\documentclass{aa}

\usepackage{color}

\usepackage{graphicx}

\usepackage{txfonts}

\usepackage{natbib}

\usepackage[dvipsnames]{xcolor}

\usepackage{subfig}

\usepackage{multirow}

\usepackage[labelfont=bf]{caption}

\usepackage[switch]{lineno}

\usepackage{placeins}

\usepackage{hyperref}
\hypersetup{
    colorlinks=true,
    linkcolor=NavyBlue,
    filecolor=NavyBlue,   urlcolor=blue,
    citecolor=NavyBlue
    }

\begin{document}

\title{The chemical enrichment histories across  the Milky Way disk}

\author{
    Valeria Cerqui\inst{1}, 
    Misha Haywood\inst{1}, 
    Owain Snaith\inst{2}, 
    Paola Di Matteo\inst{1}, 
    Laia Casamiquela\inst{1}
}

\offprints{V. Cerqui, \email{valeria.cerqui@obspm.fr}}

\institute{
    \inst{1}LIRA, Observatoire de Paris,  PSL Research University, CNRS, 92195 Meudon, France\\
    \inst{2}University of Exeter, School of Physics and Astronomy, Stocker Road, Exeter, EX4 4QL, UK}

\authorrunning{V. Cerqui et al.}

   \date{}

  \abstract
  {The variations in the production of metals with time and their dilution in the interstellar medium are a function of the star formation and gas accretion rates. Thus measuring age-chemistry relations across the Milky Way disk is our most important constraint on the gas accretion history, the past star formation history, but also provides crucial information on place of birth of the stars.}  
   {We present a new study of these relations based on a sample of almost 30 000 dwarf stars from the APOGEE survey DR17 within 2~kpc from the Sun for which we measure accurate ages.}
   {Various combinations of parameters are tested to find the best determination of stellar ages from stellar isochrones. The resulting age-chemistry relations for a selected subsample of 12 000 stars are interpreted with the help of a chemical evolution model.}
   {The data show a very well defined and tight thick disk sequence, characterized by high [$\rm \alpha$/Fe] content, subsolar metallicities and ages greater than 8~Gyr. The thin disk, characterized by a lower $\rm \alpha$-content and by ages younger than 8 Gyr shows a large metallicity spread at all ages, with apparent structures. When detailed in inner ($\rm R_{guide}<$7~kpc), intermediate (7.6 kpc<$\rm R_{guide}$<9 kpc) and outer ($\rm R_{guide}>$10~kpc) disk using guiding radius, the data show distinct chemical evolutions. We find in particular that the inner disk is typical of a monotonic, homogeneous evolution, with little dispersion, while the outer disk shows little increase in metallicity over the last 8~Gyr. The evolution at the Solar radius seems to be a mix not only because some stars have migrated from the inner and outer disk, but more importantly because of the chemical evolution of the ISM in the intermediate region is resulting from mixed gas from the inner and outer disk. In particular, we demonstrate that in the Solar neighborhood the evolution shows a decrease in the mean metallicity of the ISM that occurred 7-9 Gyr ago. One possible explanation assumes a radial inflow from the outer disk of lower metallicity gas at this epoch that diluted the gas leftover by the formation of the thick disk, giving rise to the metallicity gradient observed in this intermediate region.}
   {}
   {}

   \keywords{stars: abundances -- stars: kinematics and dynamics -- Solar neighborhood -- Galaxy: disk -- Galaxy: chemical evolution
               }

   \maketitle

\section{Introduction}

The variations in the production of chemical elements over time and their dilution in the interstellar medium (ISM) depend primarily on star formation and gas accretion rates (along with other factors such as feedback, radial gas flows, turbulence, spiral arms, bars, diffusion, etc.) and their measurements provide strong constraints on the evolution of our Galaxy. However, these constraints can be difficult to interpret, because these variations are also a function of the position in the Galaxy, and because dynamical mechanisms can reshuffle stars after their formation. Studies of the chemical enrichment in the Solar neighborhood have been struggling with this question for half a century. 
The presence of disk stars in the Solar neighborhood with metallicities significantly higher than those of the youngest local stars has been known since the 1960s, challenging the idea of a monotonic increase of the metallicity of the ISM.  For example, 51 Peg was recognized as one of the very first metal-rich stars in 1965 \citep{van_Den_bergh1965} with [Fe/H]=+0.4 (a bit higher than more recent measurements $\sim$0.2 dex,  e.g. \citealp[]{santos2003}). Comparison with the open cluster NGC188 led to the conclusion that these objects were old
(see also \citealp{Spinrad_1972}). The importance of the birthplace of stars was quickly recognized in \cite{Janes1972}, leading to the conclusion that the position within the galaxy could be a more important factor than age in determining the abundance of disk population stars. 
Super metal-rich stars in particular have been seen as coming from the regions closer to the Galactic center (see \citealp{Grenon1972}). It is only after a clear difference could be observed between the metal-poor thin disk and the thick disk using $\rm \alpha$ elements (e.g. \citealp{fuhrmann1998}, \citealp{bensby2003}, \citealp{reddy2006}) that the same could be deduced for thin disk stars coming from the outer disk \citep{Haywood2008}. 
The significant spread in metallicity (over 0.5 dex) at a given age was confirmed by the detailed study of \cite{Edvardsson1993}. However, the mechanism by which stars with such varied metallicities could reach the Solar circle remained unclear. \cite{Carraro1998} provided a comprehensive overview of the issue, highlighting that the radial epicyclic motion of stars, even when considering the secular increase in their orbital energy, was insufficient to explain the observations \citep{binney2007}.
By showing how the angular momentum of stars could be modified by transient spirals, the study by \cite{Sellwod_binney2002} has opened a Pandora's box now full of various mechanisms by which stars can loose or gain angular momentum. These processes can be categorized by the asymmetry responsible for this variation: transient spiral arms \citep{Sellwod_binney2002}, bars \citep{Halle2015,martinezbarbosa15,Khoperskov2020} coupling between bar and spiral arms \citep{minchev2010} or interaction with a satellite \citep{quillen2009}. All these processes have been invoked as possible explanation of the metallicity spread observed within the Solar vicinity \citep[e.g.][]{Haywood2008,schonrich2009,minchev2013,frankel2020,Sharma_2021,chen2023,ratcliffe2023}. 

\begin{figure*}[hbt!]
\includegraphics[width=9cm]{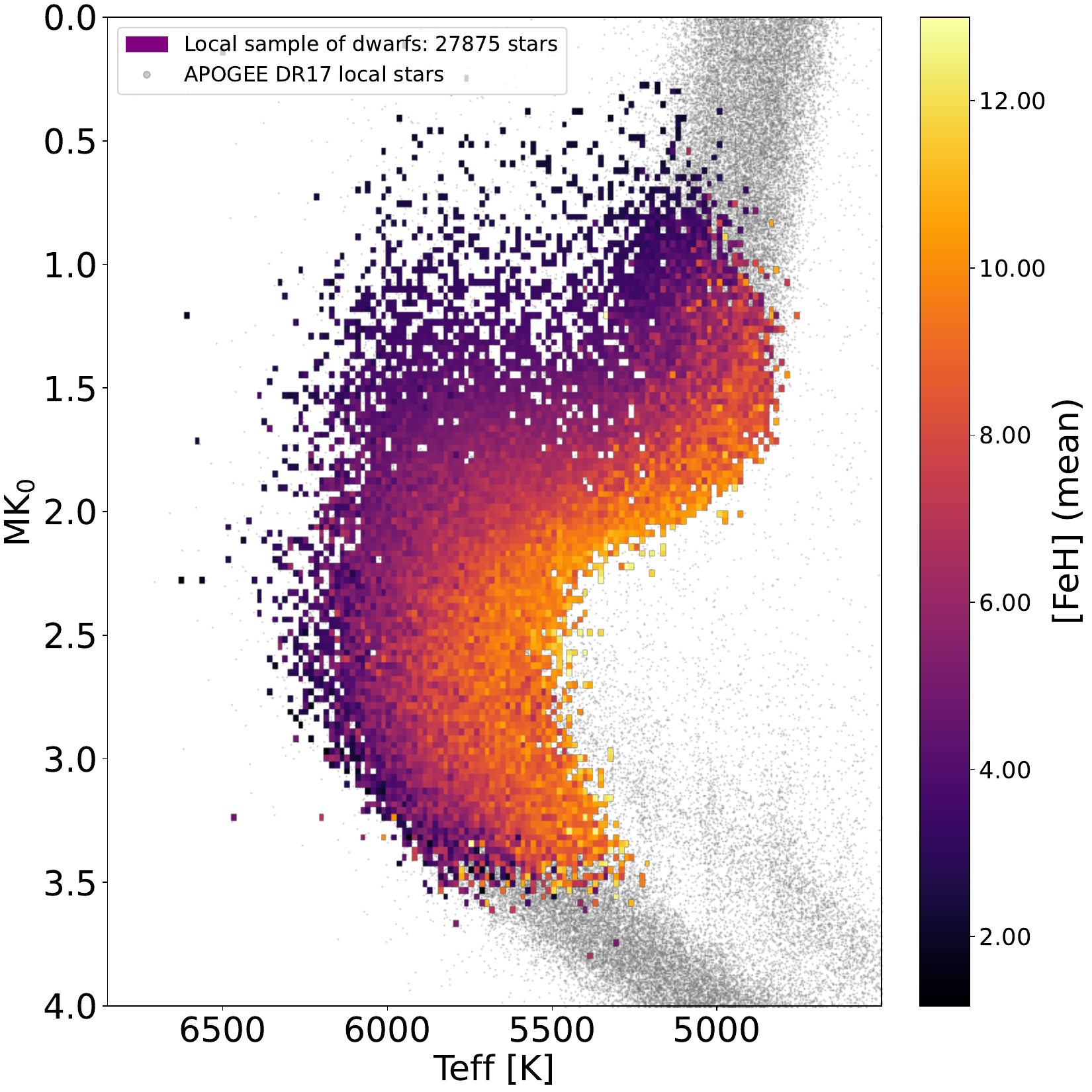}
\includegraphics[width=9cm]{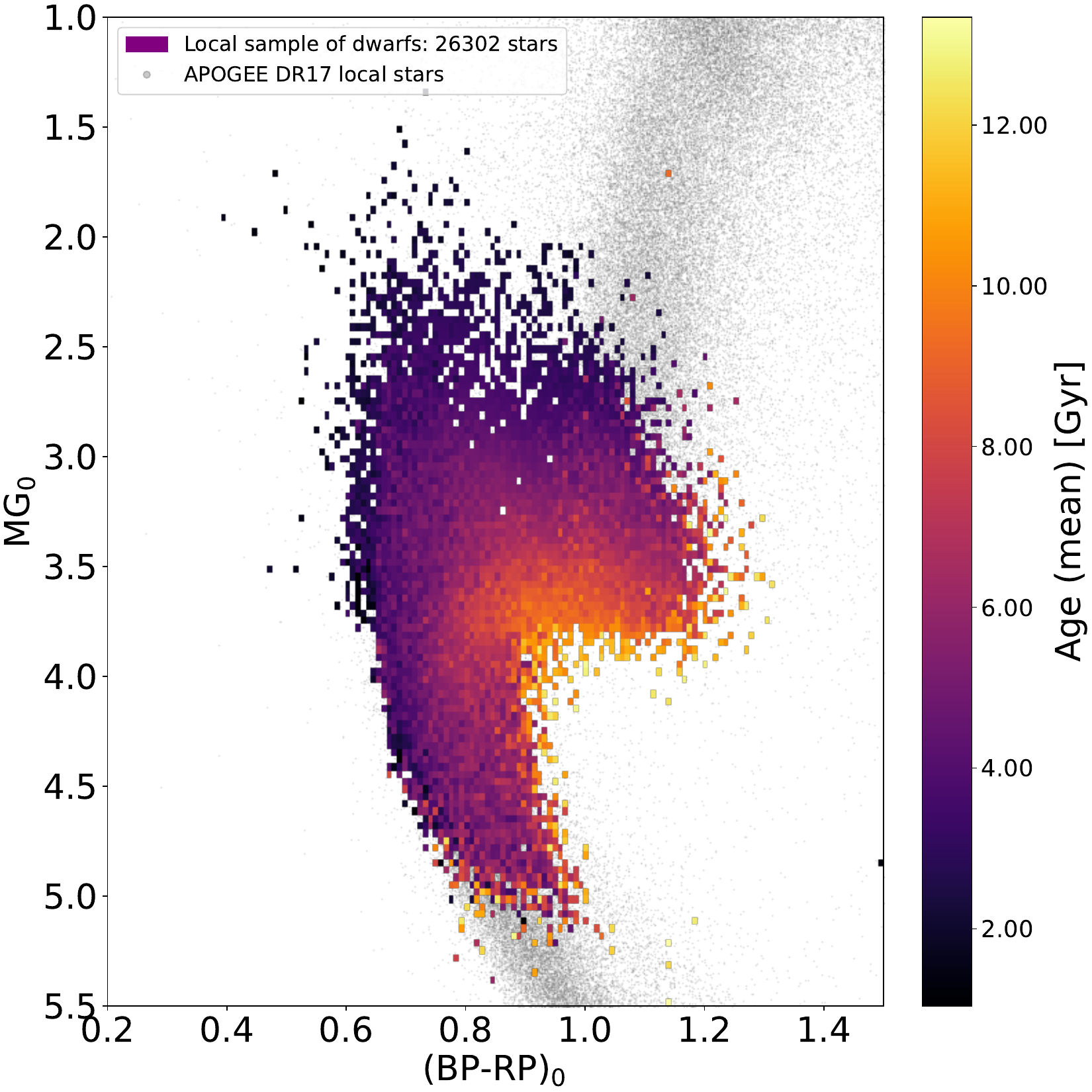}
\includegraphics[width=9cm]{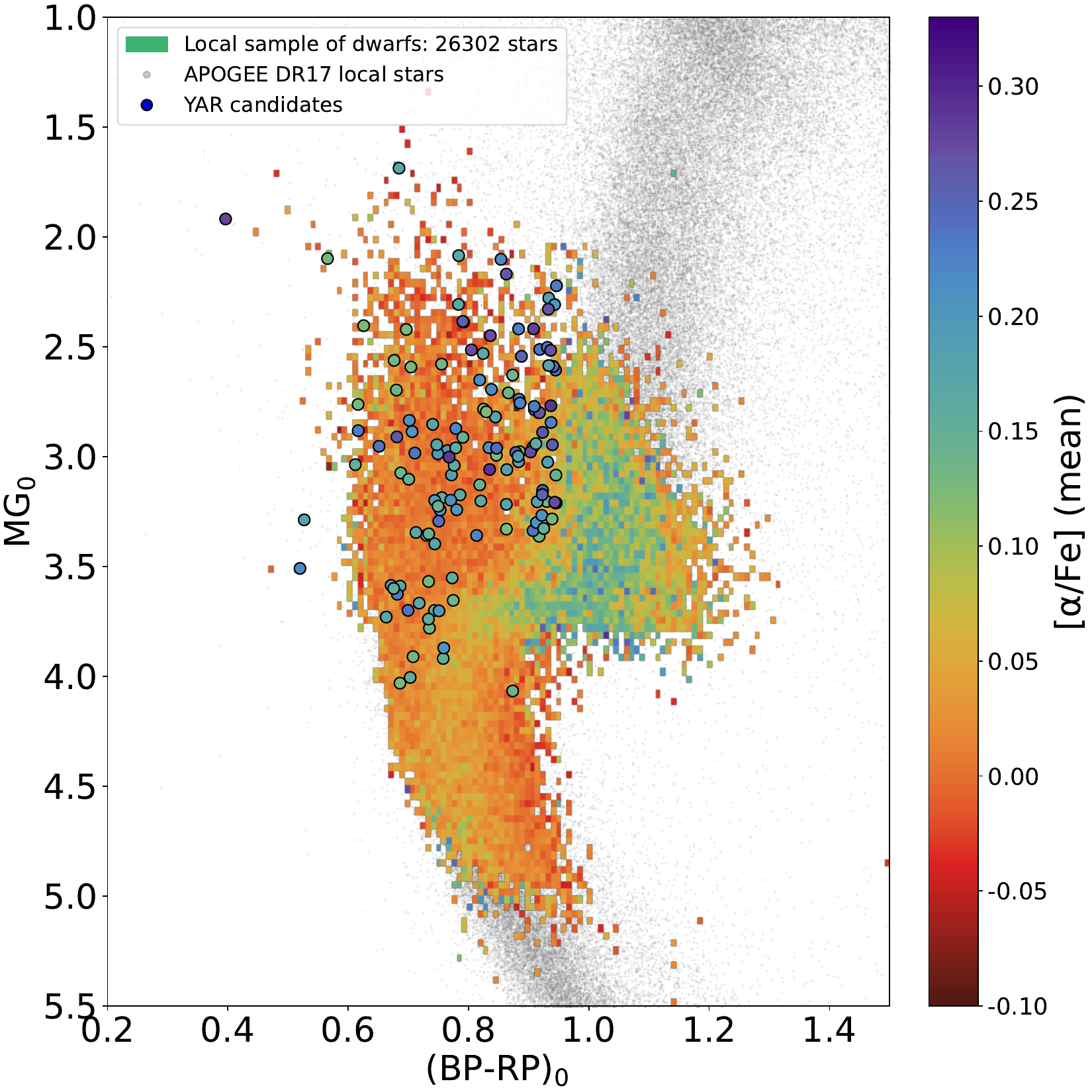}
\hspace{8pt}\includegraphics[width=9cm]{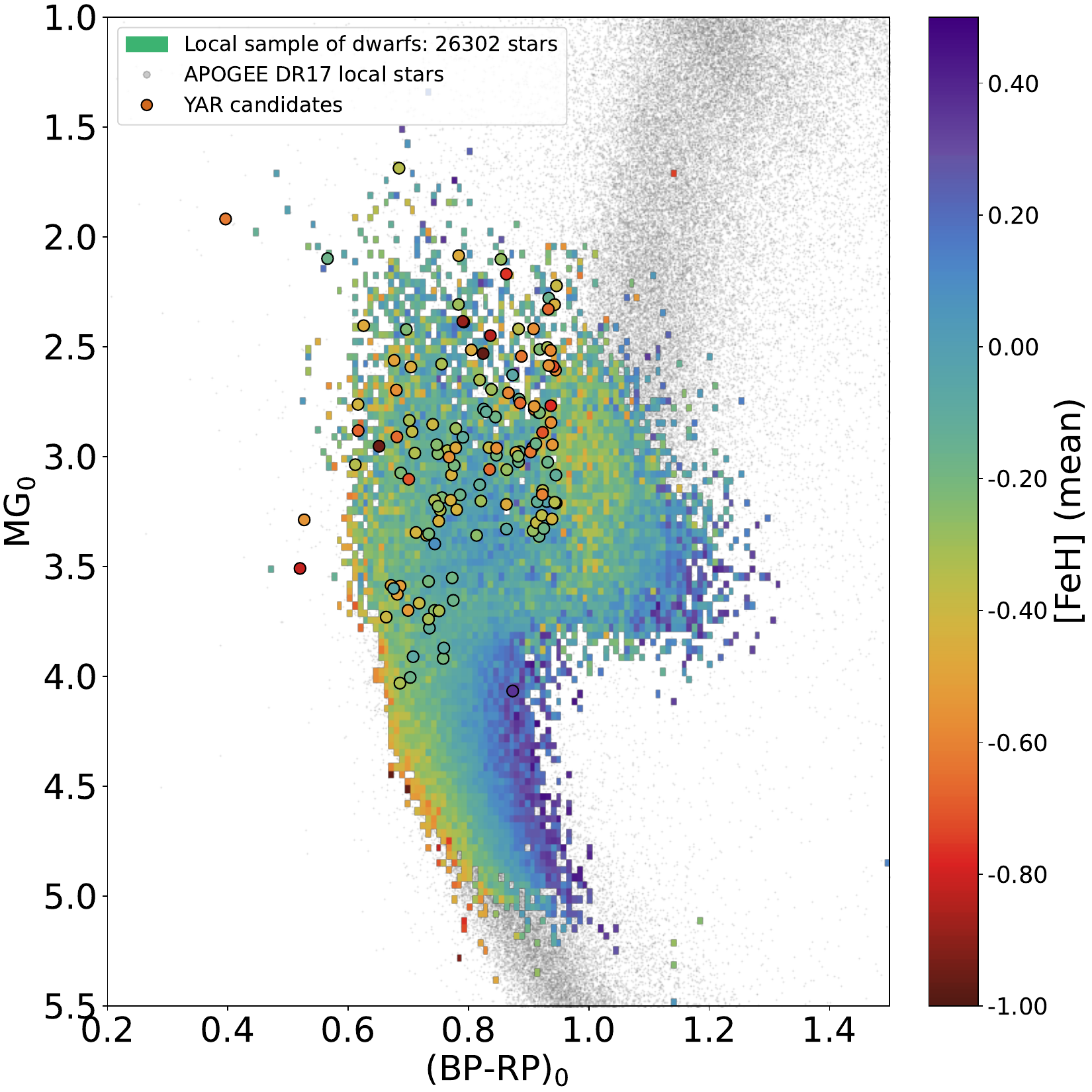}

\caption{$\rm MK_{0}-T_{eff}$ plane (upper left left) and  Gaia photometry color-magnitude diagram (upper right, bottom left and bottom right) for stars in APOGEE DR17. The local sample of dwarfs is highlighted by the color-code in age (upper panels) [$\rm \alpha$/Fe] (bottom left) and [Fe/H] (bottom right), while the rest of the stars (within 2 kpc from the Sun) is under plotted in gray. In the bottom panels we additionally over-plot young alpha-rich candidates we find in our sample (see Section \ref{section: local age-chemistry}). The details about the correction for reddening and extinction of the 2MASS and Gaia photometry are given in the text.}

\label{fig:HRD_local_sample}
\end{figure*}

The question of the disk chemical evolution is now being addressed on two fronts. The first is to better characterize both the local and large scale age-chemistry relationships. 
In this regard, huge progresses have been possible thanks to large spectroscopic surveys and Gaia \citep{Gaia_DR3}, leading a number of studies to show that local age-chemistry distributions are complex, with in particular an apparent bimodality in the thin disk regime, and/or possible clumping of stars \citep[see for example][]{nissen2020,Casali2020,jofre2021,miglio2021,Xiang_Rix_2022,Salholhdt_22_GALAHDR3,Hayden2022,Sharma2022,Patil2023,Leung2023,ciuca2023,Anders2023,queiroz2023_anders,ratcliffe24}. 
In addition, detailed maps of chemical elements as a function of Galactocentric distance have been obtained as well as detailed dissections of the metallicity gradients as a function of age \citep{Hayden2015, Anders2023, Katz2021, Imig2023,Haywood2024}.

The second front is to understand the link between the variation of chemical patterns as a function of age and distance to the Galactic center and the global disk dynamical evolution, and not only that of the radial migration. In this regard, understanding the effect of bar resonances (see \citealp{Halle2015}, \citealp{Wheeler2022} and \citealp{Haywood2024}) and the role played by spiral arms (e.g. \citealp{Khoperskov2022,Khoperskov2023}) is crucial.

In the present work, we focus on the evolution of chemical abundances with time in the Solar vicinity, as now defined by the distance out to which high resolution spectroscopic data on dwarfs is available (less than 2~kpc), and aim to investigate the following questions: can structures in the age-metallicity distribution be reliably identified, and are they the effect of a local chemical evolution?  Or are these features related to the evolution of the disk outside the Solar neighborhood? Does the Solar vicinity have its own evolution with marginal connection with the inner and outer disk? What scenario of the disk formation do these observations suggest?  

The paper is organized as follows:
in Section \ref{Section: local sample}, we describe the data utilized and the definition of our samples, highlighting the age determination performed;
in Section \ref{section: local age-chemistry}, we present the age-chemistry relations obtained for the local sample of dwarfs stars while in Section \ref{sec: age-metallicity literature} we compare our results with the literature.
We detail the kinematic properties of our sample in \ref{sec: kinematics of the populations}. In Section 
\ref{sec: chemical modelling} we present the chemical modeling of our data, and in section 
\ref{sec: discussion} we discuss our results and possible scenarios. Finally, we conclude by summarizing the key points in Section \ref{sec: conclusion}.

\section{The local samples of dwarf stars}
\label{Section: local sample}

\subsection{Definition of the samples}
\label{subsec: Data selection}

Since we want to test the robustness of the structures detected in the age-chemistry planes, we determine the ages of two independent samples, namely a sample of dwarfs from APOGEE Data Release 17 \citep[DR17,][]{apogeedr17} and another from GALAH survey Data Release 3 \citep[DR3,][]{Buder2021}. The results given in this paper are relative to the APOGEE DR17 dwarfs. The motivation for this choice and the age-chemistry relations of the GALAH DR3 dwarfs are given in Appendix \ref{Appendix: GALAHDR3}.

\paragraph{APOGEE DR17} 
For our study, we make use of APOGEE atmospheric parameters and stellar abundances from data release 17 ($\rm allStarLite-dr17-synspec\text{\textunderscore} rev1$, see \cite{apogeedr17}). 
We chose stars with reliable parameters selecting $\rm ASPCAPFLAG\, bit\, 19 == 0$ (no flag for BAD metals), $\rm ASPCAPFLAG\, bit\, 20 == 0$ (no flag for BAD $\rm [\alpha/M]$), $\rm ASPCAPFLAG\, bit\, 23 == 0$ (no flag for BAD overall for stars), $\rm ASPCAPFLAG\, bit\, 41 == 0$ (no flag for a likely composed spectrum, not suitable for ASPCAP analysis),  $\rm EXTRATARG == 0$ (main survey stars) and $\rm SNR > 50$. We defined from these objects the main sample of local dwarf stars: $\rm 3.5< \log g < 4.45$, distance from the sun below 2 kpc,  \texttt{flag\textunderscore mg = 0},  \texttt{flag\textunderscore feh = 0}, $\rm MK_{0} < 4$ ($\rm MK_{0}$ being the de-reddened absolute magnitude in K band, see below). We then excluded possible binary stars checking on: the \texttt{ruwe} parameter\footnote{\texttt{ruwe} stands for renormalized unit weight error and it is an indication of the astrometric solution quality of a
source. In fact, a star that is in an unresolved multiple system affects the goodness and the
uncertainty of the astrometric solution \citep{Lindegren2021_ruwe}.} (\texttt{ruwe} $<1.4$)
from Gaia DR3 \citep{Gaia_DR3},  the \texttt{nss\textunderscore two\textunderscore body\textunderscore orbit} catalog of the non single stars (NSS) of Gaia DR3 \citep{GaiaDR3_NSS}, the radial velocity scatter from APOGEE DR17 (\texttt{v\textunderscore scatter} < 1 km/s, as indicated in the APOGEE documentation) and the radial velocity goodness of fit from Gaia DR3 (\texttt{rv\textunderscore renormalised\textunderscore gof} $>$ 4\, \text{AND}\, \texttt{rv\textunderscore chisq\textunderscore pvalue} $<$ 0.01 \text{AND}\, \texttt{rv\textunderscore nb\textunderscore transits} $\geq$ 10, as suggested in the Gaia DR3 documentation).
The final selection, built on approximately 660 000 stars in APOGEE DR17, contains around 30 000 dwarf stars located in the Solar vicinity. Fig. \ref{fig:HRD_local_sample} represents the position of this local sample in the $\rm MK_{0} - T_{eff}$ plane (left) and in the $\rm MG_{0} - (BP-RP)_{0}$ diagram (right). The stars are color-coded for their ages (upper panels) which are determined in this work (see Section \ref{subsec: age determination})\footnote{The stars highlighted in Fig. \ref{fig:HRD_local_sample} are the objects for which we obtain an age determination, selected to have age above 1 Gyr and [Fe/H]>-1.} and for the [$\rm \alpha$/Fe] and [Fe/H] content (bottom panels).

\subsection{Age determination}
\label{subsec: age determination}
\paragraph{Method and isochrones} The age determination for the local sample of dwarf stars that we performed is based on the Bayesian statistic approach described in \cite{Age_determination_Jorgensen2005}. This method involves comparing observed data and theoretical isochrones to derive the relative posterior probability density function of age (PDF) for each star.

We assume flat priors for the age and the metallicity, while the \cite{kroupa2001} initial mass function (IMF) is adopted for the prior on stellar mass. 
The chosen age is then selected as the maximum of PDF. The $\rm \pm 1\sigma$ for Gaussian errors are located at PDF $\rm \approx$ 0.6, and are at a confidence level of 68$\%$ \citep{Age_determination_Jorgensen2005}.
The advantage of this method with respect to a conventional isochrone fitting technique (based on a direct comparison of the observed data with the theoretical isochrones) is to quantify the statistical reliability of the determined ages as a function of the observational uncertainties. 
We refer to \cite{Age_determination_Jorgensen2005} for the details of the method.
We utilize the $\rm Y^{2}$ stellar isochrones from \cite{Y2_iso} which are interpolated from evolutionary tracks with metal content in the range Z=0.00001 to 0.08 and [$\alpha$/Fe] from 0 to 0.6.  
As the age determination is based on the position of each star on the Hertzsprung Russel Diagram (HRD), we tested different atmospheric parameters to see how it affects the resulting ages. In particular we determined ages for stars in the APOGEE sample relying on four combinations of parameters: (1) the Gaia photometry, using $\rm Bp - Rp$ color and G magnitude (converting stellar models to the observational plane); (2) 2MASS K$_s$ absolute magnitude combined with spectroscopic APOGEE $\rm T_{eff}$; 
(3) searching for the optimum combination, we also derived ages using G absolute magnitude and spectroscopic effective temperature $\rm T_{eff}$ from APOGEE and finally (4) 2MASS K$_s$ absolute magnitude with  $\rm Bp - Rp$ color. The results of these last two combinations are provided in Appendix \ref{appendix: Age determinations}.

\paragraph{Conversion to magnitudes and colors} An important aspect of the age determination relying on isochrone matching is the translation of the theoretical stellar models (luminosity, effective temperature) to the observational plane of the data (magnitude, color).
The conversion of luminosity to absolute magnitude in a given photometric band is $\rm M_{\lambda} = -2.5 \times \log(L/L_{\odot}) + 4.75 - BC_{\lambda}$ where $\rm BC_{\lambda}$ is the bolometric correction. We used bolometric corrections (for both Gaia DR3 and K$_s$ photometric bands) derived and described in \cite{Casagrande_VandenBerg14}, \cite{Casagrande_2018a} and \cite{Casagrande_VandenBerg2018b}\footnote{codes and tables available at \url{https://github.com/casaluca/bolometric-corrections}}. 
Additionally,
we calculated polynomial coefficients to transform effective temperature $\rm T_{eff}$ into Gaia BP-RP color performing the calibration of the relations between these two quantities on a selected sample of 1010 dwarf stars for which effective temperature and color are obtained with high accuracy (description and details in the Appendix \ref{appendix: color-teff calibration}).

\paragraph{Extinction}  

Extinction and reddening have a significant effect on the position of a star in the HRD and hence on its age (e.g. neglecting the extinction and reddening lead to overestimated ages), but come with their own uncertainties. In order to minimize the effect of extinction, magnitudes in the K band can be used. We computed $\rm K_{0}$ (corrected for extinction) using the 2MASS Ks apparent magnitudes \citep{2MASS} and absorption in the K band given in the APOGEE catalog (\texttt{AK\_TARG} parameter). 
However, one may also want to take advantage of the excellent Gaia photometry. In the case of APOGEE sample, we used the detailed reddening map of \cite{Lallement2019} to estimate the extinction in the V band, and then considered different coefficients to convert it to the Gaia photometric magnitudes. In this work we adopted $\rm A_{G}/A_{V}=0.789$, $\rm A_{G_{BP}}/A_{V}= 1.002$ and $\rm A_{G_{RP}}/A_{V}=0.589$ from \cite{Wang_Chen_extinction}.
To correct the 2MASS K$\rm _{s}$ magnitude we utilize $\rm A_{K_{s}}/A_{V}=0.078$ taken from \cite{Wang_Chen_extinction}.. 

Due to the absence of uncertainties relative to the extinction and reddening terms, we have not included them in the calculation of the absolute magnitude error. 
Consequently, readers should be aware that our estimates of age uncertainties may represent a lower limit of the true age uncertainties.

\subsection{The choice of a dataset}
\label{subsection: the choise of a dataset}

We now compare our different datasets, starting by selecting the optimum observational plane using the APOGEE data.

\begin{figure*}[hbt!]
\centering
\includegraphics[width=9cm]{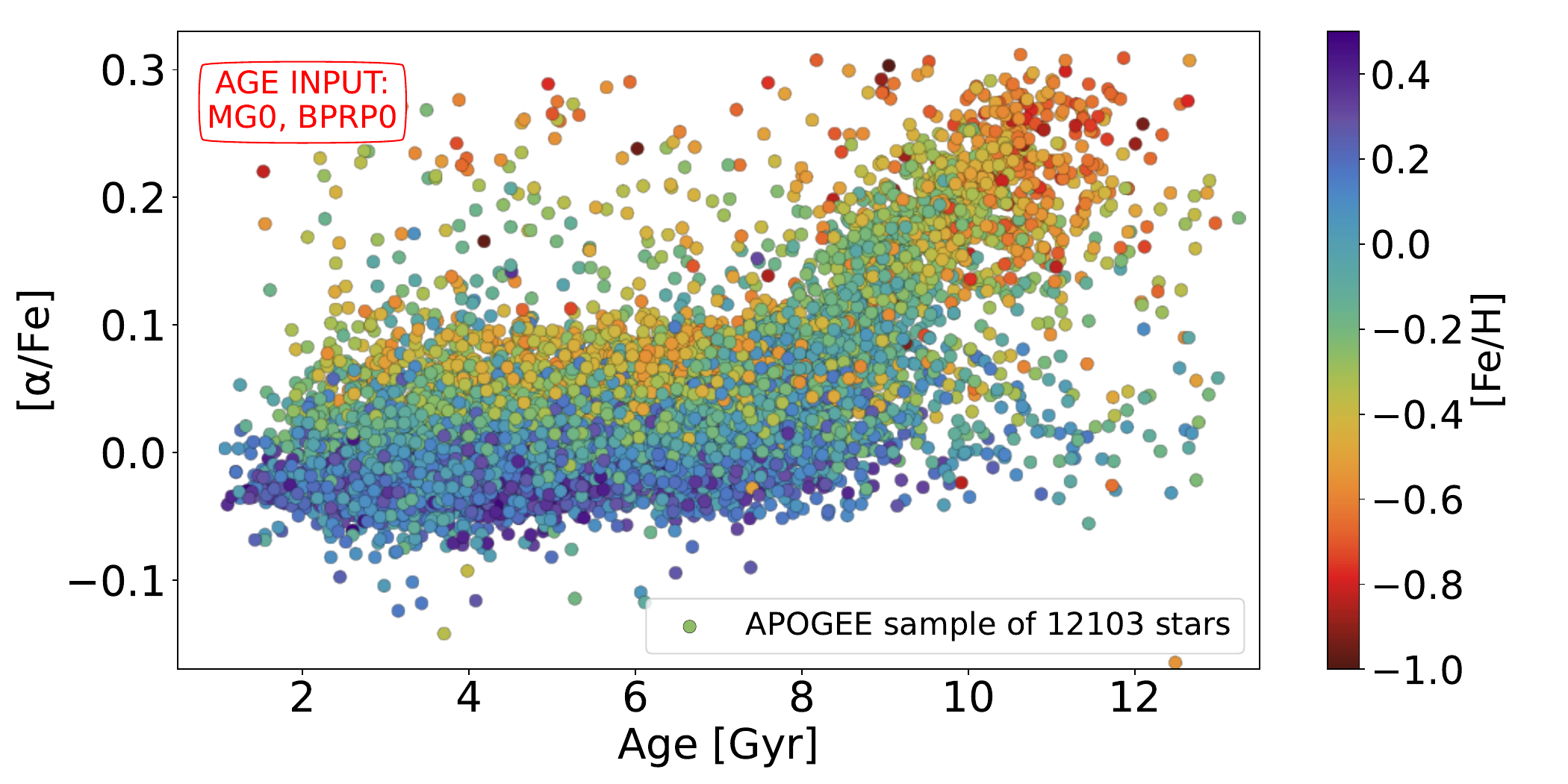}
\includegraphics[width=9cm]{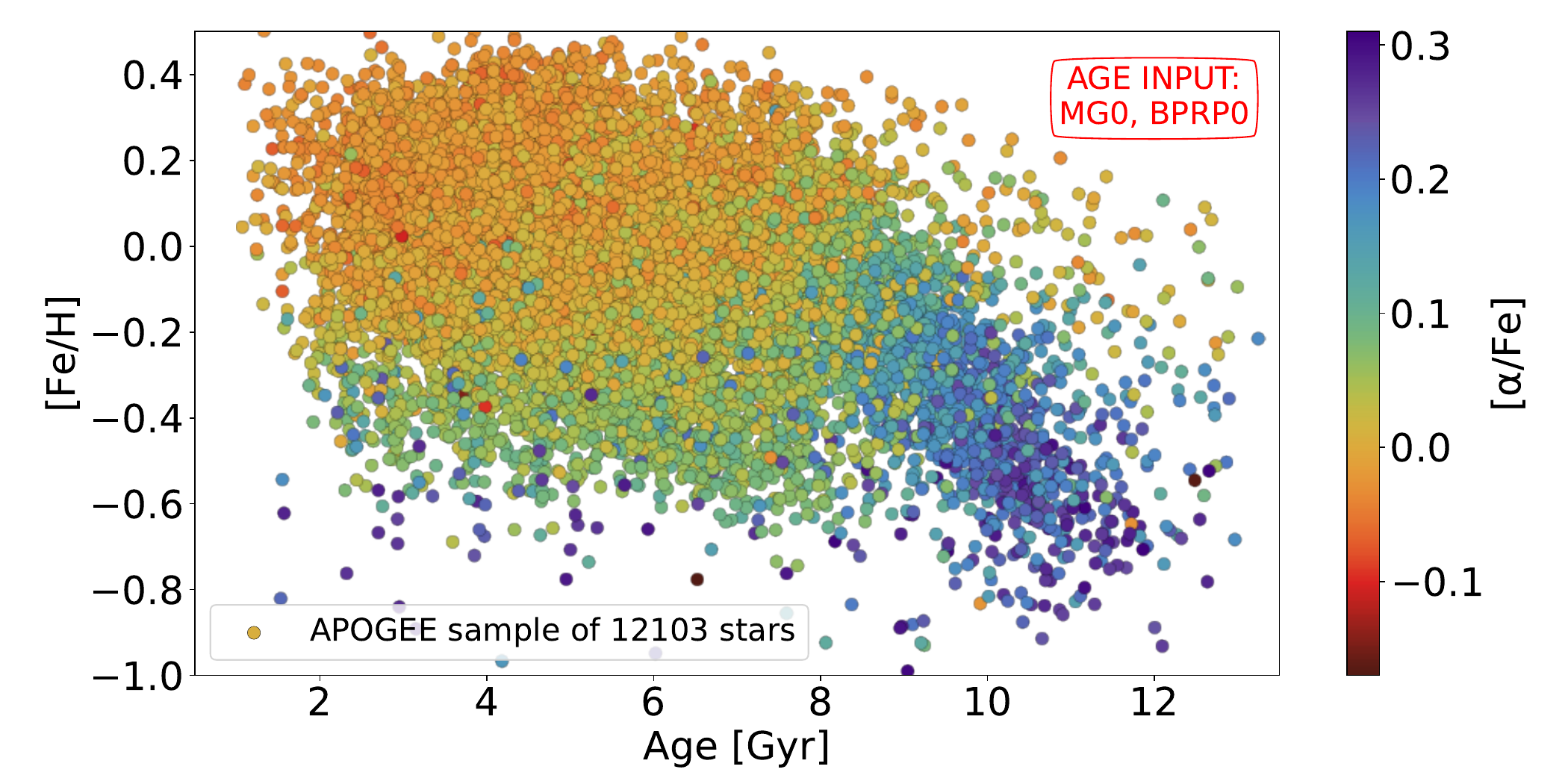}\\
\includegraphics[width=9cm]{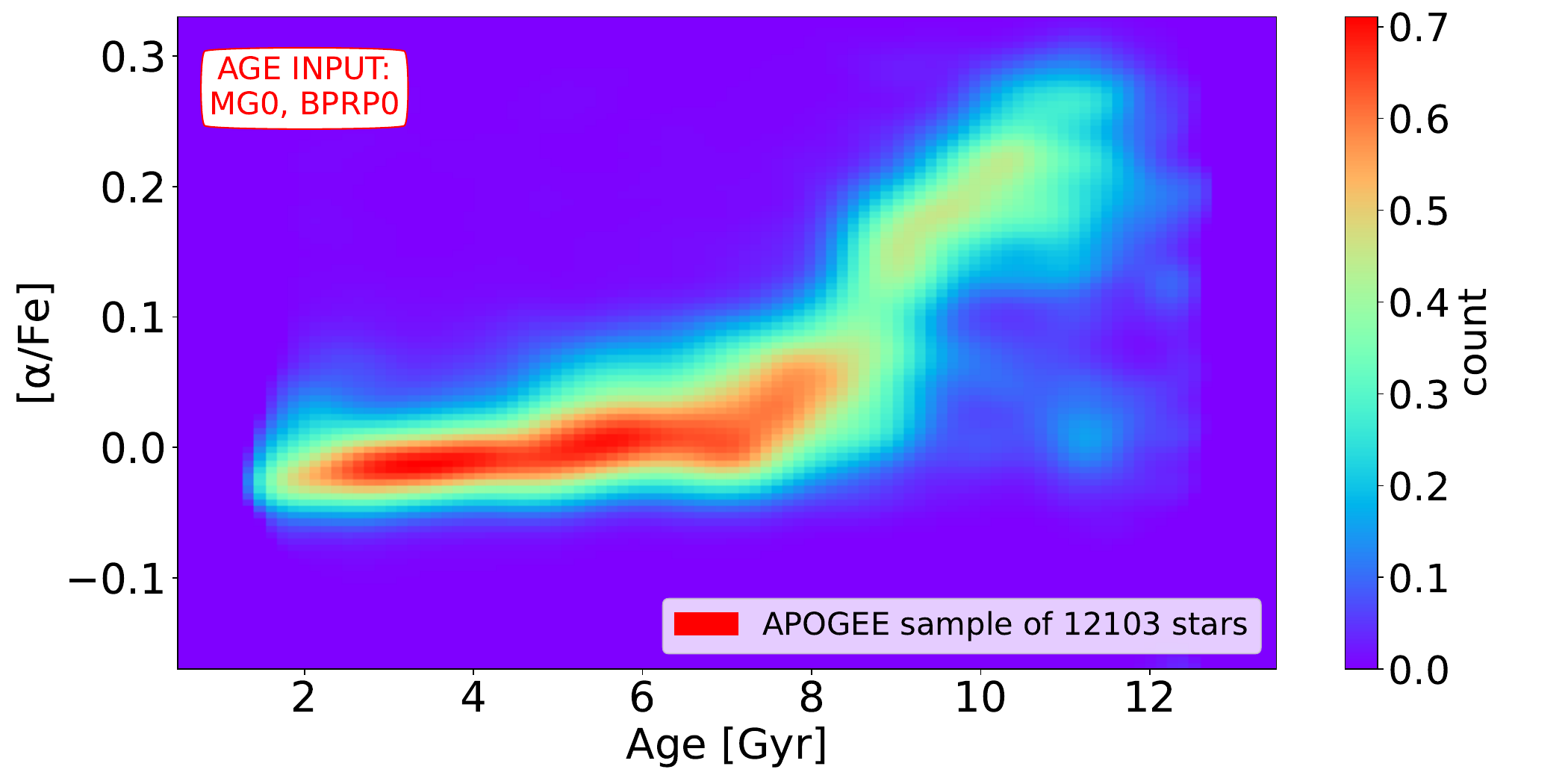}
\includegraphics[width=9cm]{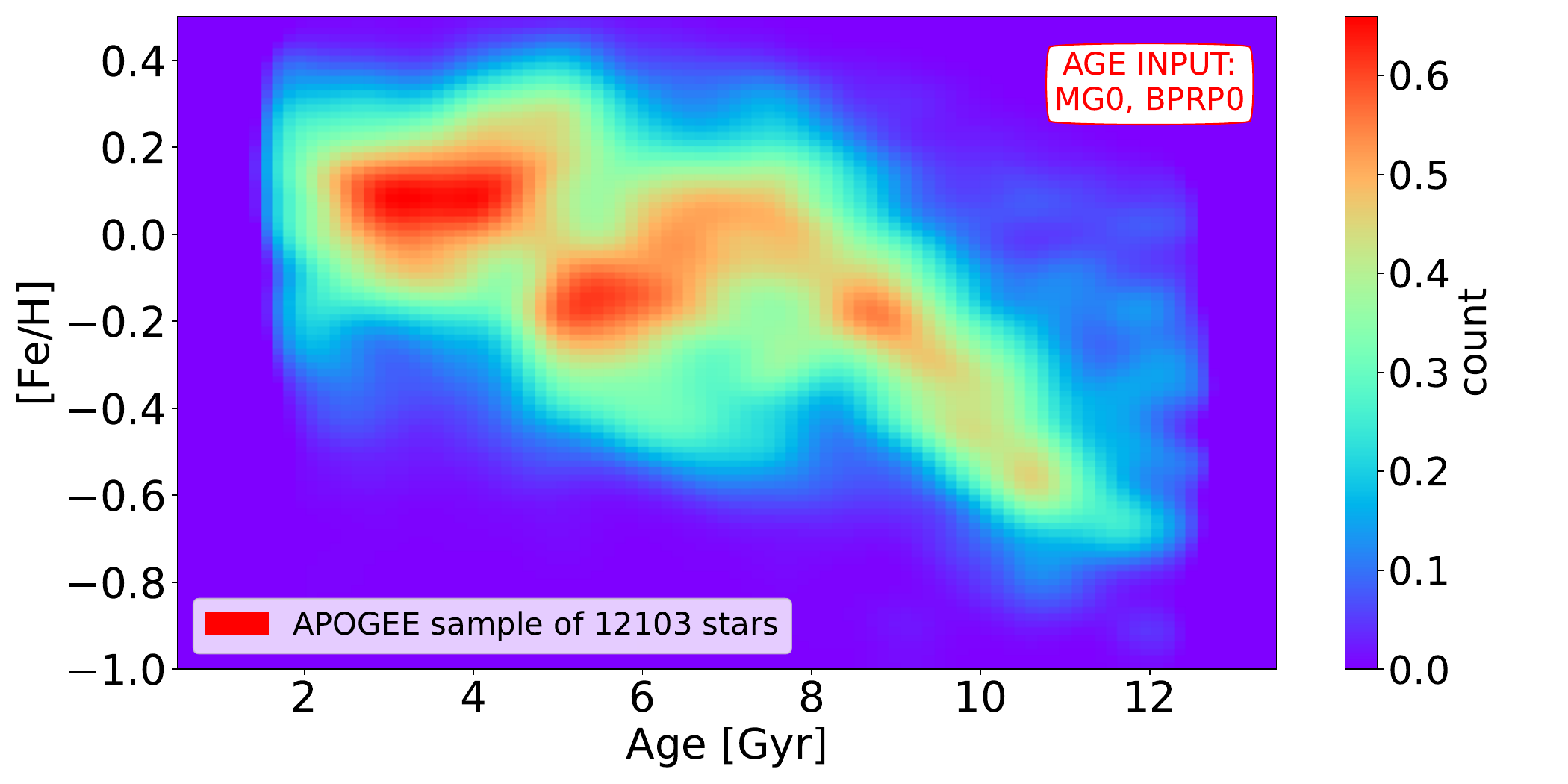}\\

\includegraphics[width=9cm]{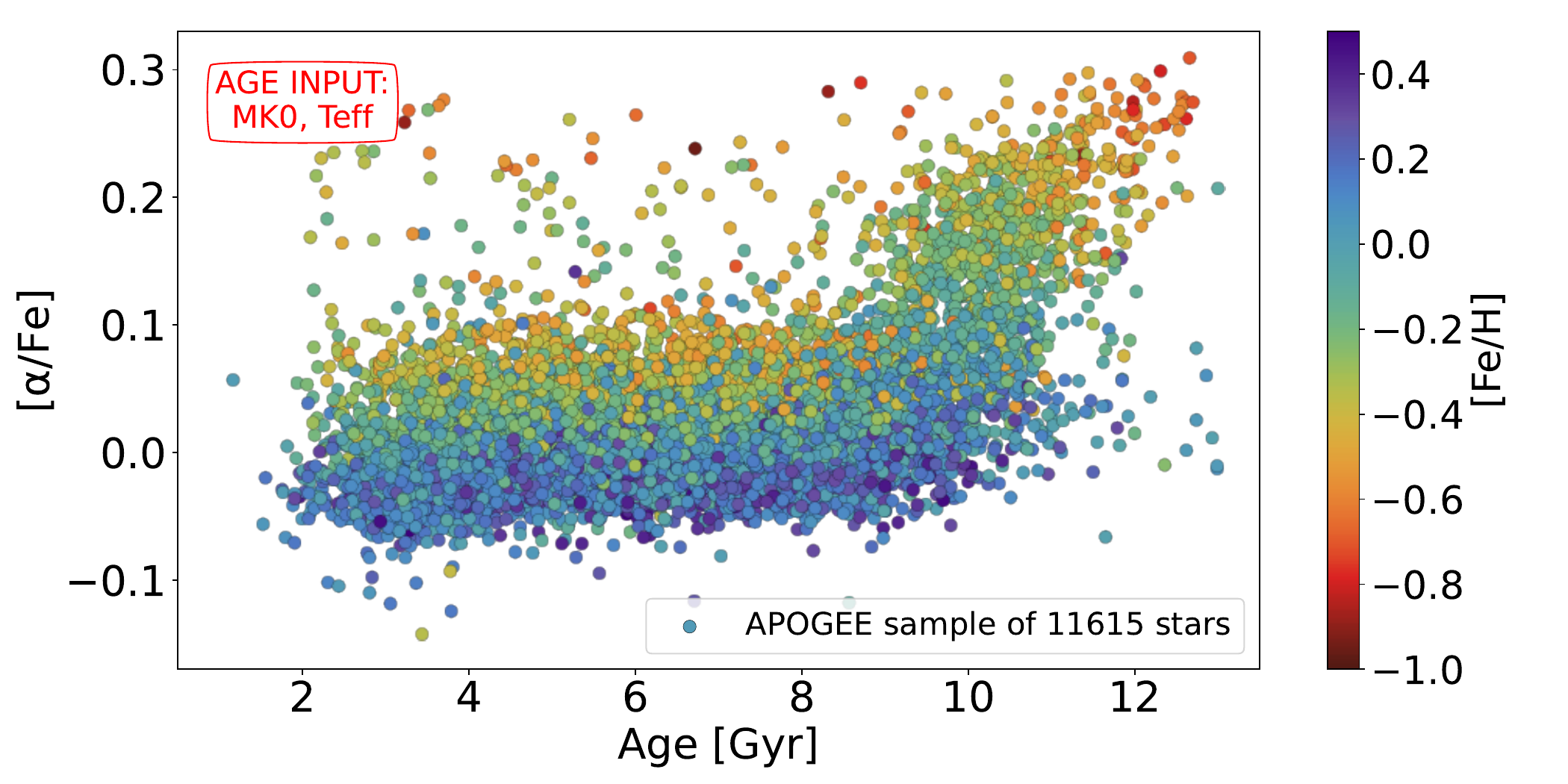}
\includegraphics[width=9cm]{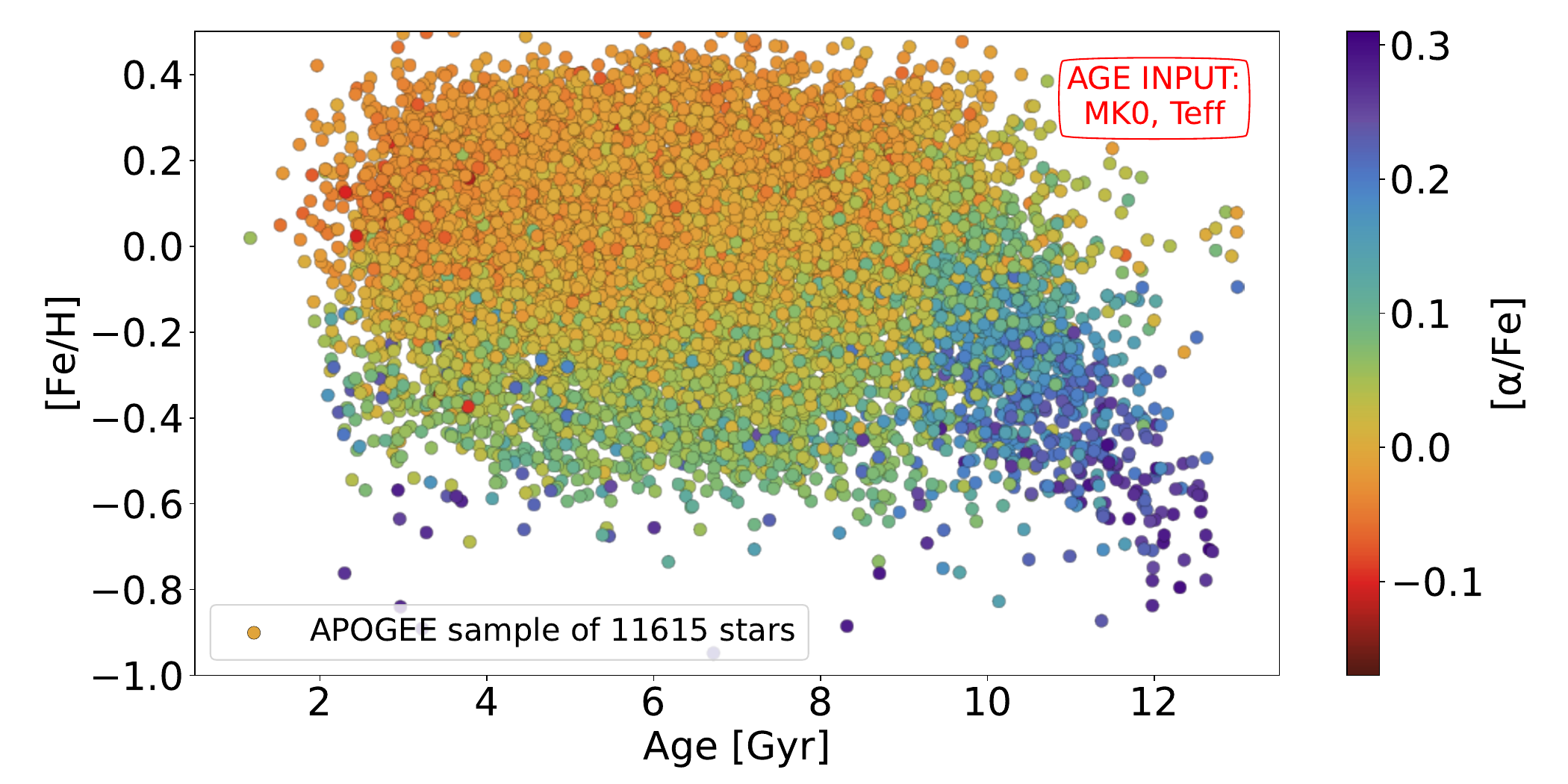}\\
\includegraphics[width=9cm]{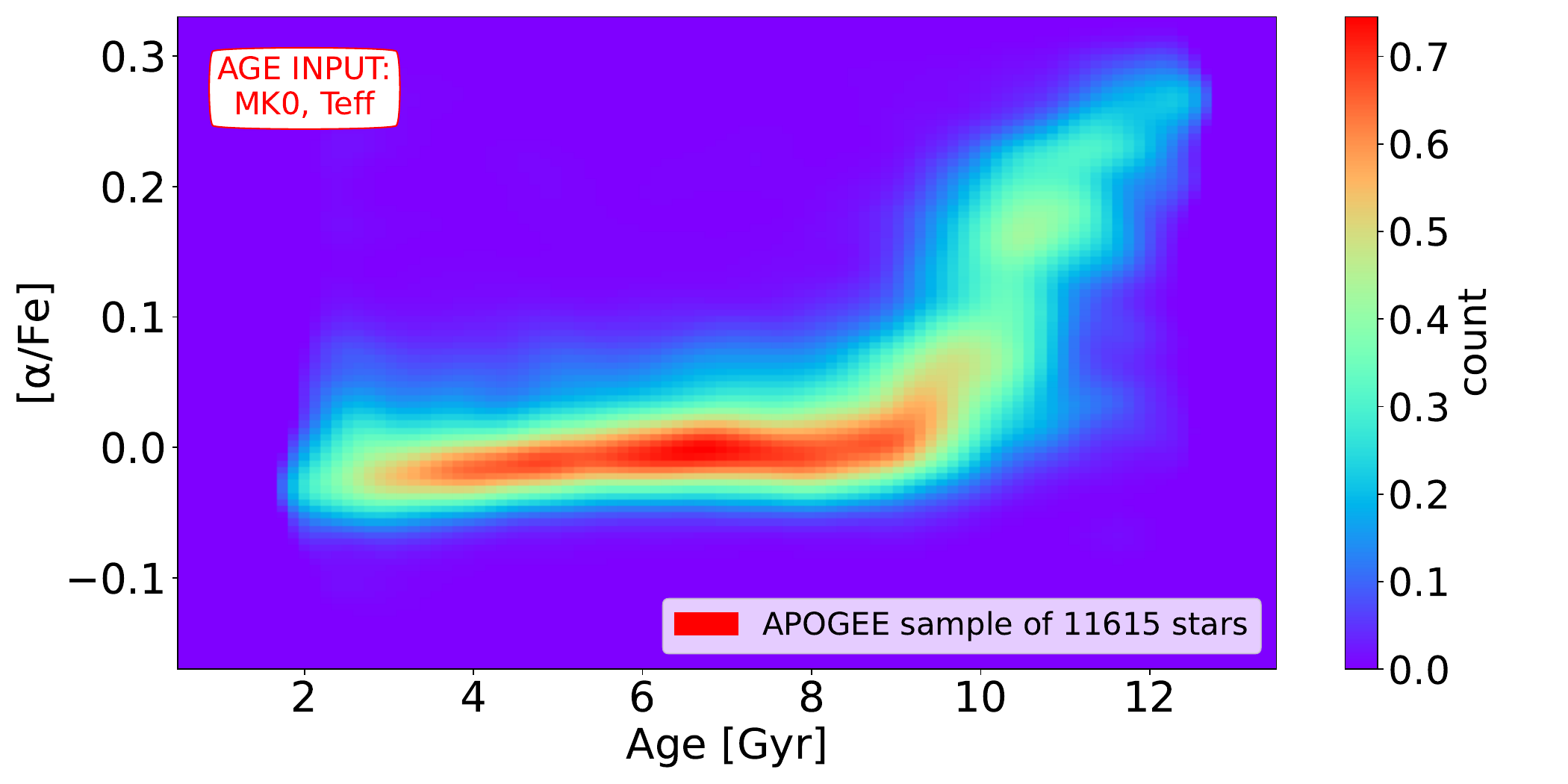}
\includegraphics[ width=9cm]{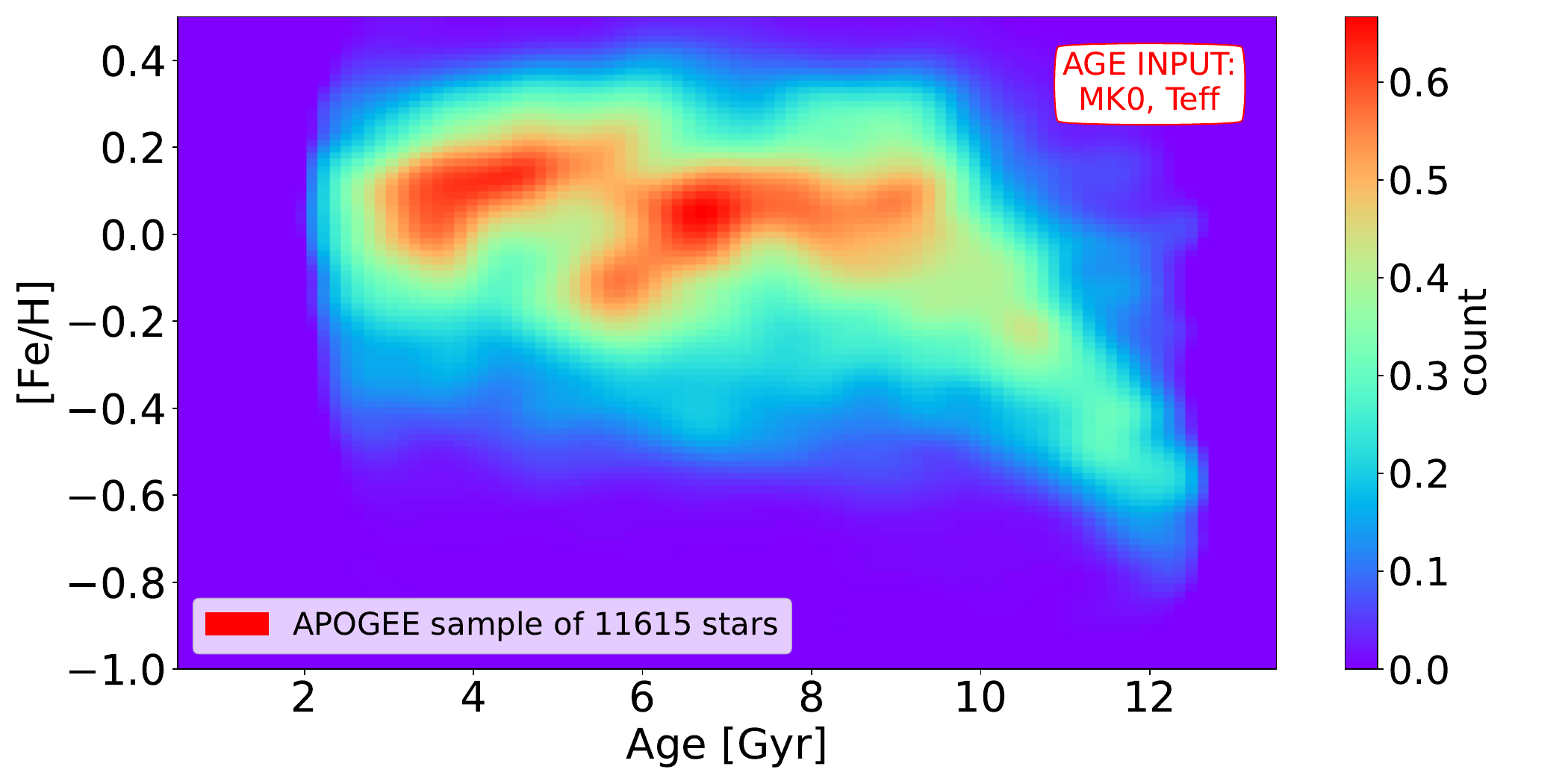}
\caption{[$\alpha$/Fe] (left) and age-[Fe/H] (right) distributions for APOGEE DR17 dwarfs in the form of scatter plots and density distributions.
In the top four plots, ages were determined using  $\rm Bp - Rp$ color and Gaia G absolute magnitude, while in the bottom four plots, the ages  were determined from K$_s$ magnitudes and T$\rm _{eff}$ effective temperatures. The number of stars in each sub-sample is indicated inside each density plot. See the text for details about the selection of the each sub-sample.
}
\label{fig:compare_age_chemistry}
\end{figure*}

Fig. \ref{fig:compare_age_chemistry} shows the age-metallicity and age-$\rm \alpha$ distributions obtained from the APOGEE DR17 datasets using Gaia photometry (first four plots) or MK0 magnitudes and spectroscopic effective temperatures from APOGEE. These two sets of parameters illustrate the maximum difference we obtain between the derived distributions. Distributions obtained using other combinations of these parameters and different cuts on the age uncertainties are given in Appendix \ref{appendix: Age determinations}. To obtain the most accurate distributions possible, in each case we display the stars with the most accurate ages (age uncertainty below 0.5 Gyr), Bp-Rp color below 0.95 magnitude and age above 1 Gyr (the PDFs of the youngest stars are generally very noisy, making a proper age determination challenging). The distributions are shown both as scatter plots, color-coded for [$\rm \alpha$/Fe] and [Fe/H], and as column-normalized density plots. The number of objects displayed is written inside each panel. 
The first apparent result is that, although there are differences, the plots show similar general distributions, demonstrating that although the atmospheric parameters from which age is determined are not the same, there is a general consistency. 
There is also an excellent behavior of the variation of the distributions as a function of $\alpha$-abundance and metallicity, as illustrated by the color coding scales  provided with the scatter plots. A number of so-called young $\rm \alpha$-rich stars are clearly visible in the age-metallicity distribution (top right plot). 
Secondly, the age determined using $\rm Bp - Rp$ color instead of the spectroscopic temperatures are a bit less fuzzy, showing in particular a tighter age-[$\alpha$/Fe] and better defined age-metallicity relationships for the thick disk population.
Thus, although the two age determinations provide very similar distributions, we prefer to adopt Gaia photometry-based ages because of the slightly better statistics on of thick disk stars and tighter thick disk sequence reflecting smaller age uncertainties.

\subsection{Comparison with an accurate external dataset: \cite{nissen2020}}
\label{Subsec: nissen comparison}

\begin{figure}[hbt!]
\centering
\includegraphics[width=\hsize]{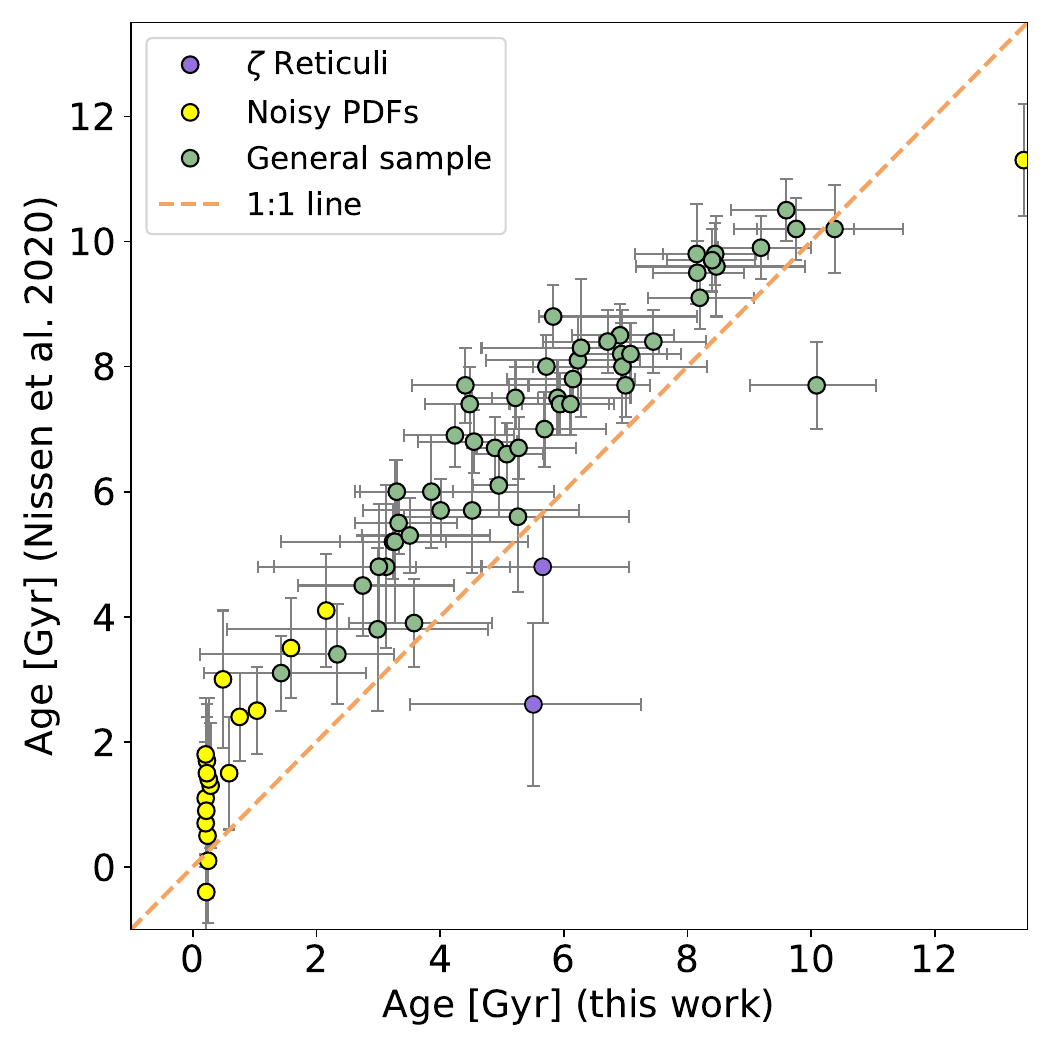}
\caption{Ages from \cite{nissen2020} compared to our determinations for the 72 stars in \cite{nissen2020}. The two violet dots (HD 20766 and HD 20807 stars) are the components of the visual binary star $\zeta$ Reticuli. 
The yellow color highlights the stars characterized by a noisy PDF that made either the determination of the $\pm 1 \sigma$ or the maximum of the  problematic. Given this issues we do not display the error bars for these objects.}
\label{fig: Nissen_this_work}
\end{figure}

Confronting our age determinations with some external dataset with age determined using different isochrones allows us to assess our age scale, both in a relative and absolute sense. 
In order to do so, we determine the ages from the sample of Solar-like stars from \cite{nissen2020}. The stars from this study have metallicities between -0.3 dex and +0.3 dex, and their stellar ages have been determined in \cite{nissen2020} using their spectroscopic effective temperatures, luminosities, and ASTEC isochrones \citep{ChristensenDalsgaard2008}. We determine the ages of these objects from Bp - Rp
color and the $\rm M_{G}$ absolute magnitude (from Gaia DR3) using our procedure. Fig. \ref{fig: Nissen_this_work} shows the comparison with the ages determined by \cite{nissen2020}. The comparison shows a systematic offset between the two age scales, in the sense that our ages are systematically younger by 1 to 2 Gyr. Additionally, some of the dated stars show noise in the PDF which causes issues in the determination of either the gaussian errors either the maximum of the PDF. The relative age values are possibly not correct and induce additional scatter at ages below 1 Gyr (yellow points in Fig. \ref{fig: Nissen_this_work}). 
Apart from these objects and from the 1-2 Gyr shift, the dispersion between the two scales is limited. 
The fact that these ages were determined from spectroscopic effective temperatures and luminosities derived from V-band magnitudes and bolometric corrections in \cite{nissen2020}, while from Bp - Rp color and Gaia $\rm M_{G}$ magnitude in this work, suggests that the relative ages are well determined. The two stars colored in violet, HD 20766 and HD 20807, are the components of the $\zeta$ Reticuli visual binary star in the reticulum constellation \citep[for details refer to][]{nissen2020}. 
To conclude, this comparison tells us that our ages are in agreement with a well determined set of ages, although we have to be aware that the absolute values may be subjected to systematic correction of about 1 to 2 Gyr.

\section{Age-chemistry relations}
\label{section: local age-chemistry}

\begin{figure*}
\centering
\includegraphics[width=0.675\columnwidth]{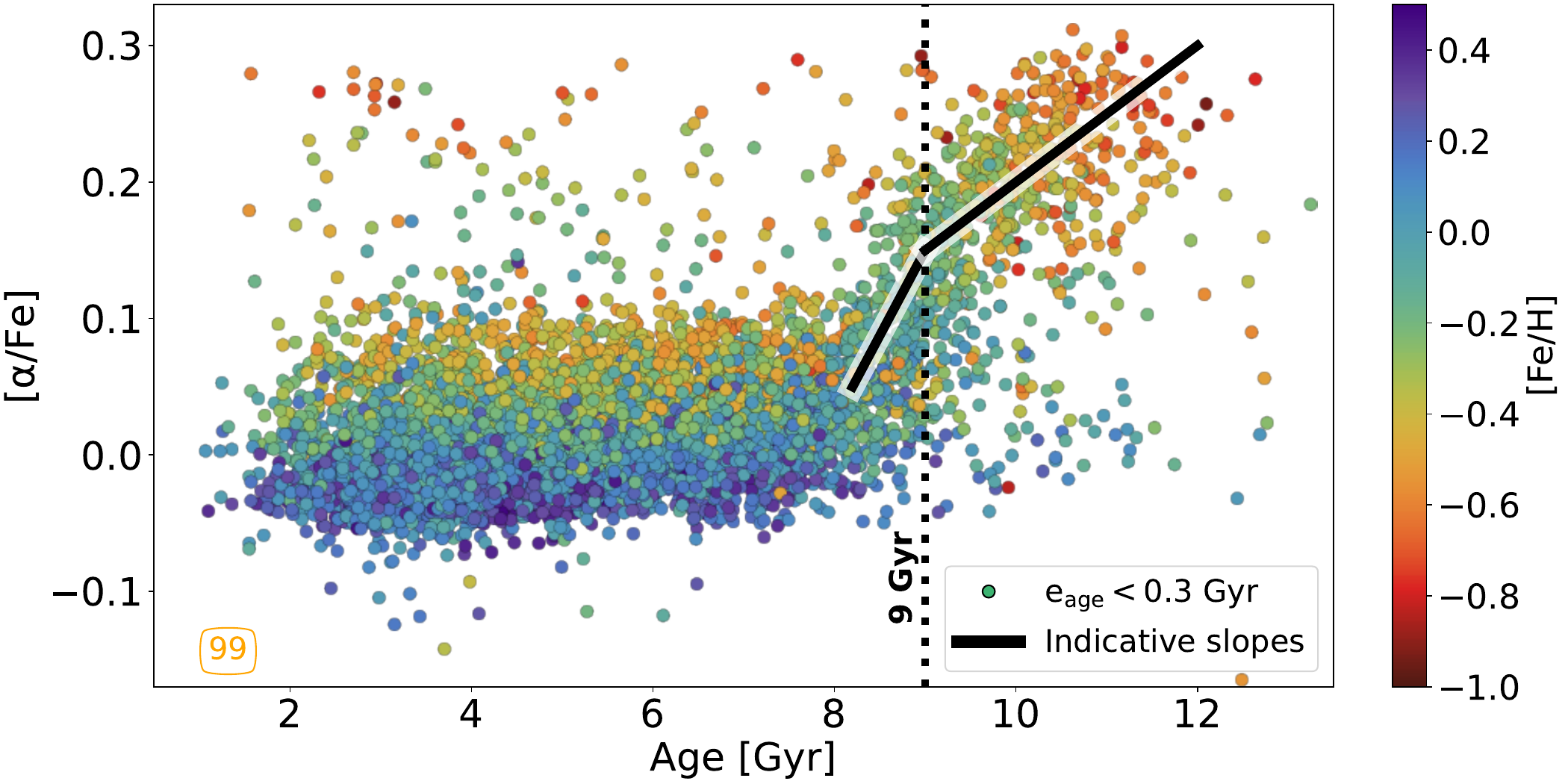}
\includegraphics[width=0.675\columnwidth]{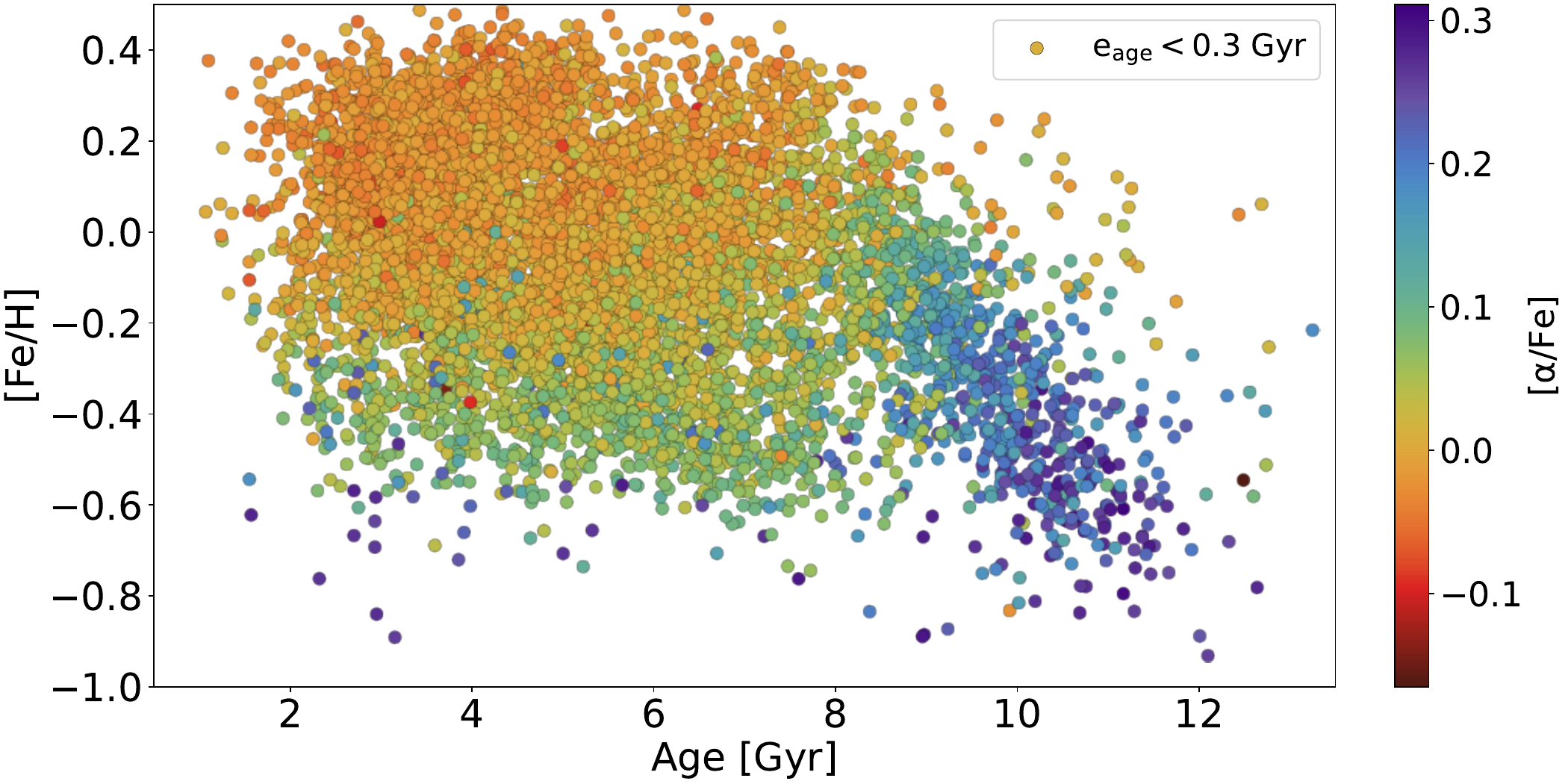}
\includegraphics[width=0.675\columnwidth]{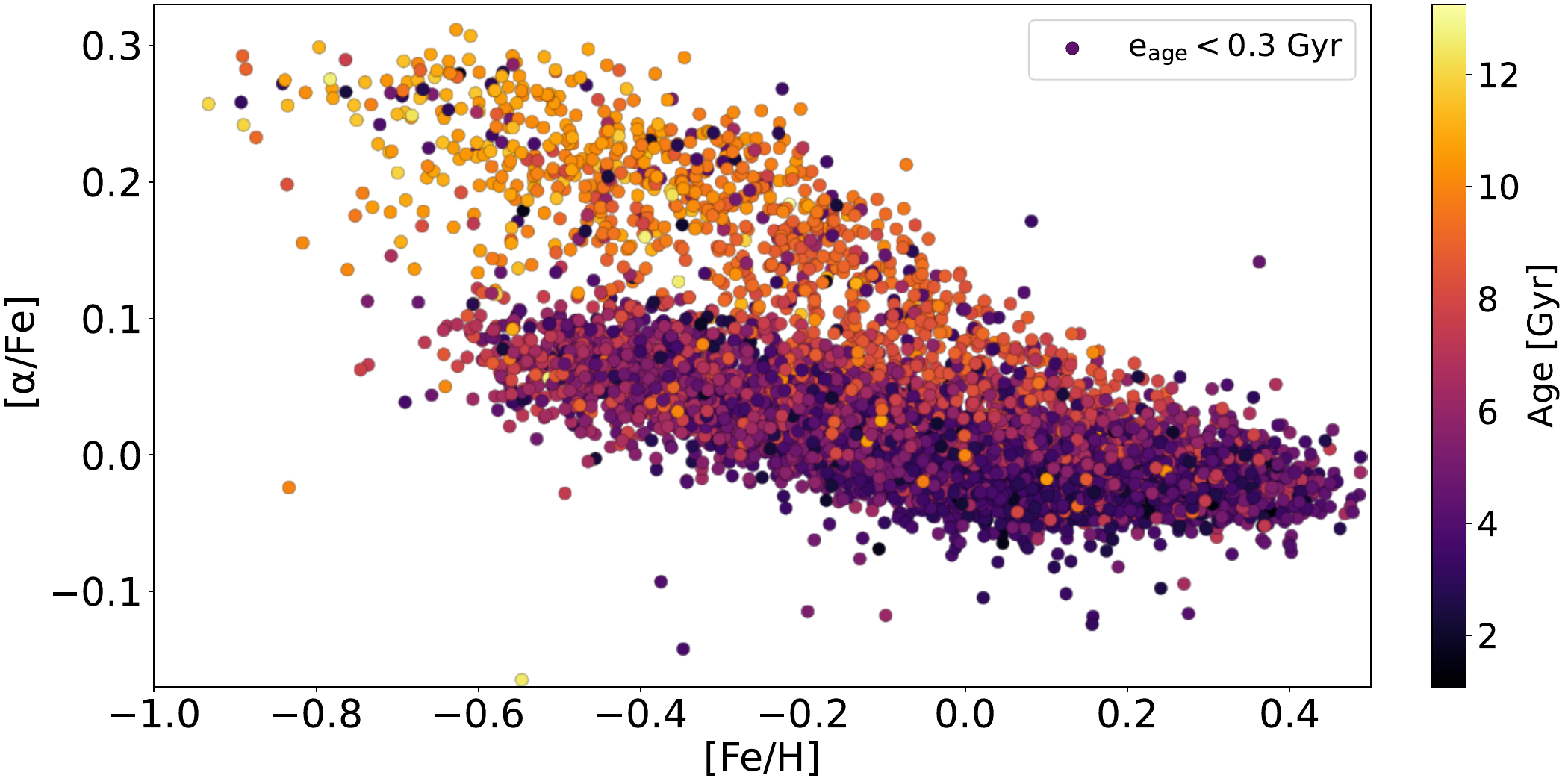}\par

\includegraphics[width=0.675\columnwidth]{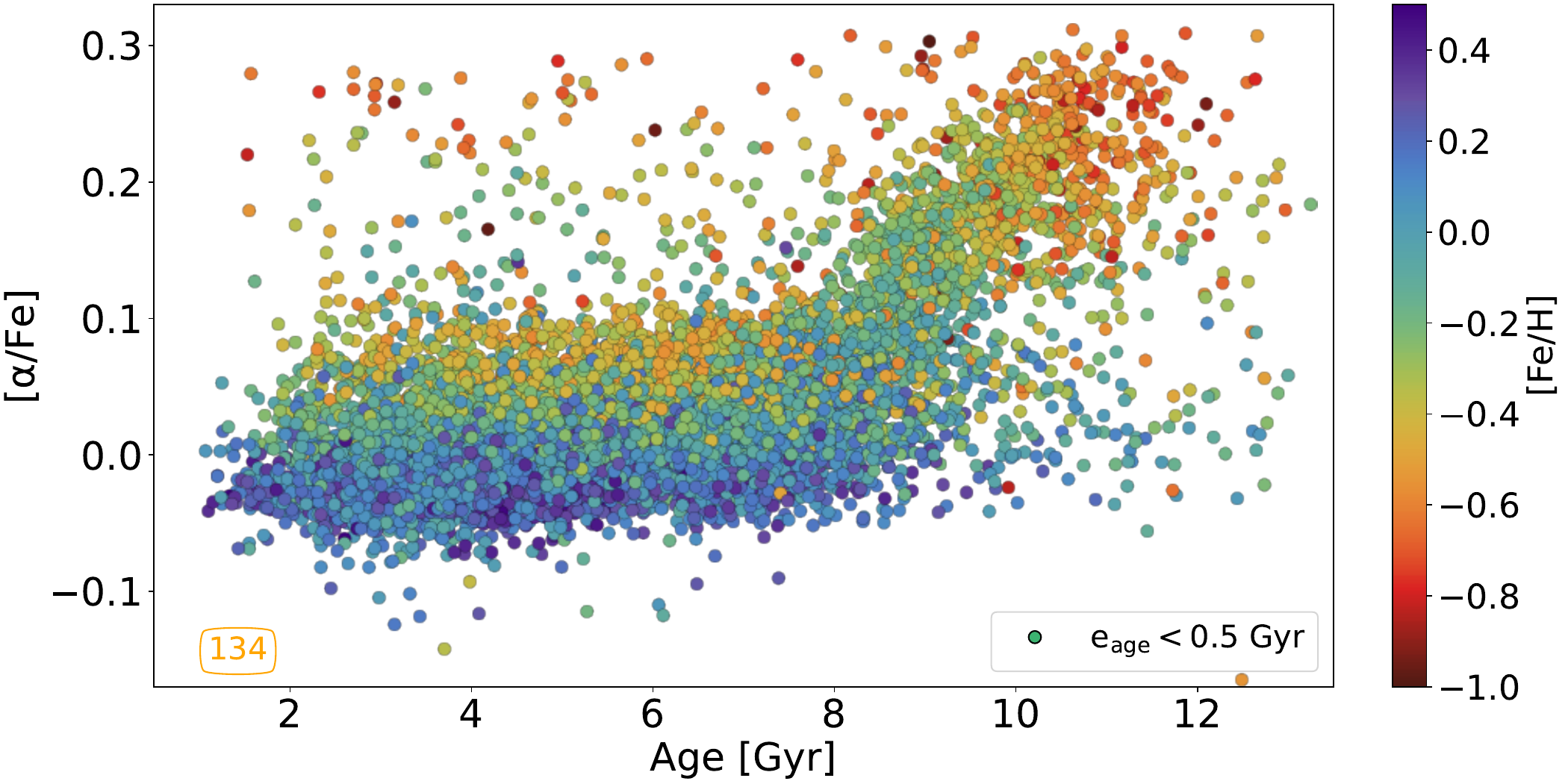}
\includegraphics[width=0.675\columnwidth]{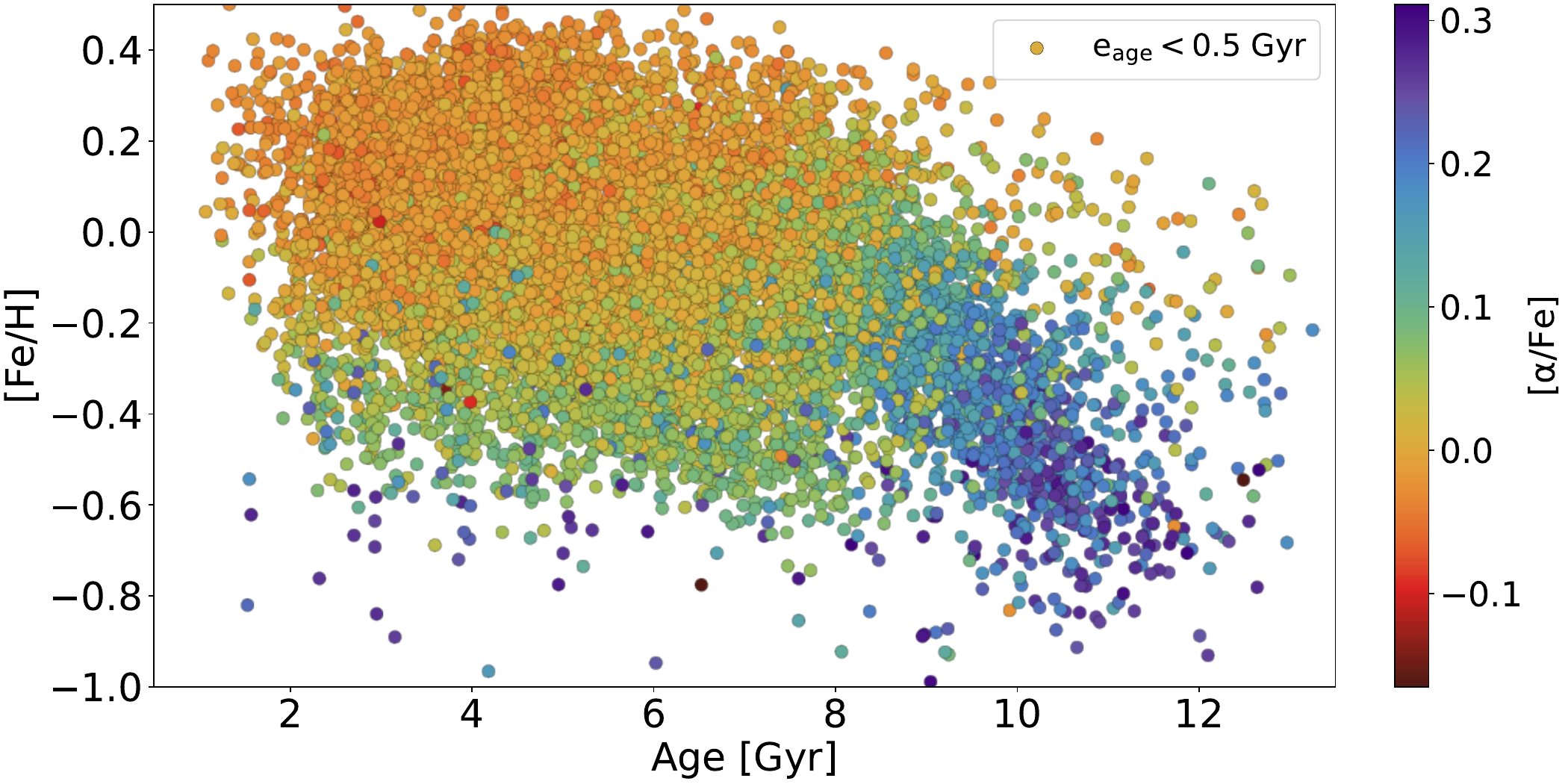}
\includegraphics[width=0.675\columnwidth]{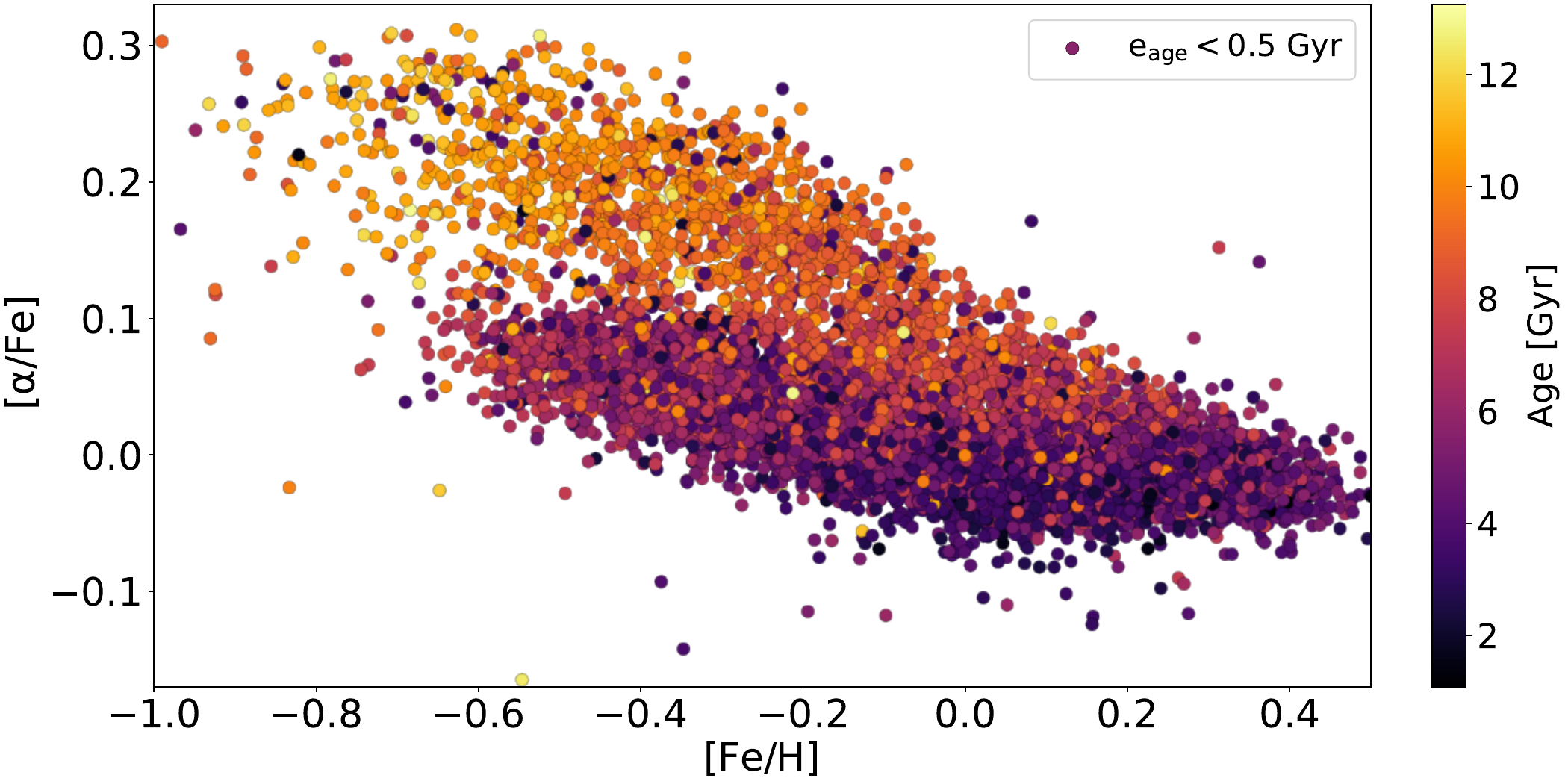}\par
\includegraphics[width=0.675\columnwidth]{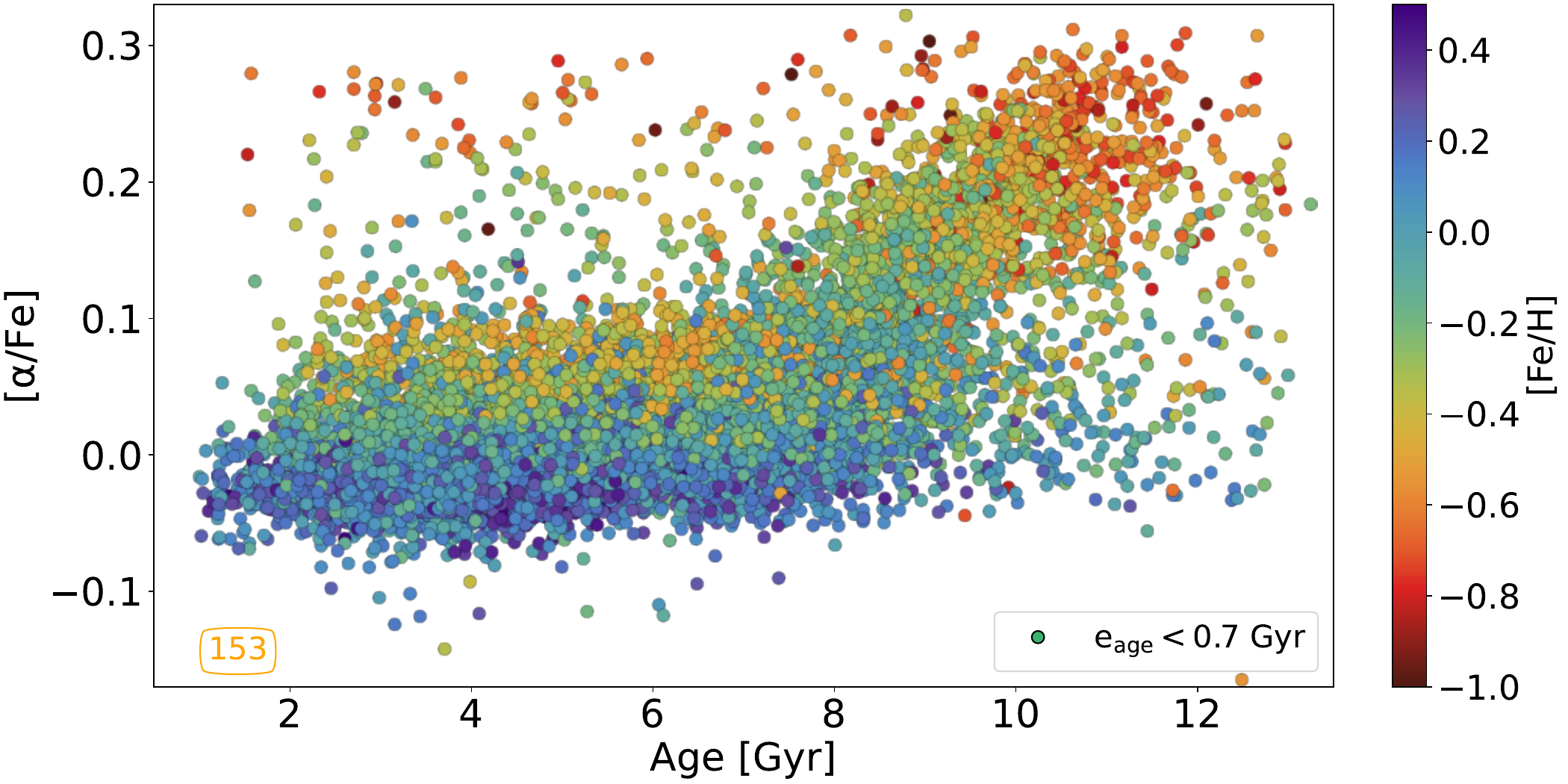}
\includegraphics[width=0.675\columnwidth]{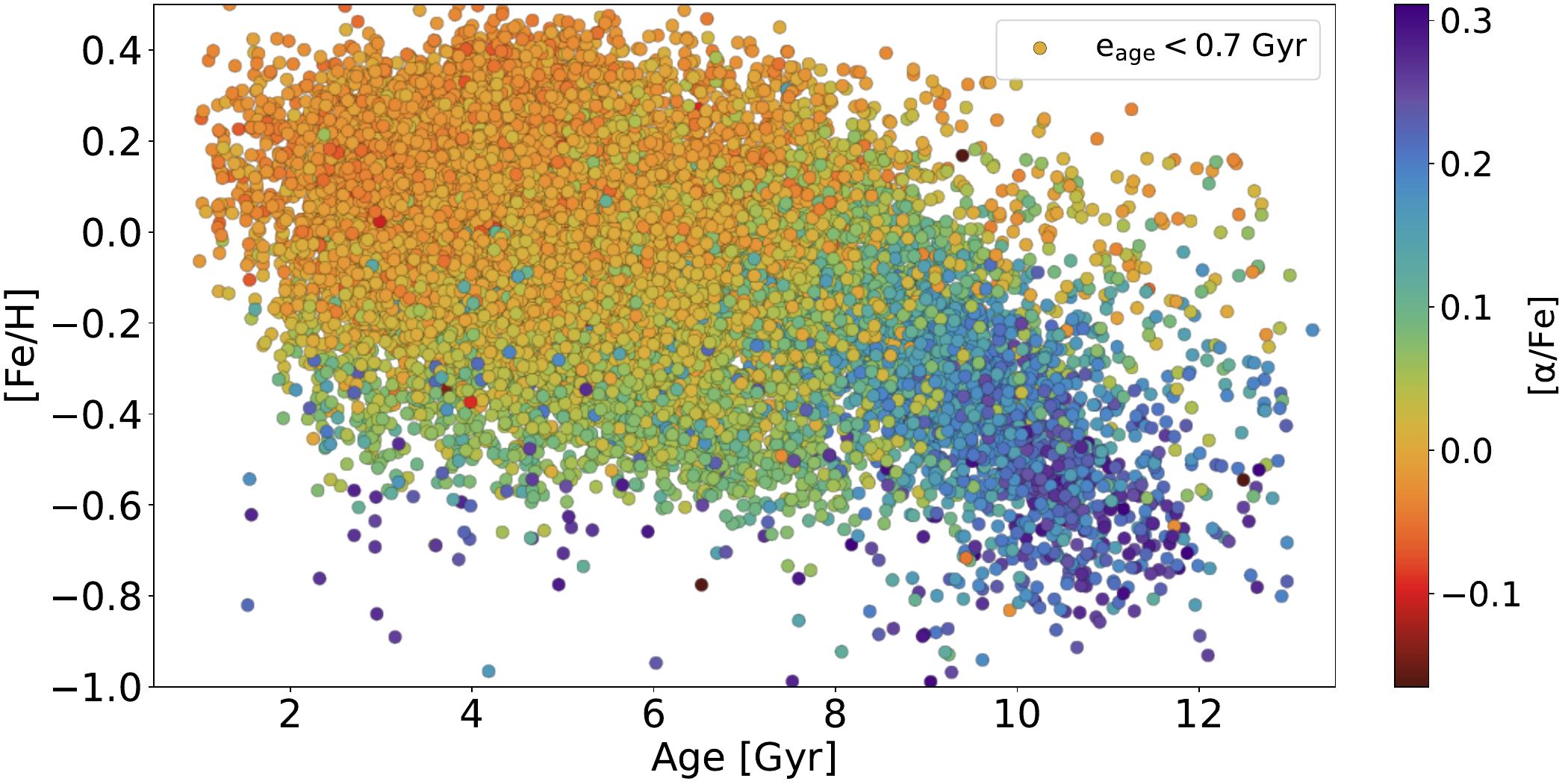}
\includegraphics[width=0.675\columnwidth]{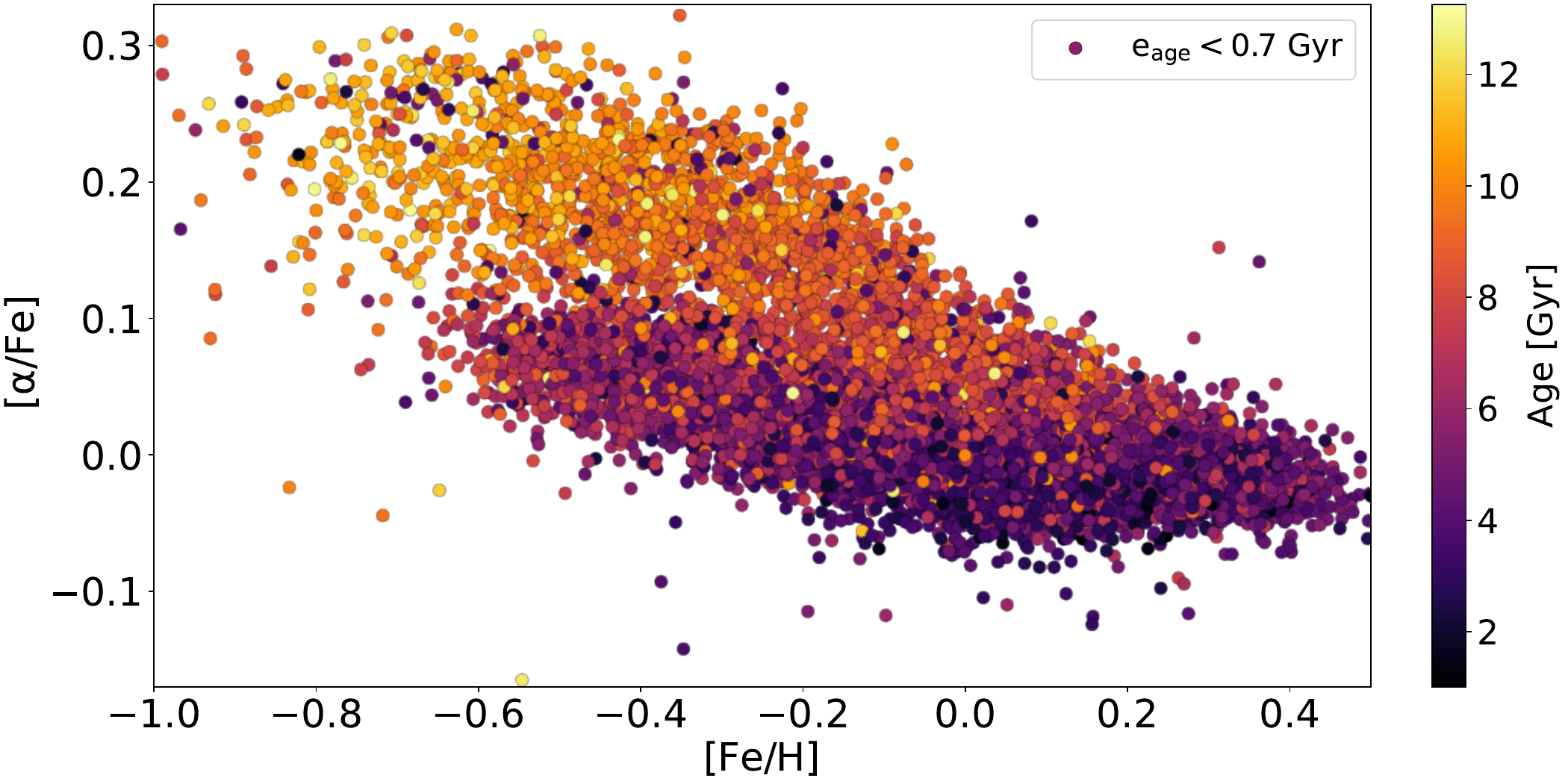}\par

\includegraphics[width=0.675\columnwidth]{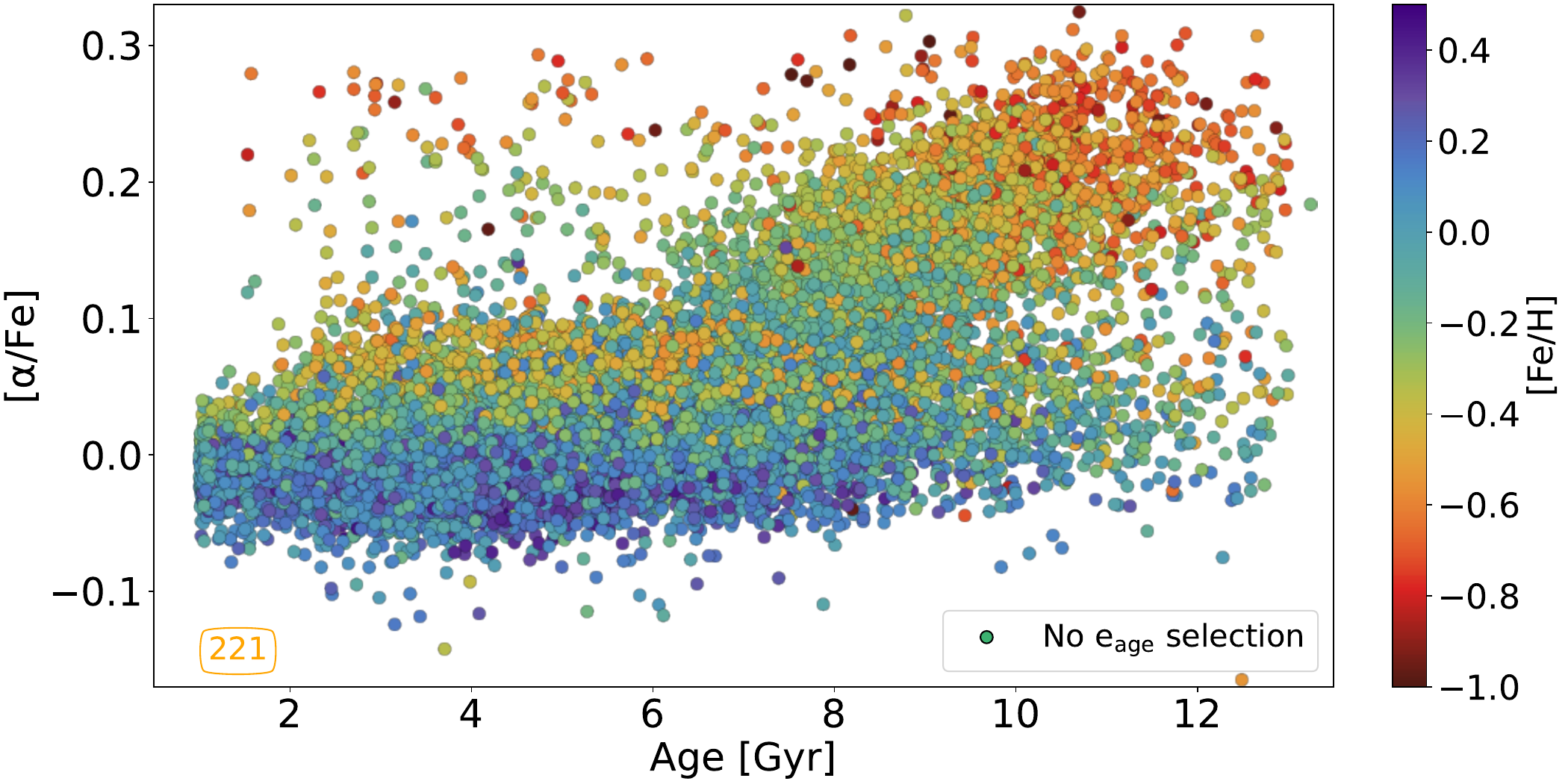}
\includegraphics[width=0.675\columnwidth]{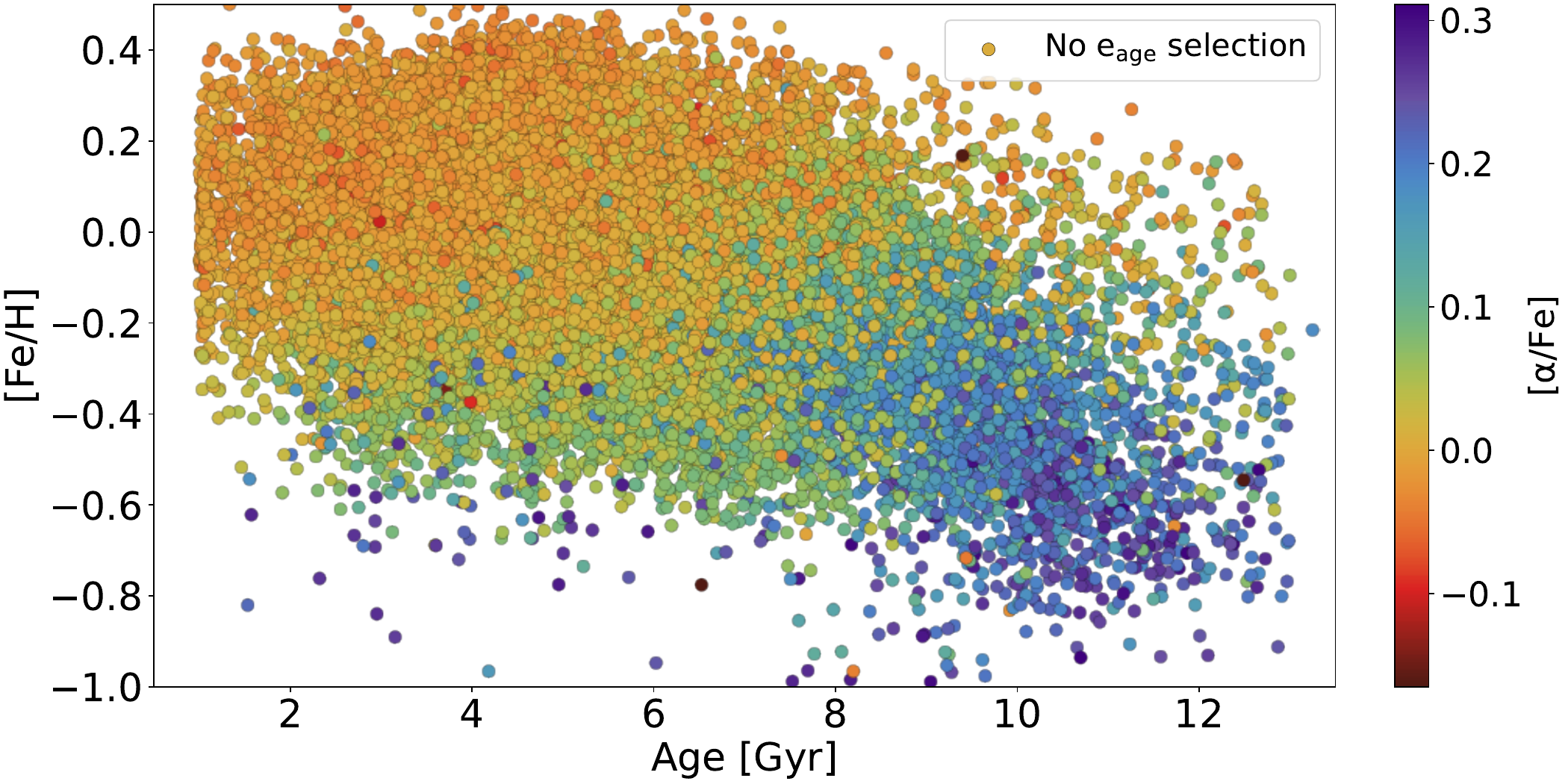}
\includegraphics[width=0.675\columnwidth]{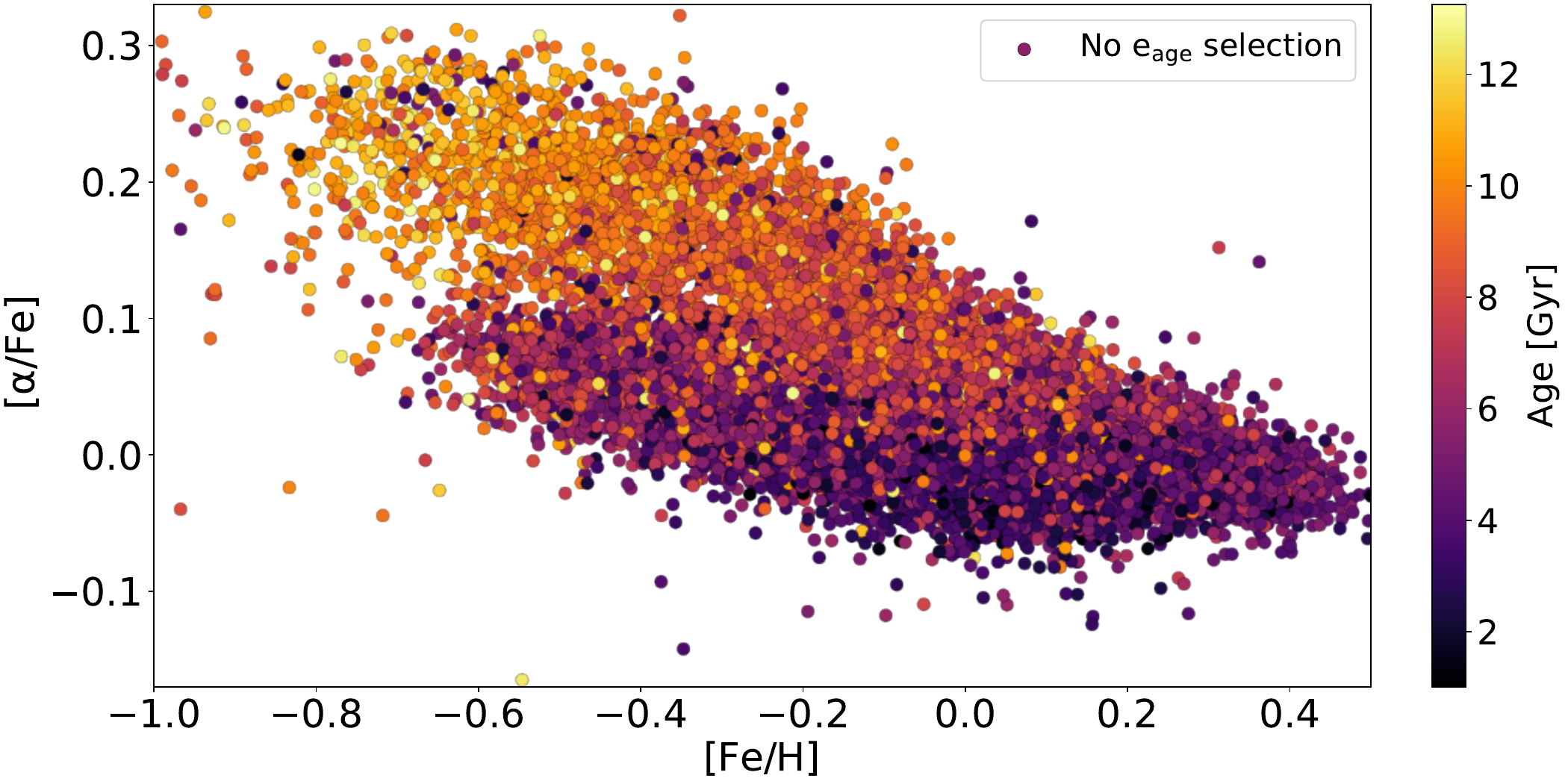}
\caption{Effect of the age uncertainties on the age-[$\rm \alpha$/Fe] distribution (left column), age-metallicity distribution (middle column) and [$\rm \alpha$/Fe]-[Fe/H] distribution (right column) for APOGEE DR17 local dwarfs. The ages in this case are obtained with the observational input G absolute magnitude, $\rm MG_{0}$, and Gaia DR3 de-reddened color $\rm (Bp-Rp)_{0}$. The stars in these plots have been selected to have: age above 1 Gyr, [Fe/H] lower than -1 dex, $\rm(BP - RP)_{0}$ lower than 0.95. We compare then three different cuts: $\rm e_{age}<0.3$ Gyr, $\rm e_{age}<0.5$ Gyr, $\rm e_{age}<0.7$ Gyr with the total distribution to which no selection on age uncertainty is applied (bottom row). The stars are color-coded for [Fe/H], [$\rm \alpha$/Fe] and age respectively. The number of candidate YAR (with age below 7 Gyr and [$\rm \alpha$/Fe] above 0.12 dex) is written in orange in each  age-[$\rm \alpha$/Fe] panel. 
Black lines are plotted in the age-[$\rm \alpha$/Fe] relation for stars with $\rm e_{age}< 0.3$ Gyr to highlight the presence of two distinct slopes in the thick disk sequence. We remind the reader that the age scale in this image may be shifted by 1–2 Gyr towards older ages if the age scale from \citet{nissen2020} is correct when making comparisons (see Section \ref{Subsec: nissen comparison} and Fig. \ref{fig: Nissen_this_work}).}

\label{fig: effect age_unc APOGEE DR17}
\end{figure*}

In this section we investigate the age-chemistry distributions in our sample making use of the ages calculated using Gaia photometry.

\subsection{General distributions}
\label{subsection: general distribution}

We begin by discussing the age-chemistry distributions based on our entire main sample. Fig. \ref{fig: effect age_unc APOGEE DR17} presents different levels of age uncertainty (stated inside each panel) and highlight that this error is well behaved: the spread in age at a given metallicity or $\rm \alpha$ abundance decreases when the uncertainty on the age decreases. Apart from that, the data show clean and well structured features.

\paragraph{Thick disk stars.}
The age-metallicity and age-$\rm \alpha$ relations for the thick disk are well defined. 
Thick disk stars are clearly separated from the thin disk, with the metallicity spread abruptly increasing from about 0.2 dex to 1 dex at about 8 Gyr. The correlation with age is tight for both metallicity and [$\alpha$/Fe] abundance as illustrated in particular by the stars with the lowest age uncertainty. The dispersion in age is continuously decreasing when the uncertainty in the stellar ages decreases, suggesting that the relationships could still be dominated by the uncertainty on ages even for stars with the lowest uncertainties (top row).

The age-$\rm \alpha$ relation in the thick disk consists of two distinct segments. In the first segment, [$\rm \alpha$/Fe] sharply increases from approximately 8 to 9 Gyr, followed by a more gradual increase after 9 Gyr. Black lines in Fig. \ref{fig: effect age_unc APOGEE DR17} are plotted to highlight the change of slope. These are indicative and the point of transition between the two slopes is determined by visual inspection. This new feature was not observed in previous studies, such as \cite{haywood2013}. The stratification in metallicity in the thick disk part of the age-$\rm \alpha$ relation is well visible, with the lowest metallicities stars having the oldest ages, as is also visible in the age-metallicity plane. 
A few so-called "young $\rm \alpha$-rich" (YAR) stars are visible as outliers in the age-$\rm \alpha$ relation (upper left part of the distribution) and as blue dots that extend to younger ages than the thick disk sequence in the age-metallicity distribution. More details on these objects are given in the next paragraph.

\paragraph{Thin disk stars.}
As already mentioned, the age-metallicity distribution part of the thin disk starts at ages below 8 Gyr, where the spread in metallicity suddenly increases to reach about 1 dex. The stratification in $\rm \alpha$-abundance is clearly visible, with the lowest metallicity stars being also the highest $\rm \alpha$-abundant objects. Due to the wide spread, it is challenging to discern any structure in the age-metallicity distribution from the scatter plots. The structure becomes clearer when examining the density plots in Fig. \ref{fig:compare_age_chemistry}.
The evolution during the thick disk phase brings the metallicity from the lowest value in our sample (about -1 dex) at ages larger than 12 Gyr to Solar metallicity at about 7 Gyr. The metallicity then decreases to about -0.2 dex at 5 Gyr. At ages younger than 5 Gyr, the stars form a separated clump of stars at a metallicity larger than Solar.

These results are in general agreement with different works in the literature \citep[e.g.][]{haywood2013, bensby2014, haywood2015, buder2019, ciuca2021, Katz2021, Sharma2022, Hayden2022, Queiroz2023, Patil2023}. 
Several studies based on stars other than dwarfs have reported a metallicity decrease consistent with our findings. \citet{Imig2023} presents age-metallicity relations for APOGEE DR17 red giant stars ($\rm 1<\log g<2$) that qualitatively resemble ours, showing a metallicity decrease after peaking at the end of the thick disk evolution in an intermediate region (see their Figure 23). Similarly, \citet{Patil2023} identify a comparable pattern for red clumps and red giants in APOGEE DR17 (see their Figures 8 and 15). \citet{Anders2023} also examine APOGEE DR17 red giant stars, finding an age-metallicity distribution in the solar vicinity (their Figure 8) that aligns with our results.

\paragraph{Young Alpha-Rich objects.}
A number of YAR stars are visible on the scatter plots on the left of Fig. \ref{fig:compare_age_chemistry}, to the left of the main thick disk branch of the age-$\rm \alpha$ relation. 
They are $\rm \alpha$-enhanced stars (usually [$\rm \alpha$/Fe] > 0.12-0.15) with ages lower than the typical high-$\rm \alpha$ objects (typically below 7-8 Gyr), located in the upper left area of the plane. These peculiar objects have been  previously investigated \citep[e.g.][]{chiappini2015, martig2015, jofre2016, yong2016,izzard2018, silva_aguirre2018, matsuno2018, sun2020, zhang2021, jofre2023, Cerqui2023, Queiroz2023, queiroz2023_anders, Grisoni2023}, and several recent works presented results in favor of an origin compatible with thick disk stars that have been rejuvenated through a mass acquisition event \citep[e.g.][]{sun2020, zhang2021, jofre2023, Cerqui2023, Grisoni2023}. 
Fig.~\ref{fig:HRD_local_sample} shows the mean trend in [$\rm \alpha$/Fe] and metallicity of our sample along the Gaia CMD, with YAR candidates highlighted with larger symbols. They are selected from the main sample to have age younger than 7 Gyr and  [$\rm \alpha$/Fe] above 0.12 dex. YAR stars located to the left and above of the thick disk sequence (light-blue and orange colors in the left and right plots respectively) and their chemical abundances show that they are both $\rm \alpha$-rich and slightly metal-poor.

\subsection{Inner, intermediate, and outer disc distributions}
\label{subsection: Rguide division}

It has been argued in \cite{Haywood2024} that the metallicity profile of the disk has been strongly shaped by the evolution of the bar. Its main resonances, the corotation and outer Lindblad resonances (OLR) in particular, define limits where the profile is affected. We define
sub-samples of the inner, intermediate, and outer disk according to these limits and using the guiding radius of our stars ($\rm R_{guide}$, provided in the astroNN catalog). Stars with a $\rm R_{guide}$ smaller than 7 kpc are considered inner disk stars. Recent measurements \citep{clarke22} indicate that the bar corotation is indeed located at 6-7 kpc from the Galactic center. In \cite{Haywood2024} it is found that this location coincides with a break in the metallicity gradient, where the metallicity profile is flatter within the corotation region. This feature is found also in the metallicity profile reported by \citet{lian2023}.
We set the inner radius limit of our outer disk sub-sample at 10 kpc to balance maintaining a sufficient number of stars and reaching radii far enough from the Solar vicinity. Ideally, we would have selected stars having guiding radii larger than 11-12~kpc, which is the inner limit of the outer disk, see \cite{Haywood2024}. Thus, our outer disk sample is in fact heavily polluted by stars originating from more inner regions.
Finally, we define a local sample in the range 7.6 kpc $< \rm R_{guide}<$ 9 kpc, positioning it approximately equidistant between the inner and outer regions and representing the intermediate region between these two. The number of stars in the inner, intermediate and outer disks defined above is 1 963, 5 351 and 567 respectively.

\begin{figure*}
\centering
\includegraphics[width=9.4cm]{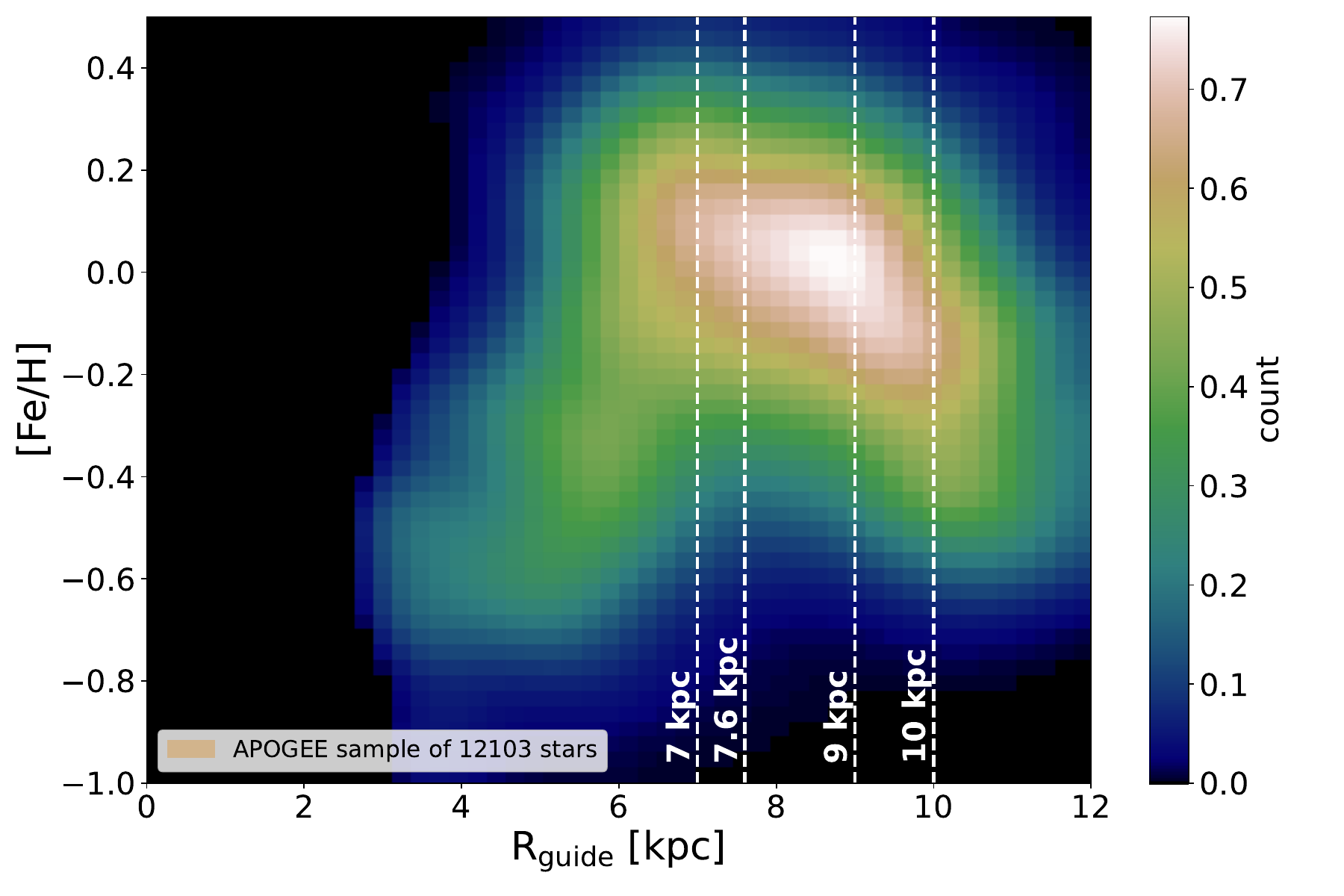}\hspace{-0.5cm}
\includegraphics[width=9.4cm]{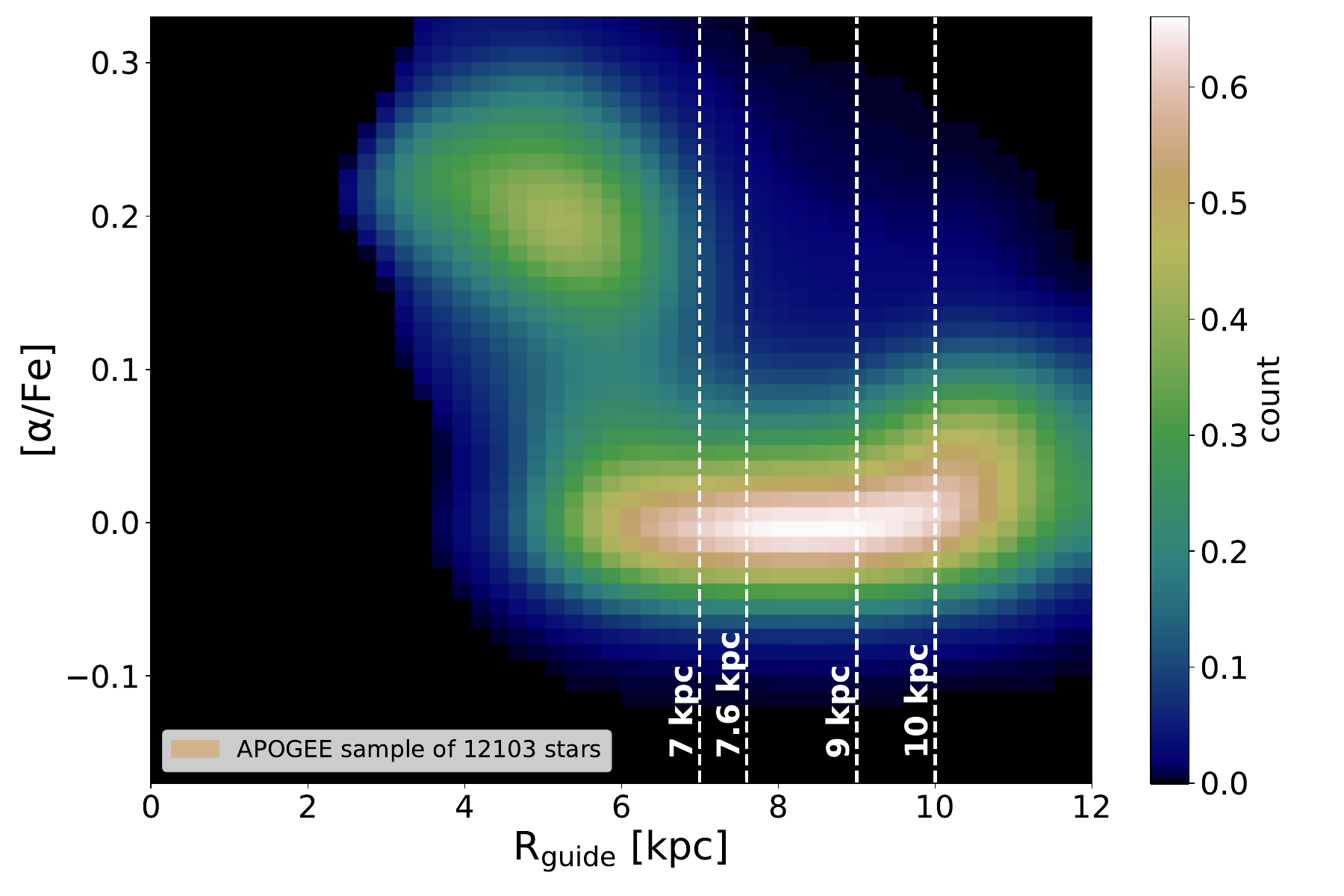}

\caption{Guiding radius - [Fe/H] (left) and [$\rm \alpha$/Fe] (right) distributions as 2D density (vertically normalized) histograms of the main sample of APOGEE DR17 local dwarfs. The vertical lines represent the thresholds we used for delimitate the inner ($\rm R_{guide}<7$ kpc), intermediate ($\rm 7.6<R_{guide}<9$ kpc) and outer ($\rm R_{guide}>10$ kpc) regions of the disk.
}
\label{fig: alpha_fe-Rguide histo}
\end{figure*}

In Fig. \ref{fig: alpha_fe-Rguide histo} we show the distributions of the main sample of dwarfs stars in the $\rm R_{guide}$-[Fe/H] and $\rm R_{guide}$-[$\rm \alpha$/Fe] planes as 2D density histograms.

\begin{figure*}
\centering
\includegraphics[width=9cm]{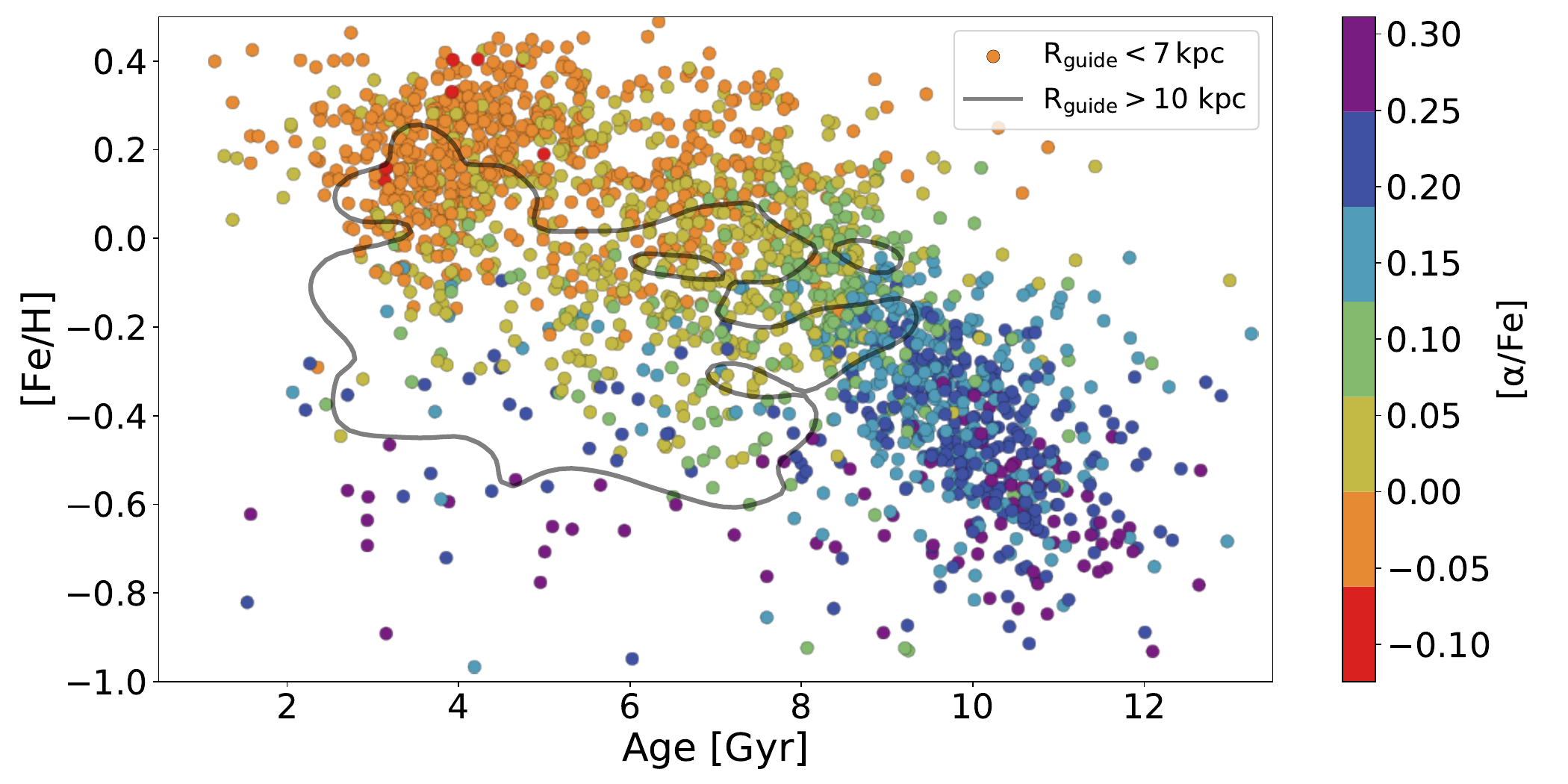}
\includegraphics[width=9cm]{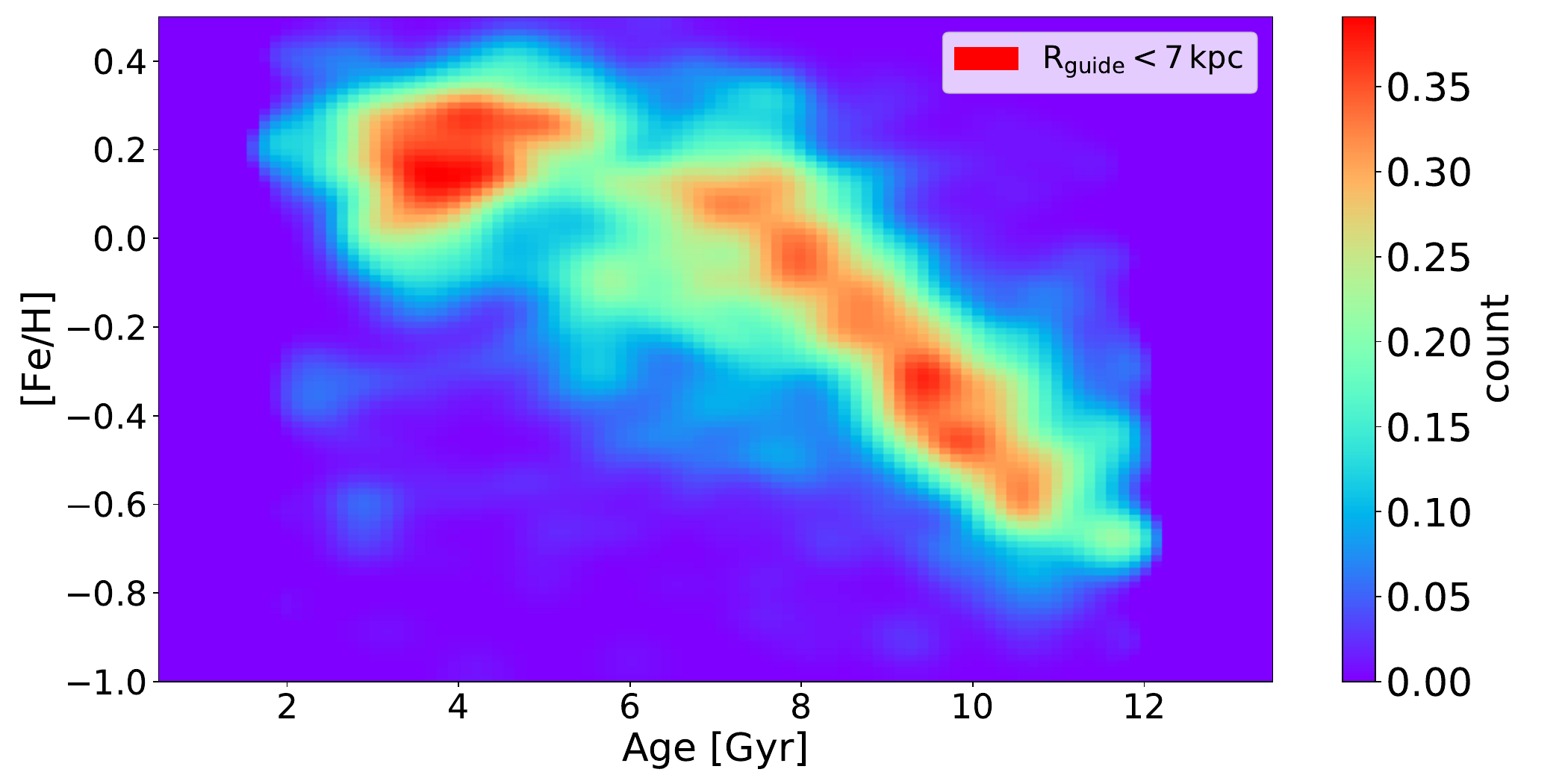}
\includegraphics[width=9cm]{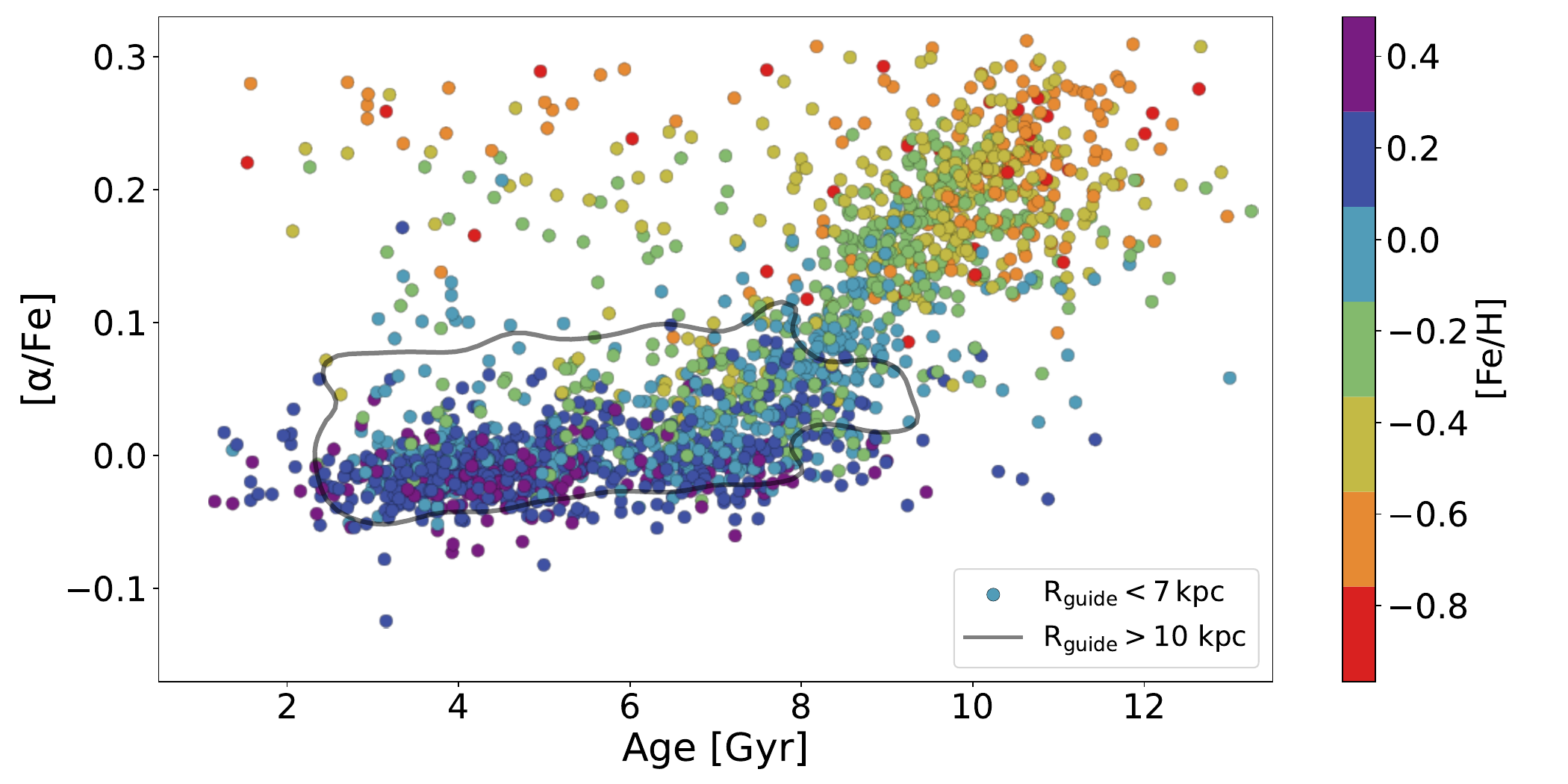}
\includegraphics[width=9cm]{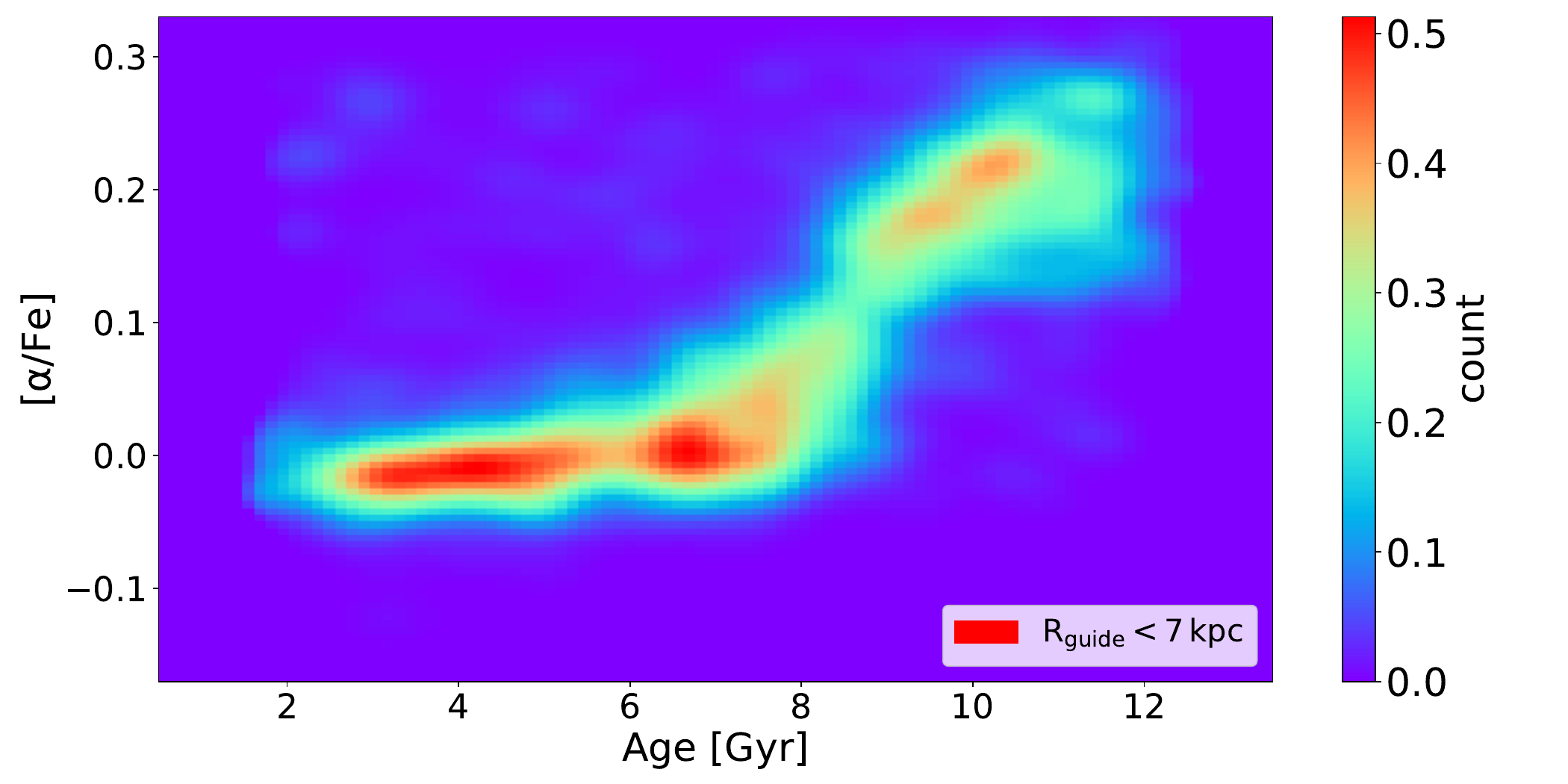}

\caption{Age-[Fe/H] and age-[$\alpha$/Fe] distributions for our sub-samples of inner disk stars, as defined by limiting $\rm R_{guide}$ below 7 kpc. Scatter plots are on the left while column normalized density plots are on the right. In the scatter plots, the stars are color-coded for their metallicity and [$\rm \alpha$/Fe] content, while with the black 30\% level iso-density contour we show the outer disk stars selected as $\rm R_{guide}$ greater than 10 kpc.}
\label{fig: age_chemistry_inner}
\end{figure*}

\begin{figure*}
\centering
\includegraphics[width=9cm]{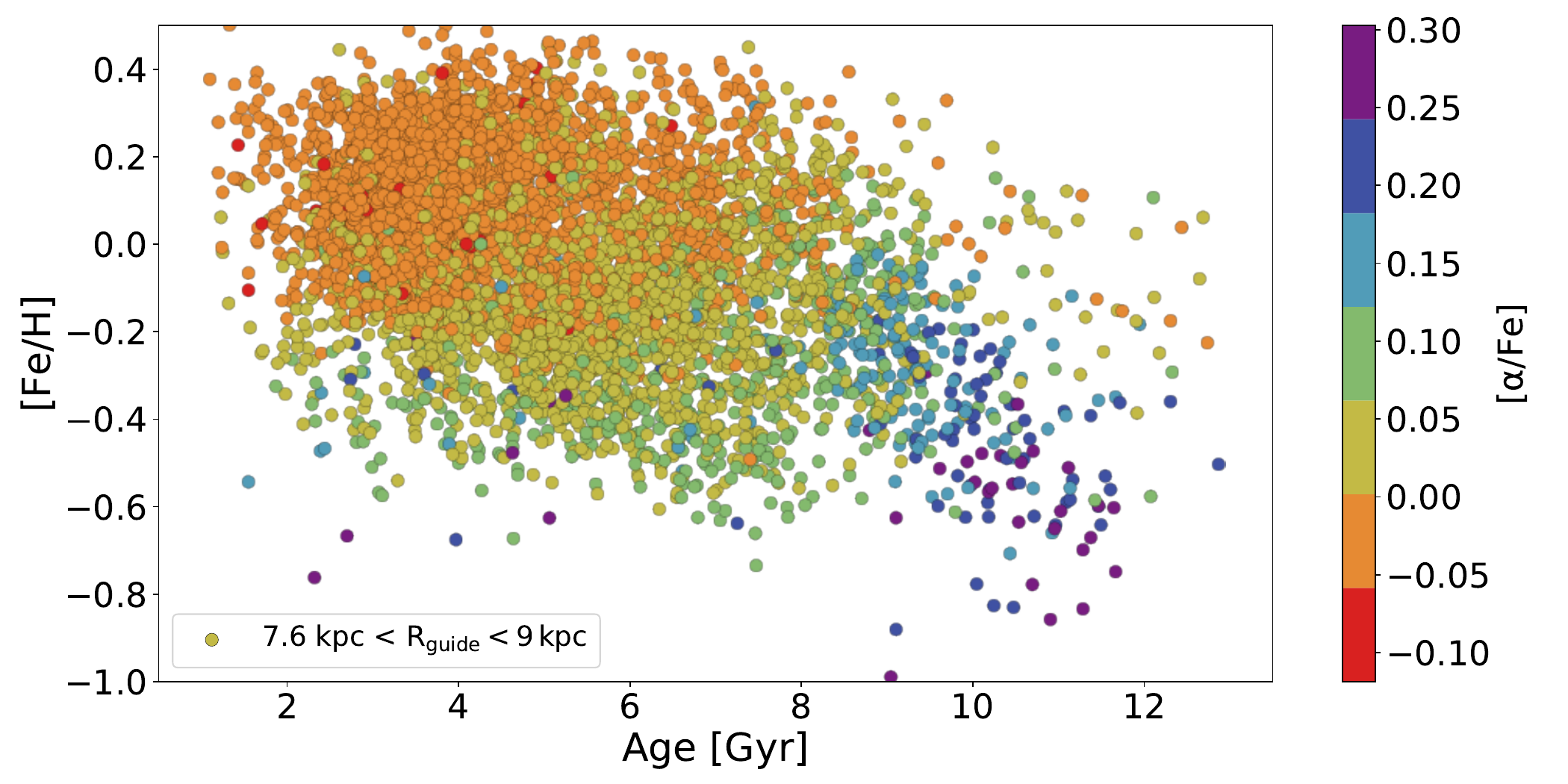}
\includegraphics[width=9cm]{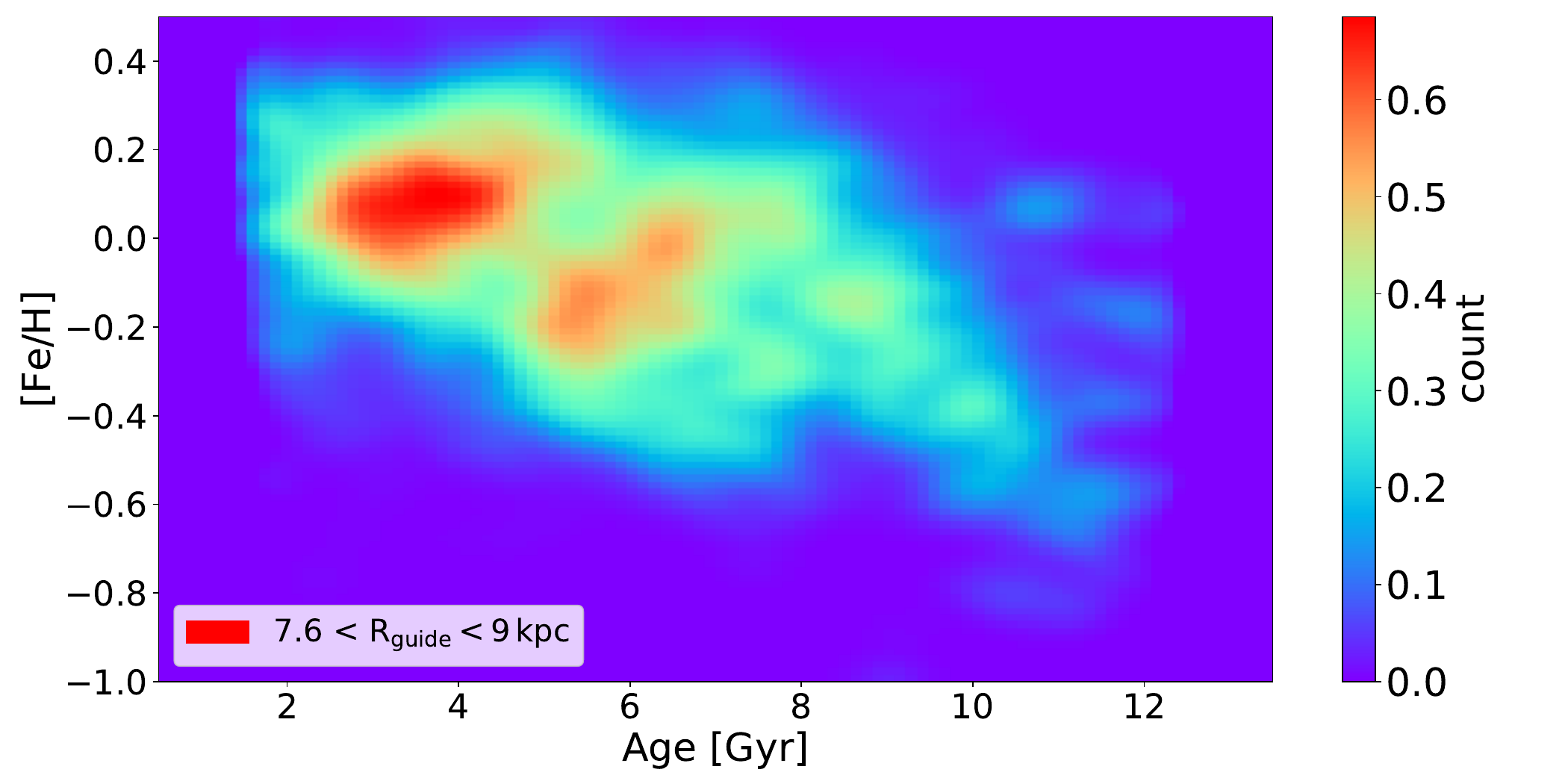}
\includegraphics[width=9cm]{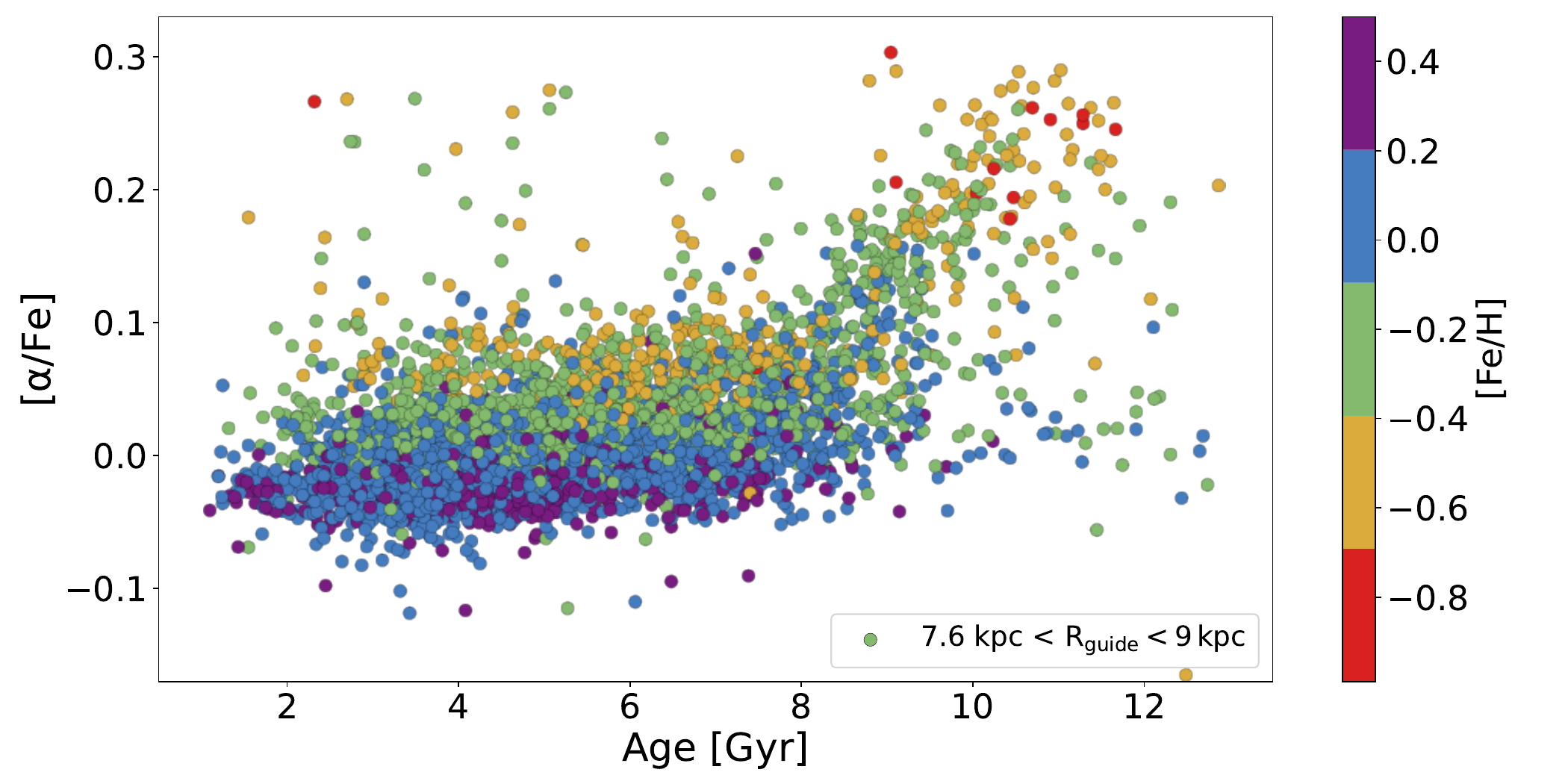}
\includegraphics[width=9cm]{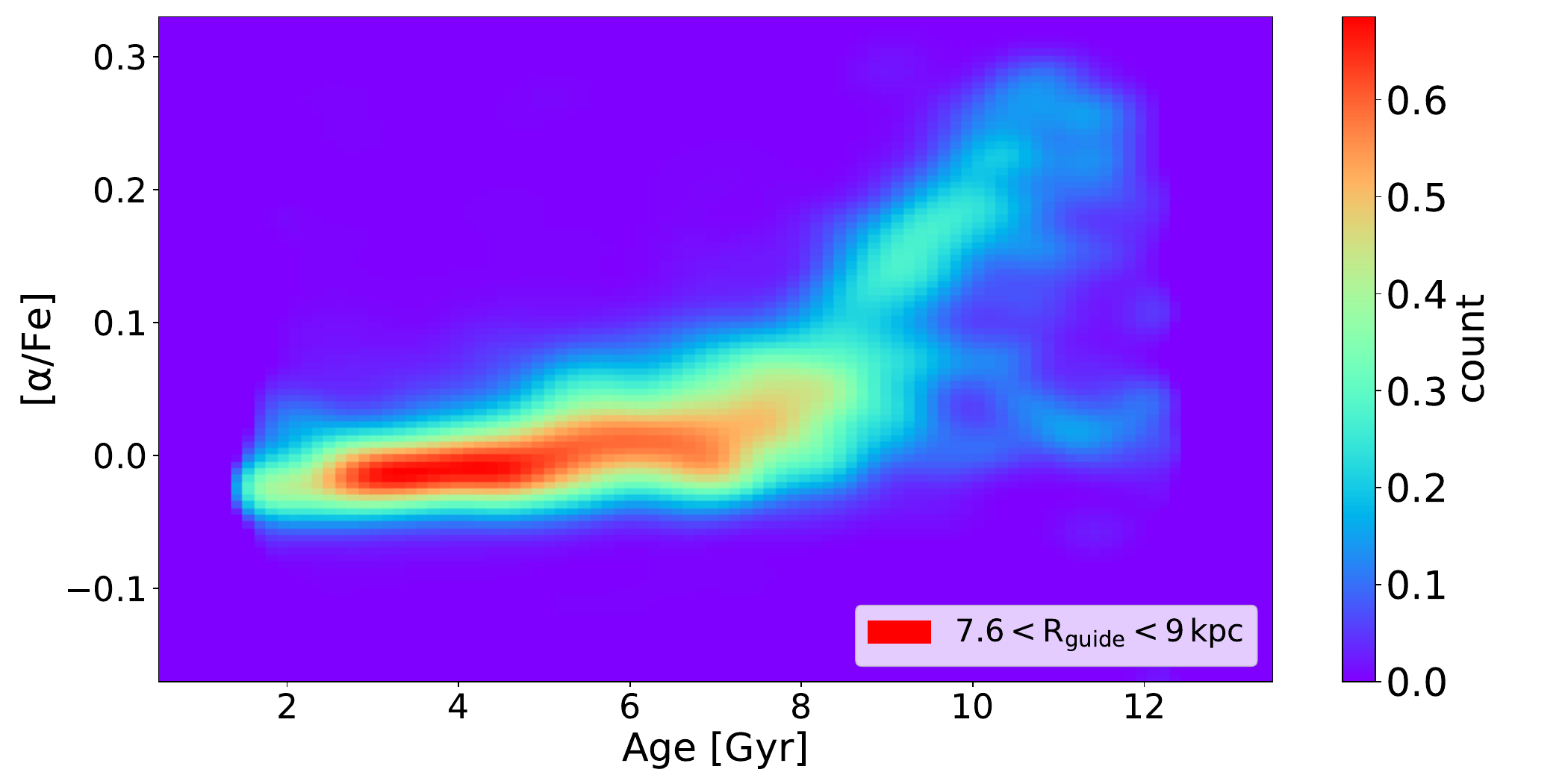}

\caption{Equal to Fig. \ref{fig: age_chemistry_inner} but for the local sub-sample: $\rm 7.6<R_{guide}<9.0$ kpc.}
\label{fig: age_chemistry_local}
\end{figure*}

\begin{figure*}
\centering
\includegraphics[width=9cm]{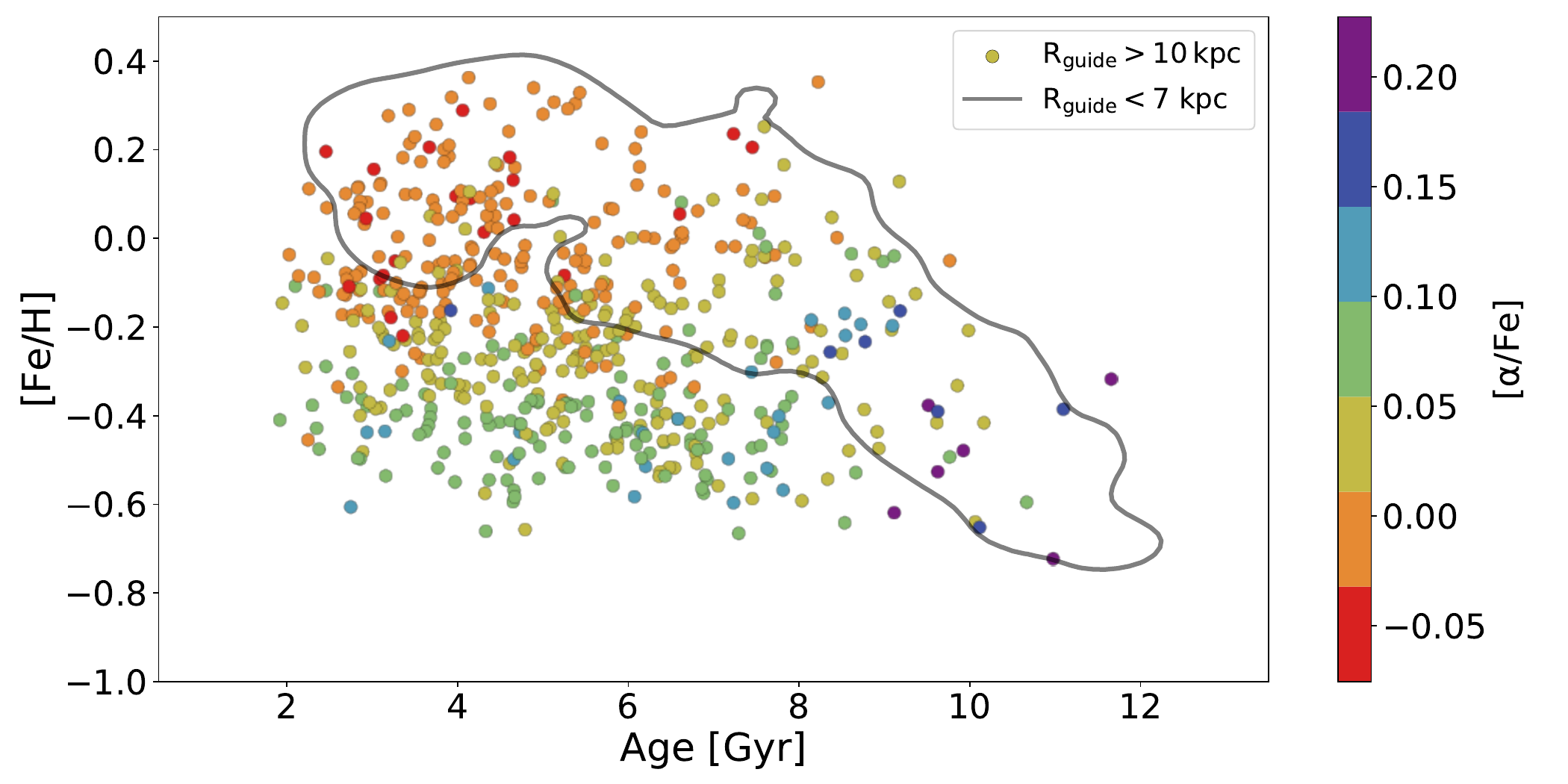}
\includegraphics[width=9cm]{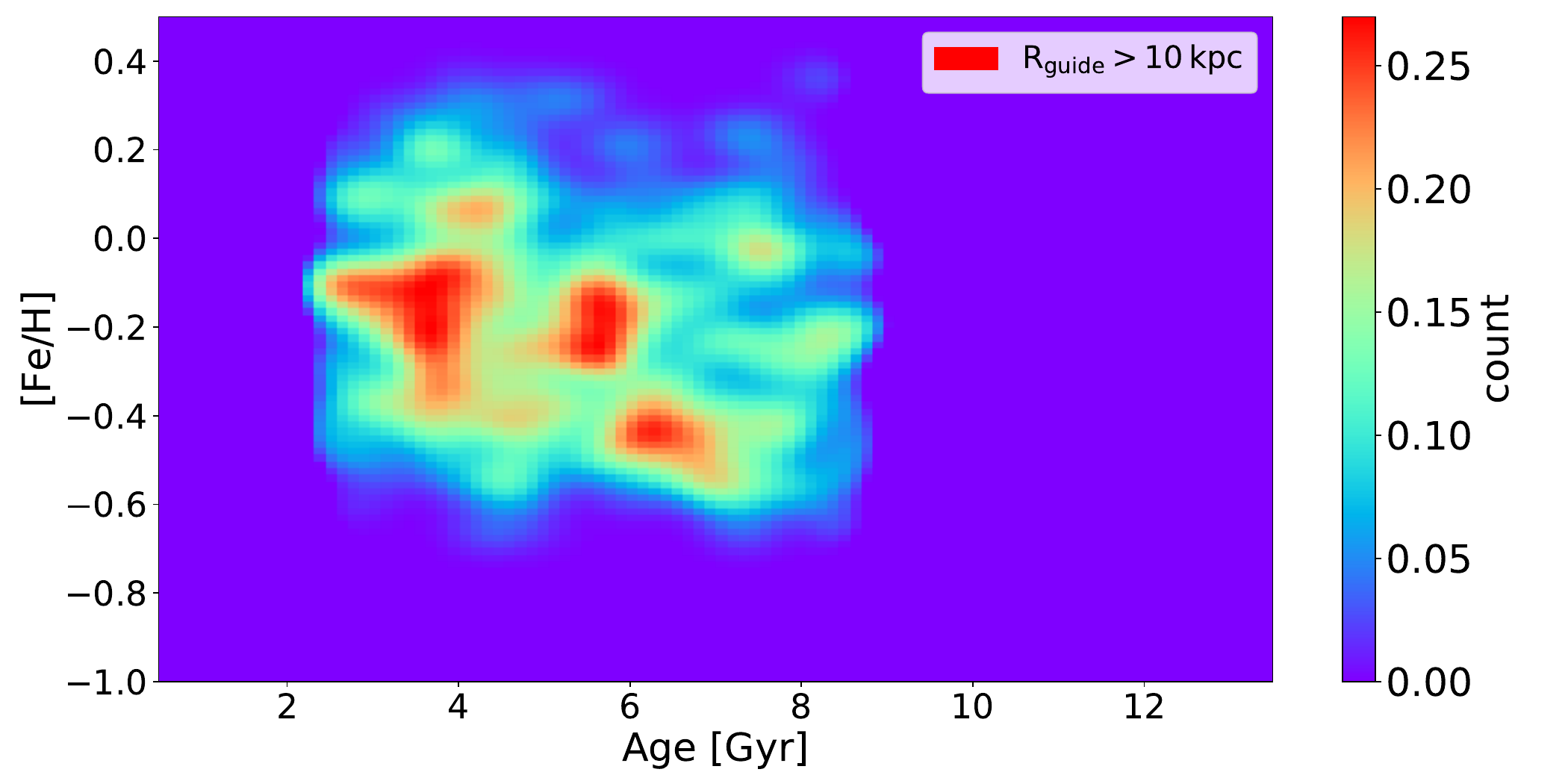}
\includegraphics[width=9cm]{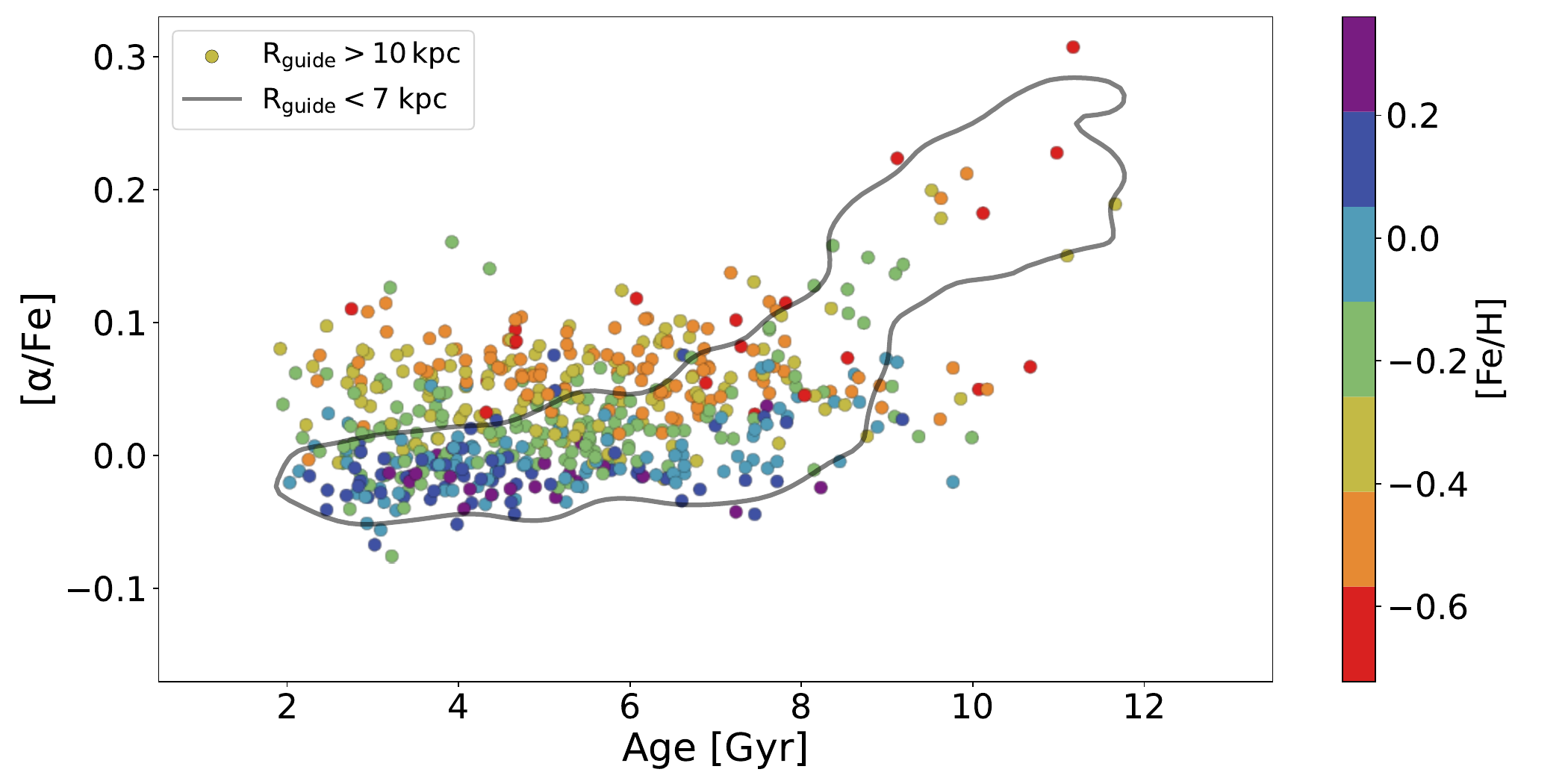}9
\includegraphics[width=9cm]{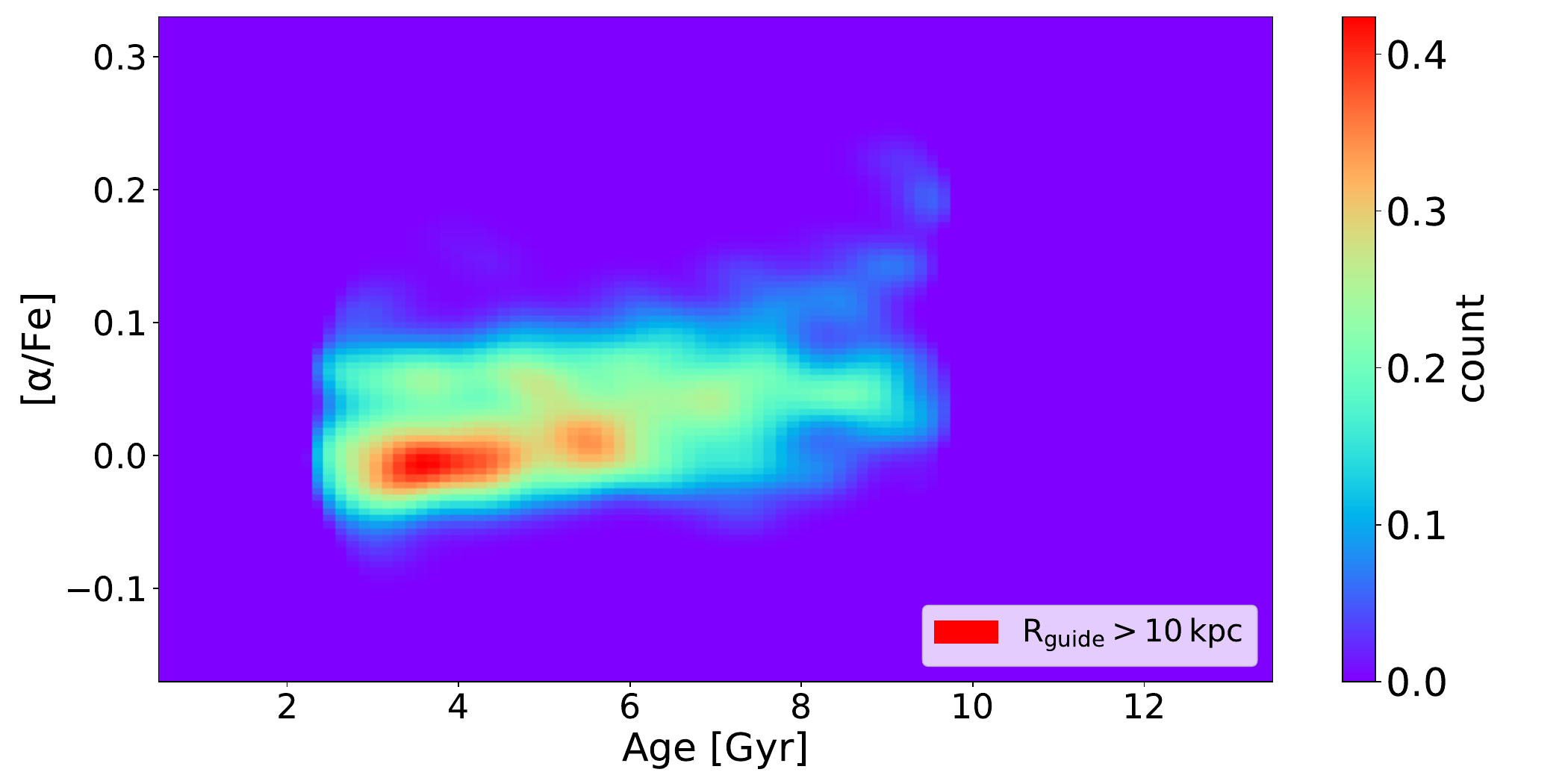}

\caption{Equal to Fig. \ref{fig: age_chemistry_inner} but for the outer disk sub-sample: $\rm R_{guide}>10$ kpc.}
\label{fig: age_chemistry_outer}
\end{figure*}

We show the age-chemistry relations of the three sub-samples in Fig. \ref{fig: age_chemistry_inner} (inner disk), Fig. \ref{fig: age_chemistry_local} (local or intermediate disk) and Fig. \ref{fig: age_chemistry_outer} (outer disk).
In each figure we highlight the scatter plot, with stars color-coded by their metallicity and [$\rm \alpha$/Fe] content, and the 2D column normalized histogram.

\paragraph{Inner disk.}
The inner disk stars, as shown in Fig. \ref{fig: age_chemistry_inner}, exhibit clearly defined relations in both the age-metallicity and age-$\rm \alpha$  distributions. The age-metallicity distribution shows a steep decrease in $\rm \alpha$-content with increasing metallicity, while the opposite trend is observed in the age-$\rm \alpha$ relation. Both relations display a distinct break at around 8 Gyr, indicating the transition between the thick and thin disk. The YAR objects are clearly visible in both age-chemistry distributions. The thick disk sequence is well-defined, and the thin disk segment is much narrower than in the global relation shown in Fig. \ref{fig: effect age_unc APOGEE DR17} and the outer disk  shown in  Fig. \ref{fig: age_chemistry_outer}, being restricted to the most metal-rich stars. Contrary to the age-metallicity trend of the main (Fig. \ref{fig:compare_age_chemistry}) and local (Fig. \ref{fig: age_chemistry_local}) sub-samples, the inner disk stars do not show an obvious decrease to lower metallicities at 5-6 Gyr (see the density histogram in the upper right panel of Fig. \ref{fig: age_chemistry_inner}).

\paragraph{Intermediate disk.}
The local disk distribution 
shown in Fig. \ref{fig: age_chemistry_local} presents similar trends and patterns to those commented in Section \ref{subsection: general distribution}. Being less populated, the thick disk sequence is less defined. The thin disk sequence of the age-$\alpha$ distribution is thicker compared to the inner thin disk (Fig. \ref{fig: age_chemistry_inner}), reflecting a wider range of metallicities. The age-[Fe/H] density distribution shows a clear decrease in metallicity beginning at 8 Gyr down to 5-6 Gyr. This feature, absent in both the inner and outer disks, is specific to the disk evolution within the 6-7 to 10-12 kpc range.

\paragraph{Outer disk.}
As visible from Fig. \ref{fig: age_chemistry_outer}, the outer disk sub-sample is almost entirely restricted to low-$\rm \alpha$ stars ([$\alpha$/Fe]$<$0.1), the high-$\rm \alpha$ stars, or thick disk stars, being essentially limited to R$<$10~kpc. The spread in metallicity at a given age is large, because this sub-sample is contaminated by more metal-rich stars belonging to the intermediate disk. No clear pattern of the metallicity and [$\rm \alpha$/Fe] as a function of age is apparent. We explore chemical evolution models for these three zones in section~\ref{sec: chemical modelling}.

\section{Comparison with age-metallicity trends from the literature}
\label{sec: age-metallicity literature}

The age-chemistry distributions of disk stars have been subject of several investigations, with the aim of understanding how the Galactic disk has formed. We now highlight the trends found by some of these works.

\subsection{Age-metallicity distribution: bimodal or not?}
\label{subsec: bimodalw}

A number of studies have pointed to an apparent bimodality of sequences in the age-metallicity distribution of the Solar neighborhood (e.g., \citealp[]{nissen2020, jofre2021, Xiang_Rix_2022, Anders2023, Patil2023, Casamiquela2024}), while others found smooth distributions \citep[e.g.][]{Hayden2022, ciuca2023}. 
Different results could be due to different disk sampling of these studies, in particular the age-metallicity distribution of the inner disk (i.e. the sequence evolving from the thick disk to the inner, metal-rich, thin disk) on the one hand, and the two clumps of stars at 5-6 Gyr and 2-4 Gyr on the other. Our distribution, similar to that of \cite{Salholhdt_22_GALAHDR3} (see Fig. \ref{fig:compare_age_chemistry}), shows more complex patterns, with a combination of sequences (as seen in the inner disk) and clumps (as observed in the Solar vicinity). We will now compare our results with this latter study in more detail.

\subsection{\cite{Salholhdt_22_GALAHDR3}}
\label{Subsec: comparison sahlhold}

An alternative paper that makes use of dwarf and subgiant stars to study metallicity trends in the Galactic disk is \cite{Salholhdt_22_GALAHDR3}. The authors exploit GALAH DR3 data and their resulting age-[Fe/H] distribution (their Figure 2) presents characteristics similar to our findings: 
starting with the clumpy configuration dominant in young objects (e.g. the main peak at Solar metallicity and at 3 Gyr) to the tight sequence that begins from the oldest and metal-poorer stars in the sample up to the most metal-rich stars at around 7 Gyr.
However, ages being determined using different isochrones, and the APOGEE DR17's and  GALAH DR3's abundance scales differing \citep[see for instance][]{hegedus2023}, small but significant differences are expected.

\begin{figure}
\centering
\includegraphics[trim= 0mm 10mm 0mm 10mm,width=\hsize]{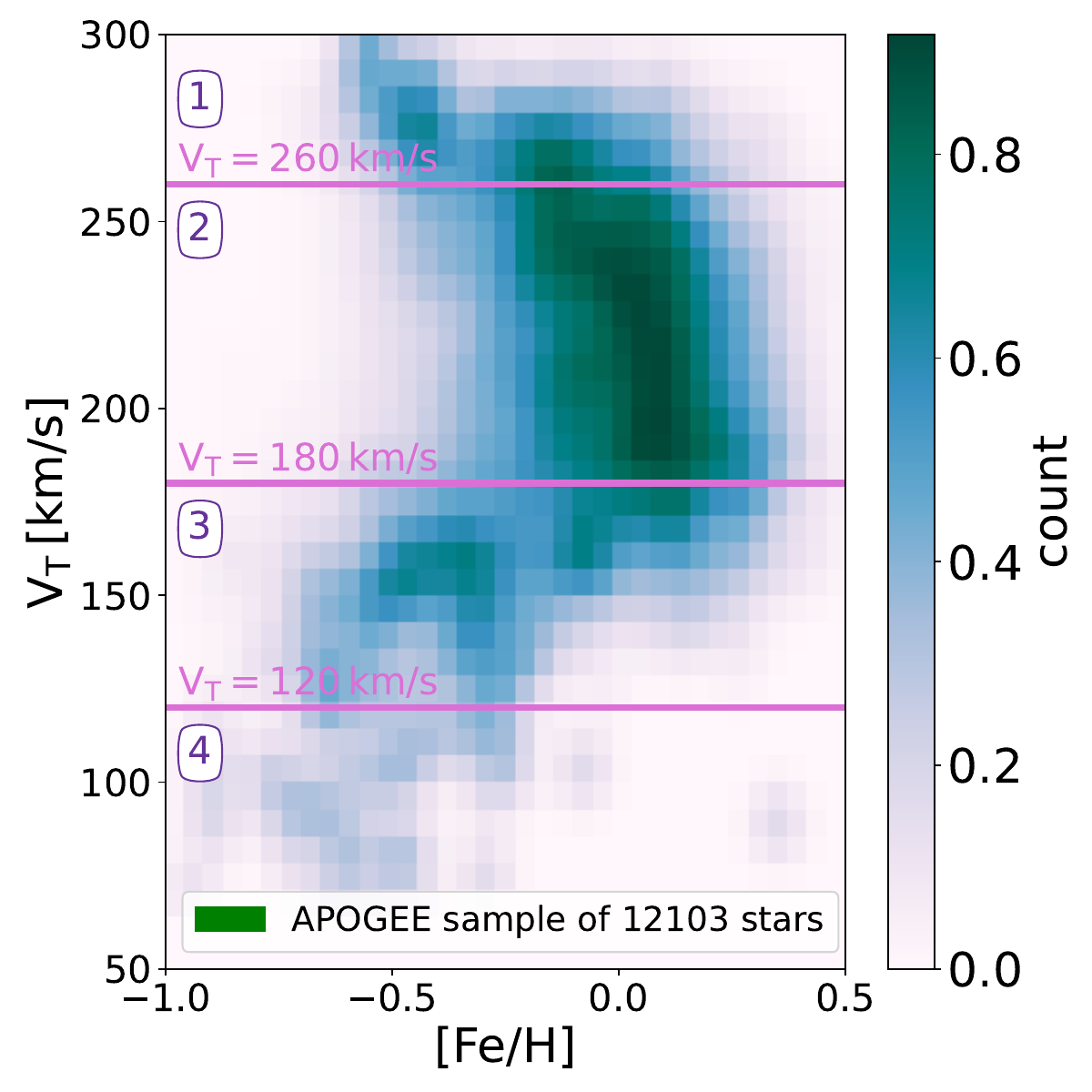}
\caption{Horizontally normalized density distribution of the Solar vicinity sample of dwarfs in the metallicity-tangential velocity plane. The three pink lines represent the threshold levels of $\rm V_{T}$ used to break the sample into smaller groups that we tag with a number from 1 to 4, as written in the diagram.}
\label{fig: Vt-Rmean dwarfs}
\end{figure}

\begin{figure}
\centering
\includegraphics[width=9cm]{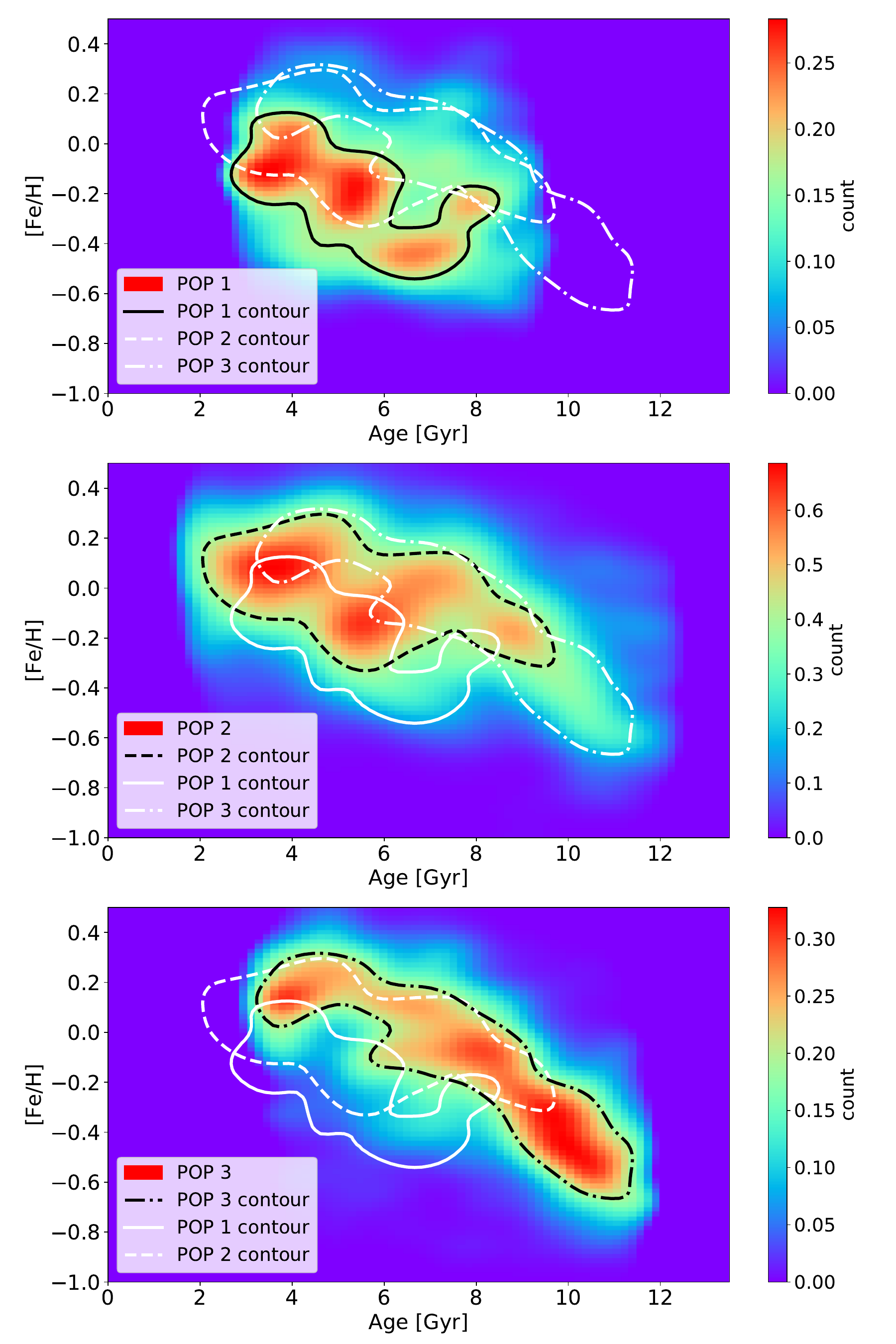}
\caption{Age-[Fe/H] distributions for each kinematically selected subsamples of local dwarfs. In each panel the 60$\%$ of the maximum density contour line is shown in black and the equivalent contour lines of the other populations are shown in white.}
\label{fig: age-feh_pop_dwarfs}
\end{figure}

For the purpose of comparison,  we decompose our main sample using the same kinematic selection as \cite{Salholhdt_22_GALAHDR3}, following a method 
inspired by \cite{Belokurov2020}: we divide our sample into four sub-samples, tagged POP 1 to POP 4, on the basis of their position in the [Fe/H]-$\rm V_{T}$ diagram, as shown in Fig. \ref{fig: Vt-Rmean dwarfs}. For the purpose of comparison, we adopt the thresholds of \citet{Salholhdt_22_GALAHDR3}\footnote{The only difference in the adopted thresholds is for the delineation of POP 1, where we set the cutoff at 260 km/s, while \citet{Salholhdt_22_GALAHDR3} use 270 km/s.}, 
but note that they are somewhat arbitrary and in particular do not take into account the obvious structures visible the metallicity-$\rm V_{T}$ plane. $\rm V_{T}$ is the azimuthal component of the Galactocentric velocity in cylindrical coordinates (namely tangential velocity), and we use the values provided by the astroNN added value catalog. Note that although we use a similar threshold to divide the four populations, POP 4 contains only 55 stars. For this reason, it is discarded in the following analysis.
The tangential velocity is correlated to the mean radius, see for example Fig. 5 in \cite{Salholhdt_22_GALAHDR3}.
However, a specific value of tangential velocity corresponds to a range of mean or guiding radii. Thus the kinematically defined samples 
are only approximately corresponding to our samples defined using guiding radii. Nevertheless, it can be seen that the age-metallicity plots of our outer, local and inner regions (Fig. \ref{fig: age_chemistry_inner}, \ref{fig: age_chemistry_local}, \ref{fig: age_chemistry_outer}) correspond well to the kinematically defined POP 1, 2 and 3 (Fig. \ref{fig: age-feh_pop_dwarfs}) and are characterized by the same features, although the small-scale structures are slightly different.

The inner disk stars (our POP 3 and population C from \citealp[]{Salholhdt_22_GALAHDR3}) show a similar overall evolution, initially following the thick disk sequence and then transitioning to more metal-rich stars at younger ages. Our POP 3 appears to include younger stars compared to the population C, but this may be due to differences in age scales. The main difference lies in the distribution patterns: our sample shows more clumpy structures, while \cite{Salholhdt_22_GALAHDR3} shows more regular patterns, a difference seen in all populations.

Population B in \cite{Salholhdt_22_GALAHDR3} shows the same characteristics as our POP 2, with a drop in metallicity at the end of the thick disk sequence - between 6 and 8 Gyr in their case, and up to about 5 Gyr in ours. While our data show increasing metallicity at younger ages, theirs remains relatively constant.

Finally, our POP 1 shows similar characteristics, with no noticeable trend in metallicity increase. Most stars have metallicities between Solar and -0.4 dex. However, POP 1 shows a dip in metallicity around 7 Gyr, reaching -0.5 dex, followed by an increase to -0.25 dex around 8 Gyr, a pattern not observed in their Pop. A. 

Thus, globally the two samples compared here show consistent features, which is encouraging given that they are based on different spectroscopic observations and the ages are based on different atmospheric parameters and stellar isochrones.

\section{Kinematics of the populations}
\label{sec: kinematics of the populations}
To complement the chemical description of our sample of APOGEE dwarfs, we now present in this section the and kinematic properties of the inner, local, and outer disk components into which the main sample can be subdivided (by a guiding radius selection, see section \ref{subsection: Rguide division}). Orbital properties are highlighted in Appendix \ref{Subsec: orbital parameters}.

\subsection{Vertical velocity dispersion}
\label{subsec: sigma_z}

\begin{figure}
    \centering
    \includegraphics[width=9.5cm]{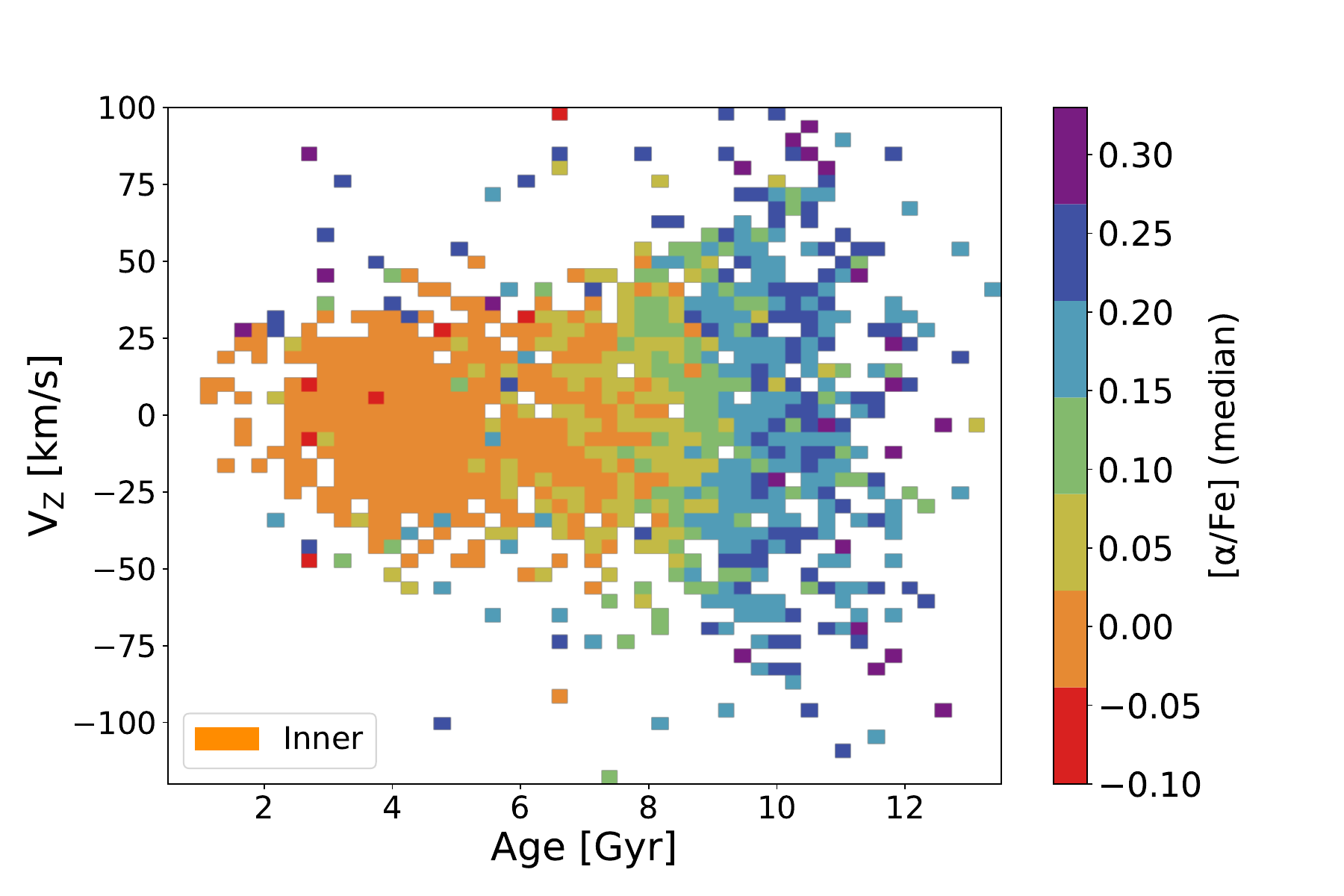}\vspace{-0.5cm}
    \includegraphics[width=9.5cm]{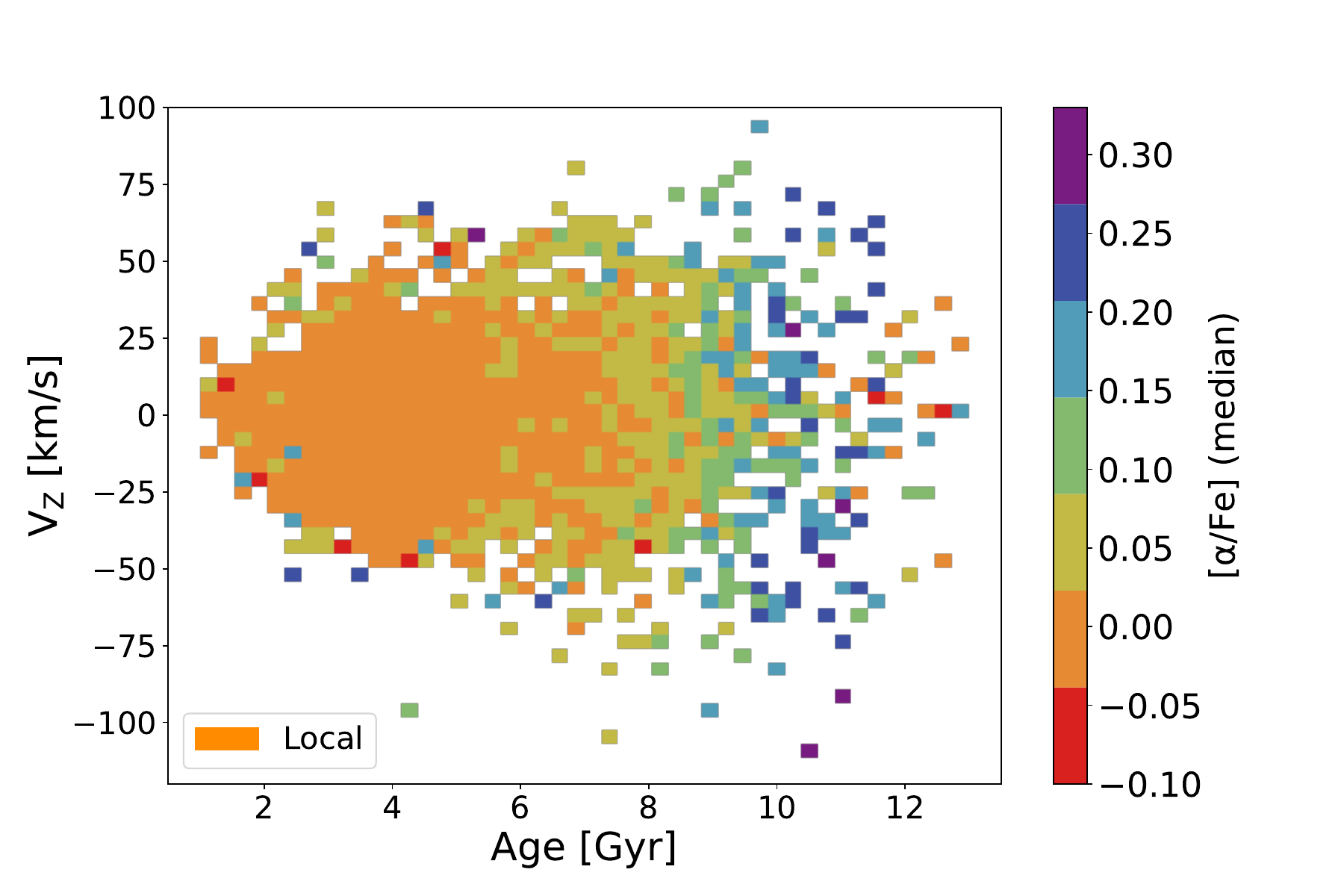}\vspace{-0.5cm}\par
    \includegraphics[width=9.5cm]{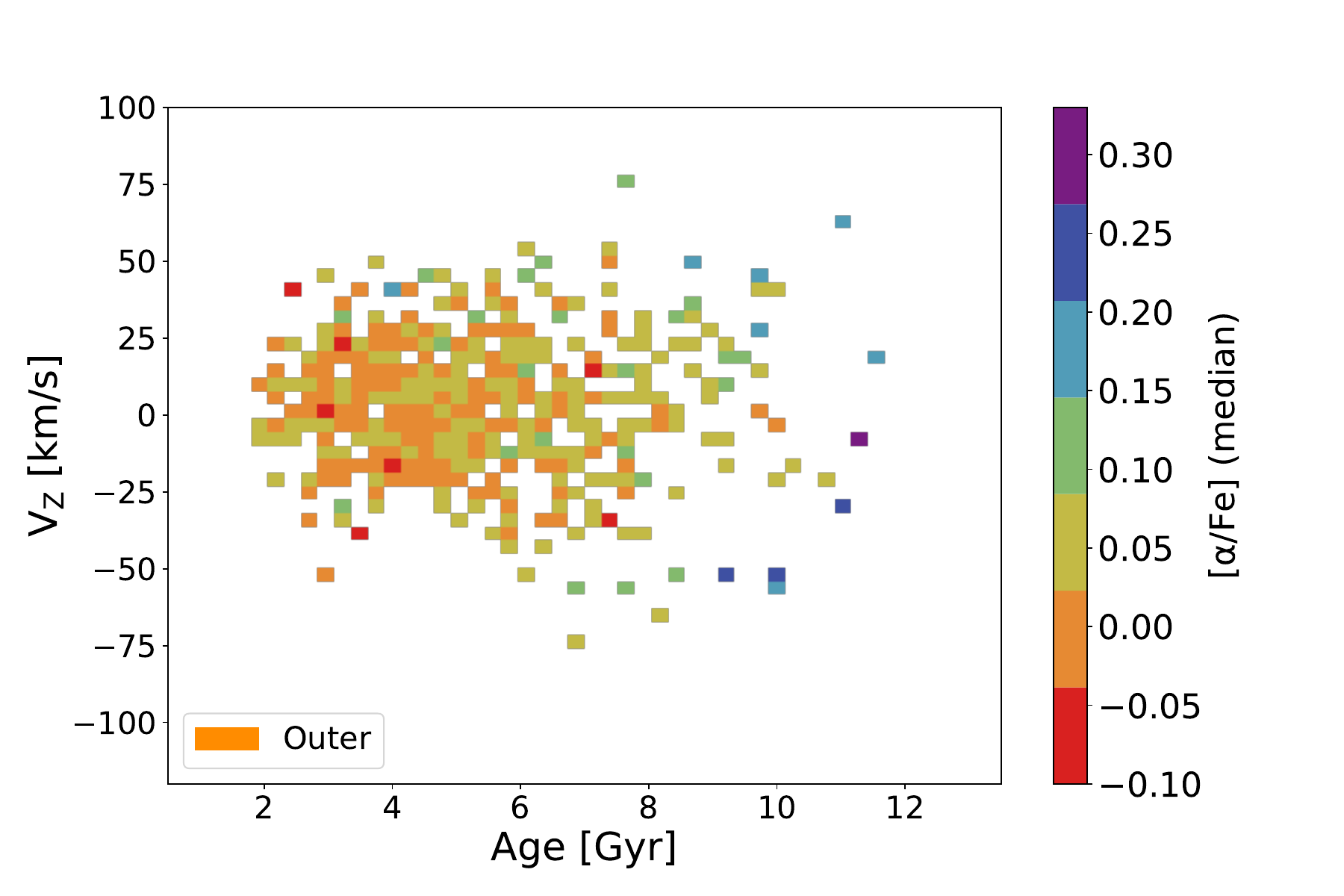}
\caption{Age-Vertical velocity distribution for the three sub-sample of inner, local and outer disk as defined in the text. Stars are color coded for $\rm \alpha$-elements.}
\label{fig: age-vz-pop}
\end{figure}

\begin{figure}
\includegraphics[width=10cm]{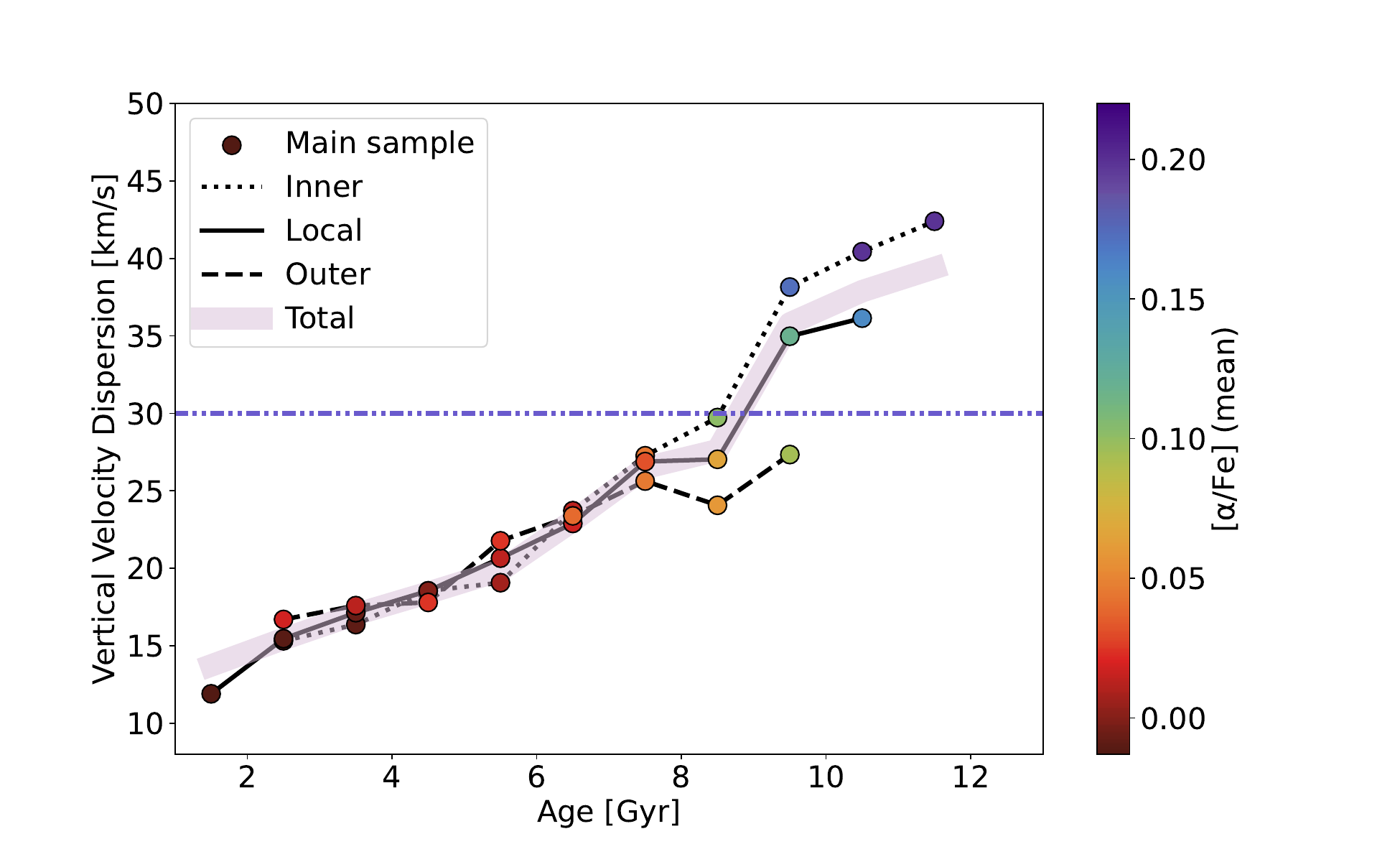}\vspace{-0.5cm}
\includegraphics[width=10cm]{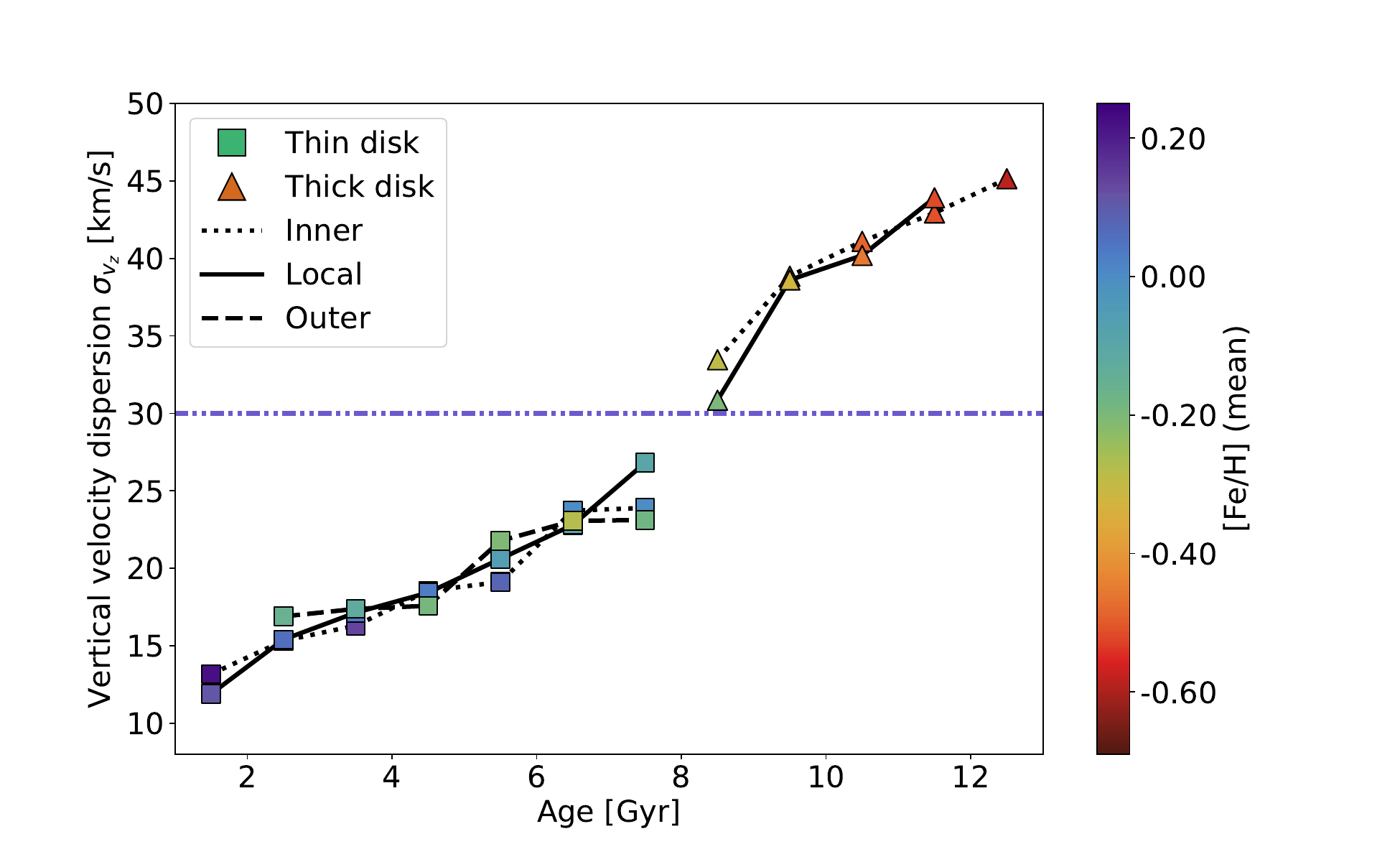}
\caption{Vertical velocity dispersion $\rm (\sigma_{z}$) of the inner, local and outer sub-samples
as function of stellar ages. In the top panel we display the relation obtained for the main sample; in the bottom panels we show the relation representative of the thick disk selected with [$\rm \alpha$/Fe] > 0.10  and age older than 8 Gyr (triangles) and the thin disk sample selected with [$\rm \alpha$/Fe] < 0.10 dex and age younger than 8 Gyr (squares). This selection allow us to avoid the region in which the thin and thick disk populations mix and the possible YAR candidates.
$\rm \sigma_{z}$ is determined for 1 Gyr age bins and the points are plotted in the middle of each bin. The color coding indicates the mean of the individual [$\rm \alpha$/Fe] (top and middle) and  [Fe/H] (bottom) of the stars in each bin. For each population, the bin with the fraction of stars exceeding 1\% of the total number of stars is shown. The pink thick line represents the global trend of the main sample. A blue line at $\rm \sigma_{z}$ = 30 km/s is also shown for reference.}
\label{fig: sigmaz_age-pop_fe_no_yar}
\end{figure}

The relation between the age and vertical velocity dispersion of stars in the Solar vicinity is a crucial probe for constraining the vertical heating history of the Milky Way's disk. This relation has been extensively explored over the years using different observable tracers \citep[e.g.][]{Wielen74, casagrande2011,raddi2022}.
In this section we explore relations representative of our inner, local and outer disk populations, previously selected through $\rm R_{guide}$. 
To calculate $\rm \sigma_{z}$, we made use of the Galactocentric cylindrical vertical velocities provided in the astroNN added value catalog \citep{Leung_Bovy_2019}. In particular, we computed the $\rm \sigma_{z}$ value for the different groups in bins of 1 Gyr each and the points are displayed in the middle of each bin.
We do not display the results for age bins that contain less than 1\% of the total number of stars in a given sample. 

The increase in the vertical velocity dispersion with age can be first seen in Fig. \ref{fig: age-vz-pop}, where we plot the vertical velocity ($\rm V_{Z}$) as a function of stellar age for the inner, outer, and local subsamples. 
This figure shows a clear stratification of [$\alpha$/Fe] with age, especially in the inner disk sub-sample (upper panel), which demonstrate the quality of our stellar ages.
The scatter in $\rm V_{Z}$ increases sharply for ages larger than 8 Gyr (this is obvious in the inner disk sample, but also noticeable in the local sample, even though the old thick disk component lacks stars at these radii).

The increase in velocity dispersion is quantified in Fig. \ref{fig: sigmaz_age-pop_fe_no_yar}.
The top panel represents the global trend, with no chemical separation between thin and thick disks.
We discard from each sub-sample the YAR candidates selected to have ages below 7 Gyr and [$\rm \alpha$/Fe] above 0.12 dex,
that contribute to the young age populations with a vertical velocity dispersion typical of older stars (having a thick disk origin).
Globally, each sub-sample shows a tendency of $\rm \sigma_{z}$ to increase at increasing ages.
Inner, local and outer disk trends overlap up to 7.5 Gyr (7-8 Gyr age bin). 
Beyond 7.5 Gyr the vertical velocity dispersions of the three different sub-samples diverge, due to the different fraction of thick disk stars, with a particular steepening of the trend appreciable both for inner and local stars above 8.5 Gyr (8-9 Gyr age bin).

These results are in agreement with previous works
(e.g. \citealp{haywood2013,Feuillet2016,Hayden17,Yu_Liu2018,silva_aguirre2018,Mackereth19,Sharma_2021,miglio2021,Hayden2022,Salholhdt_22_GALAHDR3, Anders2023, Casamiquela2024}), and reproduce the general findings in the literature.
The main feature that differentiates these results is the steepness of the relationship of the high-$\rm \alpha$ and low-$\rm \alpha$ populations considered: some works found that the vertical velocity dispersion smoothly increases with age \citep[for instance][]{Hayden17,Hayden2022}, while other authors \citep[e.g.][]{silva_aguirre2018,miglio2021, Sun2024} find an abrupt change in vertical velocity dispersion between the high-$\rm \alpha$ and low-$\rm \alpha$ disks. We believe that the difference lies in how strict the chemical definition of the thick disk is made (as suggested also in \citealp[]{Sharma_2021} and \citealp{Hayden2022}). 

In our case, we select the thin and thick populations directly from the age-[$\rm \alpha$/Fe] relation carefully avoiding the area of the "knee" where the two populations may overlap: the stars of the thin disk present [$\rm \alpha$/Fe] lower than 0.1 dex and age younger than 8 Gyr while on the contrary the thick disk objects have been selected to have [$\rm \alpha$/Fe] higher than 0.1 dex and age older than 8 Gyr. In addition, this cut allow us to discard possible YAR candidates.
The resulting relations, shown in the bottom panel of Fig. \ref{fig: sigmaz_age-pop_fe_no_yar}, highlight the continuity between the $\rm \sigma_{z}$ trends with age of the thin and thick disks. The apparent absence of an abrupt separation or "jump" between the vertical velocity dispersion of the thin and thick disks makes our results in agreement with \cite{Hayden17,Hayden2022}. However, we noticed that a more rough selection (for instance a simple horizontal cut in $\rm \alpha$-elements) could induce a separation in our trends similar to the one observed in \cite{silva_aguirre2018,miglio2021}. Thus, we underline the importance of a careful chemical selection (of the thick disk sample in particular) and its impact on the results.

Some studies in the literature suggest that the steepening or 'jump' observed in the vertical velocity dispersion around 9 Gyr in the disk stars by many works (also visible in Figure \ref{fig: sigmaz_age-pop_fe_no_yar}, top plot) could be attributed to a merger event \citep{Quillen2000,martig14}, possibly compatible with the Gaia-Sausage-Enceladus (GSE). 
Our data shows that the observed steepening is internal to the thick disk -- stars immediately younger and having lower dispersion are still thick disk objects (Figure \ref{fig: sigmaz_age-pop_fe_no_yar}, bottom plot). Thus, the sudden increase of the vertical velocity dispersion is not a marker of the transition between the thin and thick disk components. The heating occurs as the thick disk evolves from a metallicity of  about -0.4 dex to about -0.2 dex, as visible in the bottom panel of Fig. \ref{fig: sigmaz_age-pop_fe_no_yar}. At this metallicity, \cite{Dimatteo19} found a significant number of thick disk stars becoming counter-rotating, suggesting that a major accretion event happened during this period, which our findings seem to support. In this sense, the results indicate that, when defined chemically, the thick disk is not just a heated disk: it is a disk with distinct chemical characteristics which has been only partly heated ([Fe/H] < -0.3 dex).
This assessment implies that the accretion event occurred before the thick disk phase completed. \cite{martig14} observed in their simulated galaxies that those undergoing mergers exhibit jumps in their age-velocity relations (AVRs) at ages corresponding to the end of the merger. In this context, the jump that we observe in the bottom panel of Fig. \ref{fig: sigmaz_age-pop_fe_no_yar} could be interpreted as a signature of the vertical heating caused by the Gaia Sausage Enceladus (GSE) merger event, which likely occurred between 9 and 10 Gyr ago, before the thick disk phase ended. 

The evolution of the AVR has also been investigated in cosmological simulations.
\citet{Grand2016} performed cosmological simulations using the \texttt{AREPO} N-body magneto-hydrodynamics code, resulting in a set of 16 galaxies with well-defined stellar disks with a variety of bars and spiral structures. 
They observed a range of AVRs, 
with the majority of the sample exhibiting smooth trend with age up to 10 Gyr consistent with gradual disk heating from secular evolution. A quarter of the simulations show sharp jumps linked to satellite interactions and, in particular, the two most significant heated disks on their sample show bursts of heating coinciding with satellite pericentre passages (see their Figure 15 and Section 6). Their findings highlight that minor mergers drive the most significant heating episodes, even though secular bar-driven heating is more prevalent in their sample.

\citet{Leaman2017} developed models based on observed star formation histories of Local Group galaxies to describe the ISM velocity dispersion due to a galaxy's evolving gas fraction. These models have been compared with AVRs derived from stellar velocity dispersions and ages obtained through spectroscopic observations of individual stars in eight Local Group galaxies. In their Figure 3, they show that low-mass galaxies ($\rm M_{\star} \sim 10^{10}\, M_{\odot}$) have $\sigma(t)$ profiles that can be explained almost entirely by a combination of the ISM model and the equilibrium scattering model. For intermediate-mass galaxies with $\sigma \gtrsim 15$ $\rm km\, s^{-1}$, ISM-driven dispersion alone is sufficient to reproduce the $\sigma(t)$ profiles. However, the MW and M31 exhibit present-day stellar velocity dispersions exceeding predictions from ISM turbulence and equilibrium scattering. The authors suggest that additional dynamical processes, such as heating from minor mergers, contribute to this discrepancy (see their Section 7.2.1).

\citet{Bird21} use a high-resolution cosmological zoom-in simulation of a Milky Way-mass disk galaxy (\texttt{h277}), run with the parallel N-body+SPH code GASOLINE, to link local disk kinematics to its long-term evolution. They find that the AVR in the simulated solar neighborhood closely matches that of the MW (see their Fig. 1, where the observed AVR is derived from the Geneva-Copenhagen Survey, \citealp[]{Nordstrom2004}). Their Fig. 2 further shows that mono-age stellar populations in the solar vicinity undergo additional heating after birth. \citet{Bird21} strongly suggest that the observed heating in \texttt{h277} results from a combination of inside-out disk growth and scattering by molecular clouds. In contrast, \citet{Leaman2017} argue that this mechanism alone is insufficient to explain the AVR in more massive disk galaxies like the Milky Way, where additional heating from mergers or bars is required. Given \texttt{h277}’s quiescent merger history over the past 10 Gyr, \citet{Bird21} rule out mergers as the dominant heating mechanism in this system.

To summarize, previous simulations of MW-type galaxies (e.g., \citealp[]{martig14, Grand2016, Leaman2017}) suggest that "jumps" in the AVR relation
could be due to minor mergers. Based on these findings, we suggest that jump in the AVR  that we observe in the thick disk (see bottom panel of Fig. \ref{fig: sigmaz_age-pop_fe_no_yar}) could be the signature of the vertical heating caused by GSE. In contrast, other studies (e.g., \citealp[]{Bird21}) explain the observed heating in MW-like galaxies due only scattering by giant molecular clouds, without recurring to additional heating due to merger events.

\section{Chemical evolution modeling}\label{sec: chemical modelling}

In this Section we describe the modeling we performed to describe the chemical enrichment of our inner, local and outer disk. These three zones show chemical evolution with distinct features that are specific to each region, hence we do not attempt to produce a unique model with varying parameters. 
For example, the decrease of the metallicity observed in the intermediate region is not observed in the inner region, which shows a monotonic increase of the metallicity, nor in the outer regions, where no dilution is observed.

The model used in this paper is an upgrade of the one presented in \cite{Snaith2014, snaith2015}, introducing possible gas infall. We give a brief overview of the philosophy of the model in this Section but we refer the reader to \citet{Snaith2014,snaith2015,Snaith2022} for a detailed description.

We address to the reader that the goal of the modeling done in this work is not to create a sophisticated model of the data but rather to offer a basic insight into the essential components a simple model needs to reproduce the data.

\subsection{Modeling the inner disk evolution}
\label{subsec:Closed box bodel}

\citet{snaith2015} adopted a simple Galactic Chemical Evolution (GCE) model that minimizes the number of assumptions. Section 2.1 of their paper outlines these constraints and provides arguments for their plausibility. 
A key assumption, which we also adopt in our modeling of the inner disk that most of the gas accretion in the inner disk occurred early in the evolution of the Galaxy, or alternatively that continuous gas accretion maintained a sufficiently high gas fraction to make the evolution of the disk largely independent of its accretion history. 

In other words, the initial gas supply is available for star formation from the start while also serving as a reservoir of primordial gas, with inflows and outflows considered negligible as a first approximation.
The idea that a substantial gas reservoir was available during the first few billion years of the Galaxy is consistent with observations of gas-rich disks at high redshifts \citep{Tacconi2010}.
These are known to be affected by large scale turbulence, steady gas accretion, high star formation activity and feedback. When applying our chemical evolution model to the early (thick) phase of the disk formation, we assume some sort of equilibrium is achieved during several Gyr between these processes  \citep[see e.g.][]{haywood2013, Ginzburg2022}.

In contrast with other models in the literature, the model used in this work (\citealp{snaith2015}) decouples star formation from the gas mass. Here, the SFR is treated as a free parameter rather than being constrained by the Kennicutt-Schmidt law. The only constraint applied is that the total stellar mass formed (excluding recycling) equals the initial gas mass, though the actual stellar mass is lower due to gas recycling.
The purpose of modeling the inner disk is to investigate the chemical evolution of its stellar populations over time and to determine the SFH that best matches the observational data. The optimal SFH is identified by constraining it with the chemical properties of inner disk stars, allowing us to quantify chemical enrichment track that a given SFH produces.

The chemical evolution code used is composed of two parts (refer to Section 2.1 and 2.2 of \citet{snaith2015} for the complete description of the code). The first generates the chemical evolution tracks that provide the normalized mass of metals or gas released from a stellar population of given metallicity after a given amount of time. In other words it reads the chemical yields and converts them into a chemical evolution for a simple stellar population of a given initial metallicity\footnote{It should be noted that, in general, one of the largest sources of error in GCE modeling is the uncertainty in elemental yields.}. The key steps to generate the tracks are\footnote{Refer to Section 2.2 of \citet{snaith2015} and equations (1) to (7) for the detailed description of the model ingredients discussed here. Additionally, Sections 6.3, 6.5 and 6.6 of \citet{snaith2015} elaborate on the specific choices of IMF ad yields adopted in the model.}:
\begin{enumerate}
    \item Select an IMF. We adopt the \citet{kroupa2001} IMF.
    \item Define the stellar yields. We use \cite{Nomoto2006}, \cite{Iwamoto1999} and \cite{Karakas2010} stellar yields for SNII, SNIa and AGB stars, respectively. 
    \item Determine stellar lifetime. Following \citet{snaith2015}, we use the \cite{Raiteri1996} parametrization for metallicity dependent stellar lifetimes.
    \item Specify the time delay function for SNIa. We adopt the distribution from \cite{Kawata2003}.
\end{enumerate}

The second component of the code utilizes these tracks to compute the chemical evolution of the system given a particular SFH, effectively tracing the ISM enrichment over time. Further details are provided in Section 2.2 of \citet{snaith2015}. The key aspect of this approach is that the GCE code models the galaxy by tracking the evolution of the metal and gas content in the ISM. In particular, we focus on the abundance of $\rm \alpha$-elements. From this data, we can analyze the evolution of metallicity, gas fraction, abundance ratios, and other chemical properties at each time step. In our methodology, the SFH is determined exclusively by fitting the age-[$\rm \alpha$/Fe] distribution of inner disk stars, assuming no gas accretion (see Section 5.1 of \citet{snaith2015} for a full description of the fitting procedure). The fitting process follows a straightforward $\rm \chi^2$ minimization approach: (i) an initial-guess SFH is provided; (ii) the minimization process begins from this initial guess; (iii) at each step, a complete SFH is generated, and the code computes the corresponding chemical track; (iv) the $\rm \chi^2$ is calculated based on the difference between the modeled and observed [$\rm \alpha$/Fe] values at the age of each star in the dataset; (v) the algorithm iteratively searches for the optimal solution. Convergence is defined as the point where further reductions in $\rm \chi^2$ fall below a specified threshold.

In other words, the SFR as a function of time is treated as a free parameter that emerges as an outcome of the fitting process. Once an optimal SFH is obtained using the full dataset, we apply a bootstrap procedure to assess its sensitivity to the data. The resulting SFHs from multiple iterations are then averaged to obtain a smoothed final SFH.

\subsubsection{Inner disk evolution}
\begin{figure}
\centering

\includegraphics[width=9cm]{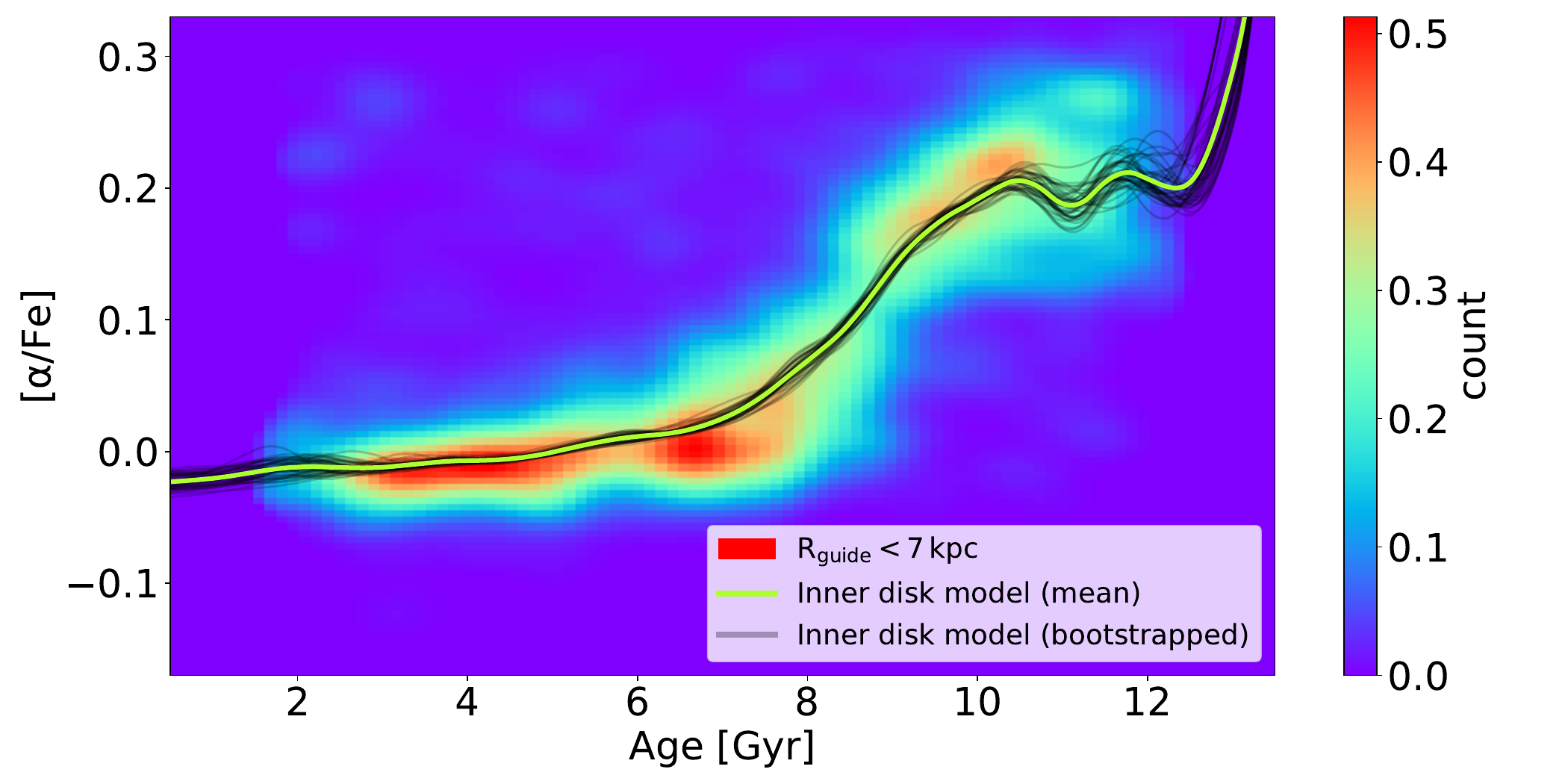}
\includegraphics[width=9cm]{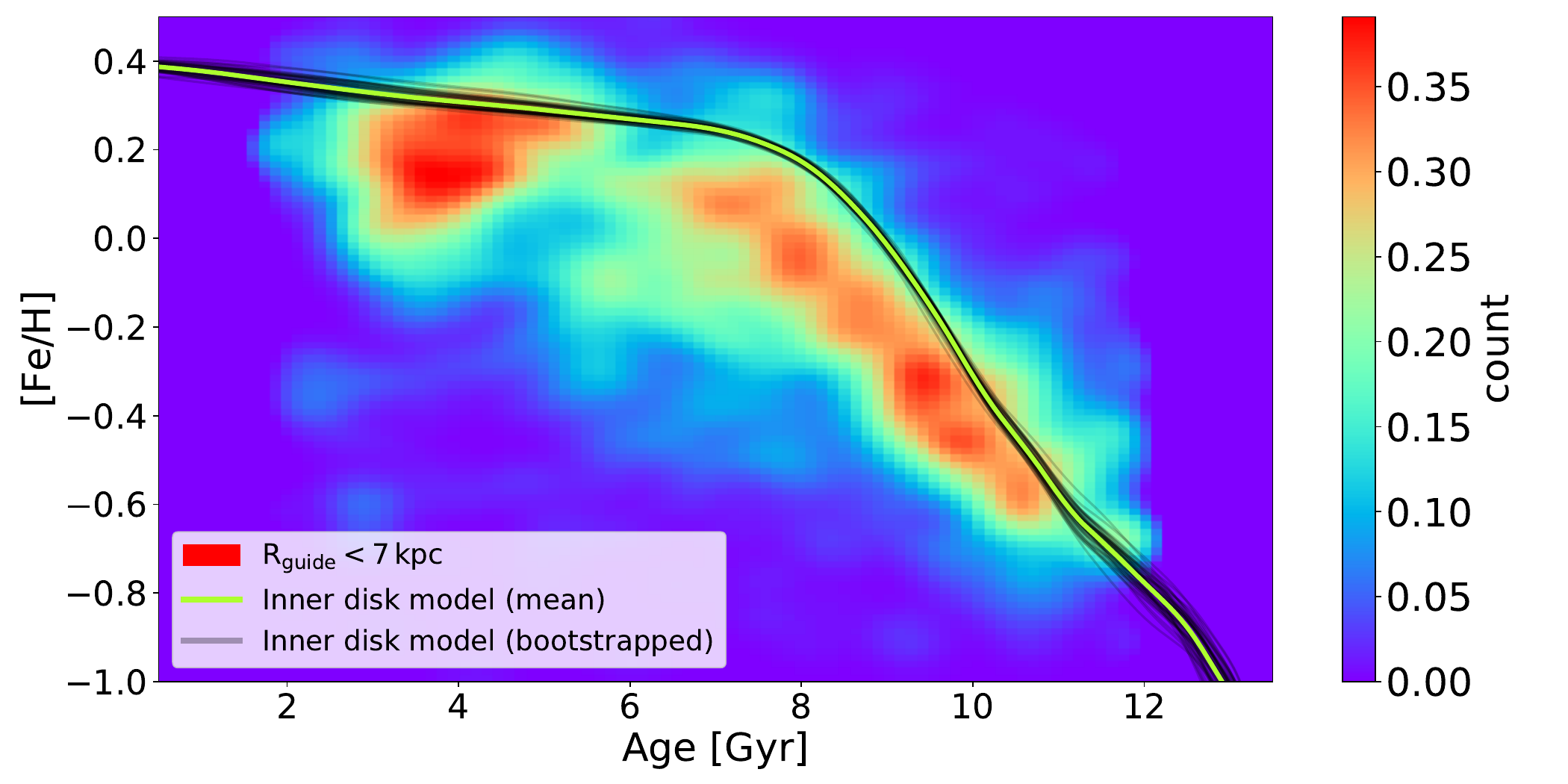}
\hspace*{-1.3cm}\includegraphics[width=7.2cm]{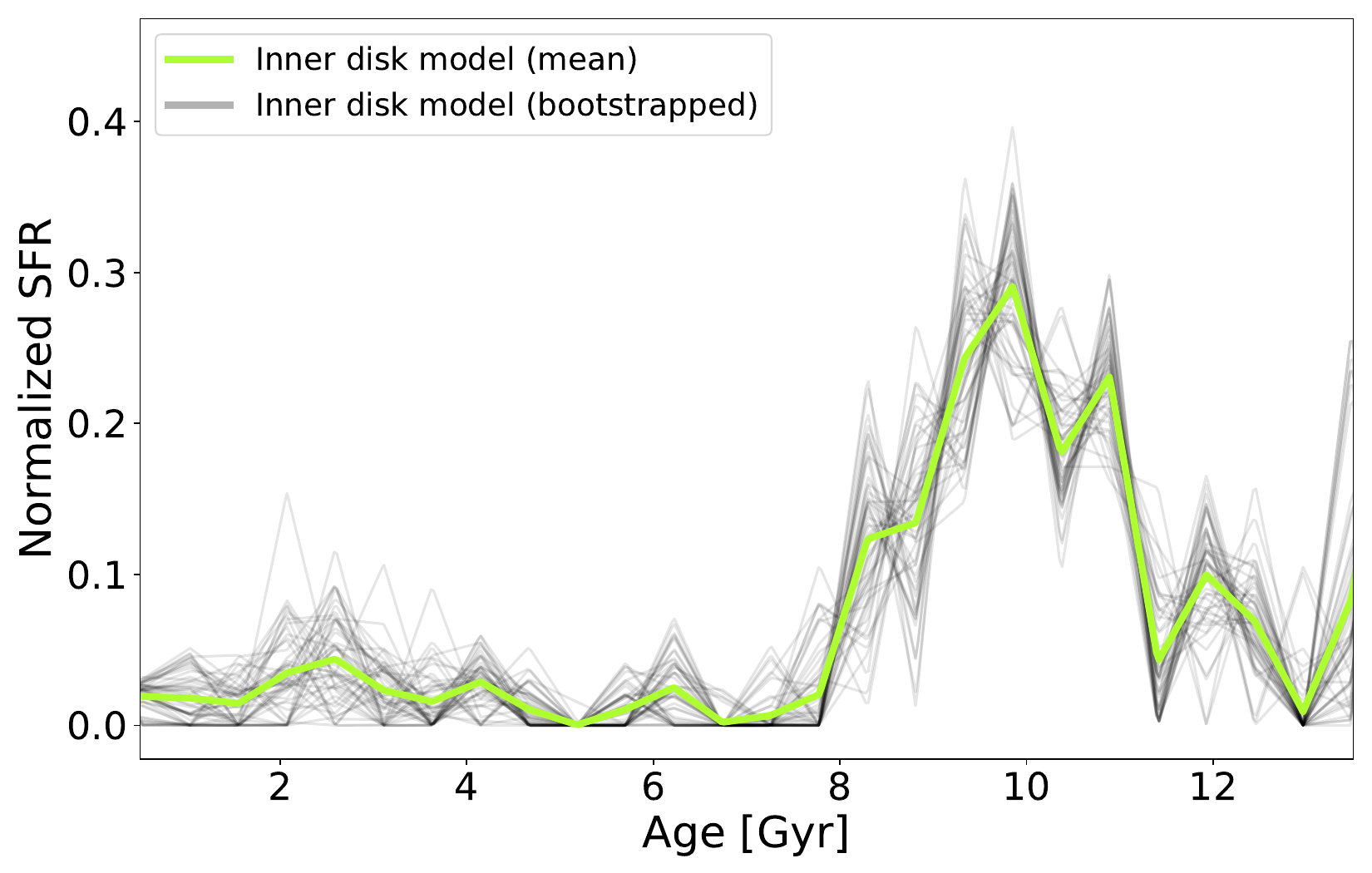}
\caption{Age-[$\alpha$/Fe] (top) and age-[Fe/H] distribution (center) for our sub sample of inner disk stars having $\rm R_{guide}$~$<$ 7 kpc together with 50 models fitted to the bootstrapped data. The mean model is in yellow-green. The bottom panel presents the corresponding 50 bootstrapped Star Formation Histories, together with the mean SFH (in yellow-green).}
\label{fig: inner_disc_model}
\end{figure}

The result of fitting the model described in Sec. \ref{subsec:Closed box bodel} to the inner disk age-[$\rm \alpha$/Fe] relation is given in Fig. \ref{fig: inner_disc_model}.
The fitted data is selected assuming $\rm R_{guide}<7$ kpc, discarding the stars that have [$\rm \alpha$/Fe] > 0.12 and age$<$7 Gyr, given that these probably are blue stragglers, and so have their age underestimated due to increased masses. 
We use a bootstrap method similar to the one used in \cite{snaith2015} to check how the fitting method is sensitive to the data. 
Figure \ref{fig: inner_disc_model} shows the result of fitting the age-[$\rm \alpha$/Fe] data with the model (top panel), the resulting age-metallicity relation for each bootstrap model and resulting star formation history.
Each plot also shows the mean of all the bootstrap models.
The resulting SFH shows a period of intense activity between 8 and 12 Gyr, reaching a maximum between 9 and 11 Gyr. 
Compared to \cite{snaith2015}, the SFH recovered in Fig. \ref{fig: inner_disc_model} is narrower by almost 1 Gyr, with most of the star formation occurring within 2 Gyr. The intensity of the SFH is decreasing at ages greater than 11~Gyr, while it remained high up to 12 Gyr in \cite{snaith2015}.
This is due to the shape of the age-$\alpha$ relation in the thick disk regime, which is a bit flatter at age $>$ 10 Gyr, and steeper below this limit, as described in Section \ref{subsection: general distribution}.

\subsection{Modeling the intermediate and outer disk evolution: introducing gas dilution}
\label{Subsec: the open box model}

The inner disk model described in Sec. \ref{subsec:Closed box bodel} and in \cite{snaith2015} is adapted to model the intermediate and outer regions of the disk.
\subsubsection{Intermediate disk evolution}

Modeling of the intermediate region is less straightforward because its evolution does not reflect a monotonic increase of its metal-content, as for the inner disk. 
The main characteristic of the intermediate region is the fact that the metallicity decreases more or less at the end of the thick disk phase. If this feature is interpreted as a dilution due to an inflow of hydrogen occurring more or less at the epoch at which the metallicity is observed to decrease, limited effect is expected on the age-$\alpha$ relation, which cannot be used to constrain detailed features of the gas inflow history. 

Rather than fitting an ad hoc model to the data, we propose here to illustrate that the feature observed in the age-metallicity distribution can be reproduced by a simple modification of the inner disk evolution. We start the model with a chemical evolution similar to the one obtained for the inner disk, keeping in particular the same high star formation rate in the 8-12 Gyr period first to allow the metallicity to rise rapidly above solar values.
Then we trigger a dilution 'by hand' at 7.8 Gyr that drops in the amount of hydrogen gas equal to 2.5 times the amount of hydrogen in the ISM at 7.8 Gyr and it does this over 100 time steps (or until 5.7 Gyr).

\begin{figure}
\centering
\includegraphics[width=9cm]{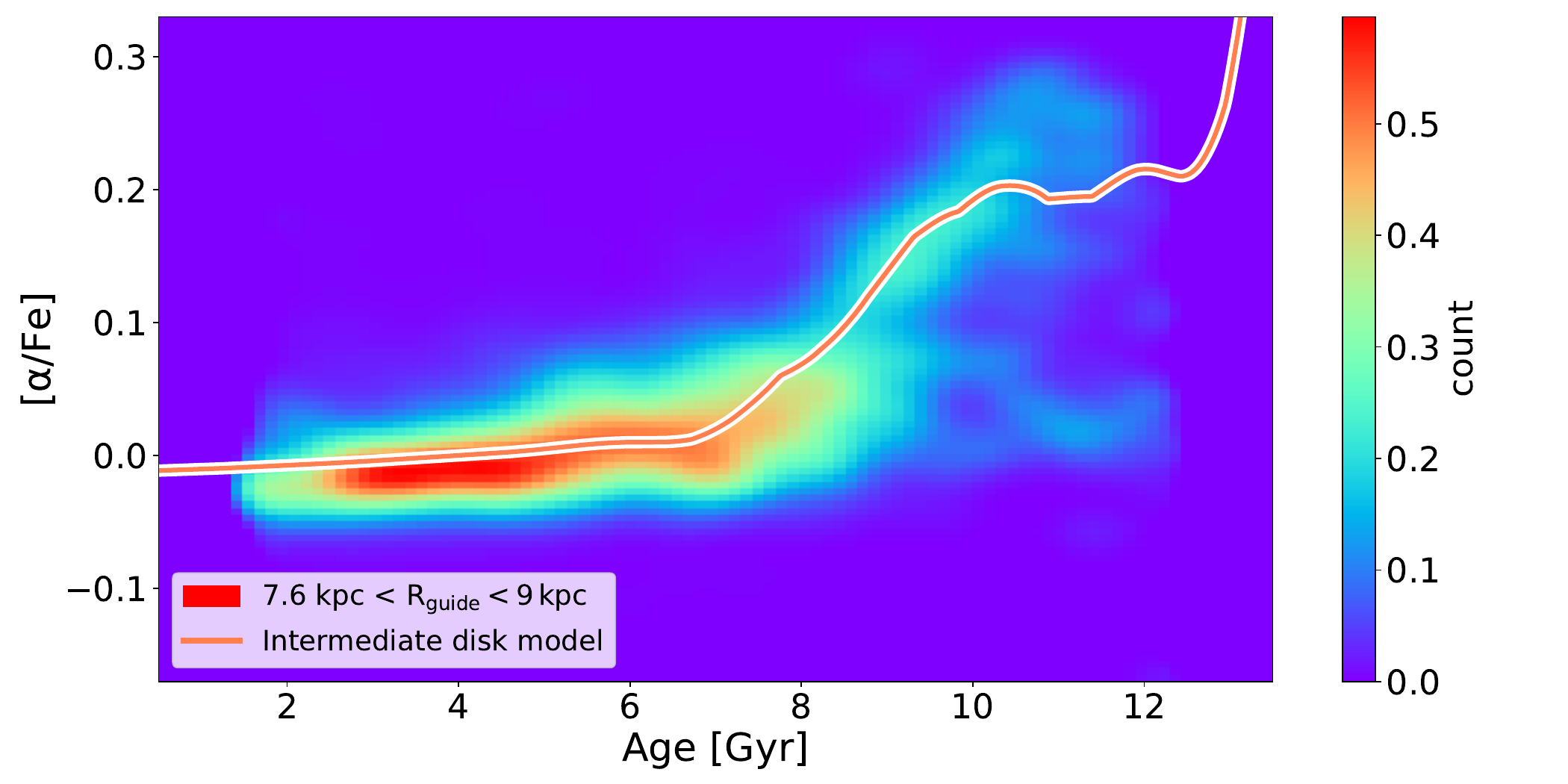}
\includegraphics[width=9cm]{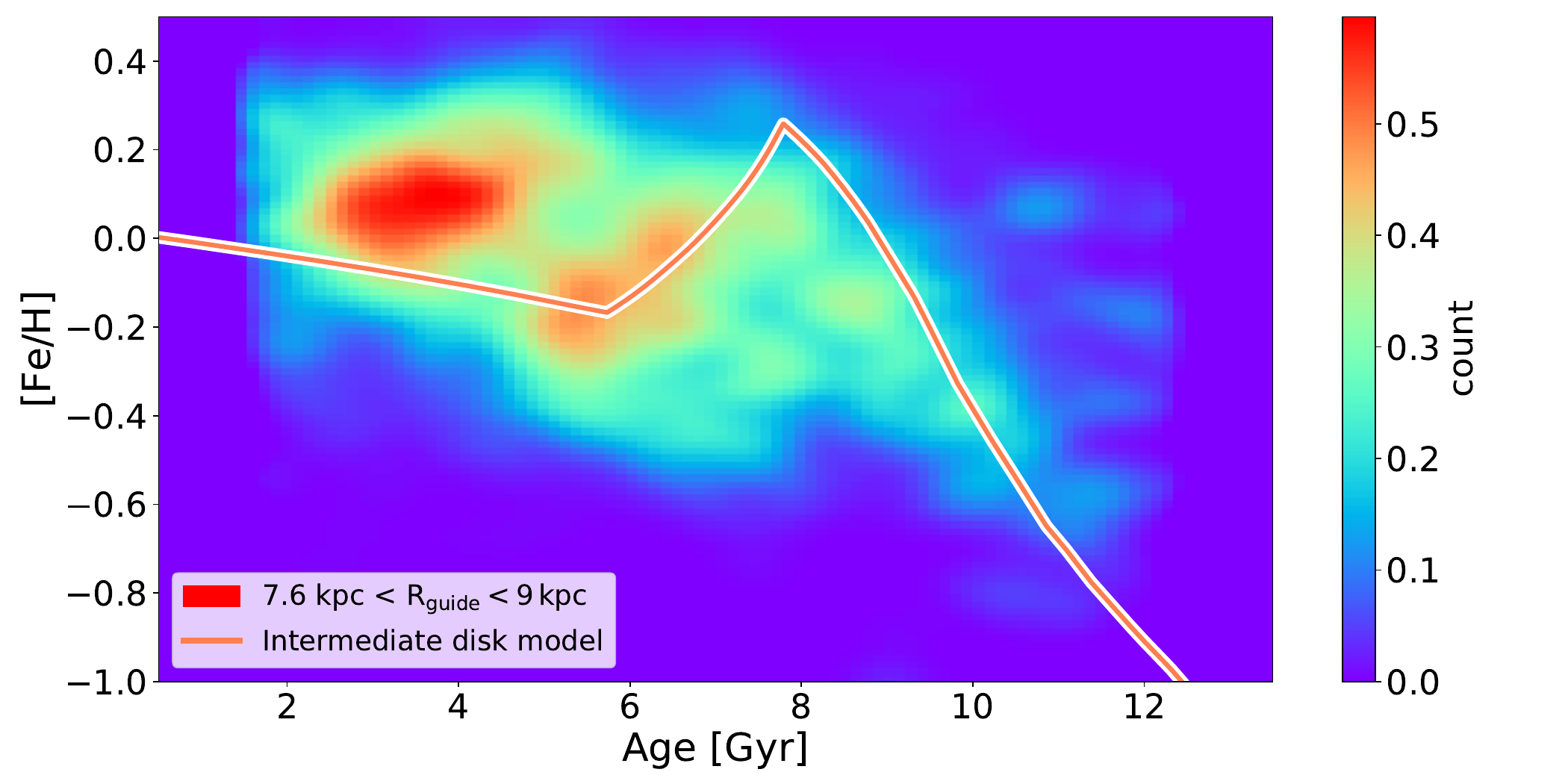}
\hspace*{-1.3cm}\includegraphics[width=7.2cm]{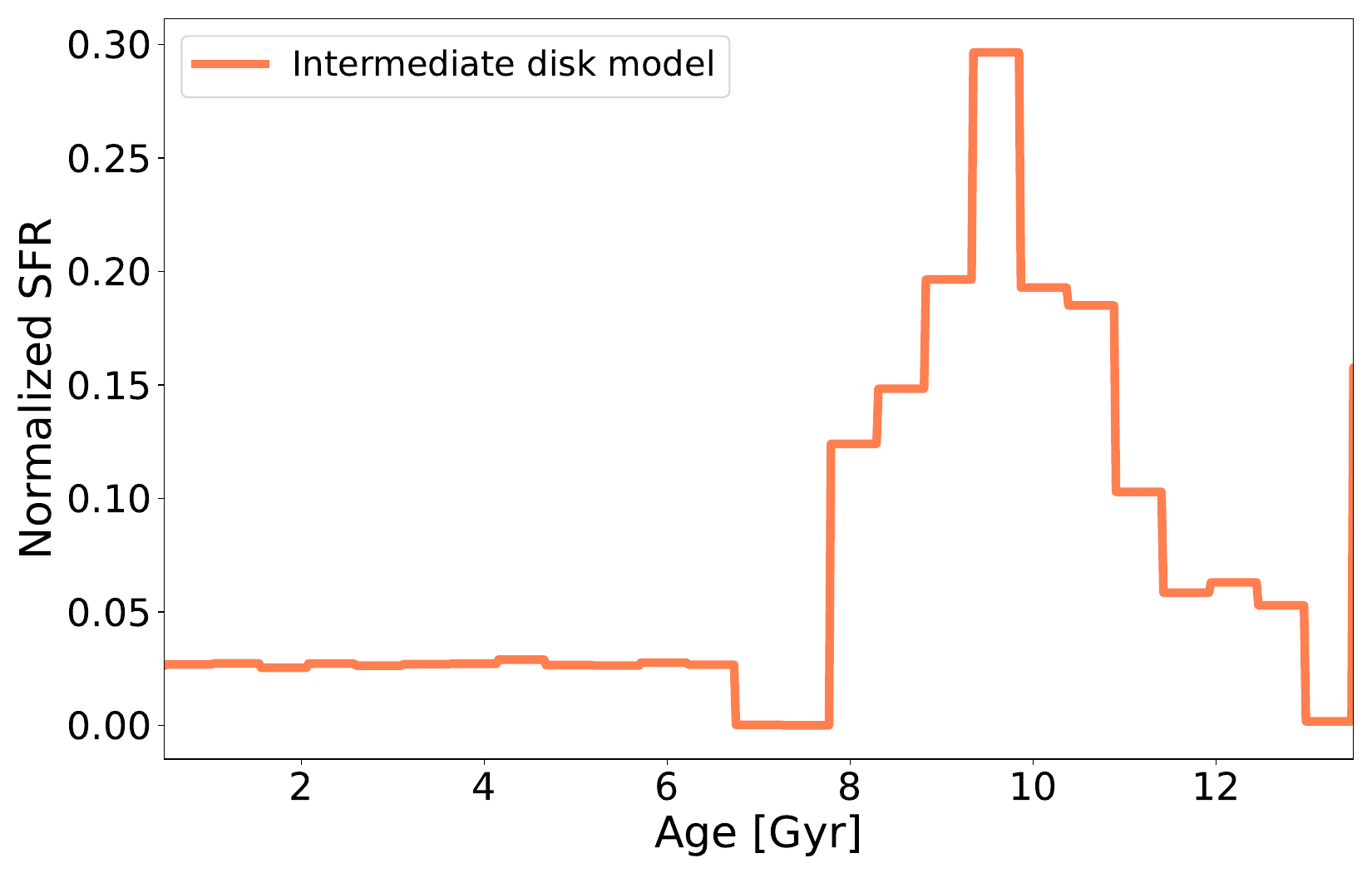}
\caption{Age-[$\alpha$/Fe] (top) and age-[Fe/H] distribution (center) for our subsample of local disk stars having  7.6 kpc $<$ $\rm R_{guide}$~$<$ 9 kpc together with the "toy" model superimposed to the data. The bottom panel presents the corresponding SFH. The SFH is normalized to 1.}
\label{fig: local_disc_model}
\end{figure}

After the end of the dilution episode, the metallicity slowly rises again from about -0.1 to reach about 0.1 at the present time, as the result of the low but continuous star formation rate imposed in the last 8 Gyr.
The aim here is not to produce a best fit, but to illustrate how the chemical trajectory here differs from the one observed for the inner disk and to roughly estimate the amount of fresh gas required to produce the observed decrease in the metallicity. The result is shown in Fig. \ref{fig: local_disc_model}.

\subsubsection{Outer disk evolution}
As already mentioned, our sample of outer disk stars is heavily polluted by stars of the intermediate and inner disk regions. Thus, 
when comparing with the model, we do not show the data as density distributions, as for the inner and intermediate disk, because only the lower metallicity fraction is representative of the outer disk.

\begin{figure}
\centering
\includegraphics[width=9cm]{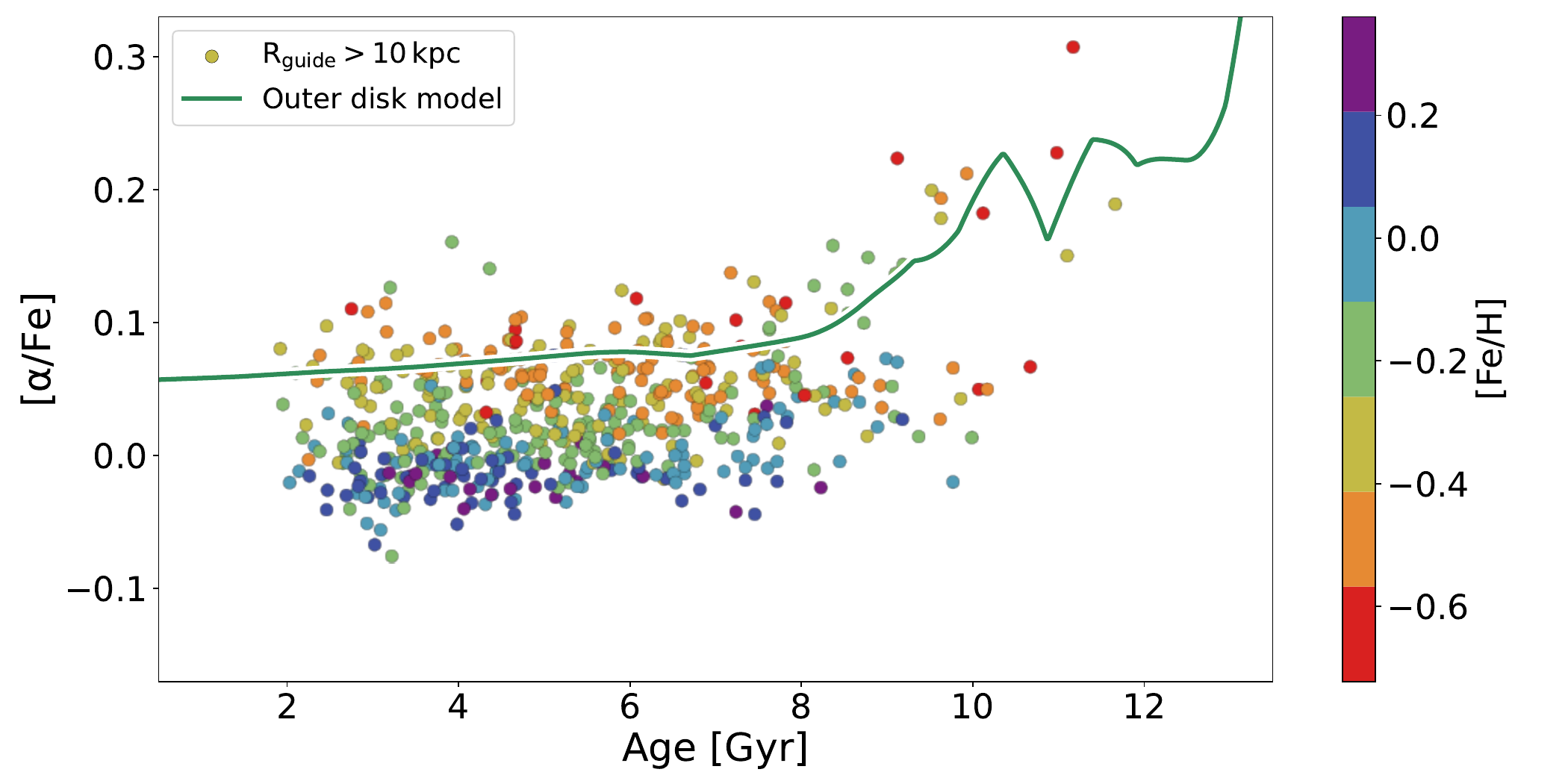}
\includegraphics[width=9cm]{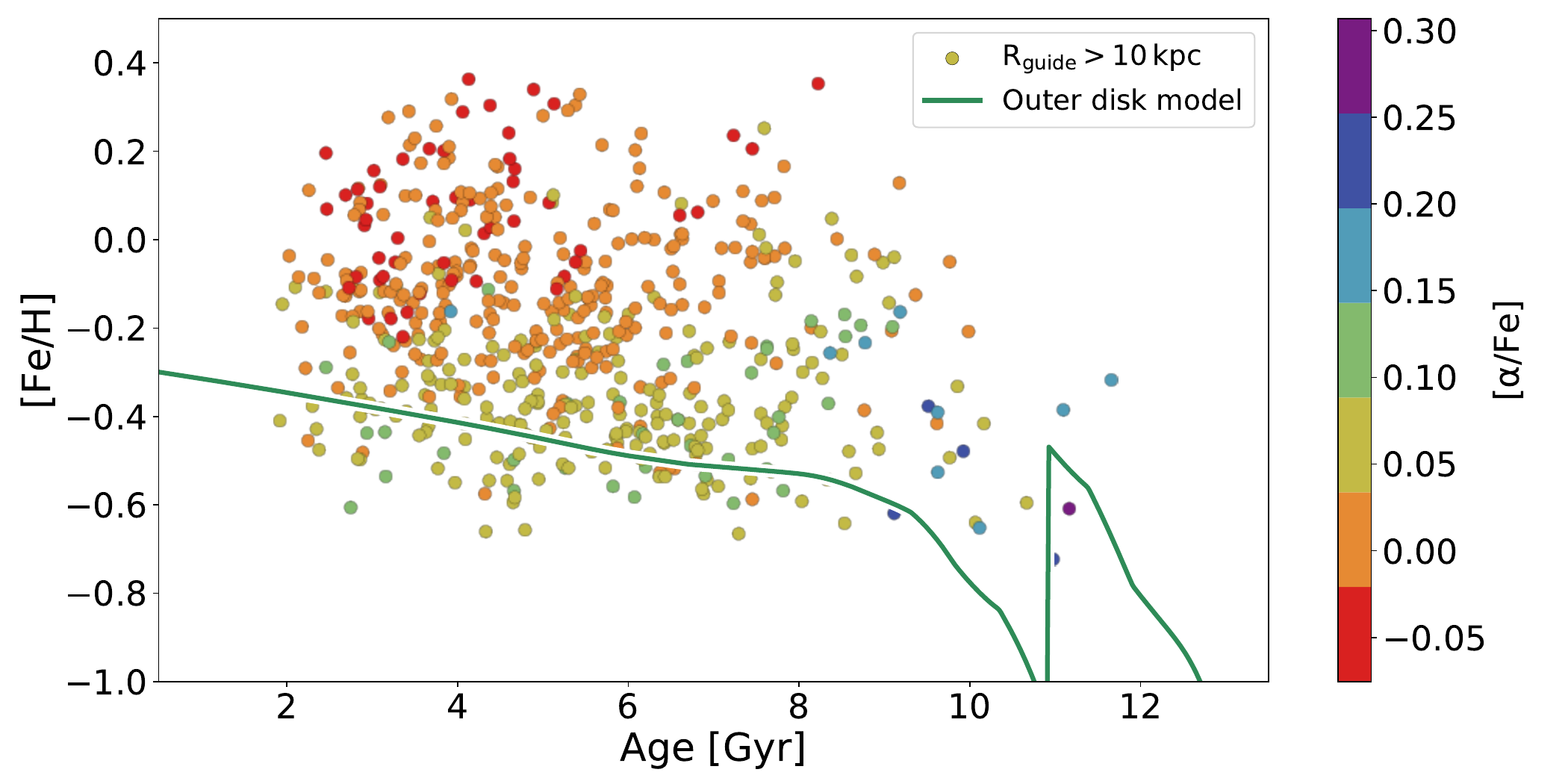}
\hspace*{-1.3cm}\includegraphics[width=7.2cm]{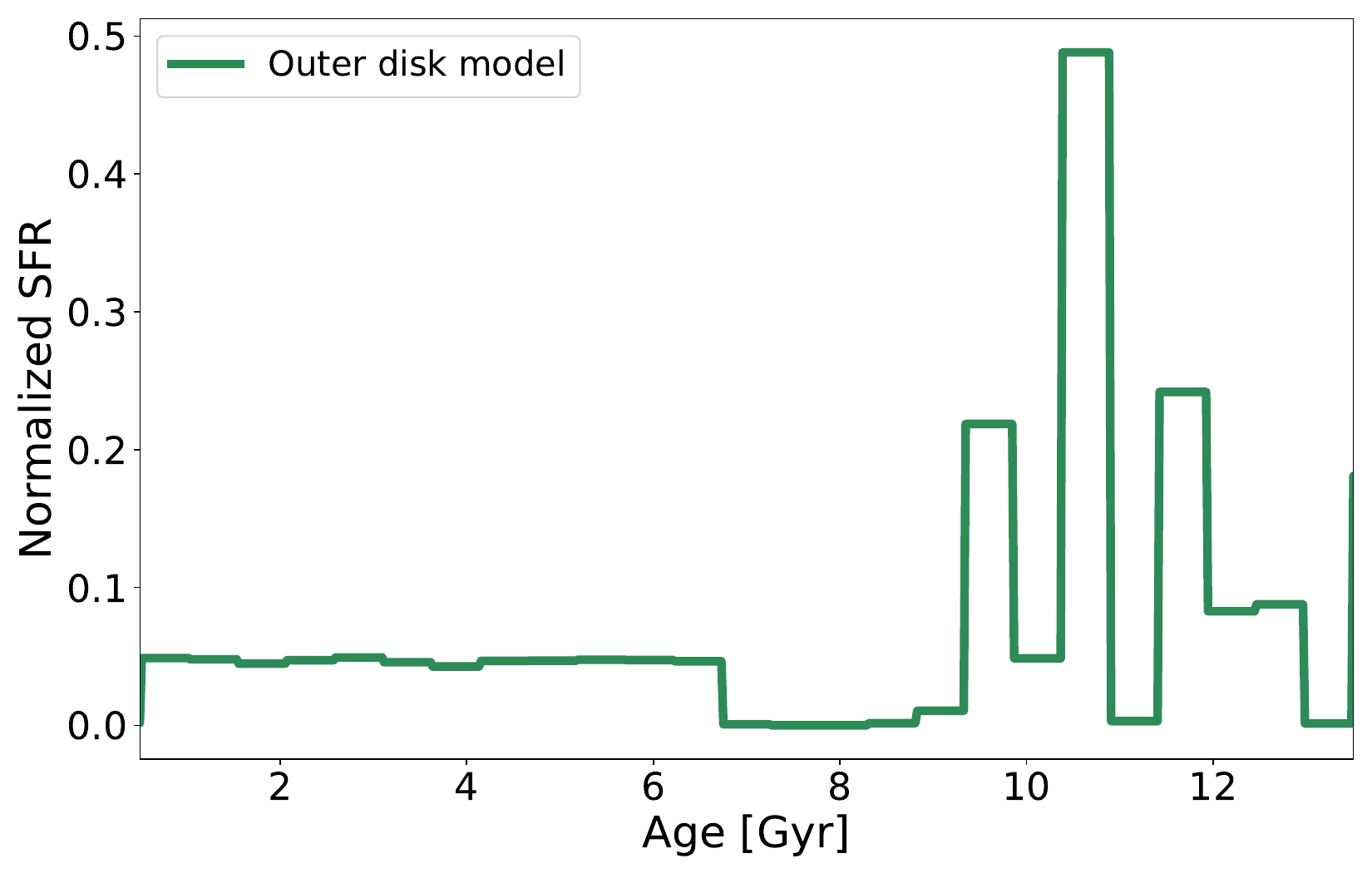}
\caption{Age-[$\alpha$/Fe] (top) and age-[Fe/H] distribution (center) for our subsample of outer disk stars having  $\rm R_{guide}$~$>$ 10 kpc together with the "toy" model superimposed to the data. The bottom panel presents the corresponding SFH. The SFH is normalized to 1.}
\label{fig: outer_disc_model}
\end{figure}

However, a clear result from the present stellar ages is that the outer disk stars first appear around 10 Gyr ago, at a metallicity of about -0.6.
Because an outer disk model cannot be fitted to the data, for the reason just mentioned, we simply illustrate that at ages below 10 Gyr, starting with the
correct initial metallicity (around -0.6 at 10~Gyr), a model with a low and flat SFH correctly represents the data.
In order to correctly initialize the metallicity, we proceed as in  \cite{snaith2015}, starting at 14 Gyr as for the chemical enrichment of the inner disk, but then dropping in at an age $>$10~Gyr (10.8 Gyr in this case) three times the amount of hydrogen as present in the disk at that time. 
This model is shown in Fig.~\ref{fig: outer_disc_model}. 
The sudden addition of hydrogen to the model does not explicitly change the [$\rm \alpha$/Fe], which decreases to about 0.1 at an age of 9~Gyr, then evolves very slowly at younger ages due to the near constant SFH, while the metallicity increases from about -0.5 at 9~Gyr to -0.3 at the present time.

\section{Discussion}
\label{sec: discussion}

Now that different regimes of chemical evolution have been identified in the data as a function of radius, the following questions are of interest. Do they form a radially continuous sequence, or are there discrete zones of evolution? Is the evolution of the Solar neighborhood separate from the inner and outer evolution, or can it only be explained in relation to them? What is the role of radial migration? Can we imagine a global scenario in which these differences can be explained? And finally, how does such a scenario compare with others discussed in the literature?

\subsection{Discrete chemical evolutionary paths?}

The intermediate region shares two characteristics with the inner region.
Like the inner disk, the intermediate region benefited from the pre-enrichment of the thick disk. As a result, the evolution of the intermediate thin disk started with a metallicity around solar, as the inner thin disk.
The second characteristic is the presence of metal-rich ([Fe/H]>0.2) stars. Metal-rich stars are endemic of the inner disk and are migrated when they are found elsewhere. 
Beyond the inner disk, metal-rich stars are found up to the OLR, at R=11-12 kpc, possibly migrated from the inner disk at the epoch of the slowing down of the bar, see \cite{Haywood2024}.
What makes the intermediate region different, however, is the fact that it was subjected to a dilution episode that was absent from the inner disk.
Finally, the metallicities at R$<$7 kpc and R$>$11-12~kpc slightly overlap, stars inside 7~kpc having [Fe/H]$>$-0.3, while stars beyond 11-12~kpc have metallicities mainly below this limit, see Fig. 11 in \cite{Haywood2024}, suggesting that there is little common evolution between these two regions, implying that there was no or limited exchange between them.
\cite{Haywood2024} noted (see their Fig. 6)  that the metallicity radial profile as deduced from APOGEE giants shows two breaks: the first one, located at about 6-7~kpc from the Galactic center, marks the transition from a very shallow gradient (-0.02 dex/kpc) in the inner disk (2-6~kpc) to a stronger gradient of about -0.09 dex/kpc to 11~kpc. After this limit, the gradient becomes flatter again, at a level of about -0.03 dex/kpc, evolving between [Fe/H] = -0.5 dex and -0.6 dex. 
They found that these breaks do in fact coincide with the positions of the co-rotation and outer Lindblad bar resonances. It was advocated that the co-rotation plays an important role in mixing the gas at and inside co-rotation radius, therefore inducing a rather flat gradient, while on the contrary, the OLR induced a barrier limiting the exchange of gas inside and outside the OLR. 
The break at 6-7 kpc was also noted previously by \citet{Haywood19} (but see also \citealp{lian2023}), and similar breaks have also been observed in the metallicity profile of gas in nearby galaxies (see for example \citealp{chen_garcia_2023}).
Other processes are known to contribute establishing or modifying gradients (inside-out process, \cite[e.g.][]{Matteucci1989}; feedback, \cite[e.g.][]{Ma2017}; mergers \cite[e.g.][]{Sillero2017}), but none of them are expected to naturally generate two breaks corresponding to three different regions. On the contrary, the bar has a direct impact on either mixing the ISM (at co-rotation) or giving rise to separate evolution as a result of the 'barrier' effect of the OLR and its consequences, as discussed in \cite{Haywood2024}.

\subsection{The role of radial migration} 

In describing the different schemes presented here, it is assumed that the peak metallicity of the stars at a given radius and epoch is still representative of the metallicity in the ISM at that same radius and epoch. 
In other words, it means that radial migration has been sufficiently limited that migrated stars never dominate stars born locally at a given radius. 
It also implies that radial migration had less effect, in structuring the disk, than the bar and its resonances did by defining regions of limited interactions once the resonances where set, as described in \cite{Haywood2024} (see also \citealp{Wheeler2022} for the role of the bar resonances).

We also describe, in \cite{Haywood2024}, how, in our view, the bar is at the origin of the episode of radial migration which spread the metal-rich stars from the inner disk up to the OLR. But this is still a minor effect, from the point of view of the number of stars being displaced, even though it is able to generate a break in the metallicity profile of stars older than about 6 Gyr, see \citet{Haywood2024}.
This point of view is different from the one presented by e.g. \cite{lu2022} or \cite{ratcliffe2023}, where migration is massive (at a given radius, migrated stars largely dominate, see for example Fig. 3 in \citealp{ratcliffe2023}). We do not follow this interpretation for several reasons. 
First, the method designed in \cite{lu2022} to recover the impact of radial migration assumes that the gradient is linear with a single coefficient covering the whole radial range under study, while in \cite{Haywood2024}, we show that this is not observed: the gradient is structured, radially and when dissected by age. As already mentioned, it is almost flat at R $<$ 6-7 kpc and R $>$ 11-12~kpc.
These radii also correspond to the location of the co-rotation and OLR of the bar respectively, and their effect on the metallicity gradient is observable as a flattening and a steep break in the metallicity gradient \citep[see also][for the flattening observed at co-rotation]{Wheeler2022}. We argue in \cite{Haywood2024} that in case of strong radial migration in the last 6-8 Gyr, this last feature would have been erased. 

Second, one of the implications of such massive migration is that, at the Solar vicinity, the vast majority of stars come inwards of 6~kpc (see Fig. 3 in \citealp{ratcliffe2023}), and the Sun itself is found to have come from 4.5 kpc. However, the Solar neighborhood is dominated by stars that have Solar metallicity and age up to 6 Gyr, while the APOGEE data (our Fig. \ref{fig: age_chemistry_inner}) show that such stars are marginal in the inner disk.  Solving this issue would require a fine-tuned mechanism to have driven the few Solar-metallicity stars of the inner disk to the Solar radius  leaving most of the metal-rich objects in the inner disk, reversing their proportions (which are the dominant type in the inner disk, but only a fraction of 10-15\% at the Solar radius). This is difficult to explain by any type of the dynamical mechanisms invoked to explain radial migration. In \cite{Haywood2024} we advocate that the data only requires a fraction of the most metal-rich and metal-poor to have migrated. Invoking that most stars in the Solar vicinity have migrated seems an unnecessarily complicated scheme, since nothing in the data prevents most stars of the Solar neighborhood from having formed at the Solar radius. 
 
Finally, this also drives \cite{ratcliffe2023} to assume a steep gradient in the thick disk phase, which is a disputed issue. The tight correlation that we find between age and metallicity (see also \citealp{haywood2013,Xiang_Rix_2022})  points to a synchronized enrichment over most of the thick disk, which would translate in an homogeneous chemical composition of the thick disk, or flat radial gradient, at a given time. On the contrary, a steep gradient implies a large dispersion at a given age (assuming a gradient of -0.15 dex/kpc over a disk of 7 kpc size implies a spread of 1 dex at a given age). Apart from the age evidence (which could be wrong since \citealp{miglio2021, Anders2023} find narrower age distributions), such spread would forbid any correlation between metallicity and $\rm \alpha$ abundances with structural parameters or kinematics.
On the contrary, it has been found that scale heights are correlated with chemistry in the thick disk \citep[decreasing with decreasing $\rm \alpha$-abundance ratio and with increasing metallicity, e.g.][]{bovy2012,mackereth2017,liam2022,Sun2024}.
Also, the kinematics of the thick disk changes at a metallicity of about -0.4, possibly due to the accretion of Gaia Sausage Enceladus \citep{Dimatteo19,Belokurov2020} 9 to 10 Gyr ago. If the growth of the ISM metallicity was not correlated with time in the thick disk, we would not expect the kinematics of this population to change at a particular metallicity.

Our description is probably more compatible with the migration estimates from \cite{frankel2020} (see also \citealp{Frankel2018}). They typically estimate that stars migrate 2-3~kpc in 6 Gyr, which is compatible with the type of migration triggered by the bar advocated in \cite{Haywood2024} to explain the presence of metal-rich stars up to the OLR.
It is beyond the scope of this paper to estimate if the metallicity spread measured here at a given age is compatible with the migration rate they measure, but this rate is probably too small to shuffle the three regions identified here.

\subsection{A possible  scenario}
\label{subsection: a possible scenario}

\paragraph{Origin of the accreted gas.}

An important component of any scenario for the chemical evolution of the disk is the gas accretion history. While many models parameterize gas infall as continuous, the data suggest two distinct accretion episodes in the Milky Way: an early, massive inflow that formed the thick disk, and a later event responsible for the dilution episode modeled in Sec.~\ref{sec: chemical modelling}.  
These episodes occurred at different epochs (early for the thick disk, later for the dilution episode and outer-disk formation) and dominated at different galactic radii. This is similar to the cold and hot accretion modes of galaxy formation in the cosmological context \citep[][]{Birnboim2003,keres2005,Keres2009,Ocvirk2008,Agertz2009,noguchi2018,stern2021}. Cold accretion feeds gas through filaments into galaxy interiors at readshifts above 2, when disks are turbulent and chaotic. Later, hot mode accretion dominates (with the time lag between hot and
cold mode accretion depending on the virialization of the circumgalactic medium), leading to the formation of thin disk through smooth gas deposit at the outskirts of the disk \citep{stern2021}. Recent studies focused on how star formation in disks is sustained by the continuous transport of gas from the circumgalactic medium to the inner disk via radial inflows.
\citep{danovich2015,Stewart2017,Peroux2020,Gurvich2023,Hafen2022,Trapp2022} through radial inflows.

\paragraph{How the gas mixed}

Once the gas is in the disk, simulations show that it may then move inwards at speeds of the order of 1-5 km/s (e.g. \citealp[]{tinsley1980}, \citealp{Trapp2024}). Note however that observational evidence of the importance of such inflows are slim in local galaxies \citep{diteodoro2021}. 

In the Milky Way, there is a striking occurrence of events about 9 Gyr ago that appears to coincide within 1-2 Gyr: (1) the end of the formation of the thick disc, (2) the formation of the bar \citep{bovy2019,sanders2024,Haywood2024}, (3) the metallicity dilution of the gas observed in the intermediate range between corotation and OLR. 
In \cite{Haywood19} the coincidence between the first two events was noted suggesting that the two could be related. This possibility was explored in depth by the study of the quenching of the star formation activity at the end of the thick disk phase \citep{khoperskov2018}.
\cite{Haywood19} inferred that a dilution episode limited in time must have occurred at an age estimated to be within 1-2 Gyr after the end of the formation of the thick disk (when the metallicity of the disk reached Solar metallicity) to explain that the mean disk metallicity was still Solar at the epoch of the Sun formation. The observations now show signatures of this dilution episode in the age-metallicity relation, see Fig. \ref{fig: local_disc_model}.
As reported in this study, the age-metallicity relation supports the fact that the dilution must have been limited in time, the dilution decreasing the metallicity below Solar at about 7-9~Gyr, but then increasing again to reach roughly the Solar value at the epoch of formation of the Sun.

Determining whether this corresponds to the long-term gas accretion (or hot mode) described in the broader context above is more challenging and requires further investigation.
\cite{Haywood19} also speculated that the dilution was triggered by the bar. Are there new indices that the bar may have play a role in regulating the dilution radially? The first one is the epoch of formation of the bar: whereas it was speculated that the Milky Way had formed its bar around redshift 1-1.5, as expected for a galaxy of its mass, recent studies \citep{bovy2019,sanders2024,Haywood2024} strongly support this scenario.
Moreover, \cite{Haywood2024} demonstrated that the metallicity profile of the thin disk is influenced by resonances, exhibiting two gradient slope changes at the positions of the bar's corotation (at 6 kpc) and the OLR (at 11-12 kpc). As argued above, these locations also correspond to regions where distinct regimes of chemical evolution are most pronounced. To understand the impact these resonances may have had, we now summarize a possible timeline in the evolution of the disk.

The thick disk forms, in a phase of high star formation \citep{Lehnert2014,snaith2015}, also corresponding to the phase of maximum cosmic star formation and high accretion, fed by an intense episode of gas accretion along cosmic filaments. The high level of turbulence expected from gas accretion, supernovae feedback and thermal instabilities flattens any possible radial gradient allowing synchronized increase of metals in the ISM. 
The bar forms with its resonances, inducing multiple possible effects. The first is quenching most of the star formation activity within the corotation \citep[see][]{Haywood16,khoperskov2018}. Because angular momentum transfer drives the gas from co-rotation to the OLR, 
it prevents the inner regions ($<$CR) to be diluted by radially inflowing gas. Hence, chemical evolution in the inner disk continues unpolluted. Outside co-rotation, the inflowing (more metal-poor) gas and the already present (metal-richer) gas left by the formation of the thick disk mix in the intermediate region between the CR and the OLR. The varying fraction of these metal-rich and metal-poor gas generated the gradient that is now observed in this region. At radii larger than 10~kpc, the influence of the thick disk is negligible for the chemical evolution of the thin disk, which is essentially fed by the incoming gas accreted at larger radii and inflowing towards the inner regions.

\subsection{Comparison with the literature}
\label{subsection: comparison with the literature}

This scenario is different from the one proposed by e.g. \cite{ratcliffe2023} or \cite{Anders2023}, who advocate, on the contrary, that the radial gradient in the thick disk was negative.

\cite{Salholhdt_22_GALAHDR3}, whose work we introduced in Section \ref{Subsec: comparison sahlhold}, proposed a scenario consistent with our statements described in the previous Section \ref{subsection: a possible scenario}. In particular, the authors support the idea of the disk being formed in separate phases: the first phase gave rise to stars on kinematically hot orbits in the inner disk; the second phase generated stars on kinematically cooler orbits, located across the entire disk with a negative metallicity gradient. We detect
the transition between the two phases in the inner disk (corresponding to a sharp decrease in the SF level) around 7-8 Gyr (in agreement with \citealp[]{Haywood19}), while \cite{Salholhdt_22_GALAHDR3} found it slightly earlier, at about 10 Gyr. This discrepancy could be due to the different age scales.
In addition to the two described main phases, \cite{Salholhdt_22_GALAHDR3} found evidences of an additional recent burst of star formation, corresponding to a prominent peak in their SAMD at around 3 Gyr seen in their kinematically young selected sub-samples. Other works in the literature found a recent peak of thin disk SFH at around 2-3 Gyr (see for example \citealp[]{Mor2019} and \citealp{johnson2021}).

Our scenario is reminiscent of the two-infall model of \cite{chiappini1997}, and its more recent versions (e.g. \citealp{Spitoni2021}), but it differs in several fundamental aspects. \cite{Spitoni2021} apply a unique two-infall model at all radii, changing the parameters of their infall law with R. It generates a radial continuity in their model predictions, with a characteristic loop in the metallicity-[$\rm \alpha$/Fe] plane, and a decrease in metallicity resulting from the second infall that occurs at all radii (see \citealp[]{Spitoni2021}, Fig.14), while we find that this feature is observed only in the intermediate region, see Section \ref{section: local age-chemistry}.
On the contrary, we argue that the disk is structured because the different regions have been affected by different processes (gas inflow and dynamical evolution). In particular, the two episodes of gas supply that we identify affect different parts of the disk. The inner disk is formed through an episode of massive infall, and no secondary infall is considered necessary to explain the formation of the thin disk in this region. 
On the contrary, no massive component formed during this epoch in the outer disk, and the thick disk contributed little to the outer disk (except possibly for increasing the $\rm \alpha$ content through pollution, see \citealp{haywood2013}). These different ingredients produced radically different evolutions in the inner, intermediate, and outer disks, as evidenced in particular by the variety of observed age-metallicity relations, while in \cite{Spitoni2021} their model predictions are very similar in all regions.

\cite{chen2023} developed a chemical evolution model characterized by one episode of smooth gas inflow and constant star formation efficiency that reproduces the observed distribution of [Fe/H] and [$\rm \alpha$/Fe] of stars at different radii and distances from the mid-plane in the Galaxy \citep[from APOGEE data, see][]{Sharma_2021}. They test four scenarios of star formation (fiducial model plus other three models with distinct prescriptions) and in each case the SFHs are essentially flat for the last 10 Gyr. 
In opposition to our view, in this model there is no separate formation phase for the thick and thin disks (no quenching episode): 
\cite{chen2023}'s star formation efficiency remains constant with time and their star formation rate does not collapse to zero to increase again during the thin disk formation. 
This is as well in contrast with the recent burst of star formation found by \cite{Salholhdt_22_GALAHDR3, Mor2019, johnson2021}.

\citet{miglio2021} used a combination of spectroscopic and asteroseismic constrains to infer ages of about 5500 red giant stars observed by Kepler and APOGEE. They find that the mean age of their chemically selected high-$\rm \alpha$ stars is about 11 Gyr, and that the width of the observed thick disk age distribution is largely due not to an intrinsic age dispersion, but rather to the relatively large uncertainties in the ages. Using a statistical model they infer the spread of the intrinsic age distribution and conclude that the thick disk population was born within $\rm 1.52_{-0.46}^{+0.54}$ Gyr. Moreover, the comparison with synthetic thin and thick disk populations \citep[generated with TRILEGAL,][]{Girardi2005} lead the authors to consider the possibility of quenching in star formation after the formation of the chemical thick disk (in agreement with our findings).

We can compare our scenario with the results provided by hydro-dynamical simulations of galaxies.
For instance both the simulations of isolated disk galaxies from \cite{Khoperskov2021} and the VINTERGATAN cosmological simulation by \cite{agertz2021, renaud2021a, renaud2021b} show an early central burst of star formation, driven by a massive early accretion of gas (see Fig. 4 of \citealp[]{Khoperskov2021}) that leads to the formation of the majority of high-$\rm \alpha$ stars, representing a significant fraction of the disk component. 
This idea is consistent with the scenario proposed in section \ref{subsection: a possible scenario} and would cause a rapid increase in metallicity during the first billion years. This is indeed observed in the thick disk sequence of Fig. \ref{fig: age_chemistry_inner} and in the lower left panel of Fig. 9 in \cite{Khoperskov2021}. 

The formation of the younger low-$\rm \alpha$ disk component is typically linked to a slower and longer period of star formation, driven by the accretion of gas.

In particular, \cite{Khoperskov2021}'s simulations agree with this work suggesting that the thin disk forms from a mixture of enriched material expelled from the thick disk and accreted gas at lower metallicity (originated from a less intense phase of gas accretion on much longer time-scales, see Fig. 4 of \citealp[]{Khoperskov2021}).

The VINTERGATAN simulation \citep{agertz2021, renaud2021a, renaud2021b} associates the formation of the low-$\rm \alpha$ sequence with a merger event. In particular, the merging galaxy does not dilute the ISM gas, but triggers the formation of stars from a metal-poor inflow of gas. The NIHAO simulation also links the formation of the thin disk stars to a merger (see \citealp{Buck2020}), but unlike VINTERGATAN, the gas that the merging galaxy brings with it dilutes the existing gas, causing a decrease in ISM metallicity. 
In this work we dated GSE accretion to be 9-11 Gyr old (see Sec \ref{subsec: sigma_z} and bottom panel of Fig. \ref{fig: sigmaz_age-pop_fe_no_yar}). If we consider that the gas took  $\approx$2 Gyr to reach the intermediate region (which, is possible, depending on the radius at which the gas from the merger was deposited), is it possible that the accretion of GSE could create the metallicity dilution we observe in the intermediate regions of the disk between 7 and 5 Gyr, even though to probe this is beyond the scope of this work.

\section{Conclusions}
\label{sec: conclusion}
In the present work we investigated the age-[$\rm \alpha$/Fe] and age-metallicity distributions of stars in the Milky Way disk using a sample of $\sim$12 000 dwarfs located in the Solar vicinity. 
The results are based on the chemical abundances of APOGEE DR17 and on stellar ages calculated using a Bayesian inference method described in \cite{Age_determination_Jorgensen2005} for which we take advantage of Gaia DR3 photometry and parallaxes.

Our sample shows the following global properties :\\
\indent 
$\bullet$ The age-metallicity distribution shows a tight and well-defined thick disk sequence for stars older than 8 Gyr, contrasting with a much wider spread in the thin disk phase.

$\bullet$ The bimodality found by previous studies in statistically less numerous samples is in fact a sequence which starts with the thick disk, then continues to Solar metallicity before a decrease at 7-9 Gyr, followed by an increase to higher metallicities. This decrease is interpreted here as the effect of the dilution due to an epoch of gas inflow, predicted to occur in particular in \cite{Haywood19}. 

$\bullet$ The existence of a well defined age-$\rm \alpha$ relation is confirmed \citep{haywood2013,bensby2014,ciuca2021}, with clear stratification in metallicity, and a number of 'young' alpha-rich stars \citep{Cerqui2023}. The part of this relation covering the thick disk phase shows two segments, the first with a steep slope between 8 and 9 Gyr, the second with a flattened evolution beyond 9 Gyr.

$\bullet$  The vertical velocity dispersion shows a sudden increase at the end of the thick disk phase at 8-9 Gyr, also corresponding to the steepening of the $\rm \alpha$ content just mentioned, suggesting a possible link between the star formation activity and this accretion event.

Following \cite{Haywood2024} where three different radial intervals were found to define the global chemical properties of the disk due to the effect of the bar resonances, we divide our sample into three guiding radius intervals.

(1) Stars having guiding radius smaller than 7~kpc define a monotonic increase of the metallicity from the oldest stars of the thick disk to the most metal-rich stars of the young inner disk.
Following \cite{Haywood16}, this evolution is well described by a model with a single gas accretion episode that gave rise to the formation of the thick disk. Fitting the age-$\rm \alpha$ relation using the model developed in \cite{snaith2015,Snaith2022}, we derived the SFH that best reflects the variation of the star formation rate in this region. The inner disk is characterized by intense star formation activity between 8 and 12 Gyr, (peaking between 9 and 10 Gyr). The steady increase of $\rm \alpha$-elements over cosmic time and the specific gradient observed in the thick disk age-chemistry relation are therefore the result of a well-mixed and turbulent ISM, coupled with an early and intense phase of star formation. The latter is then followed by an extended period of constant and rather low star formation.

(2) The age-metallicity distribution of stars with 7.6 $<$ Rguide $<$ 9.0 kpc, sampling the intermediate region (6-7 to 10-12~kpc), reveals that the local stars are marked by a more complex chemical evolution: an episode of gas dilution is required to explain the decrease in the  thin disk metallicity at 6-7 Gyr with sub- Solar metallicity, indicating a more complex chemical enrichment history than the one observed in the inner disk.

(3) The evolution of the outer disk (R $>$ 10 kpc) is dominated by the gas which, mixed with the gas left by the thick formation, gave rise to the metallicity gradient observed in the intermediate region. The low star formation efficiency at this distance from the Galactic center results in essentially a non-chemical evolution, with no or limited evolution observed as a function of time.

We emphasized that, in order to explain the evolution of the disk in the different regions considered here, two episodes of gas inflow, different in nature, are required. The first was massive and occurred early in the formation of our Galaxy. It formed the thick disk, as has long been recognized by chemical evolution studies \citep[e.g.][]{chiappini1997,Snaith2014,snaith2015,Spitoni2019,lian2020,Snaith2022}.
This gas accretion can be conceived as coming directly from filaments to the center of the Milky Way. It is reminiscent of the cold accretion mode described in simulations of galaxy formation in a cosmological context.
The second is needed to explain the outer (R $>$ 6 kpc) thin disk. We argue here that a signature of this gas inflow is readily visible in the decrease of the metallicity observed on stars born beyond the co-rotation radius (R $>$ 6-7 kpc), and in particular at the Solar vicinity, and which is absent from the inner disk. 
The metallicity gradient in the intermediate region (6-12~kpc) is the larger signature of this dilution. 
It could be due to gas that was first brought in the plane of the Milky Way and then was driven inwards, as found in recent cosmological simulations.
The dilution is visible in the age-metallicity plane up to 10 kpc, which is the distance from the Galactic center up to which the thick disk has dominated the evolution prior to the thin disk formation, raising the metallicity to Solar values. Beyond this limit of 10-12 kpc, the disk metallicity is dominated by the metallicity of the gas acquired by the Milky Way at larger radius.
In the Milky Way, the intermediate region between 6-7 kpc to 10-12 kpc is the region of overlap of two modes of gas accretion and disk formation, possibly linked to the so-called cold and hot modes of gas accretion in large disk galaxies.

Finally, we emphasize that the different chemical evolution paths individualized in this study also result from the dynamical structuring of the disk imposed by the main bar resonances. 

\begin{acknowledgements}
The authors acknowledge the support of the French Agence Nationale de la Recherche (ANR), under grant ANR-13-BS01-0005 (project ANR-20-CE31-0004-01 MWDisc).
This work has made use of data from the European Space Agency (ESA) mission Gaia (https://www.cosmos.esa.int/gaia), processed by the Gaia Data Processing and Analysis Consortium (DPAC, https://www.cosmos.esa.int/web/gaia/dpac/consortium). Funding for the DPAC has been provided by national institutions, in particular the institutions participating in the Gaia Multilateral Agreement.
This study makes use of the the astroNN catalog (\citealp{Leung_Bovy_2019}, https://github.com/henrysky/astroNN).
Funding for the Sloan Digital Sky 
Survey IV has been provided by the 
Alfred P. Sloan Foundation, the U.S. 
Department of Energy Office of 
Science, and the Participating 
Institutions. 

SDSS-IV acknowledges support and 
resources from the Center for High 
Performance Computing  at the 
University of Utah. The SDSS 
website is www.sdss.org.

SDSS-IV is managed by the 
Astrophysical Research Consortium 
for the Participating Institutions 
of the SDSS Collaboration including 
the Brazilian Participation Group, 
the Carnegie Institution for Science, 
Carnegie Mellon University, Center for 
Astrophysics | Harvard \& 
Smithsonian, the Chilean Participation 
Group, the French Participation Group, 
Instituto de Astrof\'isica de 
Canarias, The Johns Hopkins 
University, Kavli Institute for the 
Physics and Mathematics of the 
Universe (IPMU) / University of 
Tokyo, the Korean Participation Group, 
Lawrence Berkeley National Laboratory, 
Leibniz Institut f\"ur Astrophysik 
Potsdam (AIP),  Max-Planck-Institut 
f\"ur Astronomie (MPIA Heidelberg), 
Max-Planck-Institut f\"ur 
Astrophysik (MPA Garching), 
Max-Planck-Institut f\"ur 
Extraterrestrische Physik (MPE), 
National Astronomical Observatories of 
China, New Mexico State University, 
New York University, University of 
Notre Dame, Observat\'ario 
Nacional / MCTI, The Ohio State 
University, Pennsylvania State 
University, Shanghai 
Astronomical Observatory, United 
Kingdom Participation Group, 
Universidad Nacional Aut\'onoma 
de M\'exico, University of Arizona, 
University of Colorado Boulder, 
University of Oxford, University of 
Portsmouth, University of Utah, 
University of Virginia, University 
of Washington, University of 
Wisconsin, Vanderbilt University, 
and Yale University.
\end{acknowledgements}

\bibliographystyle{aa} 
\bibpunct{(}{)}{;}{a}{}{,}             
\bibliography{bibliography}

\begin{appendix}

\onecolumn
\section{$\rm Gaia\, (BP-RP)$ color - $\rm T_{eff}$ calibrations}
\label{appendix: color-teff calibration}

\begin{figure*}[h!]
\resizebox{\hsize}{!}{
{\includegraphics[width=10cm]{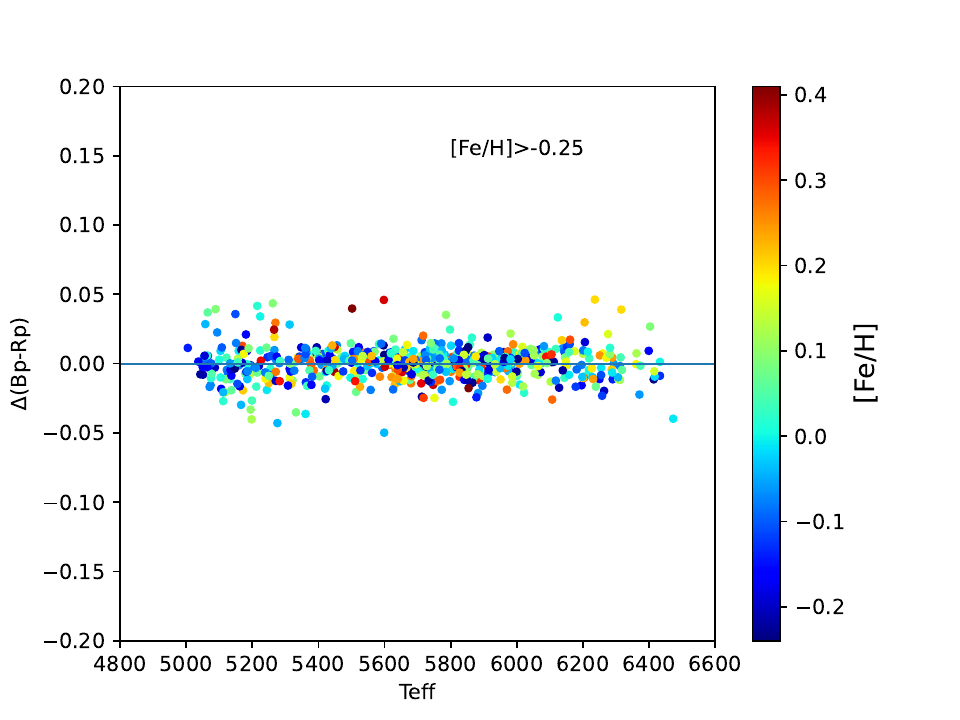}}
{\includegraphics[width=10cm]{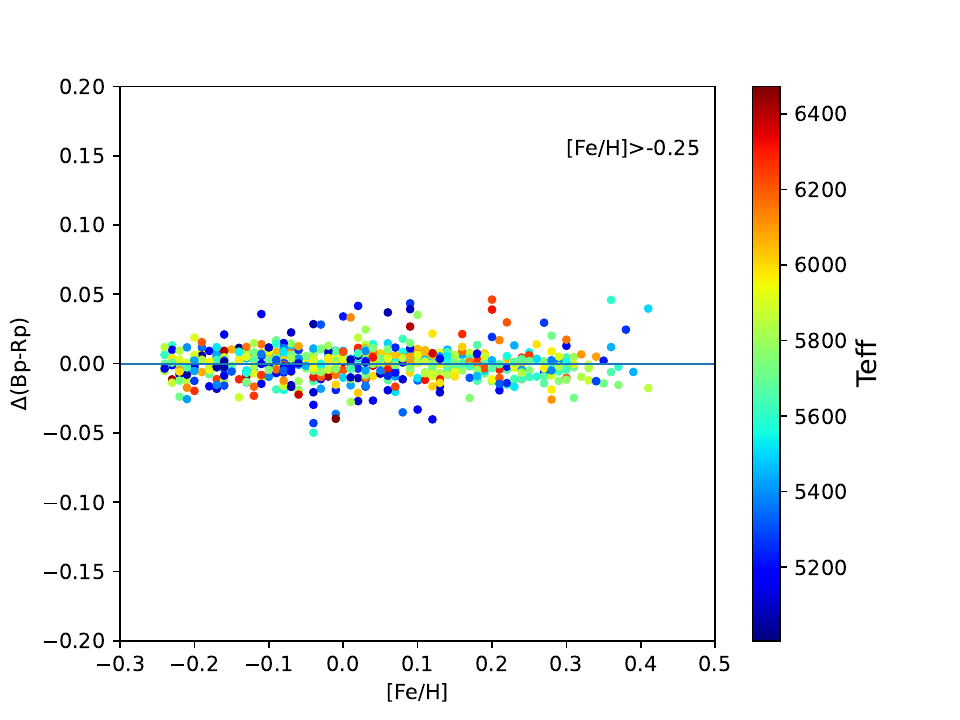}}
{\includegraphics[width=10cm]{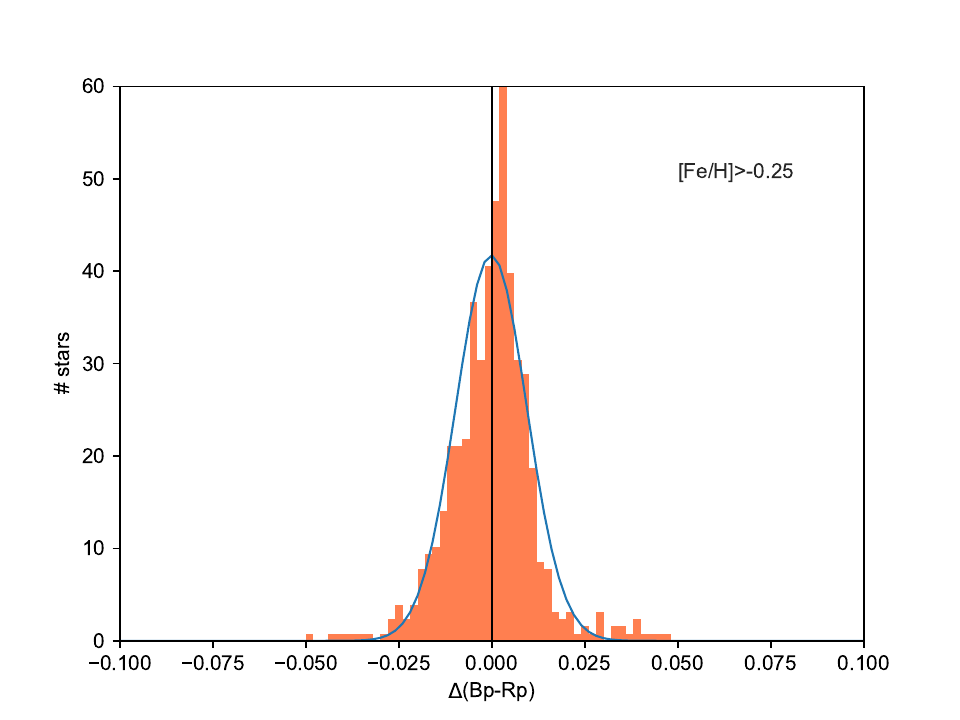}}}
\resizebox{\hsize}{!}{
{\includegraphics[width=10cm]{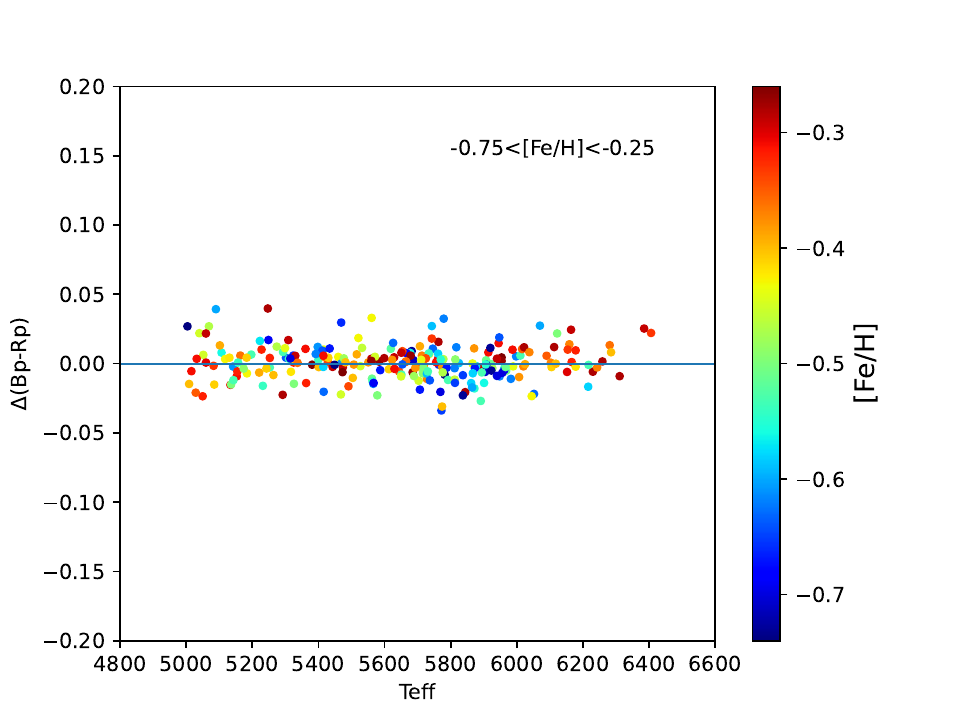}}
{\includegraphics[width=10cm]{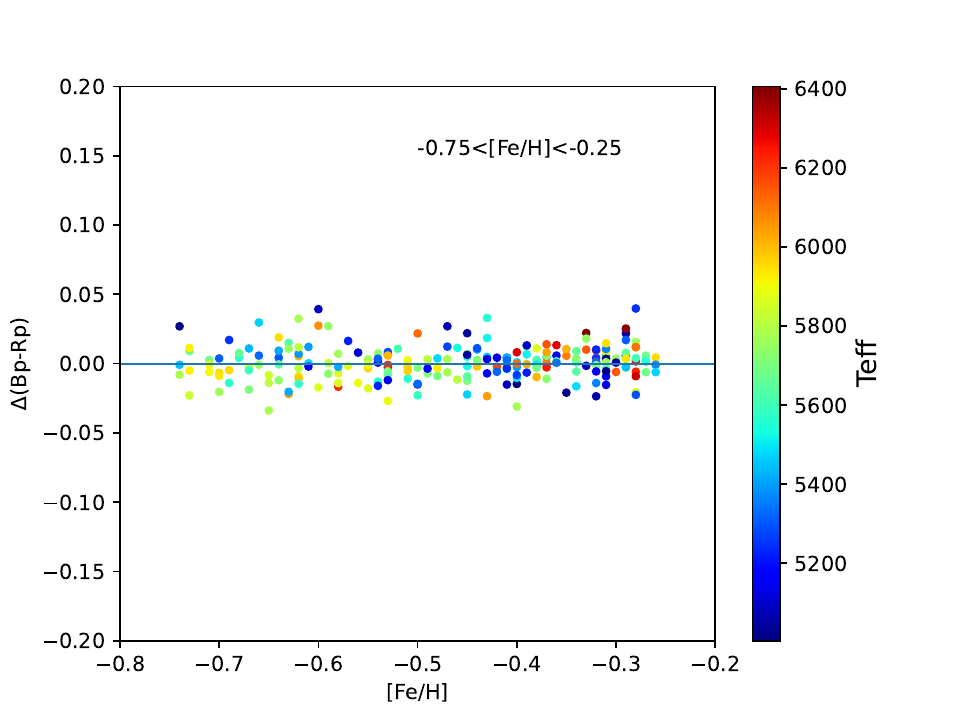}}
{\includegraphics[width=10cm]{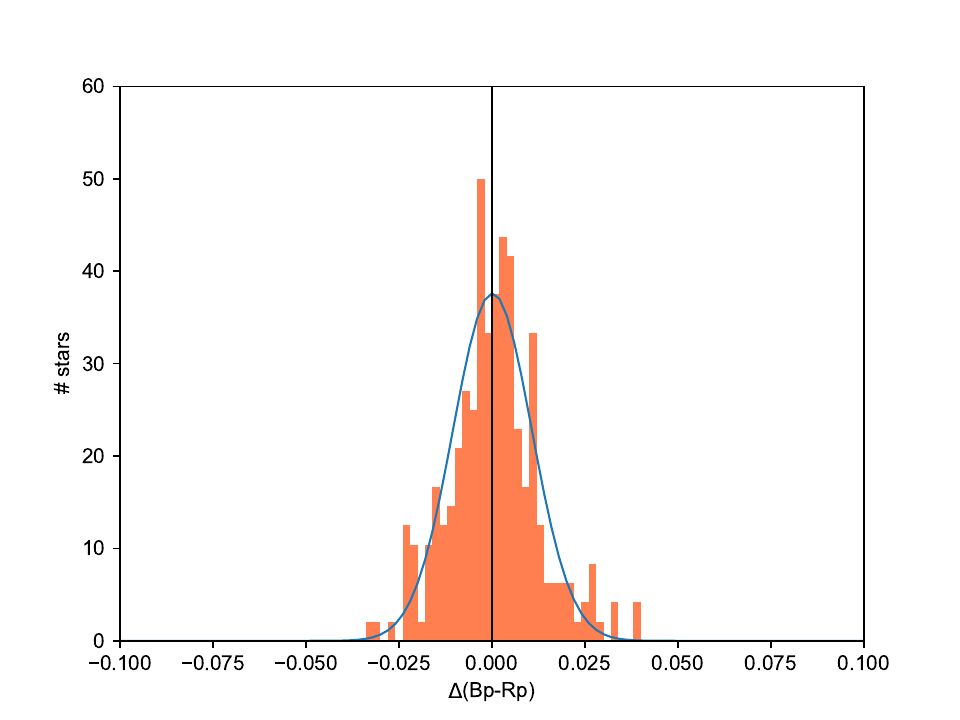}}
}
\resizebox{\hsize}{!}{
{\includegraphics[width=10cm]{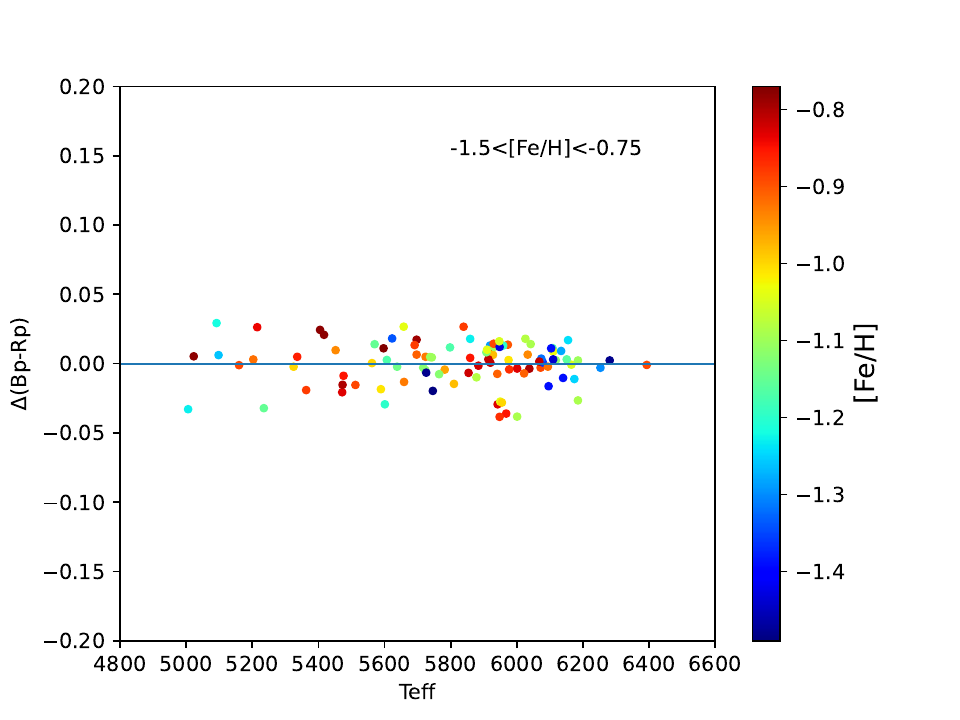}}
{\includegraphics[width=10cm]{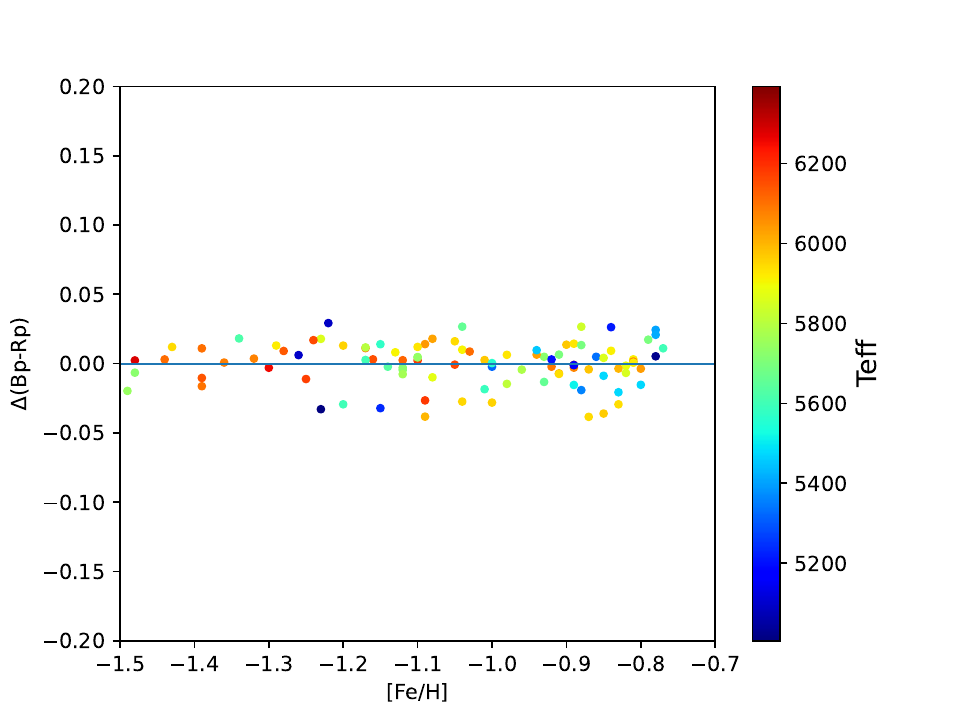}}
{\includegraphics[width=10cm]{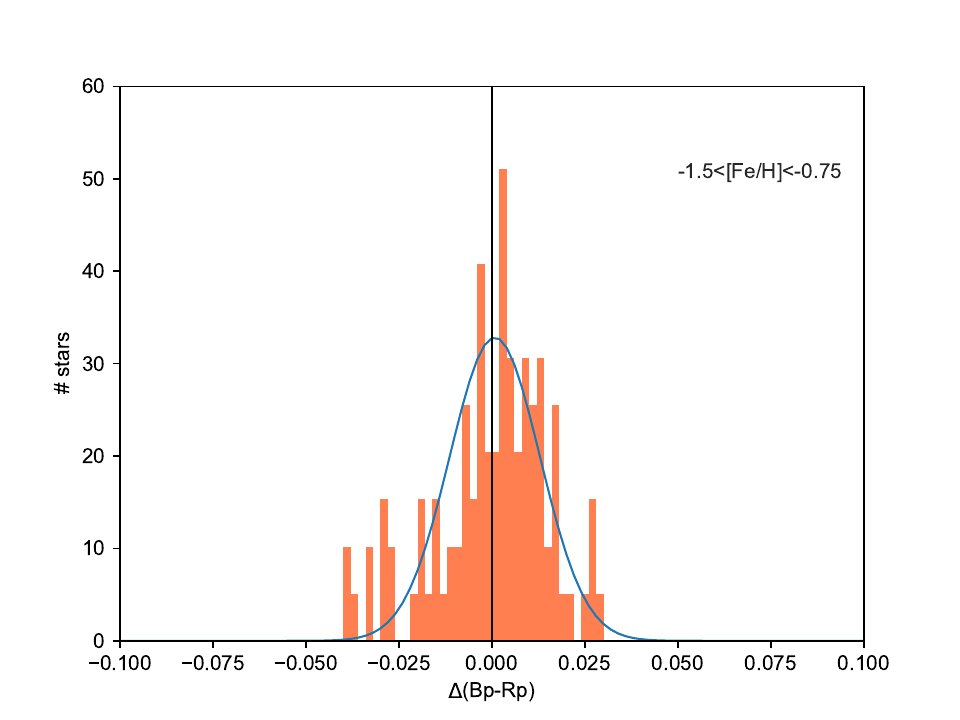}}
}
\caption{Residuals between BP-RP colors and fitted polynomials as a function of effective temperature (left), metallicity (middle), and the histogram of these residuals. Top row shows the result for stars with [Fe/H]$>$-0.25, and the middle row for stars with -0.75$<$[Fe/H]$<$-0.25 and [Fe/H]$<$-0.75.}
\label{fig: bprp_teff_calib}
\end{figure*}

The calibrations are obtained using dwarf and subgiant stars which are product of the combination of the \cite{Casagrande2010_teff} and the \cite{adibekyan2012} data-sets. The stars are selected to be within 75~pc from the Sun to avoid extinction effects and in an effective temperature interval between 5000 K and 6500~K, resulting in 1010 stars.
The distribution of the stars in the Gaia (BP-RP) color - $\rm T_{eff}$ plane is fitted with a polynomial, which is a function of the effective temperature and of the metallicity:
\begin{center}
    $\rm BP-RP = a + b \times x + c \times x^2  + d \times x^3 + e \times x^4 + f \times z + g \times z^2 + h \times z^3 $
\end{center}
 where a, b, c, d, e, f, g, h the coefficients of the polynomial, x is 5040/$\rm T_{eff}$ and z is the metallicity [Fe/H]. The coefficients are calculated for different metallicity ranges and are listed in Table \ref{tab:poly_coeff}. Figure~\ref{fig: bprp_teff_calib} shows examples of the procedure.

\begin{table*}[h!]
    \caption{Polynomial fitting coefficients for (BP-RP) - $\rm T_{eff}$ relation.}
    \label{tab:poly_coeff}
    \centering
    \begin{tabular}{c| c c c}
    \hline\hline
         & \textbf{[Fe/H] $\rm \geq$ - 0.25} & \textbf{- 0.75 $\leq$ [Fe/H] $<$ - 0.25}  & \textbf{- 1.5 $\leq$ [Fe/H] $<$ - 0.75} \\
    \hline
        \textbf{a} & - 1.26560115e+02 & - 1.10186911e+02 & - 8.33752340e+01\\
        \textbf{b} & + 5.51035166e+02 & + 4.81012003e+02 &  + 3.67918027e+02 \\
        \textbf{c} & - 8.97134410e+02 & - 7.84610828e+02 & - 6.06067325e+02\\
        \textbf{d} & + 6.49545820e+02 & + 5.68972189e+02 & + 4.44408270e+02\\
        \textbf{e} & - 1.75787123e+02 & - 1.54131155e+02 & - 1.21761773e+02\\
        \textbf{f} & + 7.75963128e-02 & - 9.98548177e-02 & + 2.56814399e-01\\
        \textbf{g} & - 3.64128996e-03 & - 2.94101629e-01 & + 2.44005287e-01\\
        \textbf{h} & - 3.44245462e-02 & - 1.87066718e-01 & + 7.81671154e-02\\
    \hline
    \end{tabular}
    \tablefoot{List of coefficients utilized to calibrate the effective temperature - Gaia (BP-RP) color relations on a sample of 927 dwarf stars. The case of [Fe/H] < -1.5 is treated as if [Fe/H] = -1.5. These values are valid only fot stars with $\rm 5000 \leq T_{eff}/K \leq 6500$.
    In the case of  $\rm T_{eff}<5000$ K or $\rm T_{eff} > 6500$ K the polynomial function depends only by effective temperature and assume the following form: BC =  $ \rm -33.869 + 0.03497 \times x - 1.295e-05  \times x^2 + 2.2885e-09  \times x^3 - 1.9641e-13  \times x^4 + 6.5913e-18  \times x^5$}
\end{table*}

\FloatBarrier

\section{GALAH DR3}
\label{Appendix: GALAHDR3}

The sample of dwarfs from GALAH DR3 \citep{Buder2021} is selected from stars with the stellar parameter quality flag \texttt{flag\_sp} == 0, the iron abundance quality flag \texttt{flag\_fe\_h} == 0, the  $\rm \alpha$-abundance quality flag \texttt{flag\_alpha\_fe} == 0, signal-to-noise for the spectrum of the camera 3 \texttt{snr\_c3\_iraf} > 65, $\rm 3.5< \log g < 4.45$, distance from the sun below 2 kpc and $\rm MG_{0}$ < 5.5. The final selection, built on approximately  589 000 stars in GALAH DR3, contains around 41 000  dwarf stars located in the Solar vicinity.

\paragraph{APOGEE $vs$ GALAH}
\label{subsev: apogee vs galah}
Having found the combination of atmospheric parameters that provide the best determination of ages using APOGEE stars (see Section \ref{subsec: age determination}), we now turn to comparing APOGEE and GALAH datasets with ages derivation also based on the Gaia photometry. To account the effect of the extinction on the GALAH sample, we took advantage of the reddening map from \citet{Lallement2019}.
We then converted the values in the Gaia photometric bands using, the same coefficients assumed for the APOGEE sample from \citet{Wang_Chen_extinction}.

\begin{figure*}[h!]
\centering
\includegraphics[width=8cm]{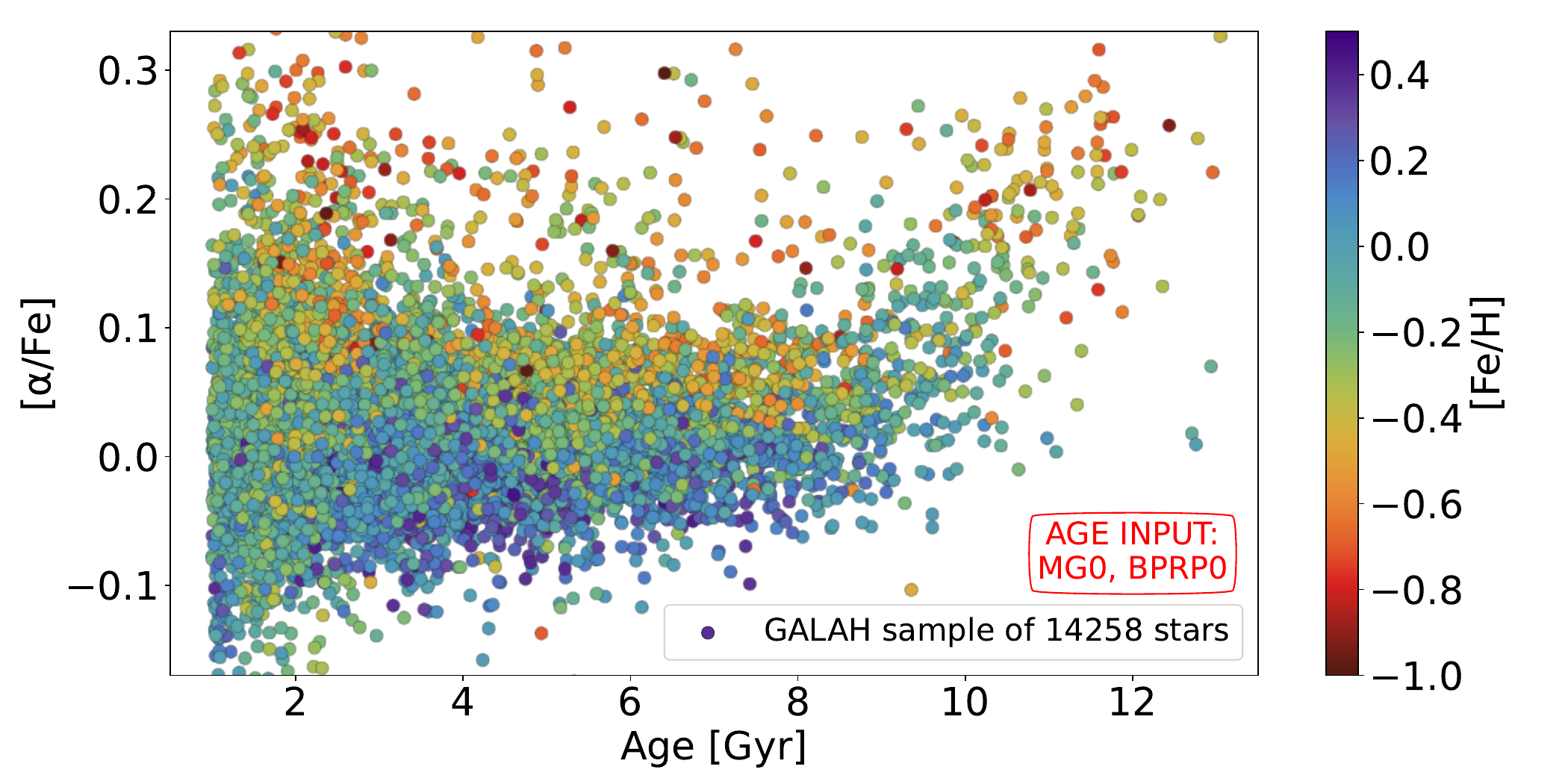}
\includegraphics[width=8cm]{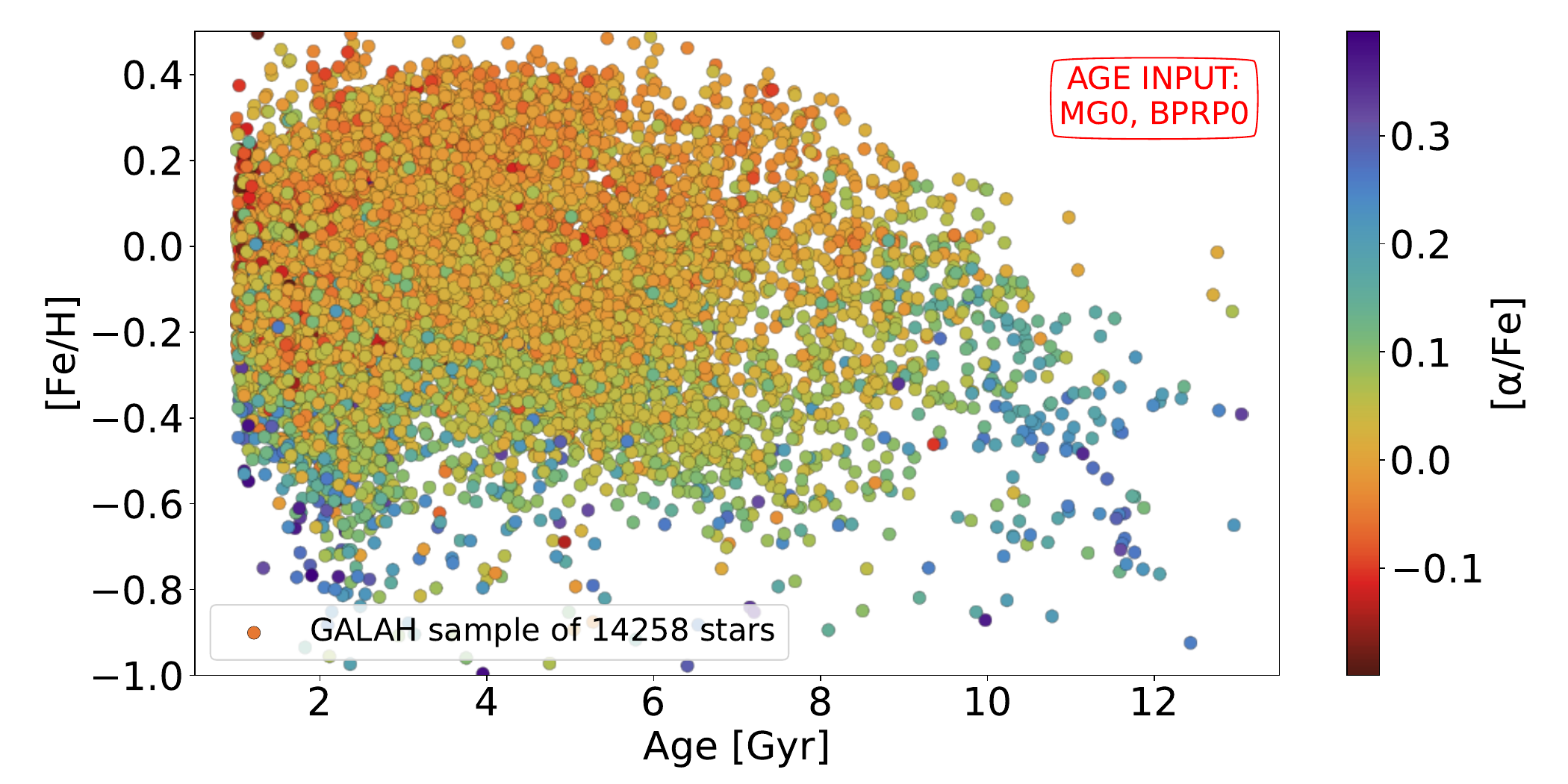}
\includegraphics[width=8cm]{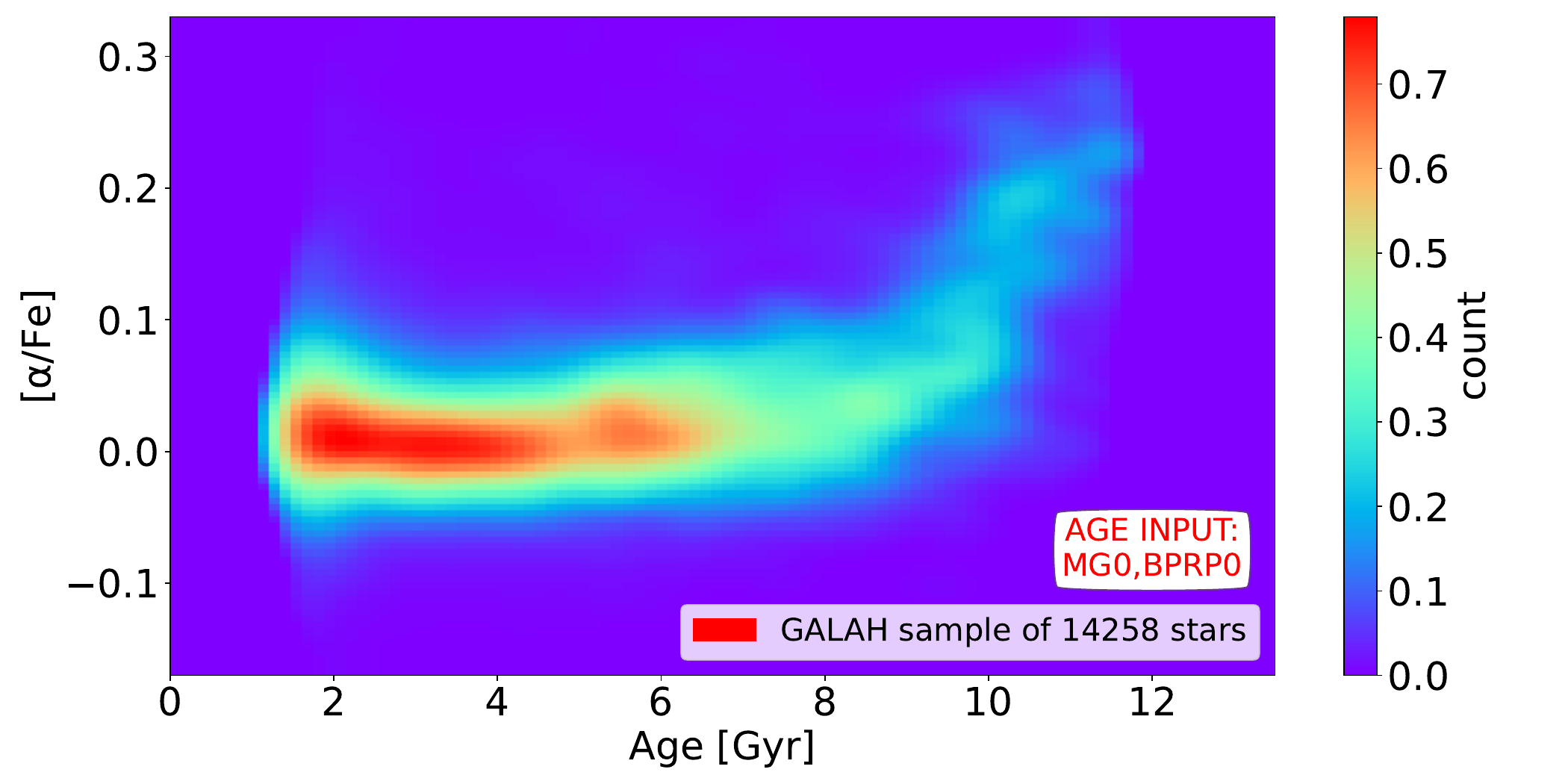}
\includegraphics[width=8cm]{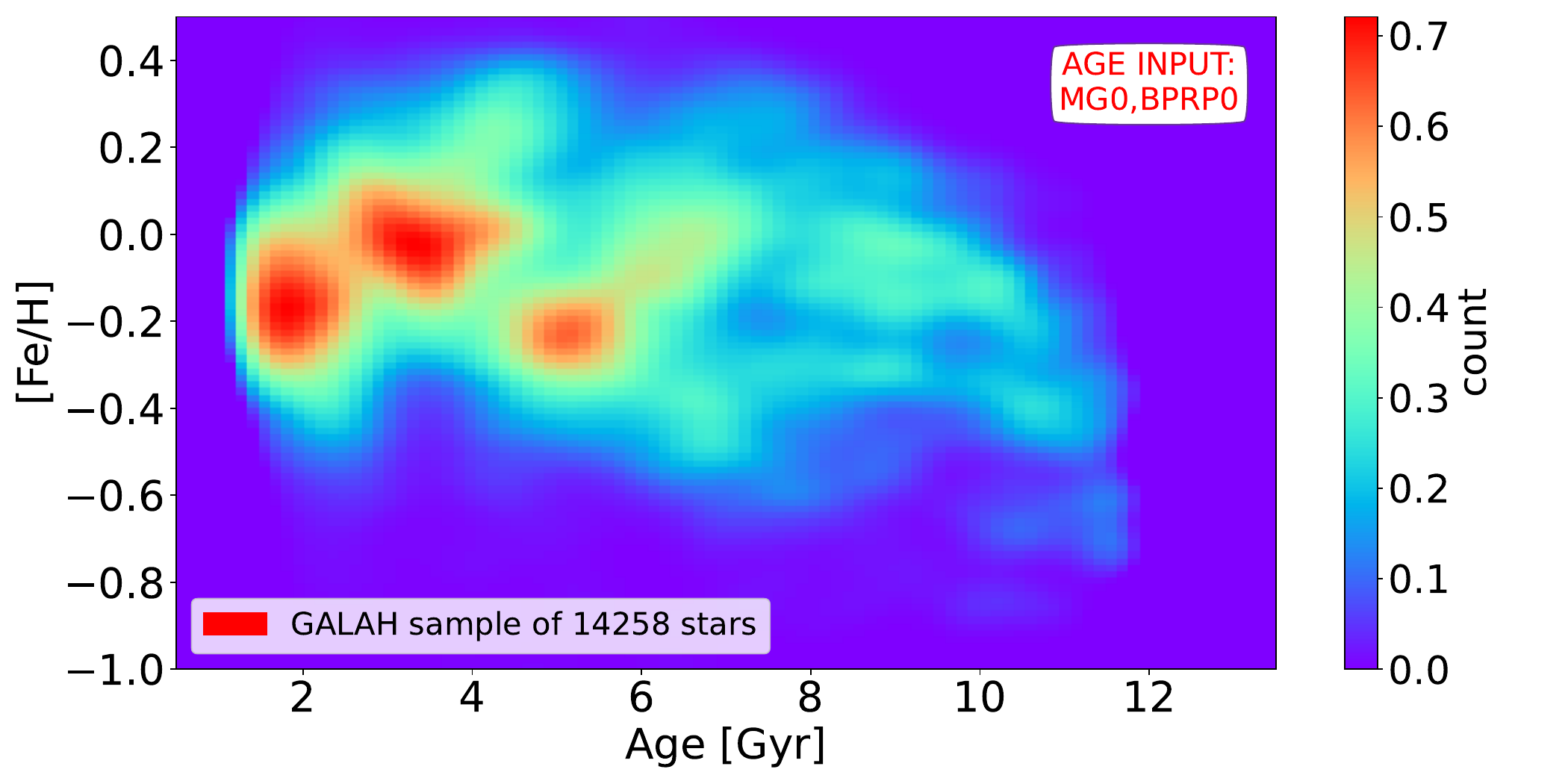}

\caption{Age-chemistry relations for the GALAH DR3 main sample of dwarf stars (see the text for details about the selection) in form of scatter plots (top) and column normalized 2D density histogram (bottom). The parameters used as input for the age determination are written inside the plots.}
\label{fig:age-chemistry GALAH and APOGEE density}
\end{figure*}

The comparison between GALAH DR3 and APOGEE DR17 is interesting because the metallicities and $\rm \alpha$-elements have been determined using entirely different spectroscopic data and procedures. 
To ensure a consistent quality of stellar parameter determinations, we adopted a more stringent lower SNR threshold for GALAH compared to APOGEE. This decision was motivated by the fact that, for a given SNR cut, the uncertainties in effective temperature and metallicity are significantly lower in APOGEE than in GALAH. Therefore, we set the SNR threshold in GALAH to retain dwarf stars with metallicity uncertainties below 0.1 dex for most cases, ultimately yielding a comparable number of stars with age estimates from both surveys.
The GALAH DR3 distributions are shown in Fig. \ref{fig:age-chemistry GALAH and APOGEE density}.

The scatter plots show apparently substantially different distributions, with an increasing dispersion towards younger ages in the age-$\alpha$ distribution, and much less thick disk stars. However the density plots shows that the main characteristics of both distributions are preserved: the age-$\alpha$ distribution shows clearly a flat segment at age below about 9 Gyr, and a steeper one above this limit. The wavy behavior of age-metallicity distribution is also observed, albeit with a less well defined thick disk sequence at ages larger than 9 Gyr in GALAH, due to poor statistics. 
Focusing on the age-metallicity density plots of both datasets, we note that all show a thick disk sequence starting at age $\sim$ 12 Gyr, increasing in metallicity to 8 Gyr, followed by an apparent decline to about 5 Gyr ago. Finally, at younger ages, a large clump of stars (two clumps, in the case of GALAH) more or less separated from the previous sequence. 

In summary, the datasets examined give consistent results, with slight differences. In particular, given the significantly better sampling of the thick disk sequence provided by the APOGEE chemical abundances and the Gaia $\rm Bp - Rp$ color and G absolute magnitude, we use the latter as the main sample in the paper.

\FloatBarrier

\section{Age determination caveats: age uncertainties}\label{appendix: Age determinations}
\label{subsec: effect_age_uncert}

\begin{figure*}[h!]
\centering
\includegraphics[width=6cm]{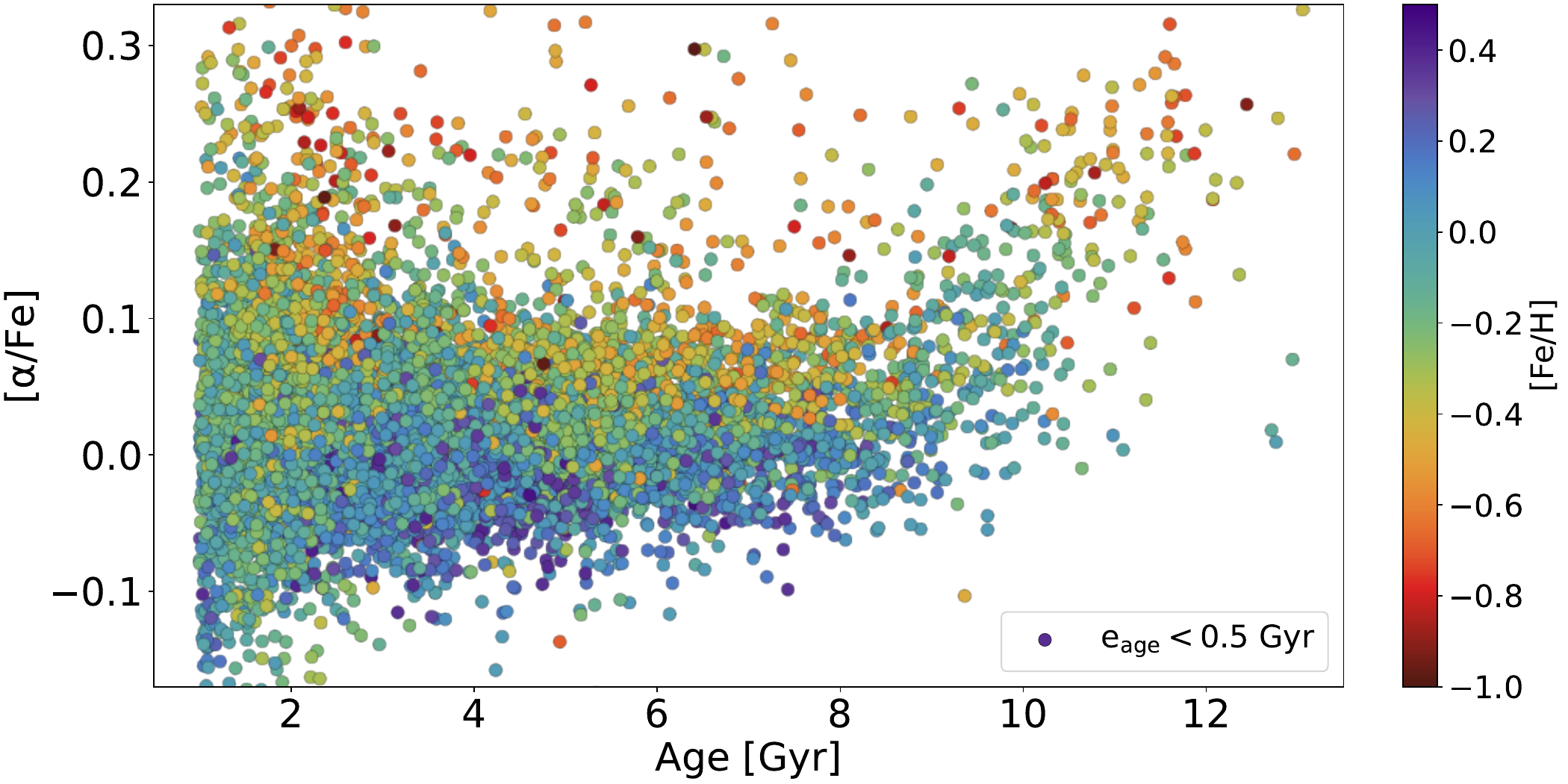}
\includegraphics[width=6cm]{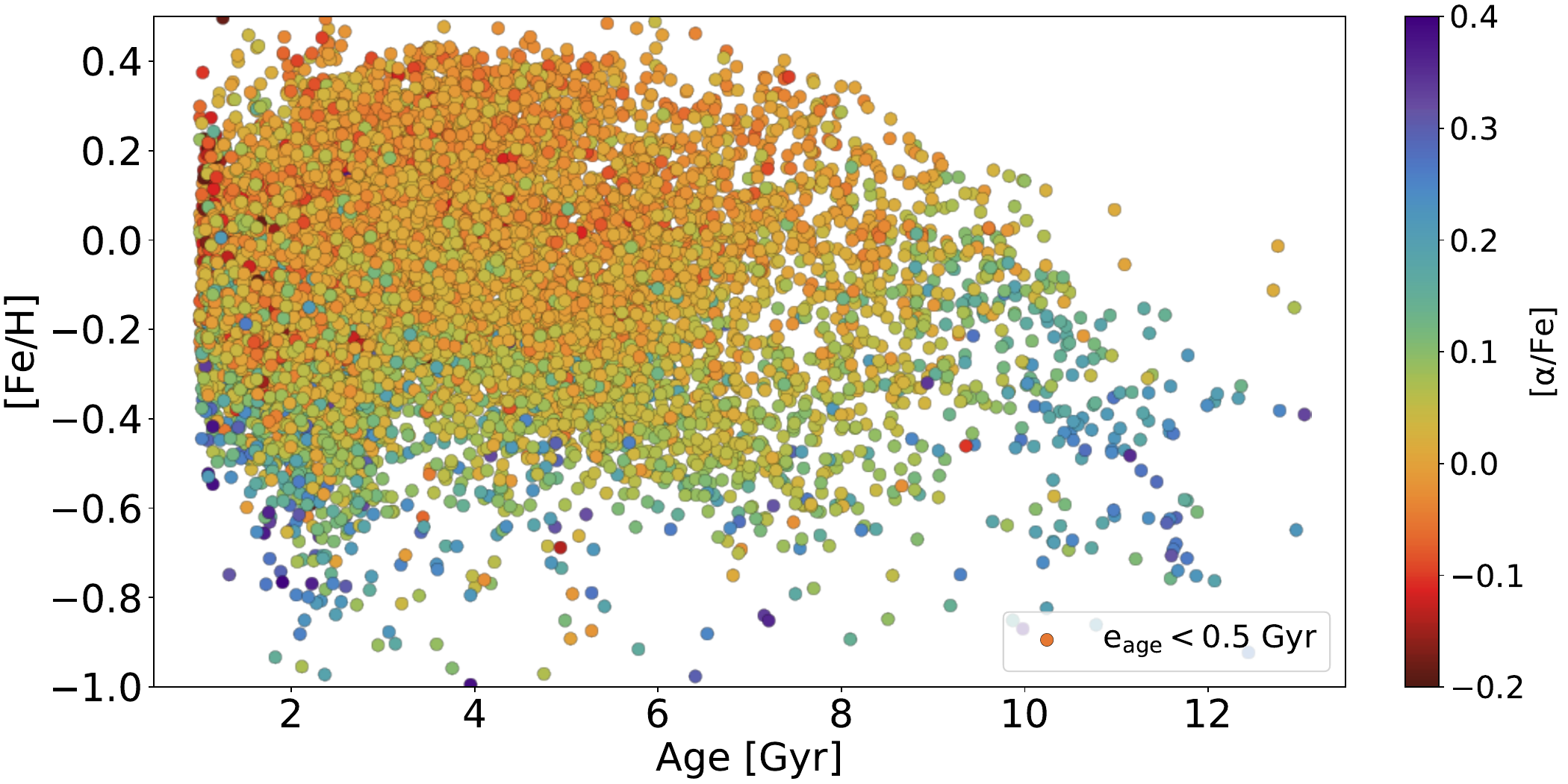}
\includegraphics[width=6cm]{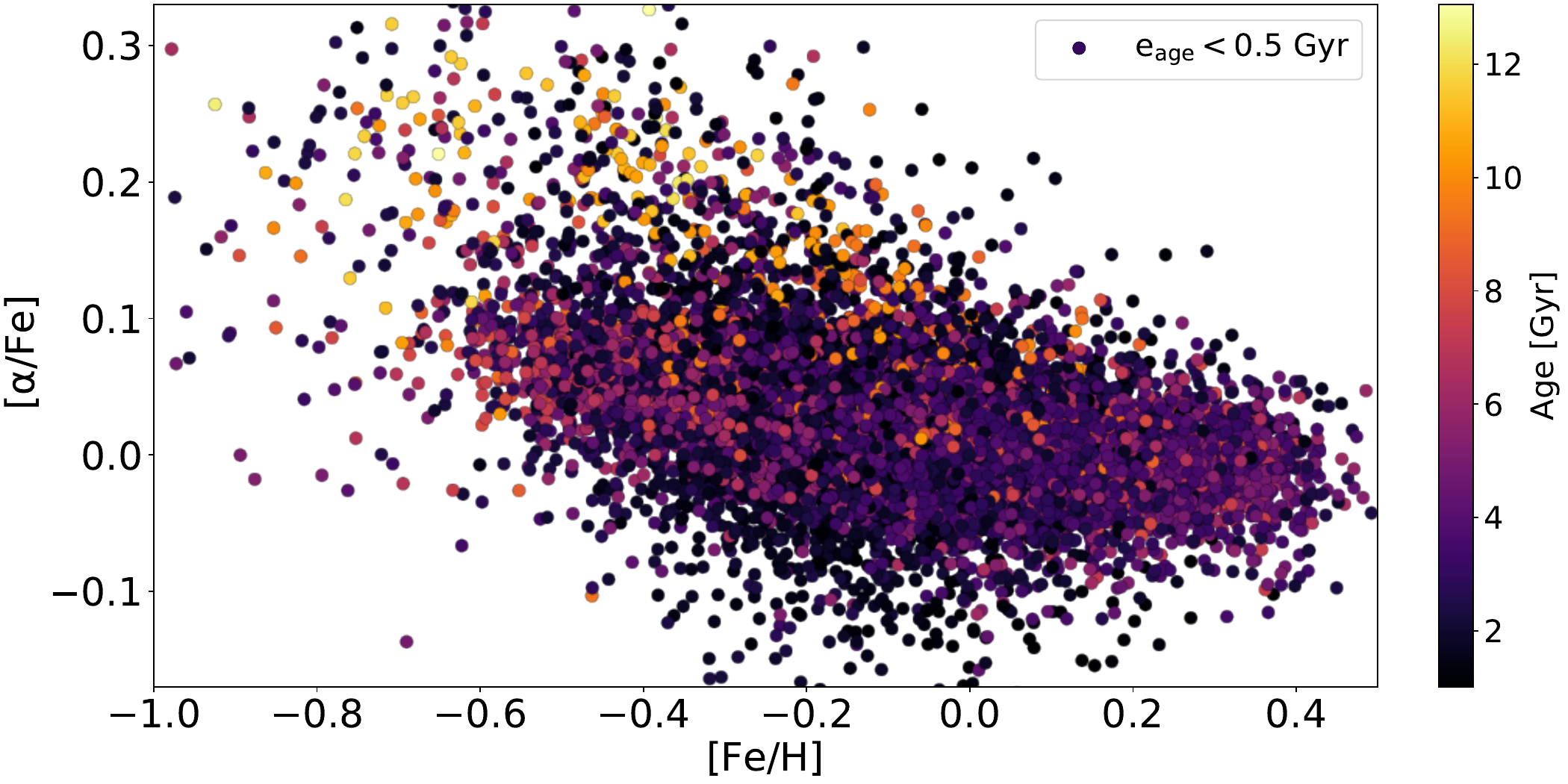}\par

\includegraphics[width=6cm]{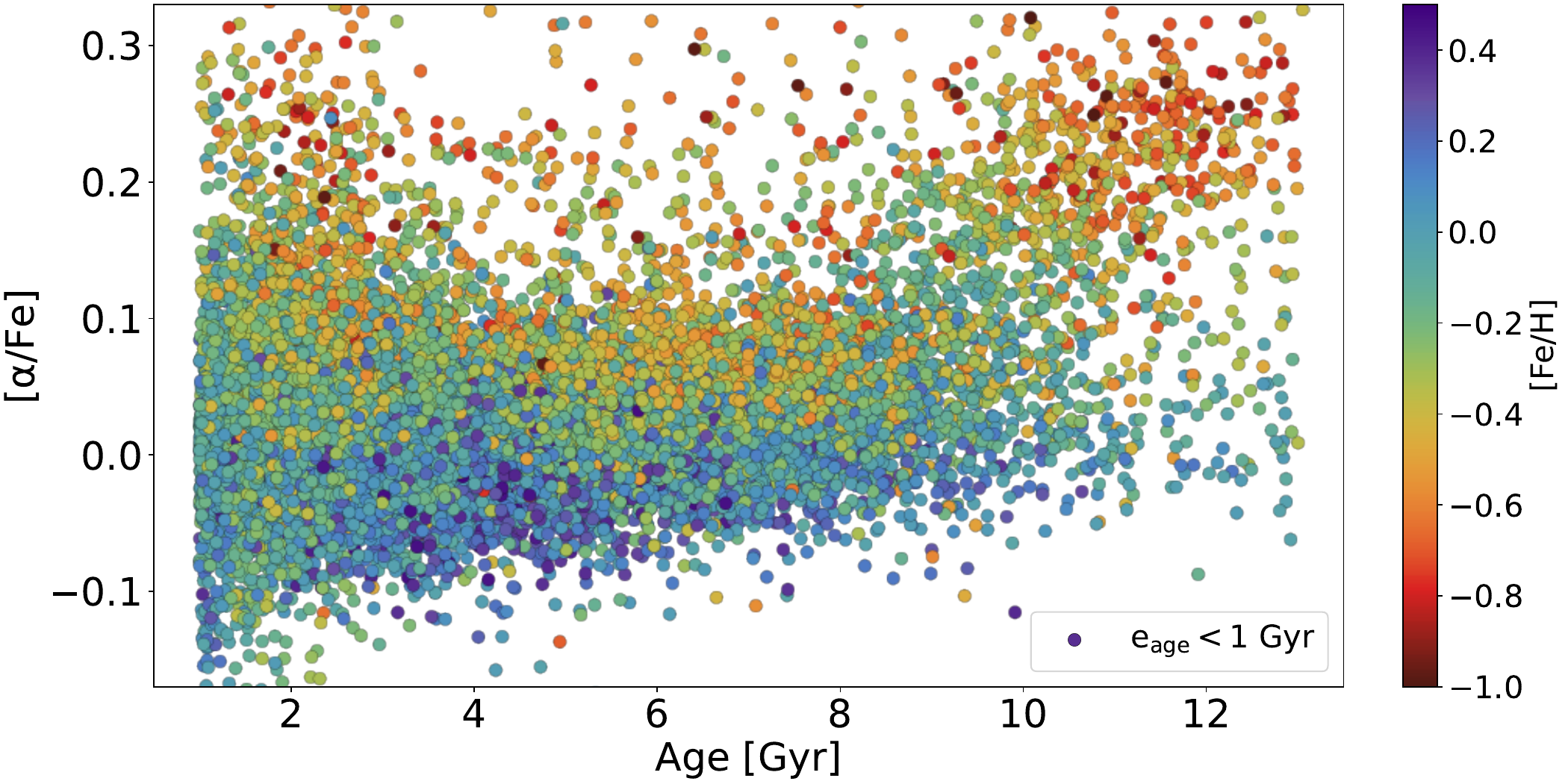}
\includegraphics[width=6cm]{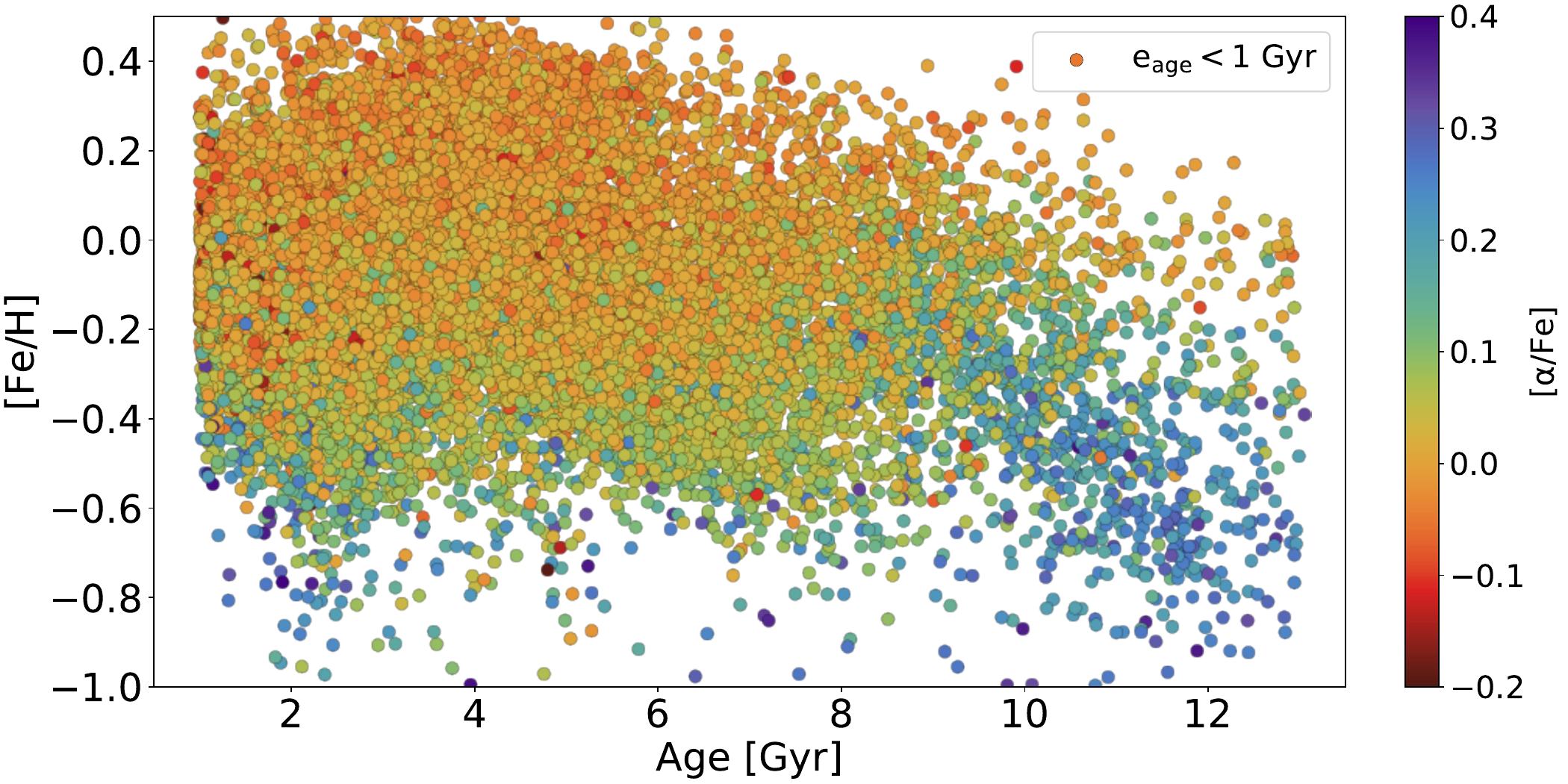}
\includegraphics[width=6cm]{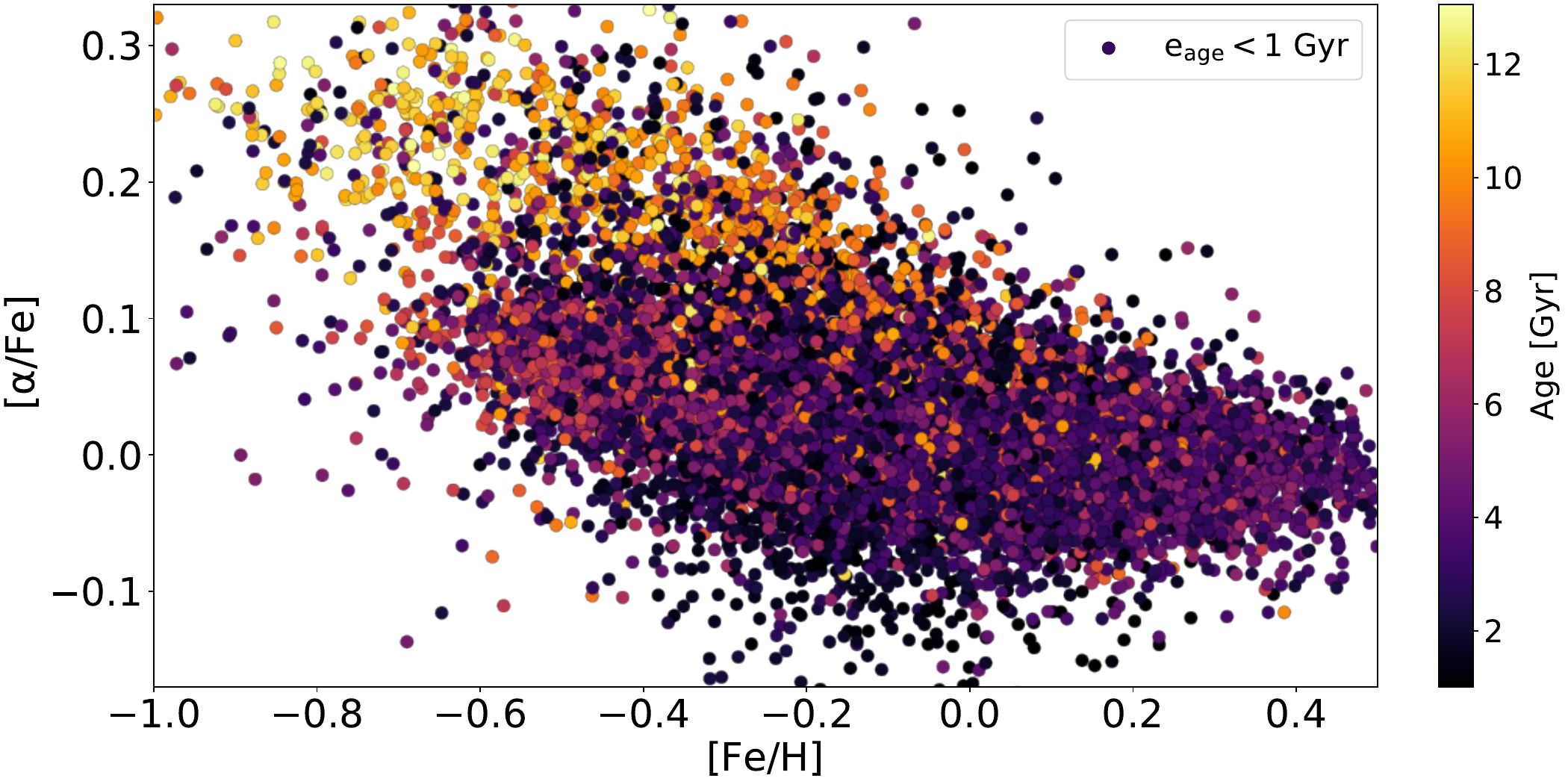}\par

\includegraphics[width=6cm]{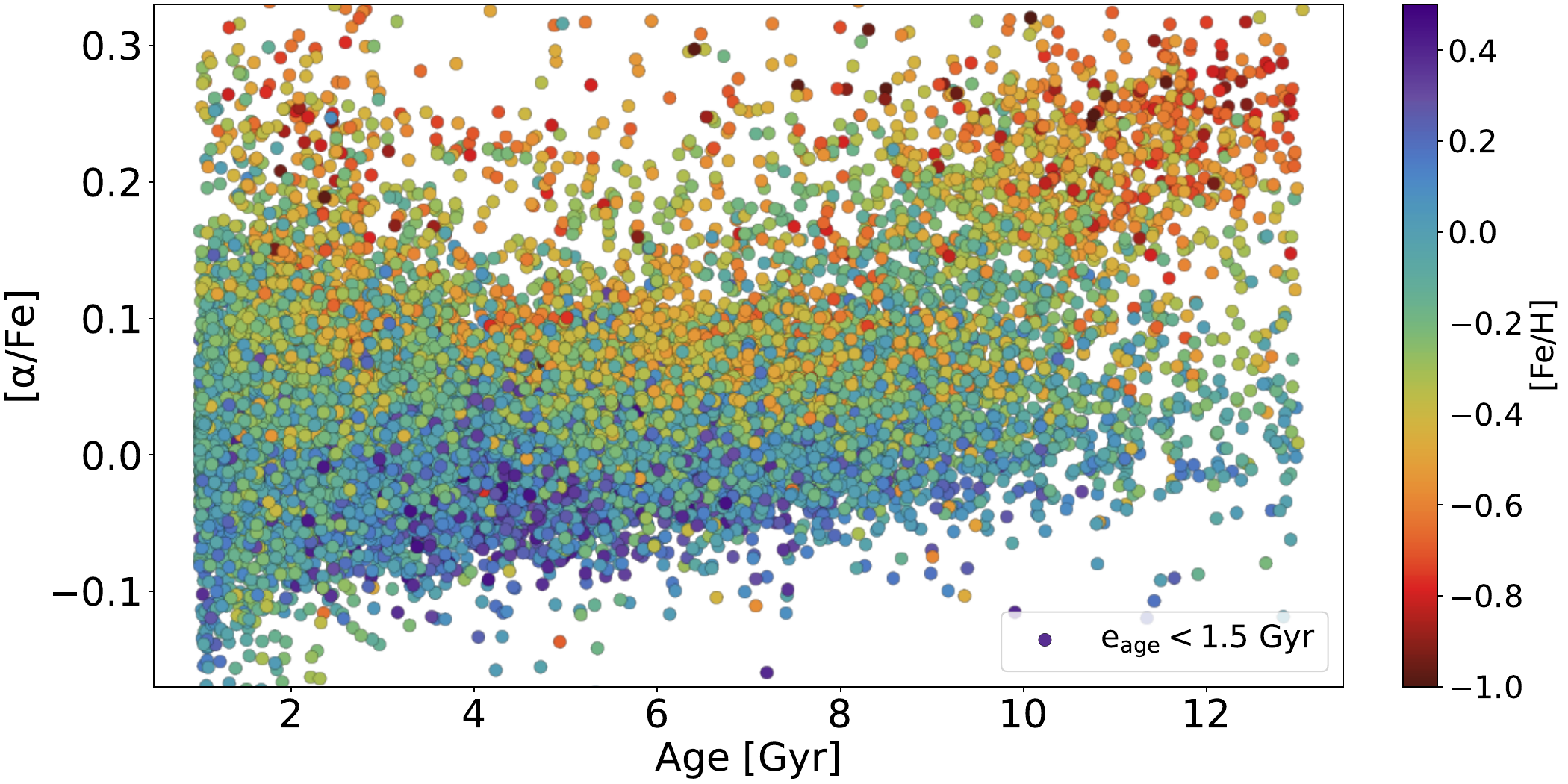}
 \includegraphics[width=6cm]{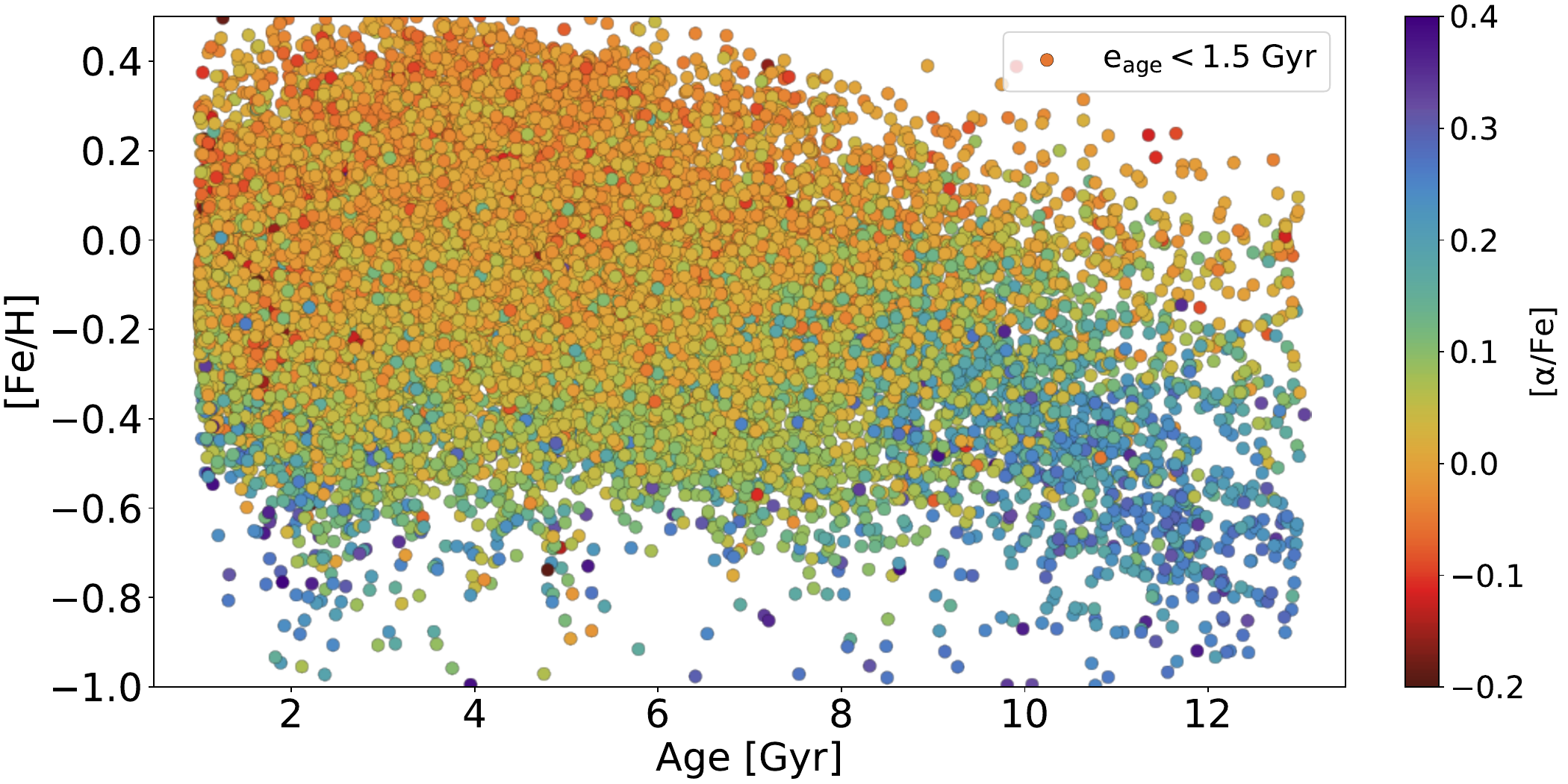}
\includegraphics[width=6cm]{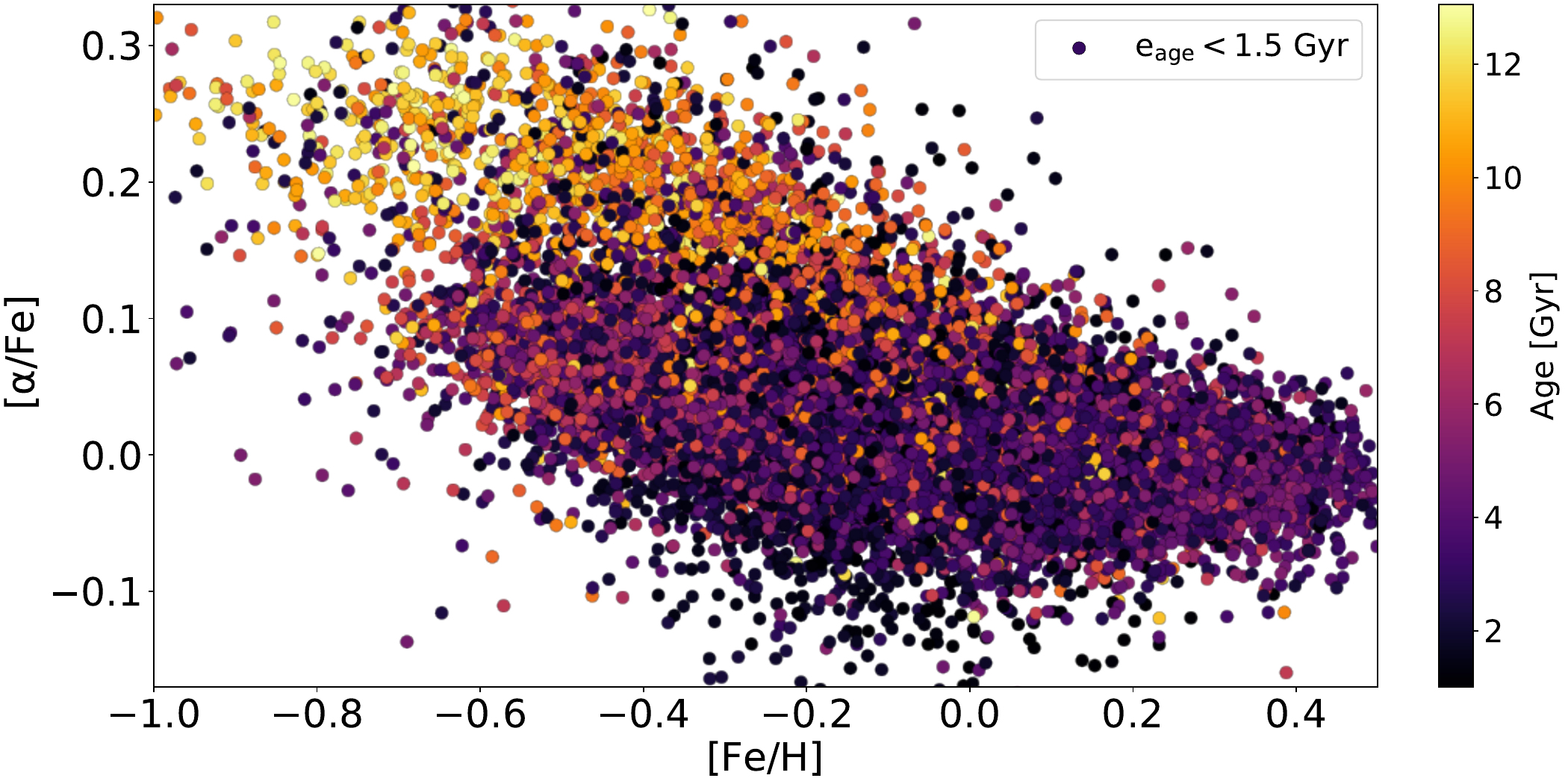}\par

\includegraphics[width=6cm]{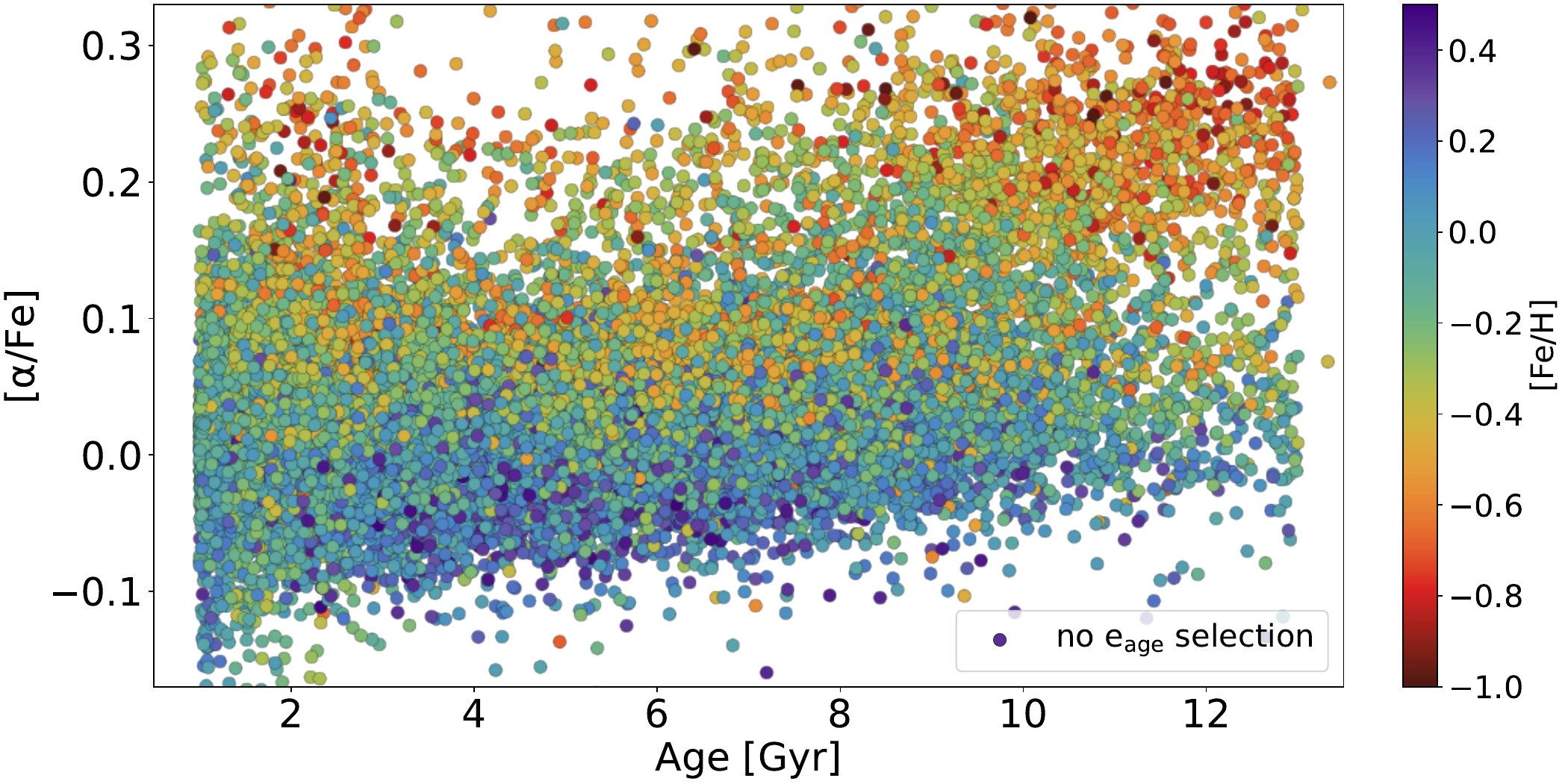}
\includegraphics[width=6cm]{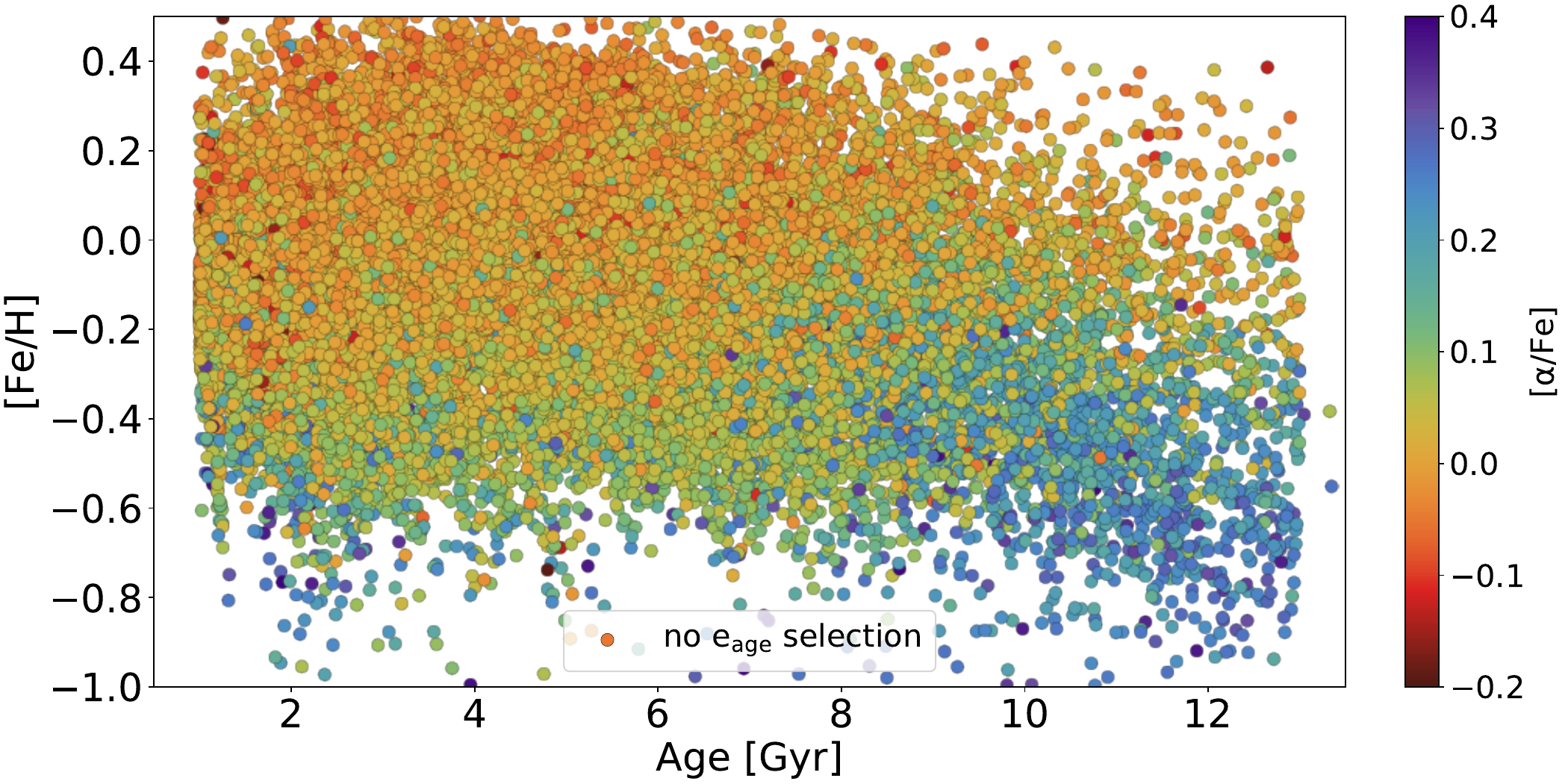}
\includegraphics[width=6cm]{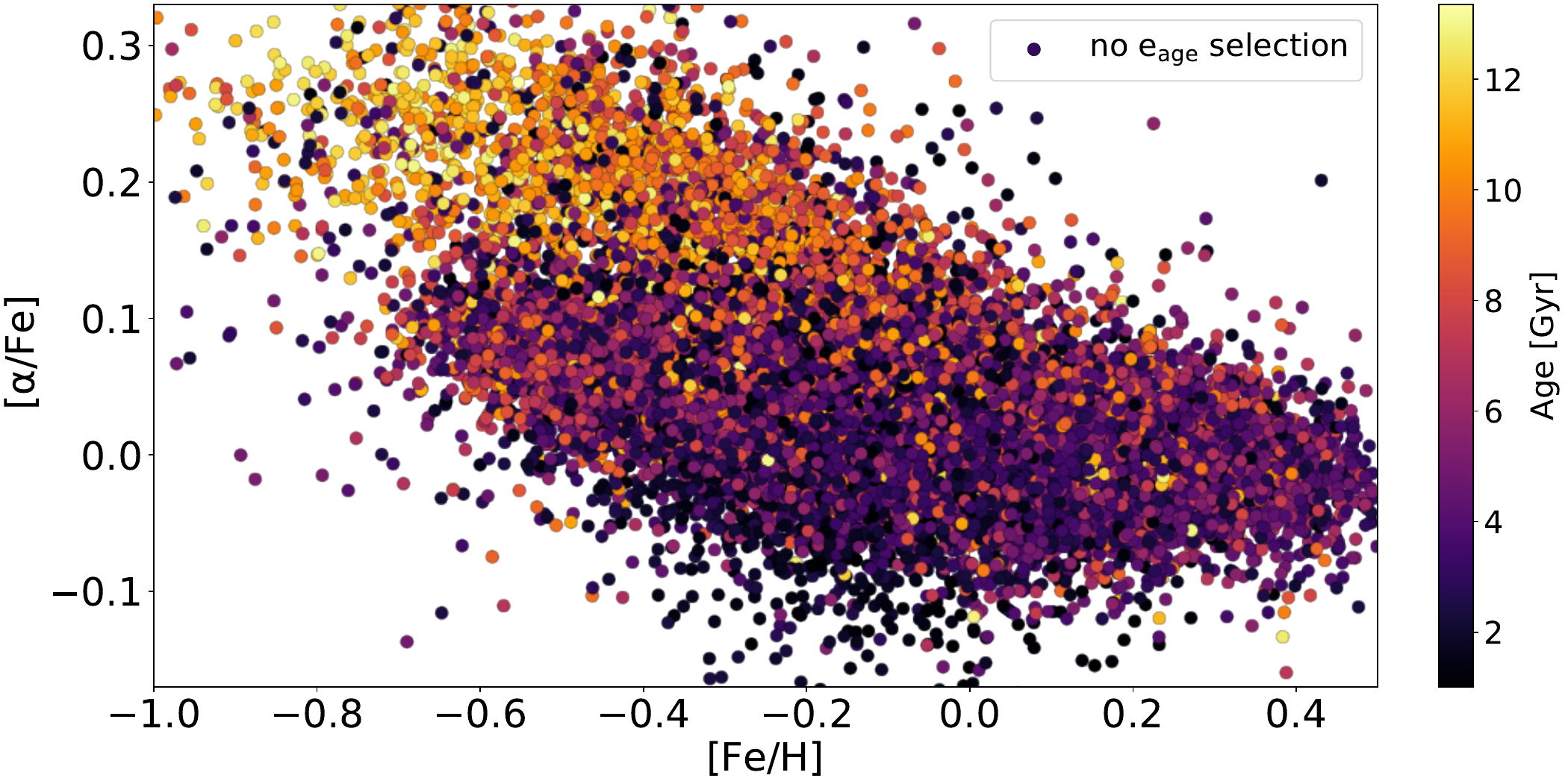}
\caption{Effect of the age uncertainties on the age-[$\rm \alpha$/Fe] distribution (left column), age-metallicity distribution (middle column) and [$\rm \alpha$/Fe]-[Fe/H] distribution (right column) for GALAH DR3 local dwarfs. The stars are color-coded for [Fe/H], [$\rm \alpha$/Fe] and age respectively. The stars in these plots have been selected to have: $\rm 3.5 < \log g < 4.45$, distance from the sun below 2 kpc, [Fe/H] > -1, age above 1 Gyr, $\rm (BP-RP)_{0}<0.95$, \texttt{snr\_c3\_iraf} above 65, \texttt{flag\_sp} = 0, \texttt{flag\_fe\_h} = 0, \texttt{flag\_alpha\_fe} = 0. We compare then three different cuts: $\rm e_{age}<0.5$ Gyr, $\rm e_{age}<1$ Gyr, $\rm e_{age}<1.5$ Gyr with the total distribution to which no selection on age uncertainty is applied (bottom row). 
}

\label{fig: effect_age_unc GALAHDR3}
\end{figure*}

\begin{figure*}[h!]
\centering
\includegraphics[width=6cm]{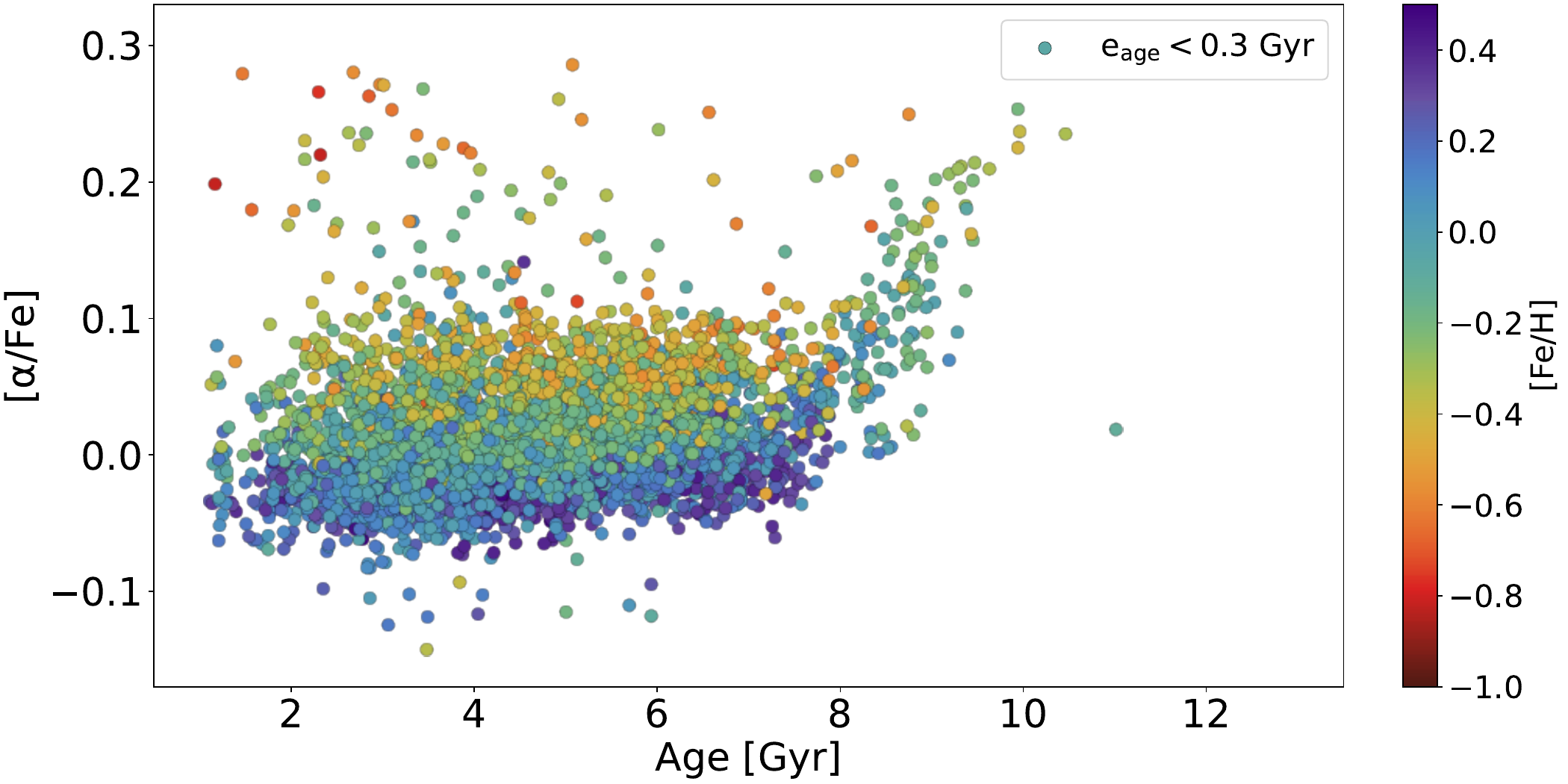}
\includegraphics[width=6cm]{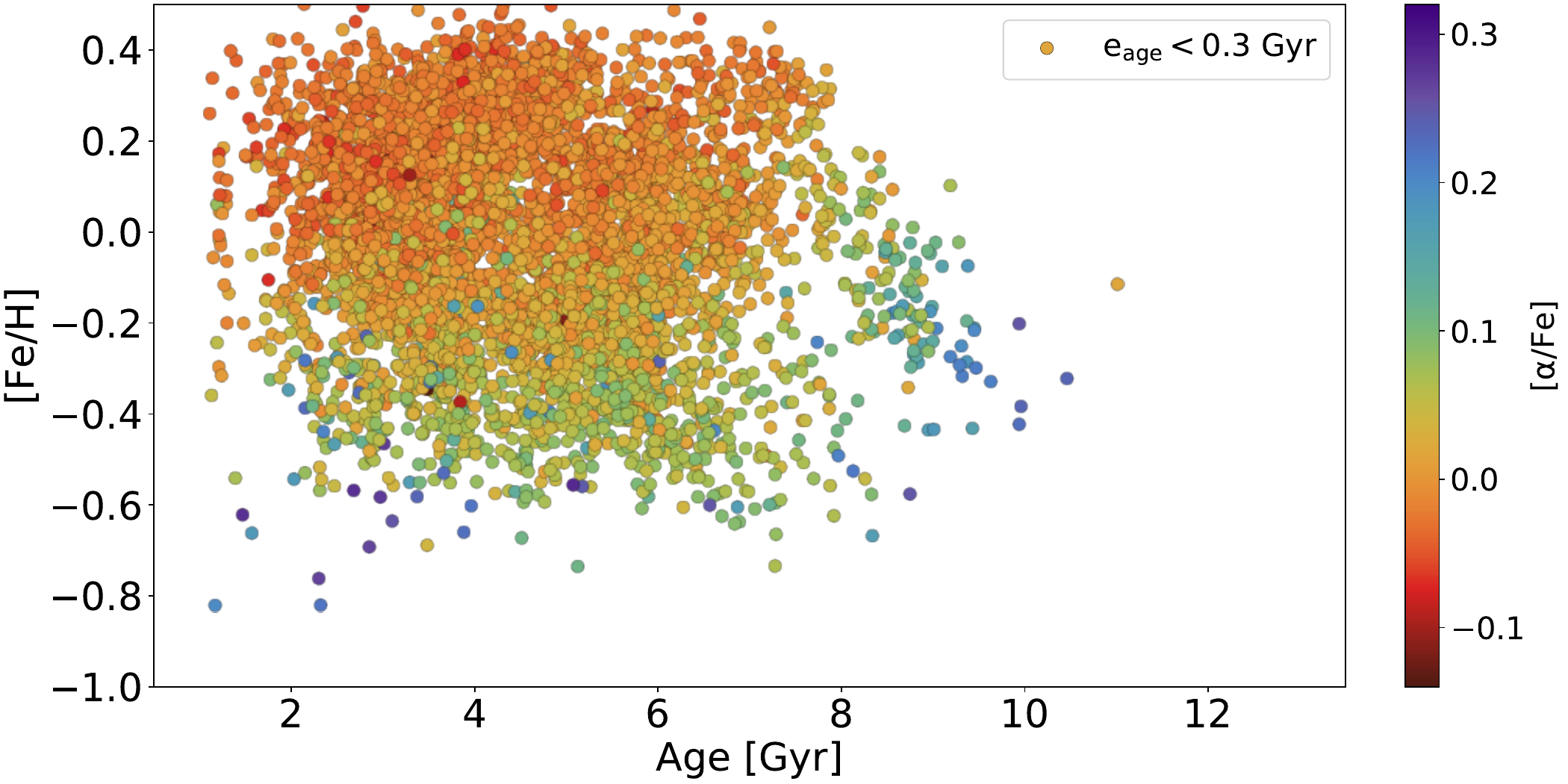}
\includegraphics[width=6cm]{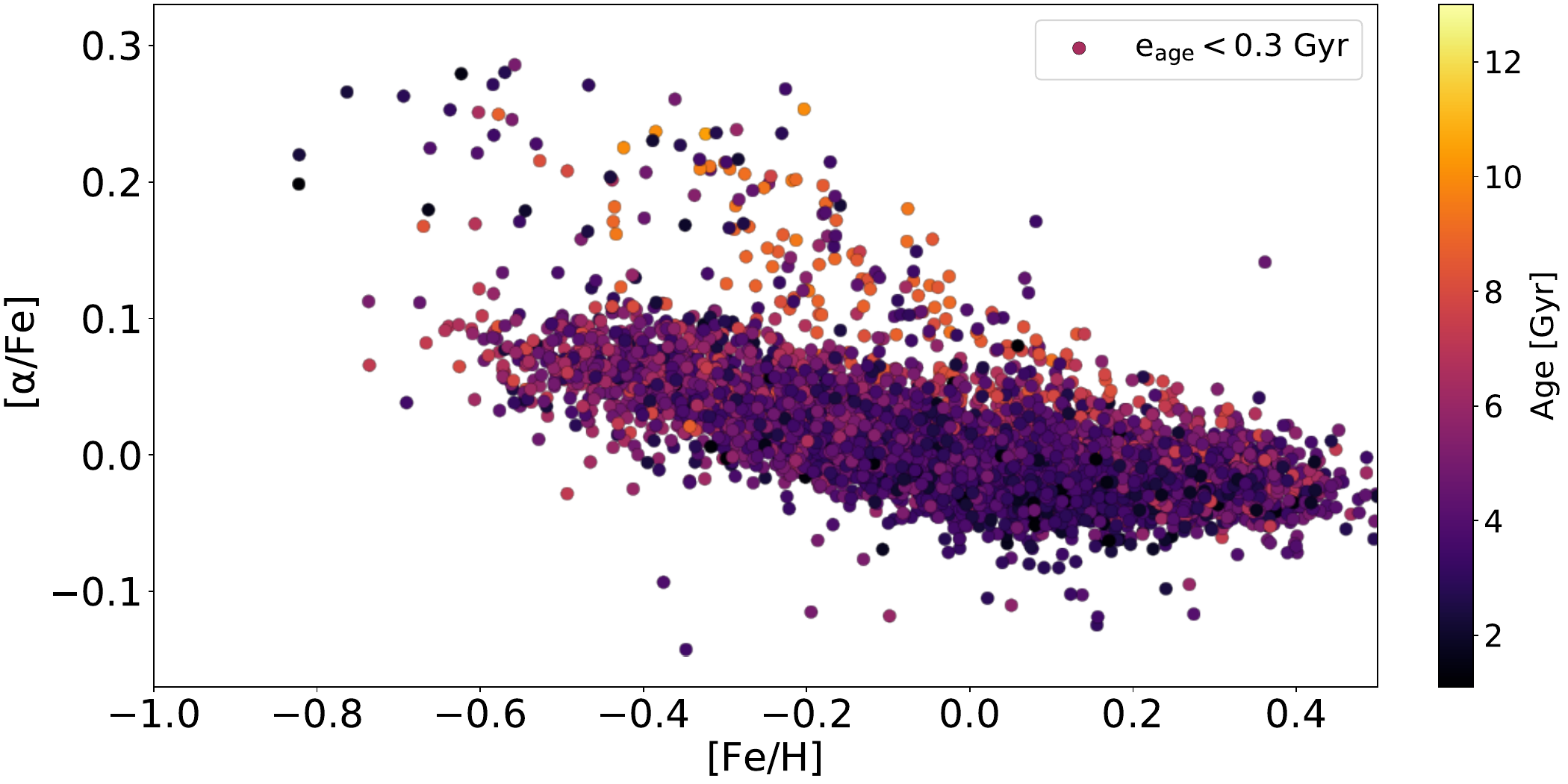}\par

\includegraphics[width=6cm]{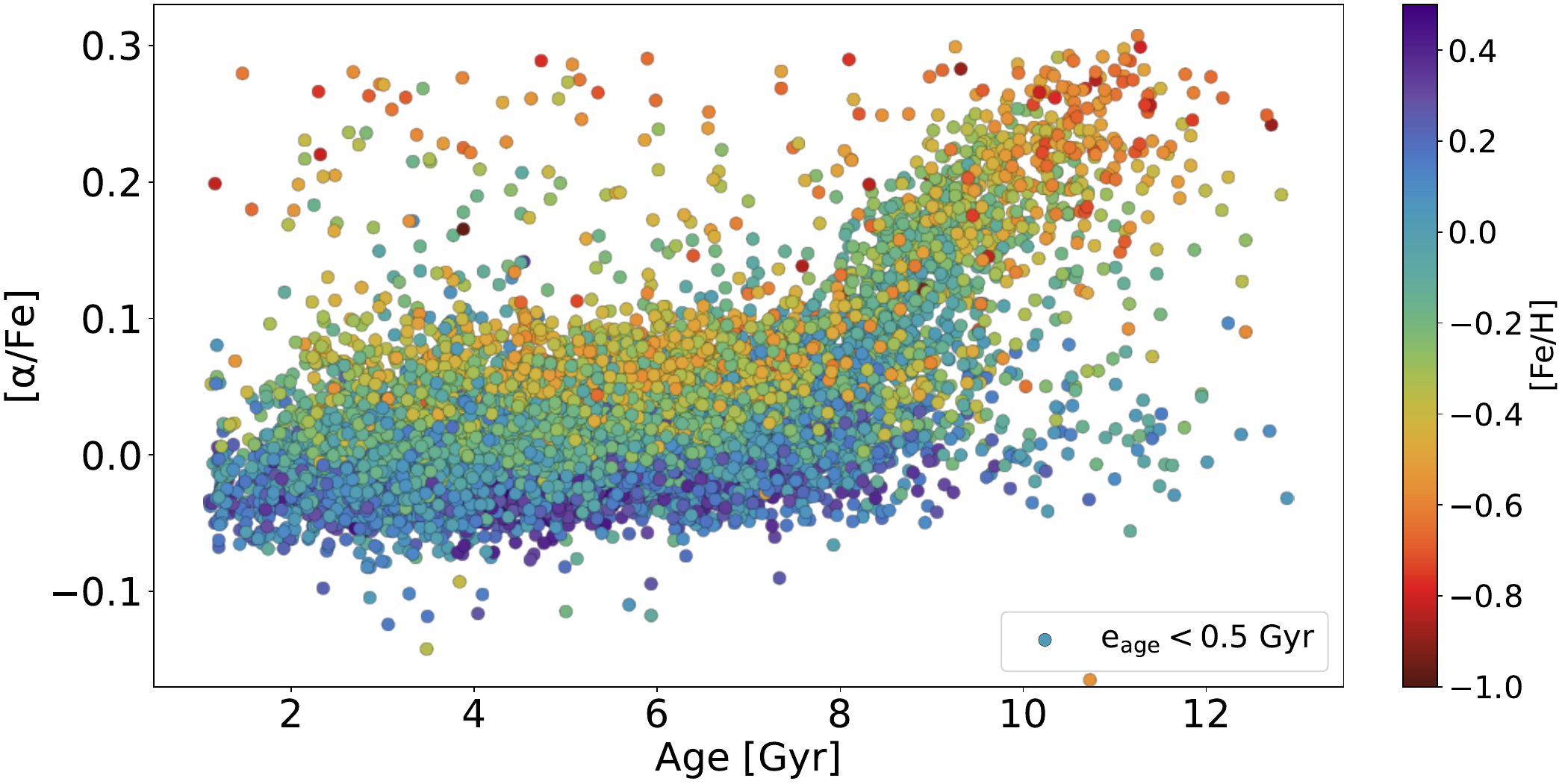}\
\includegraphics[width=6cm]{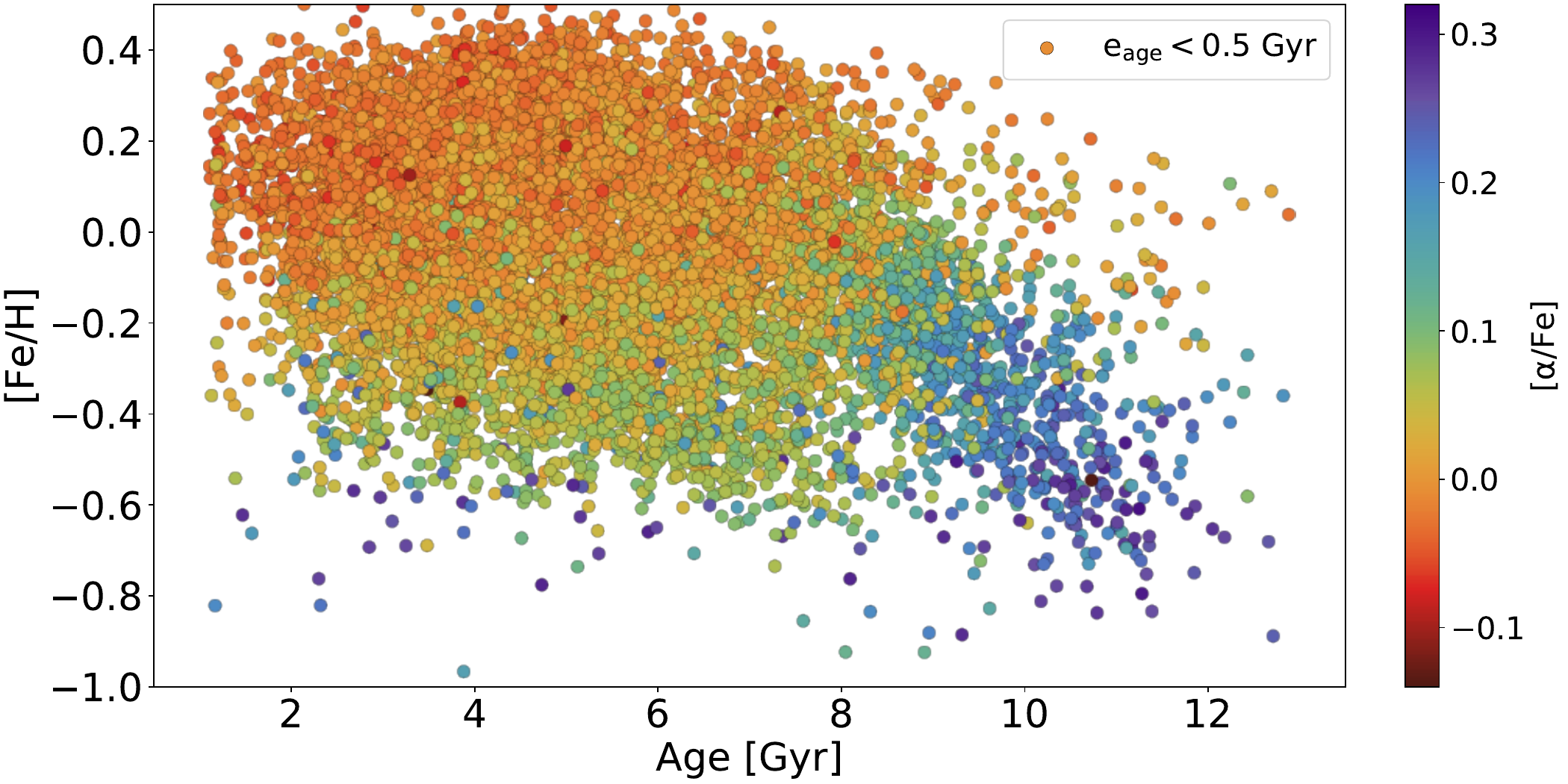}
\includegraphics[width=6cm]{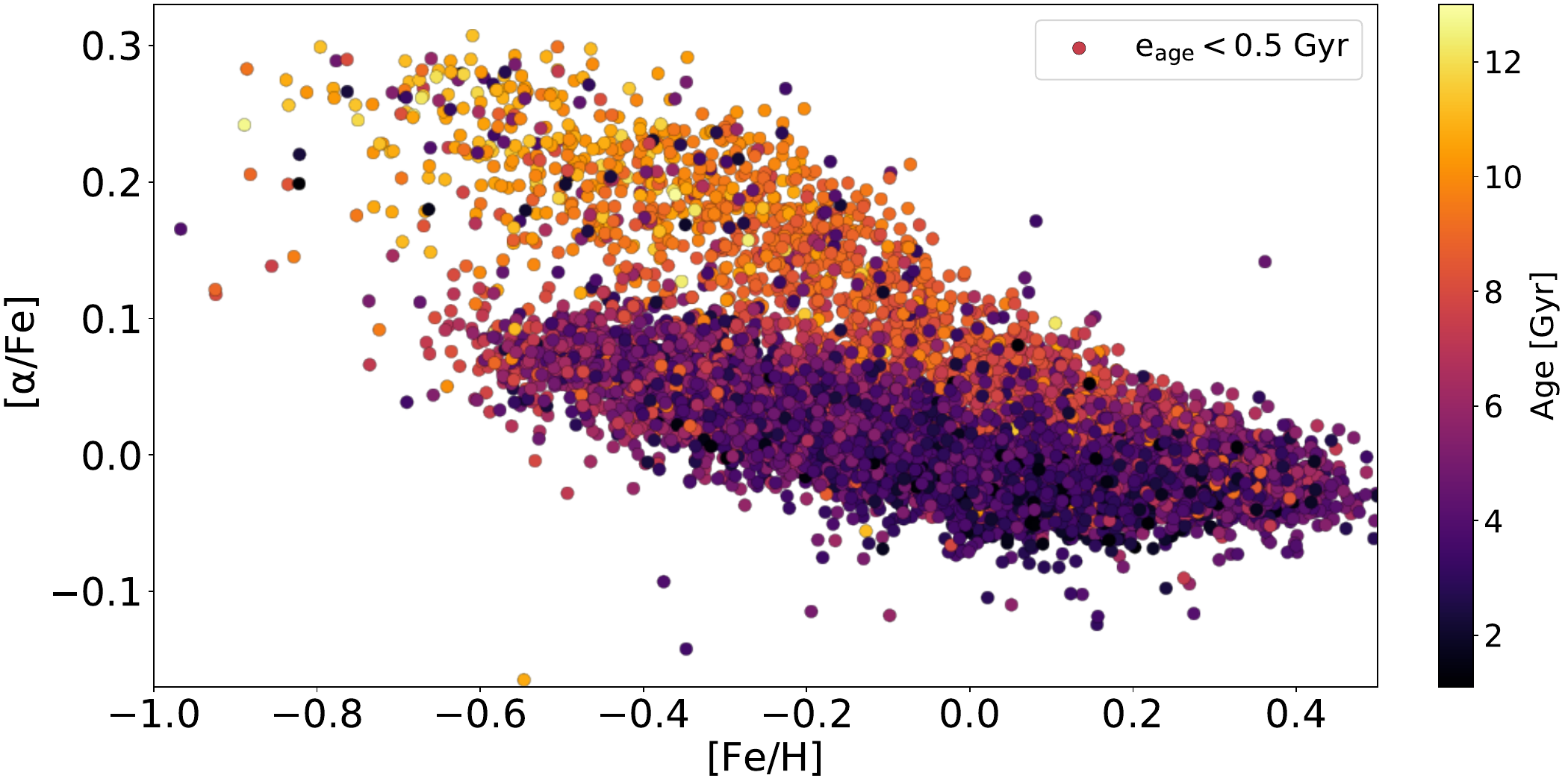}\par
\includegraphics[width=6cm]{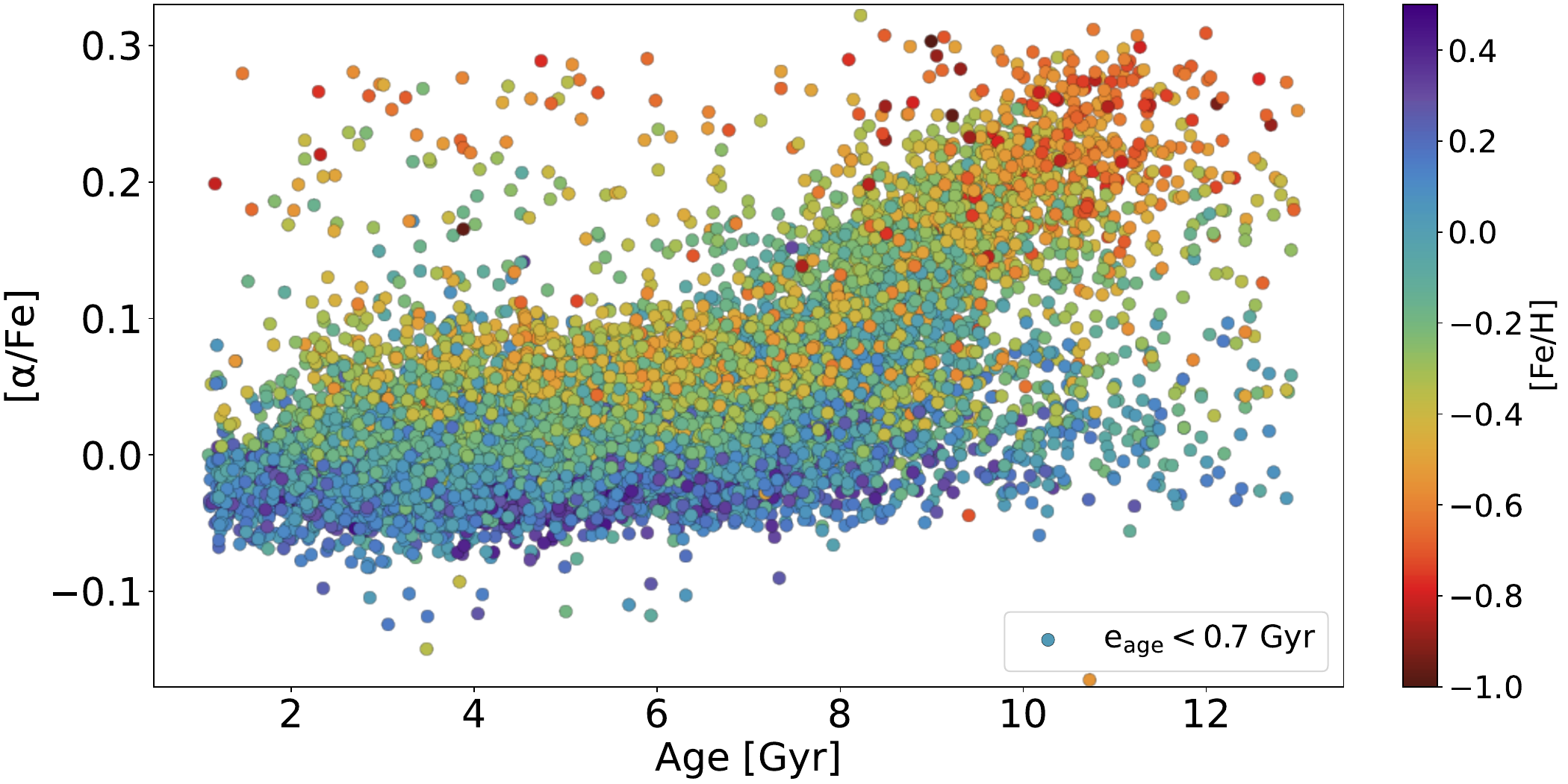}
\includegraphics[width=6cm]{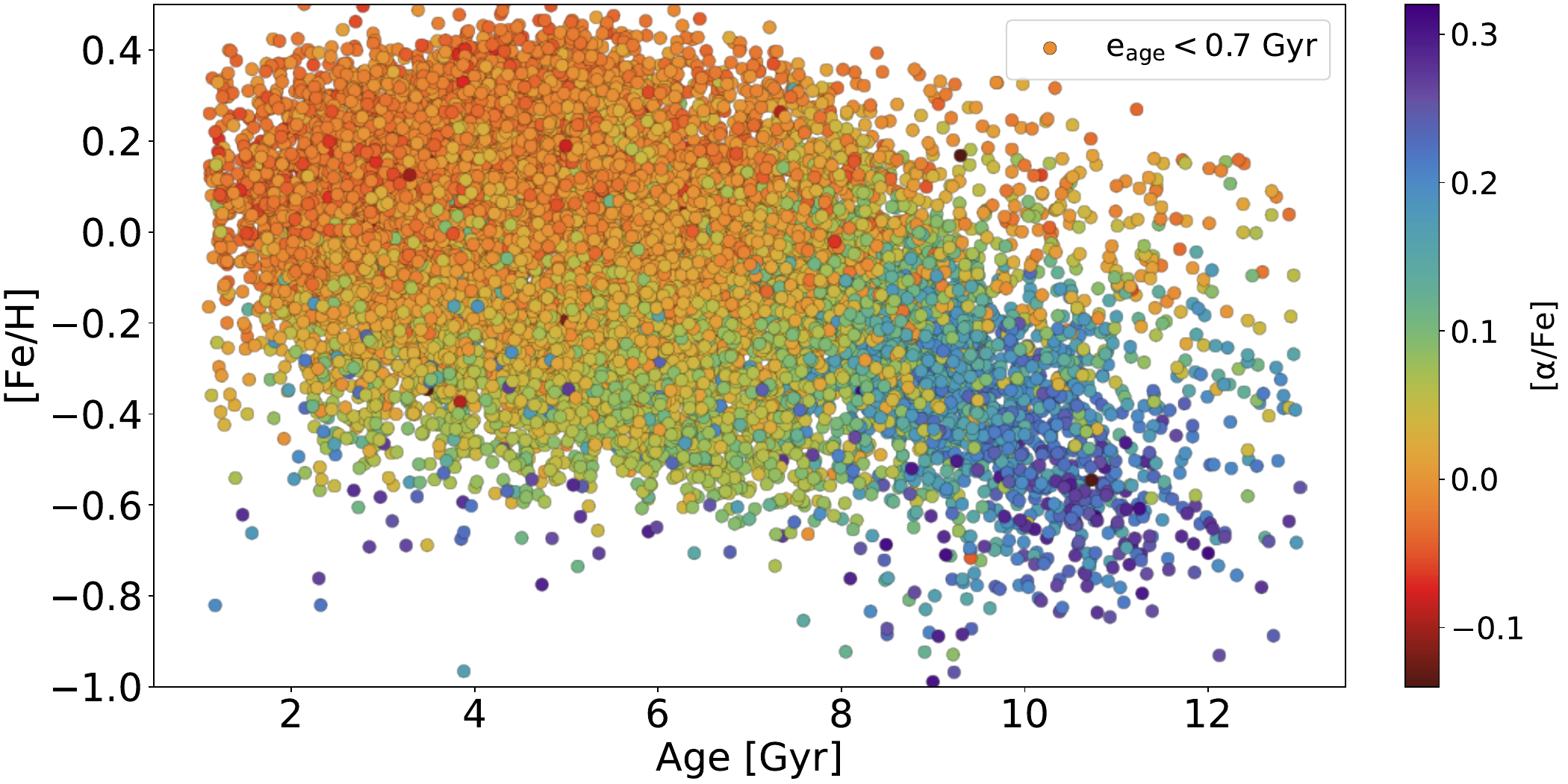}
\includegraphics[width=6cm]{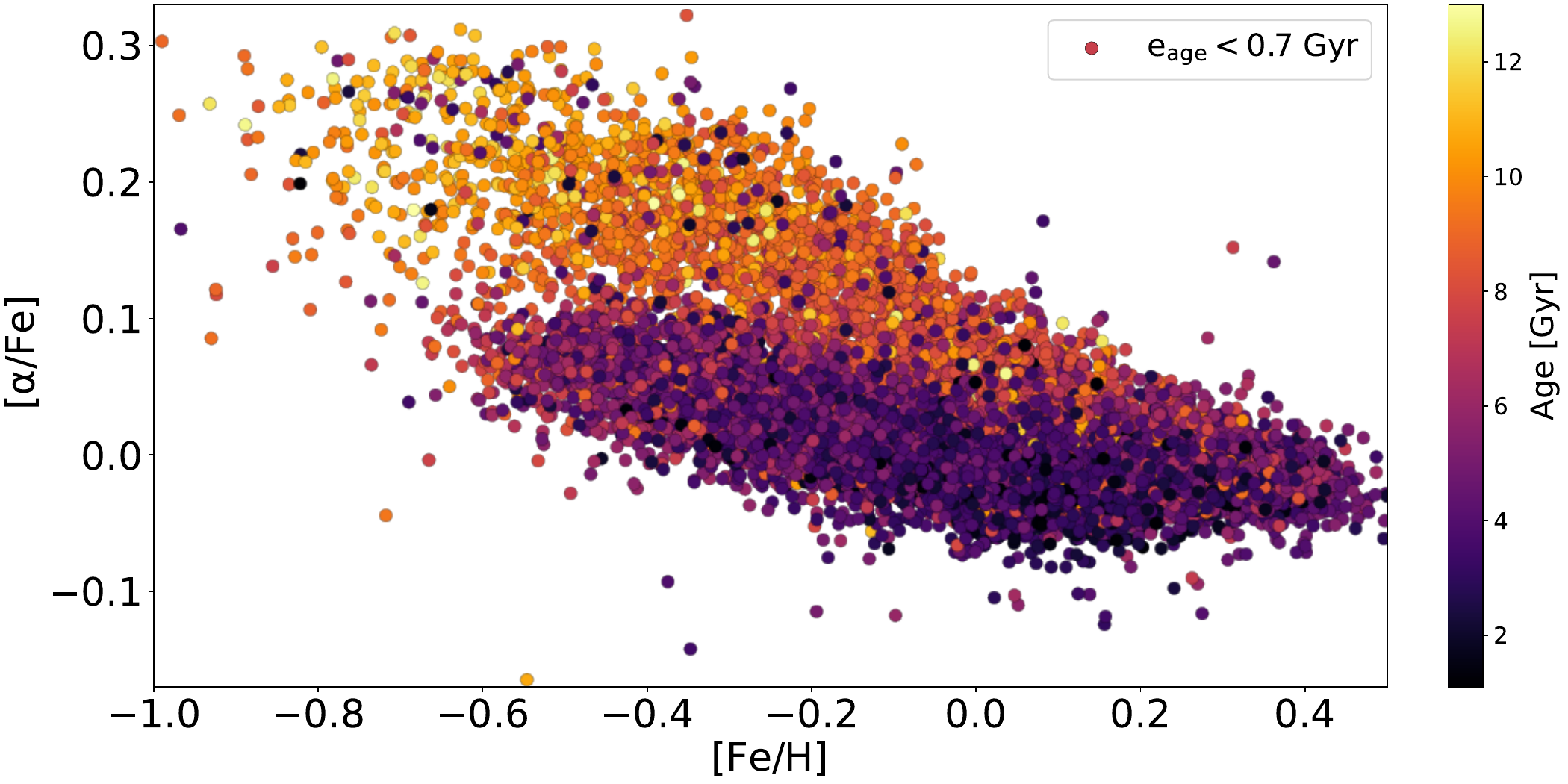}\par

\includegraphics[width=6cm]{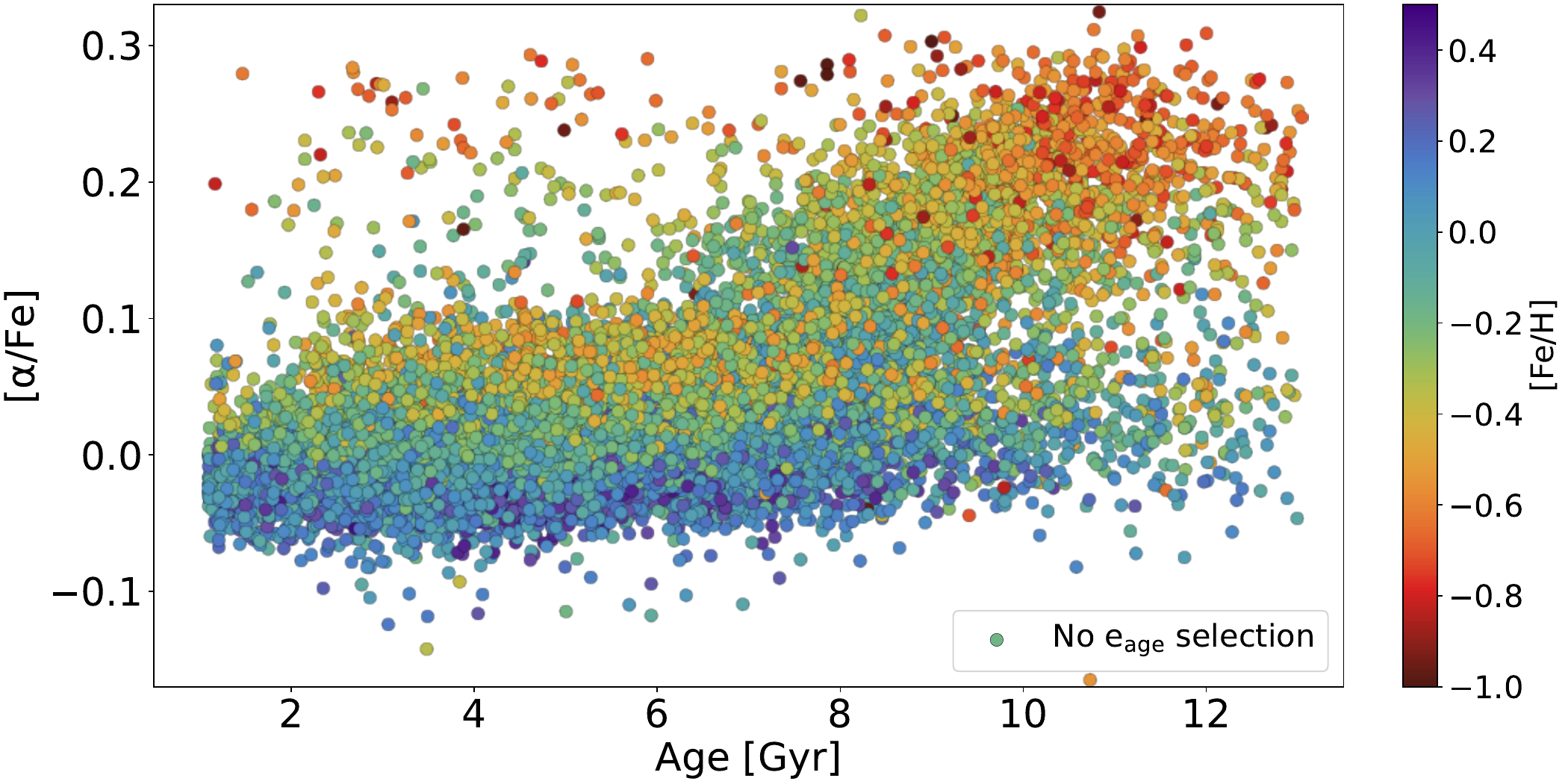}
\includegraphics[width=6cm]{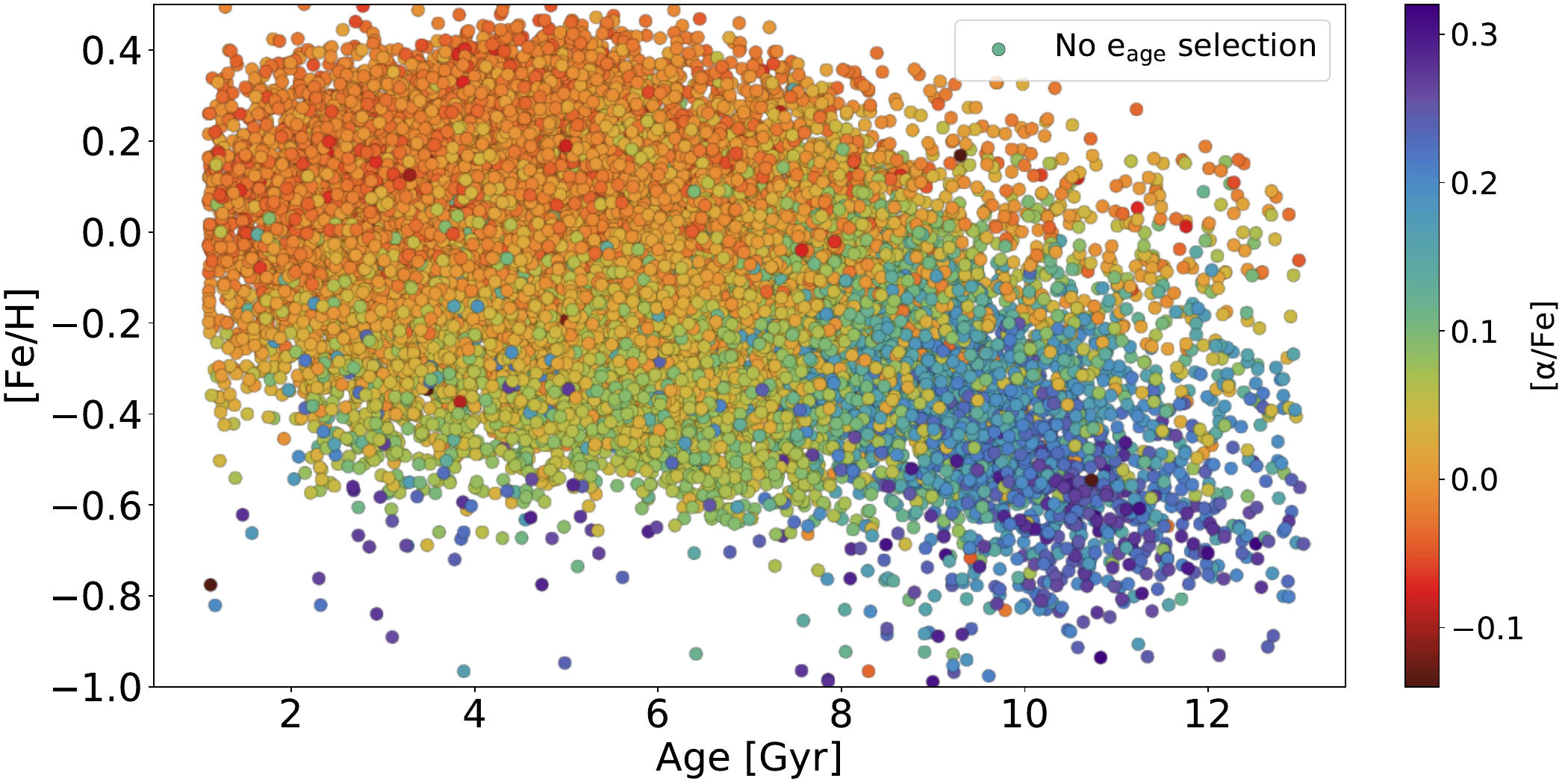}
\includegraphics[width=6cm]{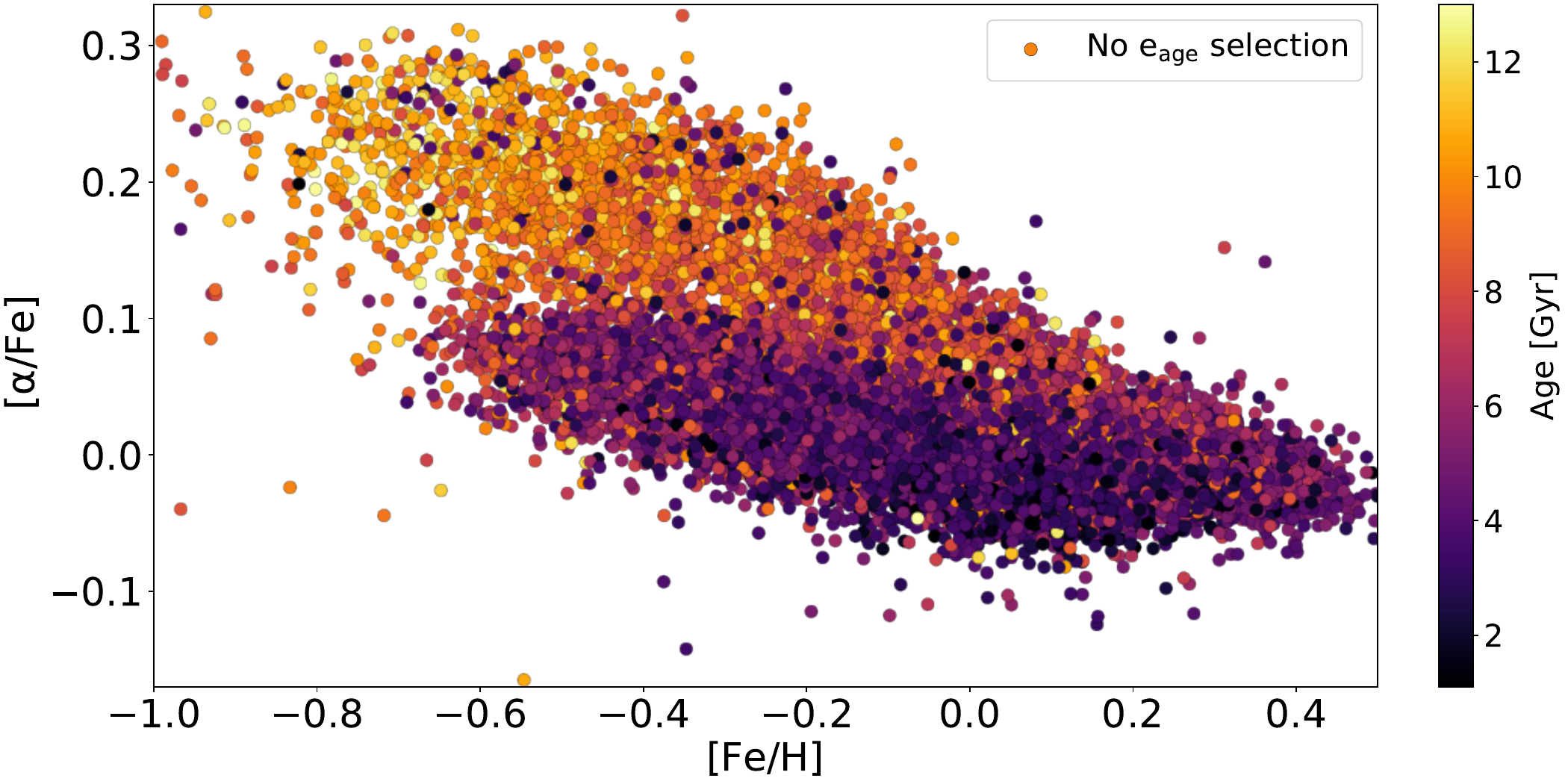}
\caption{Effect of the age uncertainties on the age-[$\rm \alpha$/Fe] distribution (left column), age-metallicity distribution (middle column) and [$\rm \alpha$/Fe]-[Fe/H] distribution (right column) for APOGEE DR17 local dwarfs. The ages in this case are obtained with the observational input K absolute magnitude, MK0, and Gaia DR3 color BPRP0.
The stars in these plots have been selected to have: age above 1.1 Gyr, [Fe/H] lower than -1 dex, $\rm(BP - RP)_{0}$ lower than 0.95. We compare then three different cuts: $\rm e_{age}<0.3$ Gyr (6 757 stars), $\rm e_{age}<0.5$ Gyr (12 035 stars), $\rm e_{age}<0.7$ Gyr (16 167 stars) with the total distribution to which no selection on age uncertainty is applied (22 233 stars, bottom row). The stars are color-coded for [Fe/H], [$\rm \alpha$/Fe] and age respectively. The number of candidate YAR (with age below 8 Gyr and [$\rm \alpha$/Fe] above 0.15 dex) is written in orange in each  age-[$\rm \alpha$/Fe] panel. 
}
\label{fig: effect age_unc APOGEE DR17 MK0 BPRP0}
\end{figure*}

\begin{figure*}[h!]
\centering
\includegraphics[width=6cm]{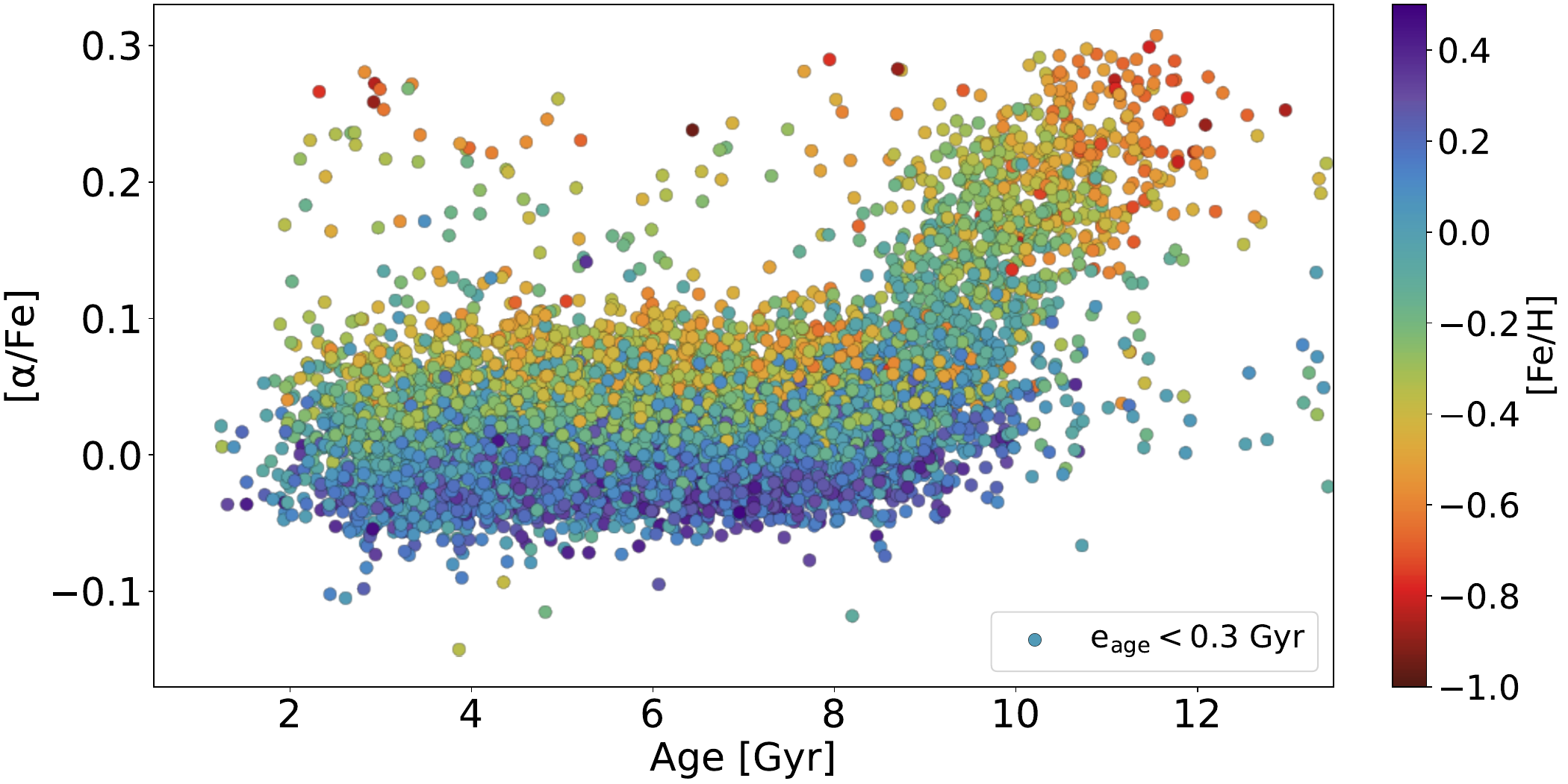}
\includegraphics[width=6cm]{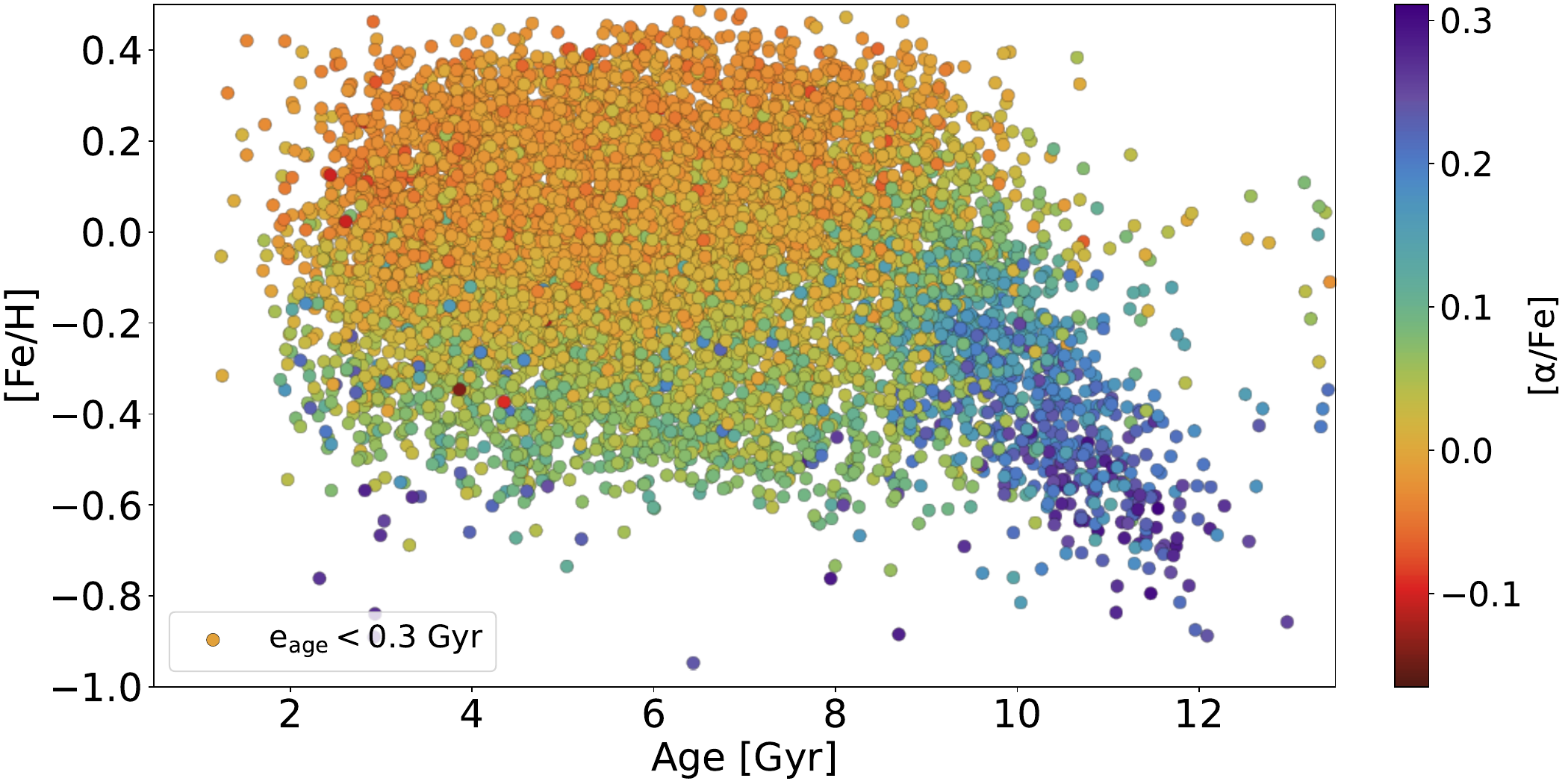}
\includegraphics[width=6cm]{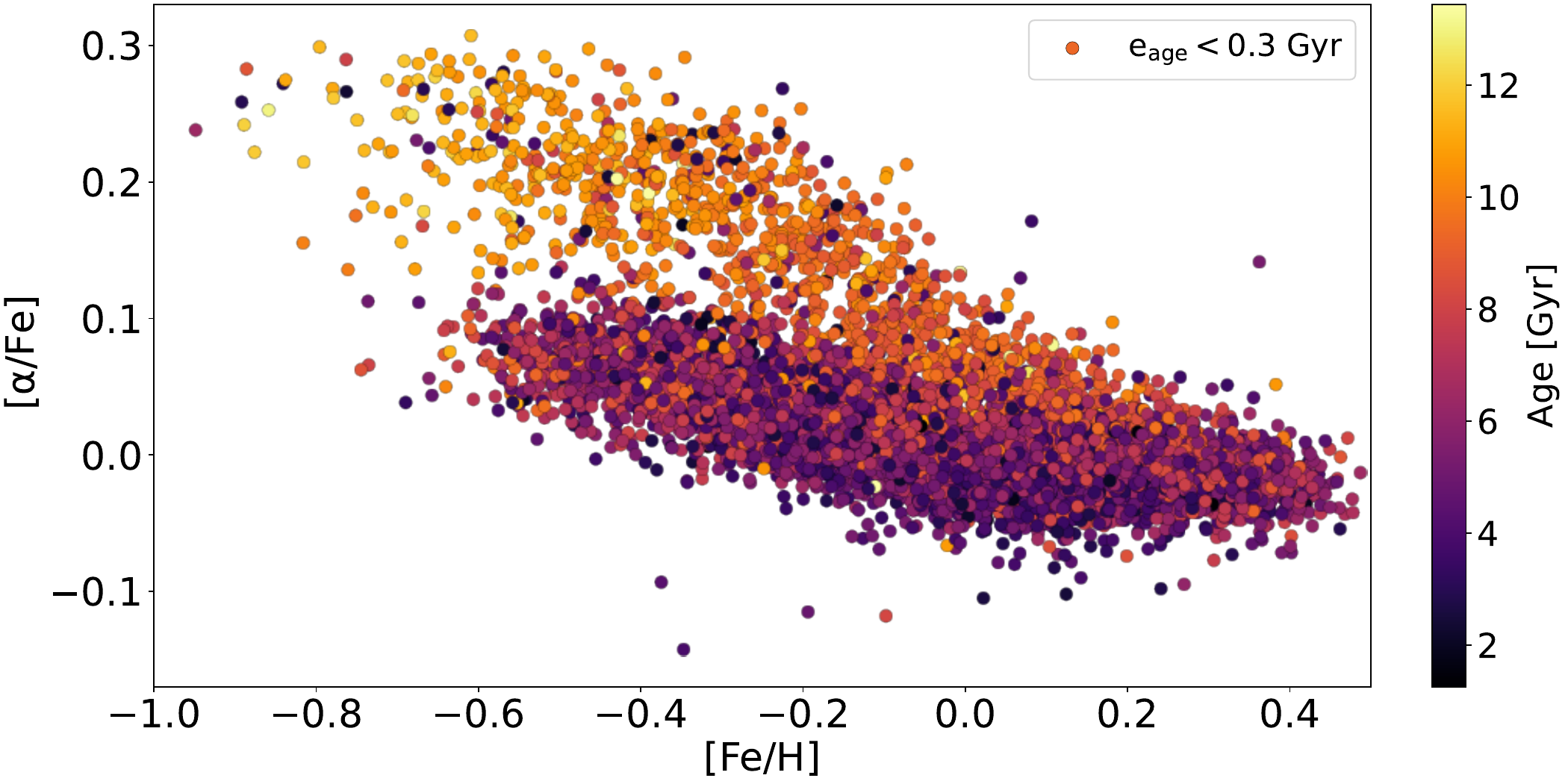}\par

\includegraphics[width=6cm]{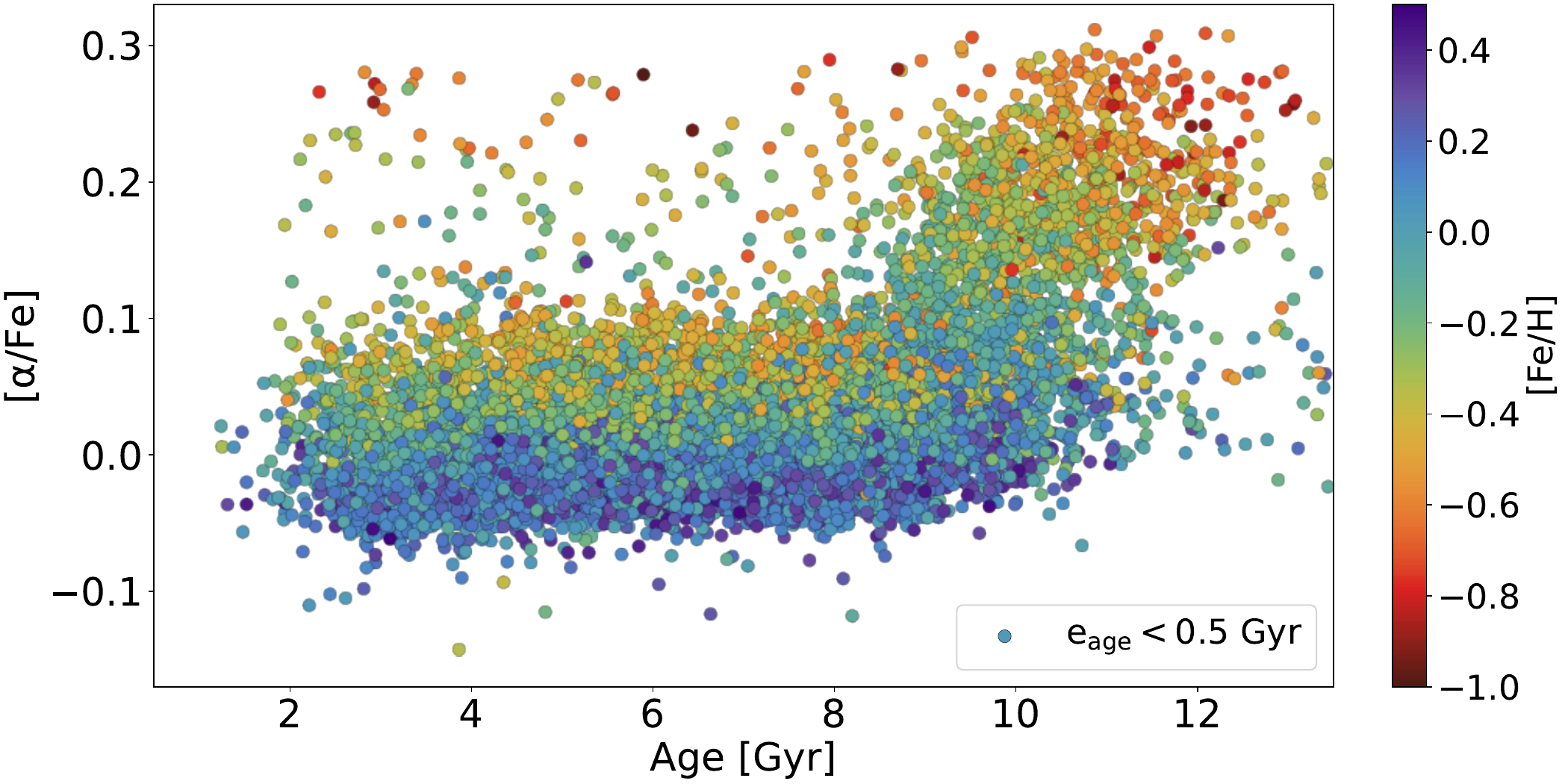}
\includegraphics[width=6cm]{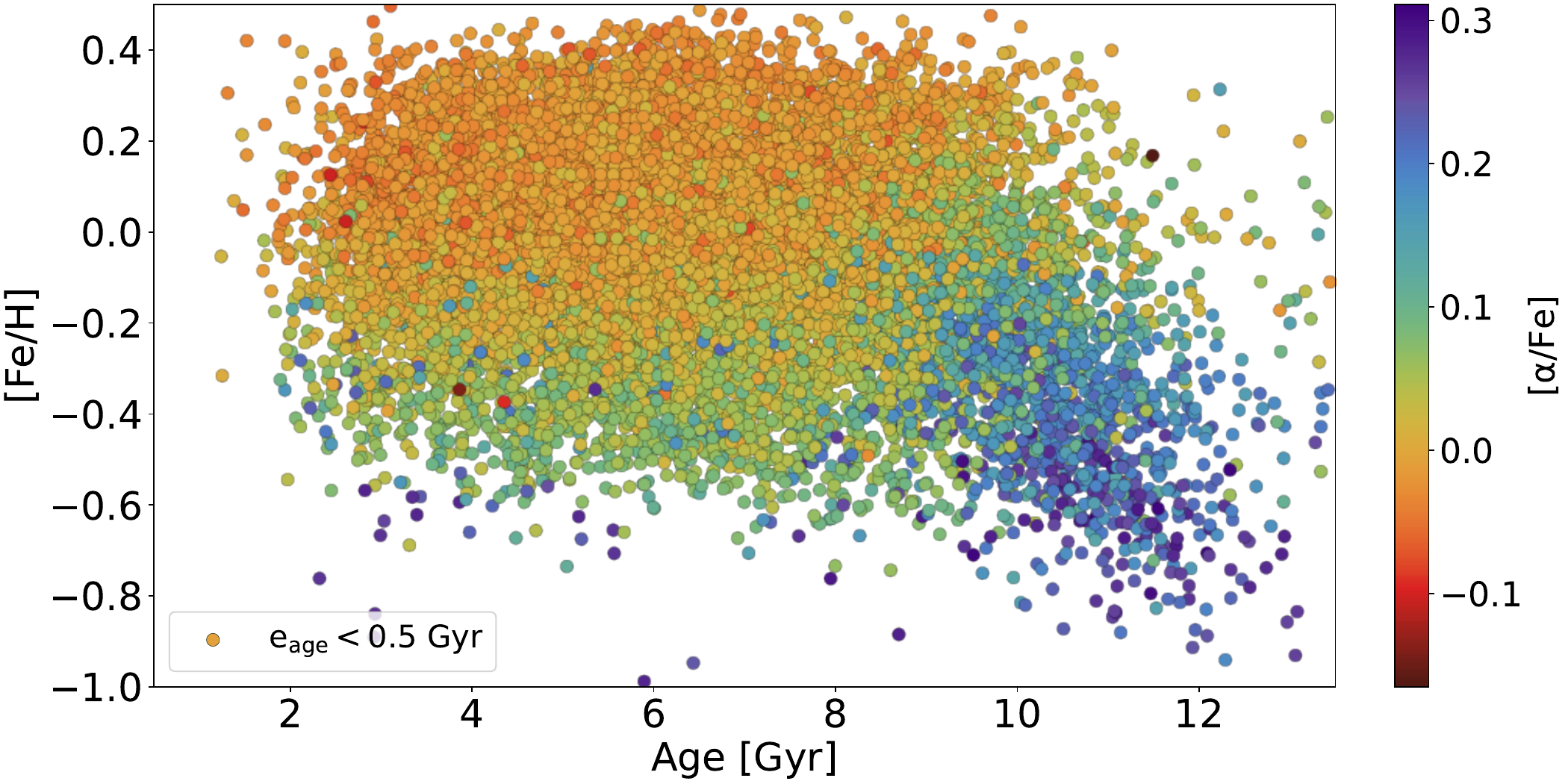}
\includegraphics[width=6cm]{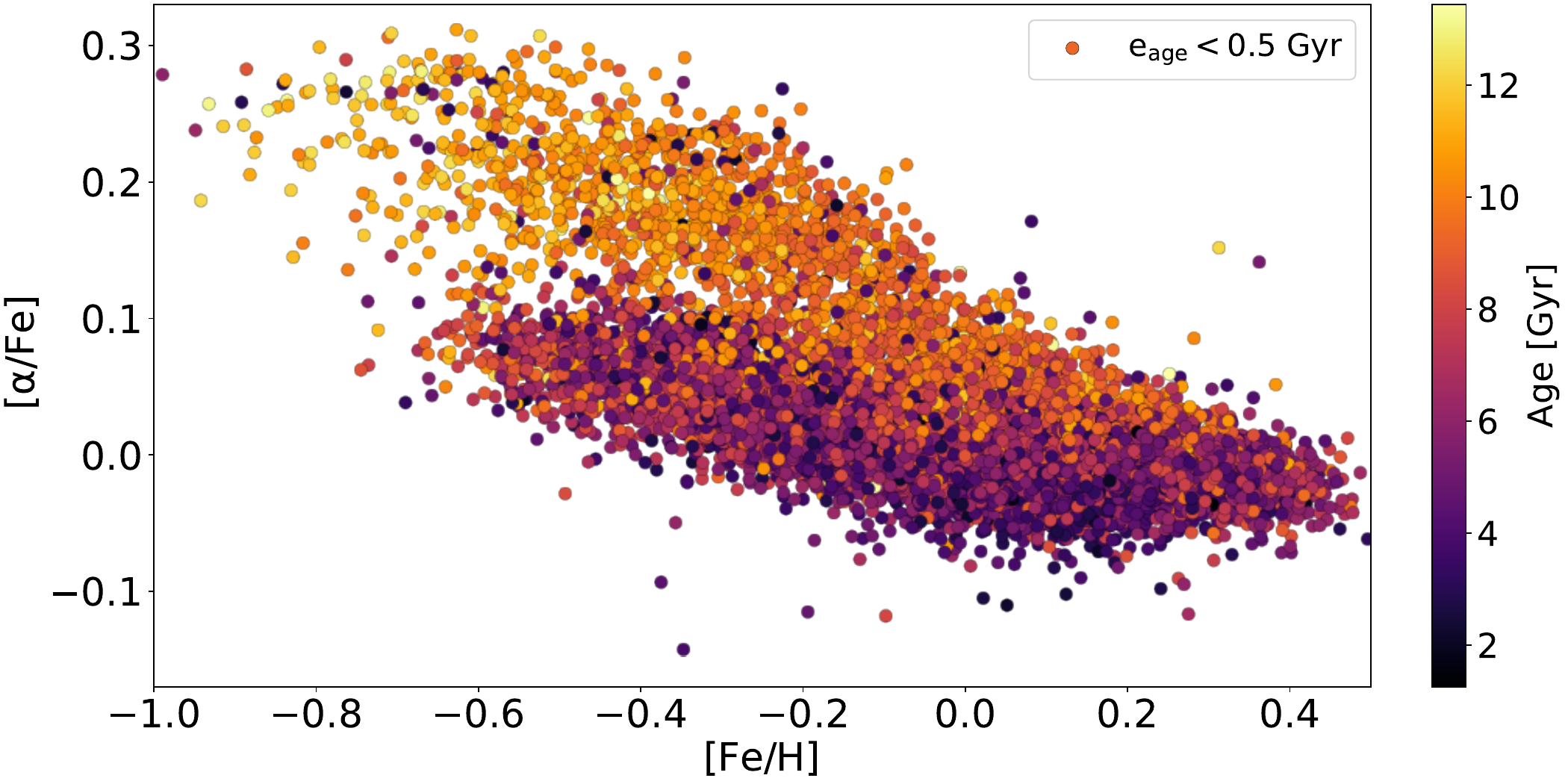}\par

\includegraphics[width=6cm]{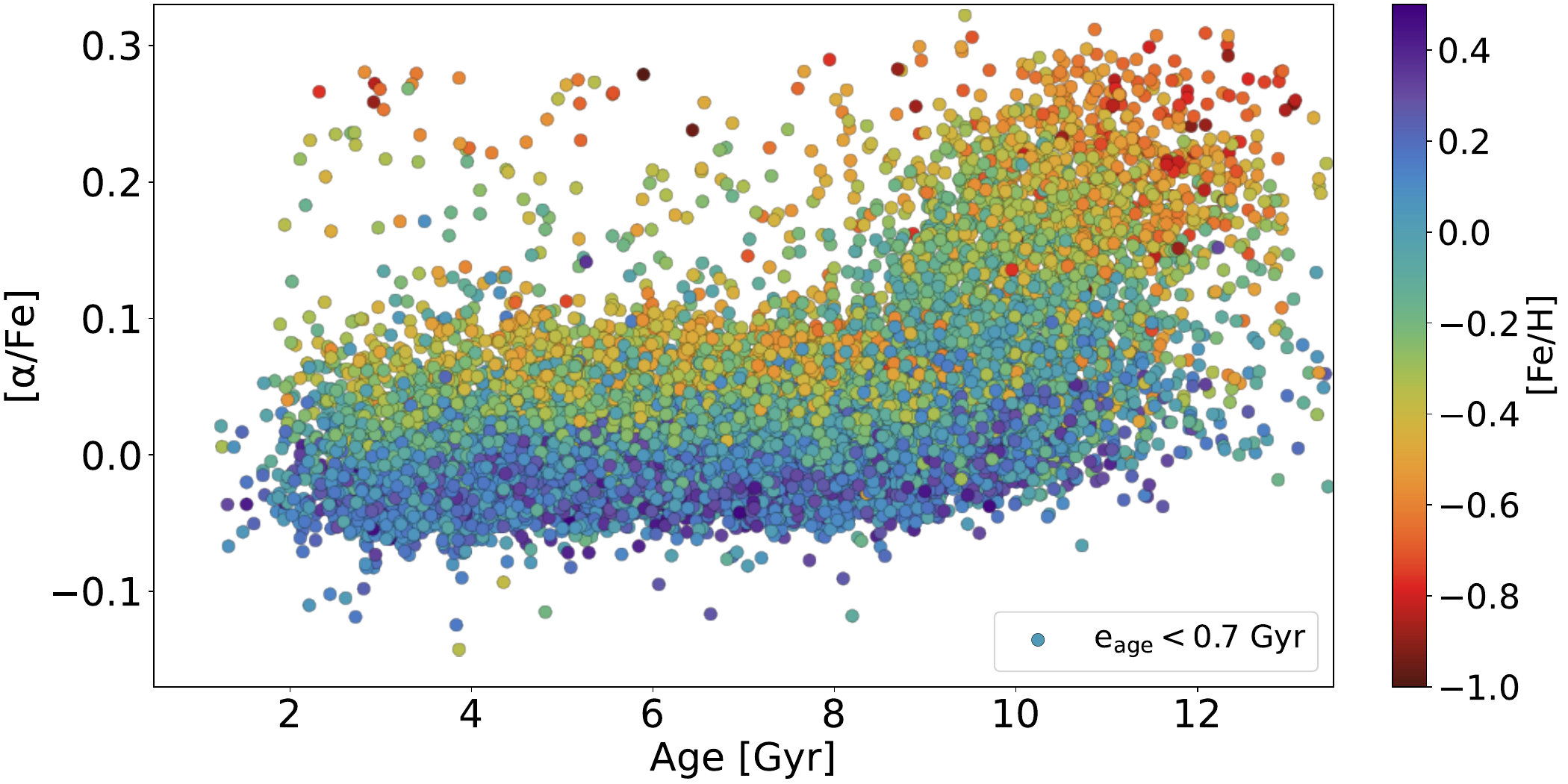}
\includegraphics[width=6cm]{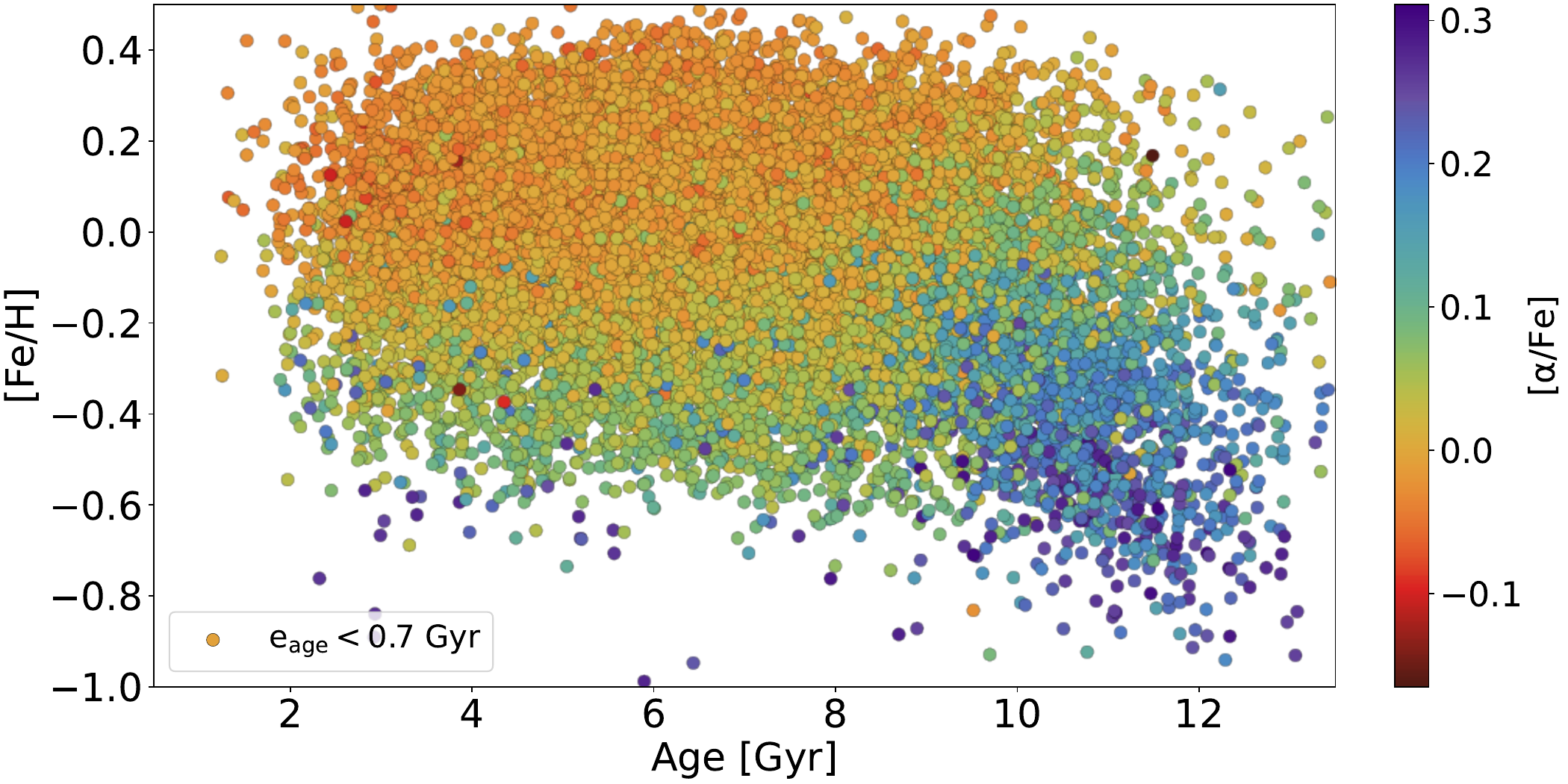}
\includegraphics[width=6cm]{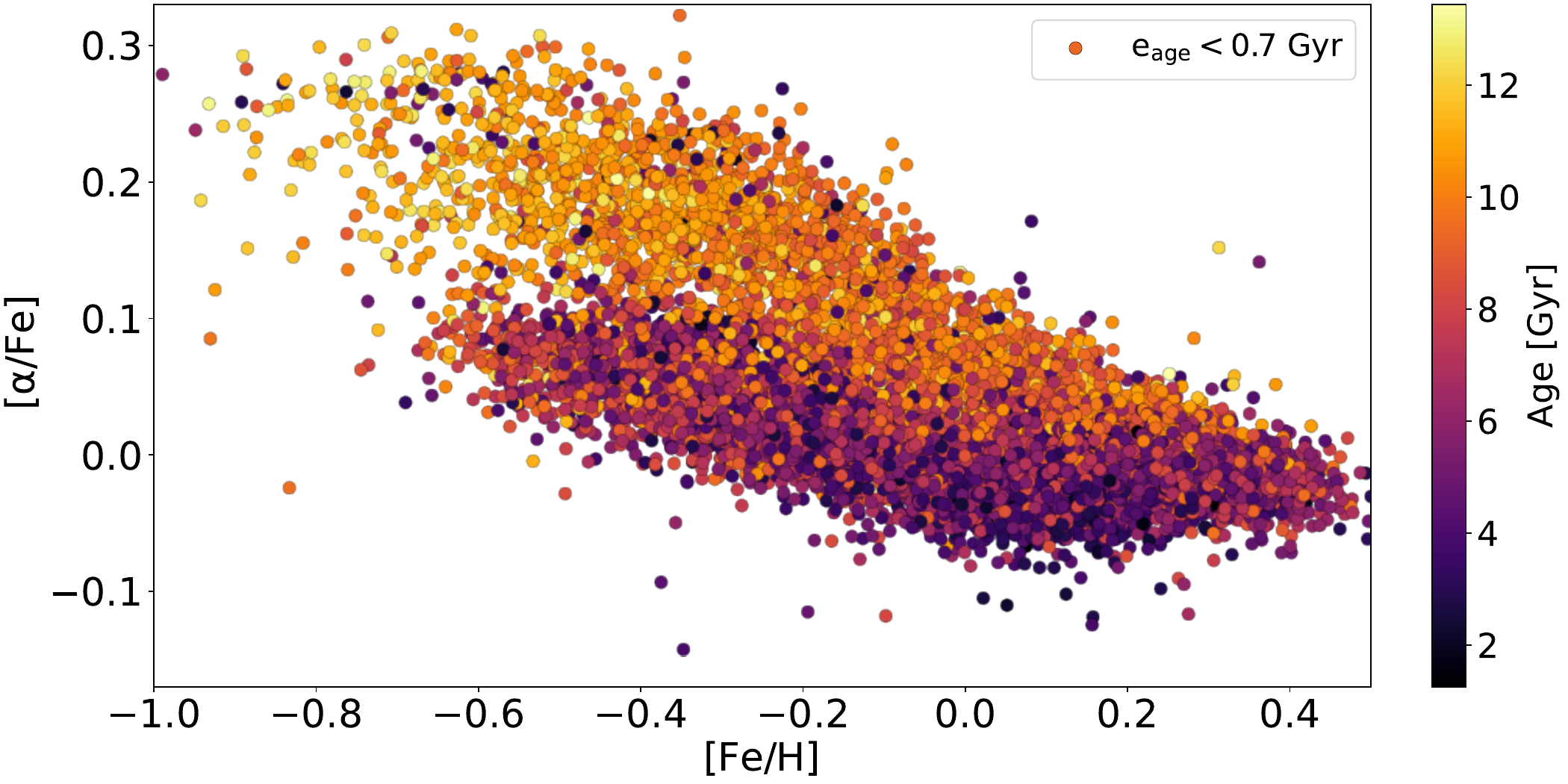}\par

\includegraphics[width=6cm]{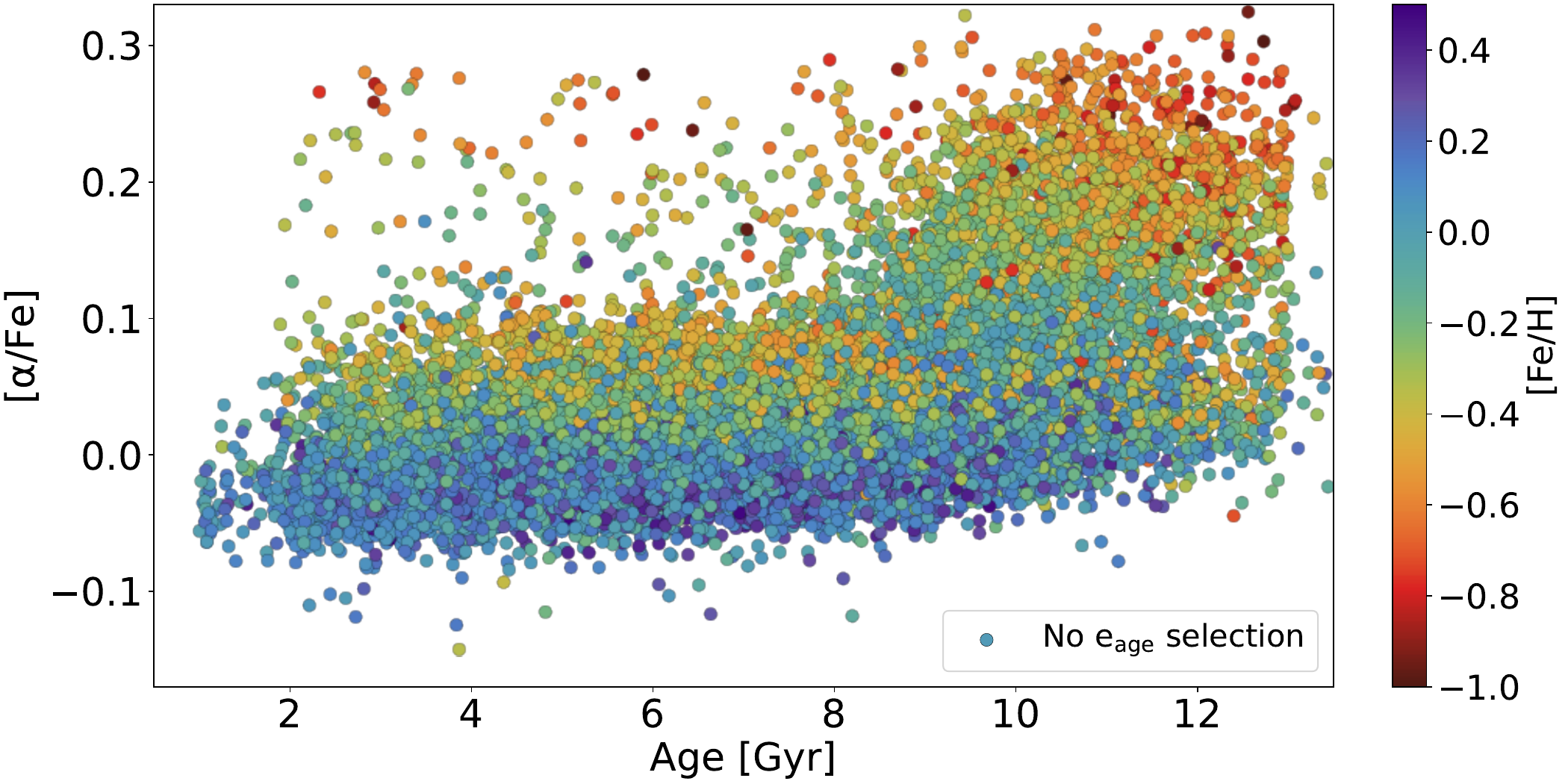}
\includegraphics[width=6cm]{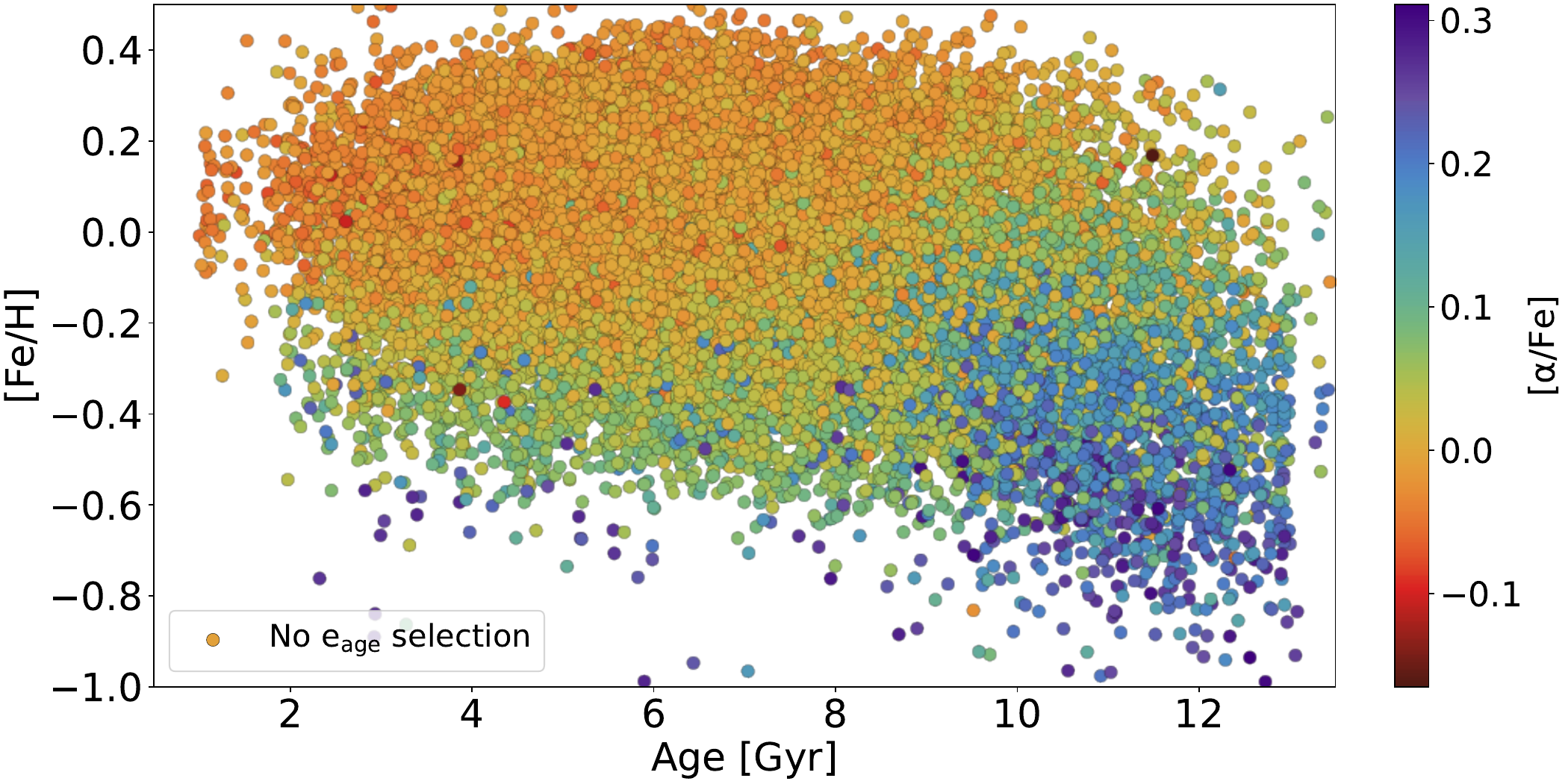}
\includegraphics[width=6cm]{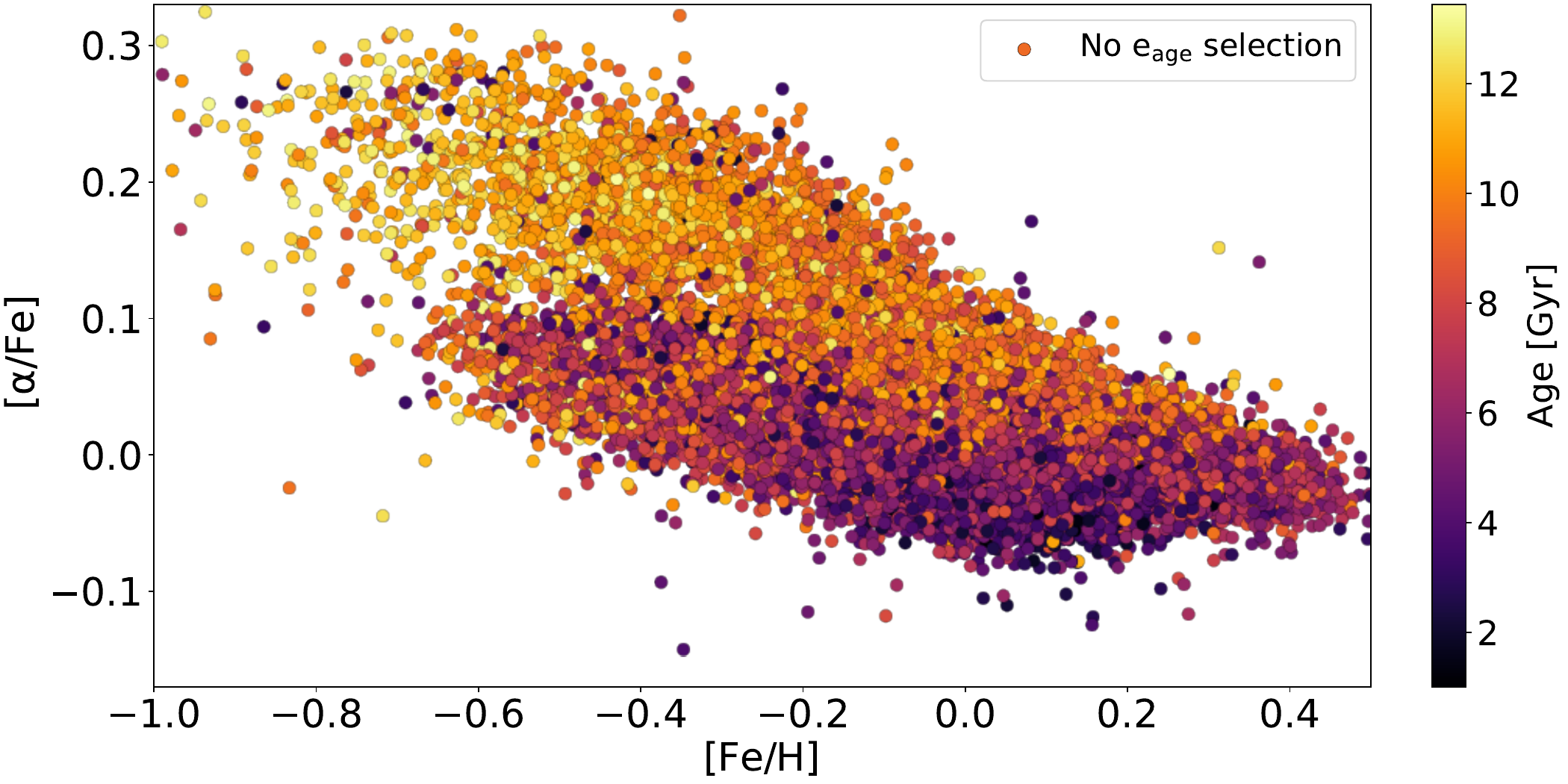}\par
\caption{As Fig. \ref{fig: effect age_unc APOGEE DR17 MK0 BPRP0} but for input parameters MG0 and effective temperature $\rm T_{eff}$.}
\label{fig: effect age_unc APOGEE DR17 MG0 TEFF}
\end{figure*}

In this section we illustrate the effect of the age uncertainties on the age-$\rm \alpha$, age-metallicity and [$\rm \alpha$/Fe]-[Fe/H] relations representative of the Solar neighborhood using the GALAH DR3 data (and Gaia DR3 photometry as input parameter, Fig. \ref{fig: effect_age_unc GALAHDR3}) and APOGEE DR17, using the 2MASS Ks magnitude and Gaia DR3 BP- RP color as observational inputs (Fig. \ref{fig: effect age_unc APOGEE DR17 MK0 BPRP0}) and MG0 and effective temperature (Fig. \ref{fig: effect age_unc APOGEE DR17 MG0 TEFF}).
In Fig. \ref{fig: effect_age_unc GALAHDR3} we highlight 3 different selection of age uncertainty: $\rm e_{age}<0.5$ Gyr, $\rm e_{age}<1$ Gyr, $\rm e_{age}<1.5$ Gyr and we compare them with the total distribution with no selection applied. Unlike the APOGEE data (for which we show $\rm e_{age}<0.3$ Gyr, $\rm e_{age}<0.5$ Gyr, $\rm e_{age}<0.7$ Gyr, see Fig. \ref{fig: effect age_unc APOGEE DR17}, Fig. \ref{fig: effect age_unc APOGEE DR17 MK0 BPRP0} and Fig. \ref{fig: effect age_unc APOGEE DR17 MG0 TEFF}) in this case the dispersion of the distribution does not seem to decrease with a stricter cut on age error. In addition, the old high-$\rm \alpha$ stars of the thick disk are removed when objects with age errors below 0.5 Gyr are selected. The thin disk component is not dramatically affected by the same quality selection.

\FloatBarrier

\section{Orbital parameters}
\label{Subsec: orbital parameters}

\begin{figure*}[h!]
    \centering
    \includegraphics[width=8cm]{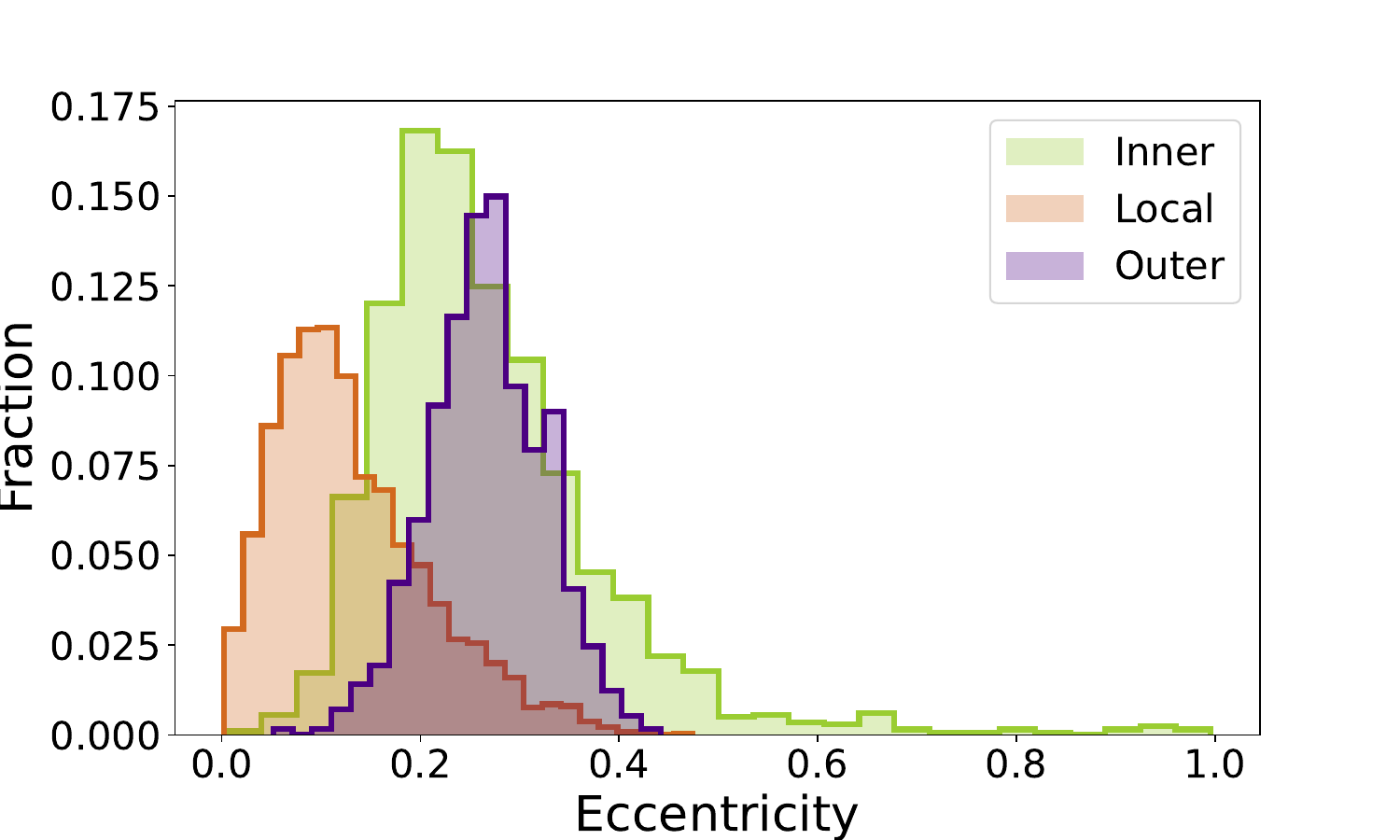}
    \includegraphics[width=8cm]{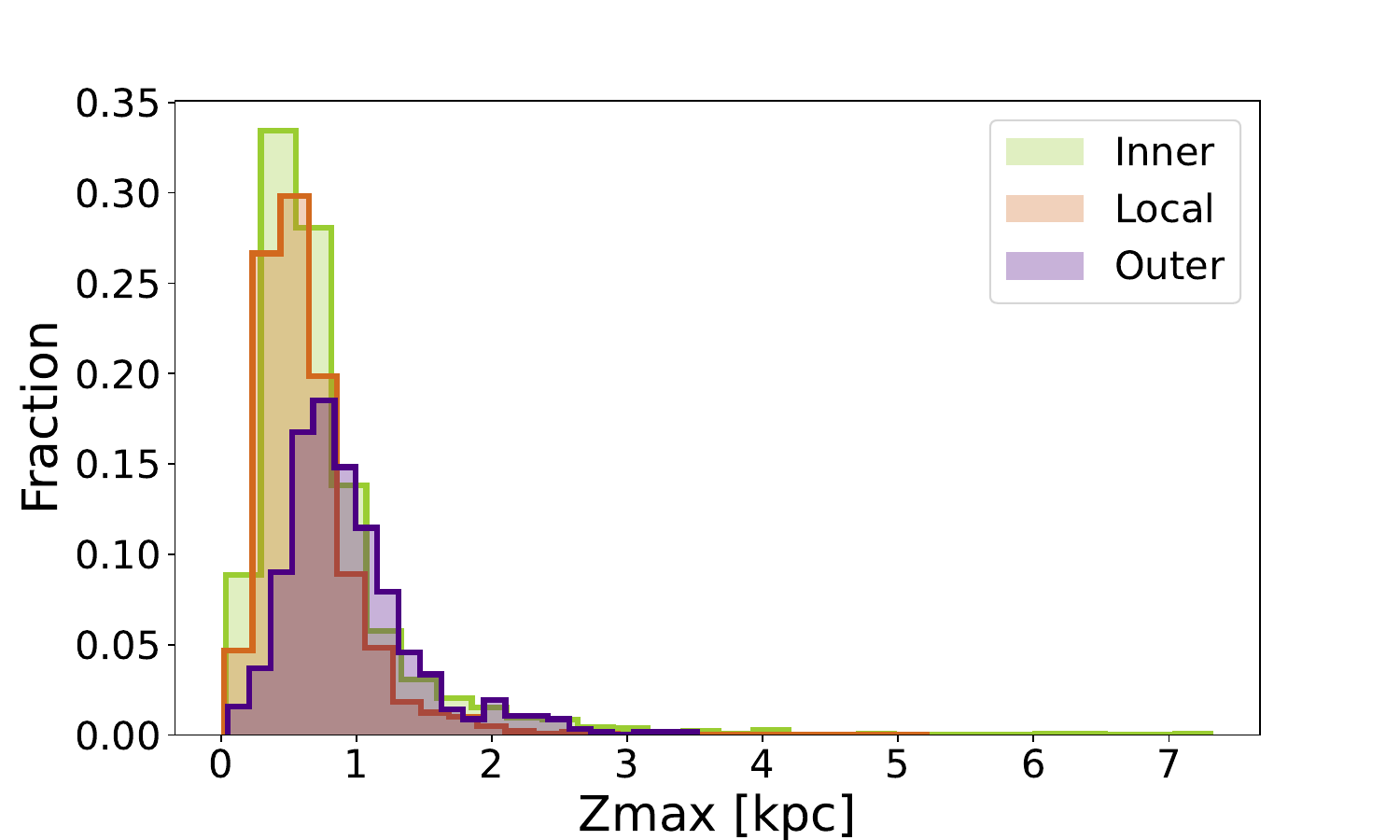}\par
    \includegraphics[width=8cm]{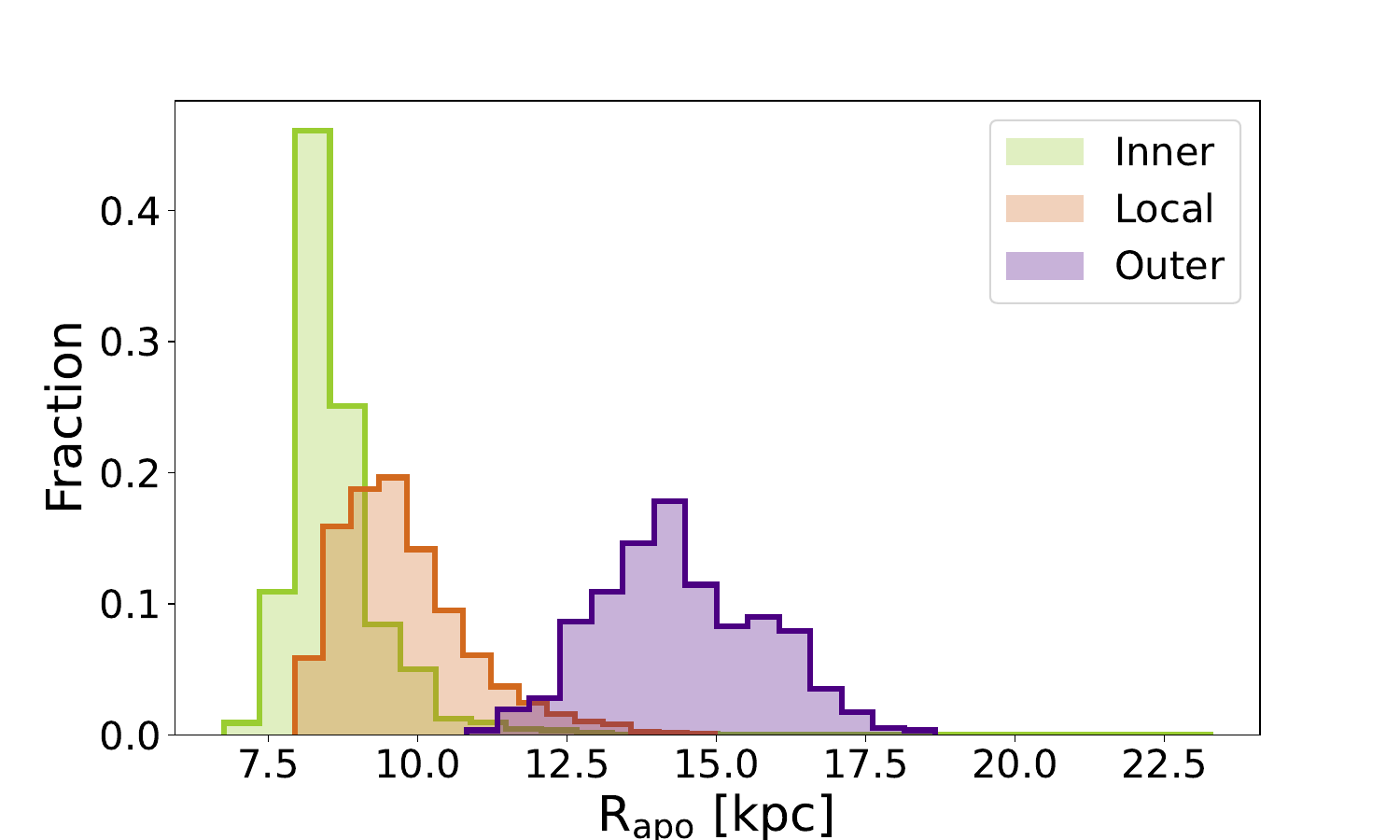}
    \includegraphics[width=8cm]{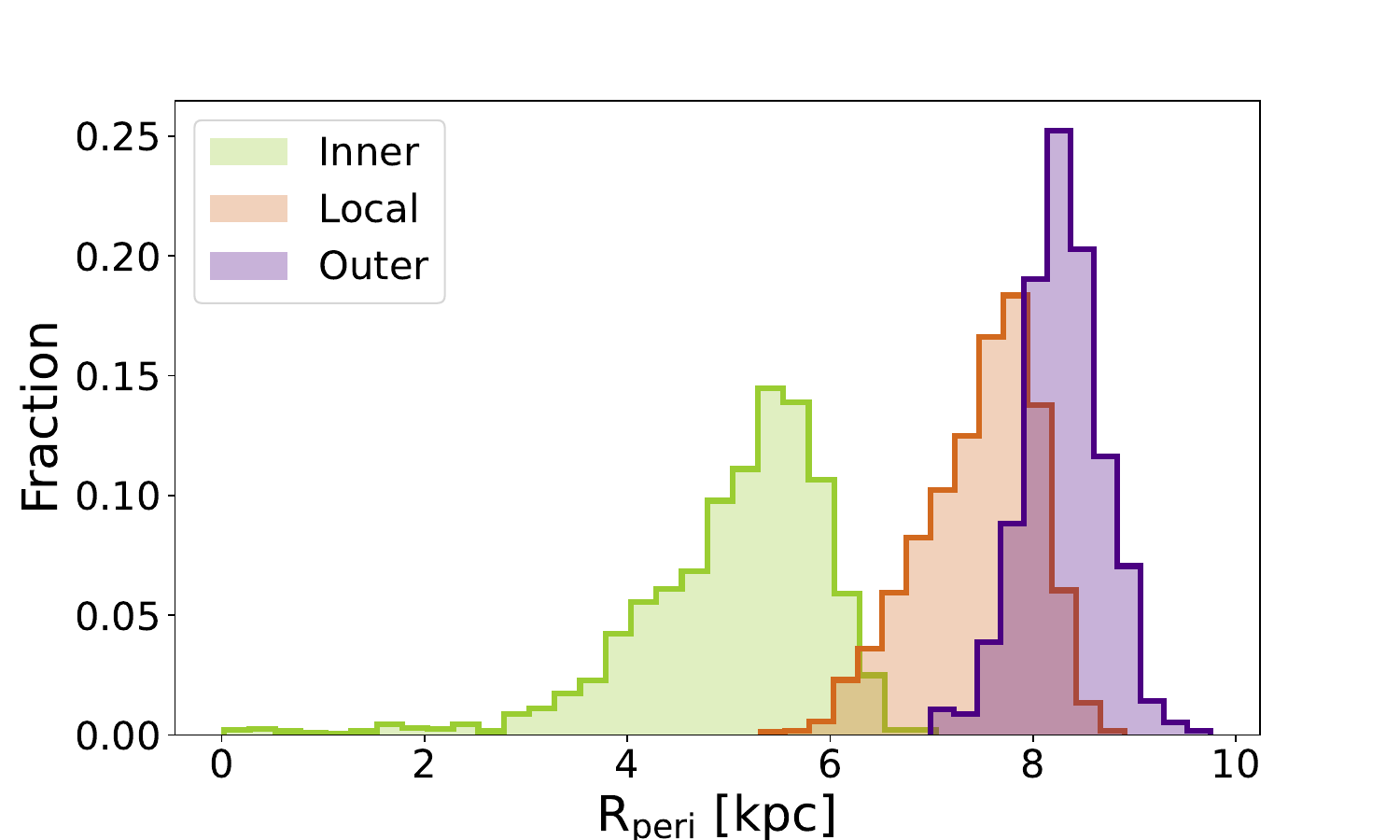}
\caption{Orbital properties of our inner, local and outer sub-samples.}
\label{fig: orbital properties 3 pop}
\end{figure*}

In Fig. \ref{fig: orbital properties 3 pop} we highlight the main orbital characteristics of the three sub-samples: eccentricity, maximum distance reached from the Galactic plane (Zmax), apocenter and pericenter distances ($\rm R_{apo}$, $\rm R_{peri}$). Inner, local and outer disk stars are highlighted with histograms of different colors.
Being sampled at the Solar vicinity, they are observed as being either at their apocentre (for inner disk stars) or their pericentre (for the outer disk stars). 

Unsurprisingly, outer and inner disks samples have higher eccentricities than the local sample. The inner disk population, in green, have eccentricities reaching high values due to thick disk stars, but with the bulk of objects peaking at 0.25.
Local stars eccentricities peak around 0.15, suggesting more circular orbits, while outer disk stars have a range of eccentricities between about 0.2 and 0.4, peaking around 0.3, suggesting more elongated orbits.

The apocenter radii distribution shows a sharp contrast: inner disk stars are tightly clustered around 9 kpc, local stars peak around 10 kpc, and outer disk stars have a broader distribution with a peak around 14 kpc, reflecting their respective orbital radii. 

Finally, the pericenter distributions show that inner stars have the smallest pericentric distances, clustering at about 5.5 kpc and spreading out to very low distances. Instead, local and outer disk stars peak near 8 and 8.5 kpc, respectively. Together, these distributions illustrate the different orbital characteristics of each sub-sample, with inner stars exhibiting more circular orbits near the Galactic Center, while outer stars occupy more extended, elliptical orbits.

Inner, local and outer objects present similar and compatible distribution in Zmax, the majority of the stars peaking at Zmax below 1 kpc.

\FloatBarrier

\end{appendix}

\end{document}